\documentclass[preprint,prd,aps,showpacs,showkeys,nofootinbib]{revtex4}
\usepackage{graphicx}
\begin{document}


\title{Feynman Integrals of Grassmannians}

\author{Tai-Fu Feng\footnote{email:fengtf@hbu.edu.cn}$^{a,b,c,d}$,
Hai-Bin Zhang\footnote{email:zhanghb@hbu.edu.cn}$^{a,b,c}$,
Chao-Hsi Chang\footnote{email:zhangzx@itp.ac.cn}$^{e,f,g}$}

\affiliation{$^a$Department of Physics, Hebei University, Baoding, 071002, China}
\affiliation{$^b$Hebei Key Laboratory of High-precision Computation and Application of
Quantum Field Theory, Baoding, 071002, China}
\affiliation{$^c$Department of Physics, Guangxi University, Nanning, 530004, China}
\affiliation{$^d$Department of Physics, Chongqing University, Chongqing, 401331, China}
\affiliation{$^e$Key Laboratory of Theoretical Physics, Institute of Theoretical Physics,
Chinese Academy of Science, Beijing, 100190, China}
\affiliation{$^f$CCAST (World Laboratory), P.O.Box 8730, Beijing, 100190, China}
\affiliation{$^g$School of Physical Sciences, University of Chinese
Academy of Sciences, Beijing 100049, China}

\begin{abstract}
We embed Feynman integrals in the subvarieties of Grassmannians through homogenization
of the integrands in projective space, then obtain GKZ-systems
satisfied by those scalar integrals. The Feynman integral can be written as
linear combinations of the hypergeometric functions of a fundamental solution system in
neighborhoods of regular singularities of the GKZ-system, whose
linear combination coefficients are determined by the integral on an ordinary point
or some regular singularities. Taking some Feynman diagrams as examples, we elucidate in detail
how to obtain the fundamental solution systems of Feynman integrals in neighborhoods of regular
singularities. Furthermore we also present the parametric representations of Feynman integrals
of the 2-loop self-energy diagrams which are convenient to embed in the subvarieties
of Grassmannians.
\end{abstract}

\keywords{Feynman integral, Grassmannian, GKZ-hypergeometric system}
\pacs{02.30.Jr, 11.10.-z, 12.38.Bx}

\maketitle

\section{Introduction\label{sec1}}
\indent\indent
Although the discovery of Higgs with mass around $125\;{\rm GeV}$~\cite{CMS2012,ATLAS2012} indicates
a great success of the Standard Model (SM), the updated experimental
data on the anomalous magnetic moment (MDM) of muon~\cite{ADM2021} and mass of the charged
gauge boson~\cite{Wmass} seem to herald the signals of new physics beyond.
Before the collision energy of the center of mass of the running collider
reaches the threshold energy of new physics, the SM theoretical
predictions of various physical observations should be accurately calculated~\cite{Heinrich2021}.
Through comparison  between the experimental data and corresponding SM evaluations,
the energy scale of new physics and strength of the coupling of new physics
particles to fields of the SM can be roughly estimated.

To obtain the SM evaluations of physical observables accurately,
one has to compute the scalar integrals of corresponding Feynman diagrams
involved at first, which is a challenging work in radiative corrections in quantum field theory.
In dimensional regularization, any scalar integral can be expanded as a Laurent series
around $\varepsilon=0$ where the time-space dimension $D=4-2\varepsilon$.
For example, the Feynman integral of the 1-loop triangle diagram $C_{_0}$ is expanded as
\begin{eqnarray}
&&C_{_0}=C_{_{0}}^{(0)}+\sum\limits_{i=1}^\infty C_{_0}^{(i)}\varepsilon^i\;,
\label{C1m-1}
\end{eqnarray}
around $D=4$, and the well-known result of $C_{_{0}}^{(0)}$ is presented in the
literature\cite{tHooft1979}. However, evaluating $C_{_0}^{(i)}$
precisely is also necessary to obtain high order radiative corrections
to the amplitude self-consistently when the virtual interactions
originate from the counter terms\cite{Collins1984}.
Considering Feynman integrals as the generalized hypergeometric functions~\cite{Regge1967},
one finds that the $D-$module of a Feynman diagram~\cite{Nasrollahpoursamami2016} is isomorphic to
Gel'fand-Kapranov-Zelevinsky (GKZ) $D-$module~\cite{Gelfand1987,Gelfand1988,Gelfand1988a,Gelfand1989,Gelfand1990}.
Correspondingly Feynman integrals satisfy indeed the systems of
holonomic linear partial differential equations (PDEs)~\cite{Kashiwara1976} whose singularities
are determined by the Landau singularities.

During the past decades, some progresses have been made on the subject.
Under the assumption of zero virtual masses, the Feynman integral of the 1-loop triangle
diagram is given as a linear combination of the fourth kind of Appell functions~\cite{Davydychev1} whose
arguments are the dimensionless ratios among the external momentum squared, and is simplified
further as a linear combination of the Gauss functions $_2F_1$ through the quadratic
transformation~\cite{Davydychev1993NPB} in the literature~\cite{Davydychev2000}.
Taking some special assumptions on parameter space, the authors of Ref.~\cite{Davydychev3}
obtain the multiple hypergeometric expressions of the scalar integral $C_{_0}$ through
Mellin-Barnes representations. An algorithm to evaluate the scalar integrals of
1-loop vertex-type Feynman diagrams is presented in Ref.~\cite{Davydychev1992JPA},
and a geometrical interpretation of the analytic expressions of the scalar integrals
of 1-loop $N-$point Feynman diagrams is also given in Ref.~\cite{Davydychev2006}.
Certainly the expressions of $C_{_0}$ function can also be derived from expressions of the
1-loop massive $N-$point Feynman diagrams~\cite{Davydychev1991JMP,Davydychev1992JMP}.
In addition Feynman integrals of ladder diagrams with 3 or 4 external lines
can be evaluated through the parametric representation and Mellin-Barnes contour
integrals~\cite{Davydychev1993}.
Actually the 1-loop 2-point function $B_{_0}$ can be expressed as a linear combination
of the Gauss functions $_2F_1$ through the recurrence relations respecting time-space
dimension, the 1-loop 3-point function $C_{_0}$ similarly is formulated as a linear
combination of the Appell functions $F_{_1}$, and the 1-loop 4-point function $D_{_0}$
is formulated as a linear combination of the Lauricella functions
$F_{_s}$~\cite{Tarasov2000,Tarasov2003}, respectively.
The expression of $C_{_0}$ is convenient
for analytic continuation and numerical evaluation because continuation of $F_{_1}$
has been analyzed thoroughly. However, how to perform continuation
of the Lauricella function $F_{_s}$ outside its convergent domain is still a challenge.
Expressing Feynman integrals as generalized hypergeometric functions in dimension regularization,
the authors of Ref.~\cite{Kalmykov2009} analyze the Laurent expansion of these hypergeometric
functions around $D=4$. The differential-reduction algorithm to evaluate those hypergeometric
functions can be found in Refs.~\cite{Bytev2010,Kalmykov2011,Bytev2013,Bytev2015,Bytev2016}.
A hypergeometric system of linear PDEs is given through the corresponding Mellin-Barnes
representation~\cite{Kalmykov2012}.
Some irreducible master integrals of sunset and bubble Feynman diagrams
are explicitly evaluated through Mellin-Barnes contour in Ref.~\cite{Kalmykov2017}.

GKZ-hypergeometric systems of Feynman integrals with codimension$=0,\;1$
are presented in Refs.~\cite{Cruz2019,Klausen2019}
through Lee-Pomeransky parametric representations~\cite{Lee2013}.
To construct canonical series solutions with suitable independent variables,
one should compute the restricted $D$-module of GKZ-hypergeometric system originating from
Lee-Pomeransky representations on corresponding hyperplane in the parameter
space~\cite{Oaku1997,Walther1999,Oaku2001}.
Certainly GKZ-hypergeometric systems of some Feynman diagrams
are obtained from Mellin-Barnes representations~\cite{Feng2018,Feng2019,Feng2020} through Miller's
transformation~\cite{Miller68,Miller72}.
Ref.\cite{Loebbert2020} explores the idea of bootstrap Feynman integrals using integrability,
Ref.\cite{Klemm2020} applies the GKZ description of periods to solve the
l-loop banana amplitudes with their general mass dependence,  and Ref.~\cite{Bonisch2021} completely clarifies
the analytic structure of all banana amplitudes with arbitrary masses, respectively.
The updated summary on some classical and modern aspects of hypergeometric
differential equations is given in literature~\cite{Reichelt2020}.
A mathematical package for integrating families of Feynman integrals
order-by order in the dimensional regulator from their GKZ-systems
is given in literature~\cite{Hidding2021}, and
a specialized integration algorithm for parametric Feynman integrals
with tame kinematics is also presented in literature~\cite{Borinsky2020}.
Actually Cohen-Macaulay Property of Feynman Integral indicates
that the process of finding a series representation of these
integrals is fully algorithmic~\cite{Tellander2021}, and regular
singularities of Feynman Integral are determined by its Landau Discriminant~\cite{Mizera2021}.
Ref.~\cite{Arkani-Hamed2022} introduces a class of polytopes to analyze
the structure of UV and IR divergences of general Feynman integrals in Schwinger parameter space.
Furthermore, Ref.~\cite{Chestnov2022} elaborates on the connection among GKZ-systems, de
Rham theory for twisted cohomology groups, and Pfaffian equations for Feynman Integrals.
Ref.~\cite{Ananthanarayan2022} presents a Mathematica package to find linear transformations
for some classes of multivariable hypergeometric functions. In literature~\cite{Berends1994}
the class of $N$ loop massive scalar self-energy diagrams with $N+1$ propagators
is studied, and the new convergent series representations for the 2-loop sunset diagram
with 3 different propagator masses and external momentum are obtained in
Ref.\cite{Ananthanarayan2019}.  The relationship between Feynman diagrams and hypergeometric functions
is reviewed in the Ref.~\cite{Kalmykov2021}.

Since the fundamental solution systems of neighborhoods of different regular
singularities are different, and many equivalent fundamental solution systems
can be constructed corresponding to neighborhoods of same regular singularity.
When a Feynman integral is expressed as a linear combination of
hypergeometric functions of the fundamental solution system,
there are obvious differences among the equivalent representations accordingly.
Finding a framework for understanding these rambling results is undoubtedly
important on both theoretical and phenomenological aspects.
The fundamental solution systems of neighborhoods of different regular
singularities can be regarded as analytic continuations of each other.
Thus the analytic expression of Feynman integral can be obtained in the whole parameter space.

Nevertheless, one only obtains the fundamental solution systems in neighborhoods of the
regular singularities $0,\;\infty$ using the GKZ-hypergeometric system of a general compact manifold.
Taking the Feynman integral as a linear combinations of hypergeometric functions from
the fundamental solution systems around different regular singularities and embedding
it in subvarieties of Grassmannians, one finds that the dual space of the GKZ-system
satisfied by the Feynman integral in the splitting coordinates determines the integer
lattice matrices, then writes the exponent matrices compatible with the integer
lattice matrices. A pair of integer lattice and exponent matrices uniquely determines
a generalized hypergeometric function where the geometric representation~\cite{Oxley2011}
of its exponent matrix can clearly tell us whether or not the hypergeometric
function of a different pair of integer lattice and exponent matrices is
proportional to it in the nonempty intersection of their convergent regions.

Some hypergeometric functions are defined on the Grassmannian $G_{_{k,n}}$~\cite{Gelfand1986-1}
in a natural way, where the Grassmannian $G_{_{k,n}}$ is the manifold of all
$(k-1)$-dimensional projective subspaces of $(n-1)$-dimensional projective
space ($k<n$)~\cite{Schwartz1968}. In order to describe Grassmannian exactly,
one can understand the manifold $Z_{_{k,n}}$ of $k\times n$ matrices of rank $k$
being a bundle space over the base space $G_{_{k,n}}$ whose projection
$Z_{_{k,n}}\rightarrow G_{_{k,n}}$ assigns to each matrix $Z_{_{k,n}}$
the $k$-dimensional projective subspace spanned by the row vectors of this matrix.
We consider an arbitrary smooth function on the manifold $Z_{_{k,n}}$
satisfying the homogeneous condition
\begin{eqnarray}
&&f(\chi_{_1}z_{_1},\cdots,\chi_{_n}z_{_n})=\prod\limits_{i=1}^n
\chi_{_i}^{\alpha_{_i}-1}f(z_{_1},\cdots,z_{_n})\;,\chi_{_i}\neq0\;,
\label{Grass1}
\end{eqnarray}
where $z_{_i}$, $(i=1,\cdots,n)$ denotes the $i$-th column vector of the $k\times n$
matrix $Z_{_{k,n}}$, and
\begin{eqnarray}
&&\sum\limits_{i=1}^n\alpha_{_i}=n-k\;.
\label{Grass2}
\end{eqnarray}
Let $t=(t_{_1},\cdots,t_{_k})$ denote the local coordinates of the
$k$-dimensional projective subspace, and the volume element
\begin{eqnarray}
&&\omega(t)=\sum\limits_{i=1}^k(-)^{i-1}dt_{_1}\wedge\cdots dt_{_{i-1}}
\wedge dt_{_{i+1}}\wedge\cdots\wedge dt_{_k}
\label{Grass3}
\end{eqnarray}
thus the integral
\begin{eqnarray}
&&F(z)=\int\limits_{S}f(t\cdot z)\omega(t)
\label{Grass4}
\end{eqnarray}
satisfies the homogeneous conditions
\begin{eqnarray}
&&F(g\cdot z)=|\det g|^{-1}F(z),\;\;g\in GL(k,C),\nonumber\\
&&F(z\cdot\chi)=\prod\limits_{i=1}^n\chi_{_i}^{\alpha_{_i}-1}F(z),\;\;
\chi={\rm diag}(\chi_{_1},\cdots,\chi_{_n})\;.
\label{Grass5}
\end{eqnarray}
Here $S$ is an arbitrary $(k-1)$-dimensional hypersurface enclosing the origin.
In this approach elements of the matrix $Z_{_{k\times n}}$ are named as the splitting
local coordinates.
Obviously the first condition is equivalent to the system of partial differential
equations (PDEs)~\cite{Gelfand1980,Gelfand1986-2,Gelfand1990+a,Gelfand1992}
\begin{eqnarray}
&&\sum\limits_{j=1}^nz_{_{ij}}{\partial F\over\partial z_{_{i^\prime j}}}=
-\delta_{_{i,i^\prime}}F,\;\;\;i\in[1,k],\;i^\prime\in[1,k],
\label{Grass6}
\end{eqnarray}
with $\delta_{_{i,i^\prime}}$ denoting the Kronecker symbol,
and the second condition of Eq.(\ref{Grass5}) is equivalent to the system of PDEs
\begin{eqnarray}
&&\sum\limits_{i=1}^kz_{_{ij}}{\partial F\over\partial z_{_{ij}}}=
(\alpha_{_j}-1)F,\;\;\;j\in[1,n],
\label{Grass7}
\end{eqnarray}
respectively. Using the GKZ-system presented above, we can construct the fundamental
solution system which is composed of the hypergeometric functions of local splitting
coordinates $z_{_{ij}}$~\cite{Gelfand1990+a,Gelfand1992}.

After embedding parametric representations of Feynman integrals in projective space,
we formulate the Feynman integrals as hypergeometric functions over Grassmannians.
Taking Feynman integrals of the 1-loop self-energy,
the massless 1-loop triangle, and the 2-loop vacuum diagrams as examples, we illuminate
how to construct canonical series solutions from those relevant GKZ-systems~\cite{Gelfand1990+a,Gelfand1992},
whose convergent regions of those series can be derived by Horn's study~\cite{Horn1889}.
Finally parametric representations of the 2-loop self energy diagrams with 3, 4 and 5 propagators
are presented respectively. Those Feynman integrals are embedded in
the subvarieties of Grassmannians by the parametric representations directly.
The adopted algorithm here is applied to obtain the fundamental solution systems of the GKZ-systems
associating with the subvarieties of the Grassmannian $G_{_{3,6}}$ in Ref.~\cite{Gelfand1990+a},
and all techniques in detail of this approach can be found in the long review article~\cite{Gelfand1992}.
In principle, we can embed the parametric representation of a given Feynman
integral into infinite subvarieties of Grassmannians.
However, the dimension of dual space of the GKZ-systems of most subvarieties
is much larger than the number of independent dimensionless ratios between the
square of external momentum and the square of virtual masses.
Correspondingly convergent regions of the obtained hypergeometric series solutions are empty.
In addition, we don't state those mathematical concepts and theorems that have been used in our analyses here,
because they can be found in some well-known mathematical textbooks~\cite{M.Saito2000,M.E.Taylor12,
L.J.Slater66,J.Leray1959,Cox1991,Cox1998,Sturmfels1995,Eisenbud1995,Coutinho1995}.

The generally strategy for analysing Feynman integral includes following steps
in this approach. First we embed parametric representation of Feynman integral in
the subvarieties of proper Grassmannians, and write out the GKZ-system satisfied by the
Feynman integral. Second the fundamental solution systems are constructed in
neighborhoods of regular singularities of the GKZ-system.
The integration constants, i.e. the combination coefficients, are determined
from the Feynman integral with some special kinematic parameters~\cite{Feng2018,Chetyrkin1980,Gu2020}.

Our presentation is organized as following. After embedding the Feynman integrals
in the subvarieties of Grassmannians appropriately, we derive the fundamental solution systems of Feynman
integrals of the 1-loop vacuum and 1-loop self energy in section \ref{1SE}, respectively.
Then we present in detail how to embed parametric representation of the 1-loop massless triangle
diagrams in section \ref{1Triangle}, and several equivalent fundamental solution systems
are presented in different neighborhoods of regular singularities.
In section \ref{2Vac}, we embed Feynman integral of the 2-loop vacuum in the subvarieties
of Grassmannian $G_{_{4,8}}$, then present the hypergeometric series in neighborhoods
of regular singularities. Finally parametric representations of the 2-loop self energy diagrams
with 3, 4 and 5 propagators are presented respectively in section \ref{2LSE},
and the conclusions are summarized in section \ref{summary}.

\section{The 1-loop self-energy\label{1SE}}
\indent\indent
We first use the well-known 1-loop bubble diagram to illustrate how to embed
the corresponding Feynman integral into the subvarieties of Grassmannians, then construct its
fundamental solution system. The Feynman integral of the 1-loop bubble diagram is written as
\begin{eqnarray}
&&A_{_{1VA}}(m^2)=\Big(\Lambda_{_{\rm RE}}^2\Big)^{2-D/2}\int{d^Dq\over(2\pi)^D}
{1\over q^2-m^2}
\nonumber\\
&&\hspace{1.8cm}=
-{i\Gamma(1-D/2)\over(4\pi)^{D/2}}{\Big(\Lambda_{_{\rm RE}}^2\Big)^{2-D/2}
\over\Big(m^2\Big)^{1-D/2}}\;,
\label{1Vac-1}
\end{eqnarray}
with $\Lambda_{_{\rm RE}}$ denoting the renormalization scale.
After performing the Wick rotation and integrating over angles,
\begin{eqnarray}
&&A_{_{1VA}}(m^2)=-{i\Big(\Lambda_{_{\rm RE}}^2\Big)^{2-D/2}\over(4\pi)^{D/2}\Gamma({D\over2})}
\int_0^\infty{dt_{_1}t_{_1}^{D/2-1}\over t_{_1}+m^2}\;.
\label{1Vac-2}
\end{eqnarray}
Introducing the homogeneous coordinate $t_{_2}=1$, one reformulates the scalar integral as
\begin{eqnarray}
&&A_{_{1VA}}(m^2)=-{i\Big(\Lambda_{_{\rm RE}}^2\Big)^{2-D/2}\over(4\pi)^{D/2}\Gamma({D\over2})}
\int{\omega(t)t_{_1}^{D/2-1}t_{_2}^{-D/2}\over t_{_1}+t_{_2}m^2}\;,
\label{1Vac-3}
\end{eqnarray}
with the volume element of projective line $\omega(t)=t_{_2}dt_{_1}-t_{_1}dt_{_2}$.
Eq.(\ref{1Vac-3}) can be taken as an integral on the Grassmannian $G_{_{2,3}}$
with splitting local coordinates as
\begin{eqnarray}
&&A^{1V}=\left(\begin{array}{ccc}\;\;1\;\;&\;\;0\;\;&\;\;1\;\;\\
\;\;0\;\;&\;\;1\;\;&\;\;m^2\;\;\end{array}\right)\;,
\label{1Vac-4}
\end{eqnarray}
whose geometric representation is 3 non-overlapping points on a projective line.
In the matrix $A^{1V}$, the first row corresponds to the integration variable $t_{_1}$,
the second row  corresponds to the integration variable $t_{_2}$, the first column represents
the power function $t_{_1}^{D/2-1}$, the second column represents the power function $t_{_2}^{-D/2}$,
and the third column represents the power function $(z_{_{1,3}}t_{_1}+z_{_{2,3}}t_{_2})^{-1}
=(t_{_1}+t_{_2}m^2)^{-1}$, respectively.
In the above splitting coordinates, the Feynman integral of the 1-loop bubble
satisfies the following GKZ-system
\begin{eqnarray}
&&\Big\{\vartheta_{_{1,1}}+\vartheta_{_{1,3}}\Big\}A_{_{1VA}}=-A_{_{1VA}}
\;,\nonumber\\
&&\Big\{\vartheta_{_{2,2}}+\vartheta_{_{2,3}}\Big\}A_{_{1VA}}=-A_{_{1VA}}
\;,\nonumber\\
&&\vartheta_{_{1,1}}A_{_{1VA}}=({D\over2}-1)A_{_{1VA}}
\;,\nonumber\\
&&\vartheta_{_{2,2}}A_{_{1VA}}=-{D\over2}A_{_{1VA}}
\;,\nonumber\\
&&\Big\{\vartheta_{_{1,3}}+\vartheta_{_{2,3}}\Big\}A_{_{1VA}}=-A_{_{1VA}}\;,
\label{1Vac-4a}
\end{eqnarray}
where the Euler operator $\vartheta_{_{i,j}}=z_{_{i,j}}\partial/\partial z_{_{i,j}}$.
The dual space of the above GKZ-system is the null space $\{0\}$. Thus
the exponent matrix is written uniquely as
\begin{eqnarray}
&&\left(\begin{array}{ccc}\;\;{D\over2}-1\;\;&\;\;0\;\;&\;\;-{D\over2}\;\;\\
\;\;0\;\;&\;\;-{D\over2}\;\;&\;\;{D\over2}-1\;\;\end{array}\right)\;,
\label{1Vac-5}
\end{eqnarray}
and the matrix of integer lattice is the zero matrix $0_{_{2\times3}}$.
Using the above matrices of exponent and integer lattice, one obtains the unique solution of
the GKZ-system satisfied by $A_{_{1VA}}(m^2)$ as
\begin{eqnarray}
&&A_{_{1VA}}(m^2)={C_{_{1VA}}^{(12)}\over\Gamma(1-{D\over2})\Gamma({D\over2})}
\Big(m^2\Big)^{D/2-1}\propto{1^{-D/2}\over\Gamma(1-{D\over2})\Gamma({D\over2})}
\Big(m^2\Big)^{D/2-1}\;,
\label{1Vac-6}
\end{eqnarray}
with the integration constant
\begin{eqnarray}
&&C_{_{1VA}}^{(12)}=-{i\Gamma^2(1-{D\over2})\Gamma({D\over2})\over(4\pi)^{D/2}}
\Big(\Lambda_{_{\rm RE}}^2\Big)^{2-D/2}\;.
\label{1Vac-7}
\end{eqnarray}
The above solution is obtained where we take the 1st and 2nd column vectors
as the basis of the projective line. Because
\begin{eqnarray}
&&\det A_{_{13}}^{1V}=\det\left(\begin{array}{cc}\;\;1\;\;&\;\;1\;\;\\\;\;0\;\;&\;\;m^2\;\;\end{array}\right)=m^2\neq0\;,
\nonumber\\
&&\det A_{_{23}}^{1V}=\det\left(\begin{array}{cc}\;\;0\;\;&\;\;1\;\;\\\;\;1\;\;&\;\;m^2\;\;\end{array}\right)=-1\neq0\;,
\label{1Vac-8}
\end{eqnarray}
we can also choose $\{(1,\;0)^T,\;(1,\;m^2)^T\}$ or $\{(0,\;1)^T,\;(1,\;m^2)^T\}$ as the basis of
the projective line respectively. As we choose the vectors $\{(1,\;0)^T,\;(1,\;m^2)^T\}$
as the basis of the projective line, the splitting local coordinate matrix is written as
\begin{eqnarray}
&&\left(\begin{array}{ccc}\;\;1\;\;&\;\;-{1\over m^2}\;\;&\;\;0\;\;\\
\;\;0\;\;&\;\;{1\over m^2}\;\;&\;\;1\;\;\end{array}\right)\;.
\label{1Vac-9}
\end{eqnarray}
Using the unique matrix of exponents
\begin{eqnarray}
&&\left(\begin{array}{ccc}\;\;{D\over2}-1\;\;&\;\;-{D\over2}\;\;&\;\;0\;\;\\
\;\;0\;\;&\;\;0\;\;&\;\;-1\;\;\end{array}\right)\;,
\label{1Vac-9a}
\end{eqnarray}
and the integer lattice matrix $0_{_{2\times3}}$, one obtains the single solution as
\begin{eqnarray}
&&A_{_{1VA}}(m^2)={C_{_{1VA}}^{(13)}\over\det(A_{_{13}}^{1V})\Gamma(1-{D\over2})\Gamma(1)}
\Big(-{1\over m^2}\Big)^0\Big({1\over m^2}\Big)^{-D/2}
\nonumber\\
&&\hspace{1.8cm}=
{C_{_{1VA}}^{(13)}\over\Gamma(1-{D\over2})}
\Big(m^2\Big)^{D/2-1}\;.
\label{1Vac-10}
\end{eqnarray}
Matching with Eq.(\ref{1Vac-1}), we get the integration constant as
\begin{eqnarray}
&&C_{_{1VA}}^{(13)}=-{i\Gamma^2(1-{D\over2})\over(4\pi)^{D/2}}
\Big(\Lambda_{_{\rm RE}}^2\Big)^{2-D/2}\;.
\label{1Vac-11}
\end{eqnarray}
Similarly taking the vectors $\{(0,\;1)^T,\;(1,\;m^2)^T\}$
as the basis of the projective line, we write the splitting local coordinate matrix as
\begin{eqnarray}
&&\left(\begin{array}{ccc}\;\;-m^2\;\;&\;\;1\;\;&\;\;0\;\;\\
\;\;1\;\;&\;\;0\;\;&\;\;1\;\;\end{array}\right)\;.
\label{1Vac-12}
\end{eqnarray}
Using the matrix of exponents in Eq.(\ref{1Vac-9a}) and the
integer lattice $0_{_{2\times3}}$, we write the solution as
\begin{eqnarray}
&&A_{_{1VA}}(m^2)={C_{_{1VA}}^{(23)}\over\det(A_{_{23}}^{1V})\Gamma({D\over2})\Gamma(1)}
\Big(-m^2\Big)^{D/2-1}
\nonumber\\
&&\hspace{1.8cm}=
-{(-)^{D/2-1}C_{_{1VA}}^{(23)}\over\Gamma({D\over2})}
\Big(m^2\Big)^{D/2-1}\;.
\label{1Vac-13}
\end{eqnarray}
From Eq.(\ref{1Vac-1}), the integration constant is given as
\begin{eqnarray}
&&C_{_{1VA}}^{(23)}={i(-1)^{1-D/2}\Gamma(1-{D\over2})\Gamma({D\over2})\over(4\pi)^{D/2}}
\Big(\Lambda_{_{\rm RE}}^2\Big)^{2-D/2}\;.
\label{1Vac-14}
\end{eqnarray}
Note that on the Grassmannian $G_{_{2,3}}$, there is only one choice of
the exponent matrix compatible with the integer lattice $0_{_{2\times3}}$. This is the reason why
three different choices on the basis of the projective line induce the same solution
of the GKZ-system which is proportional to $(m^2)^{D/2-1}$.

It should be emphasized that the matrix of splitting local coordinates in Eq.(\ref{1Vac-4})
is not uniquely determined. In fact, the columns of the matrix represent 3 distinct points on the
projective line. Multiplying each column by a non-zero constant does not change the
corresponding geometric representation of the matrix Eq.(\ref{1Vac-4}). Correspondingly those
fundamental solutions are completely consistent with each other up to a constant scalar multiple.

In the following we apply this method to present the subvarieties of
Grassmannians embedded by the Feynman integral of the 1-loop self-energy
diagram, then construct the fundamental solution systems which are
composed of linear independent hypergeometric functions in
neighborhoods of the regular singularities. As mentioned above,
the analytic expression of the Feynman integral is expressed as the linear combinations
of these hypergeometric functions of the fundamental solution systems
in different coordinate neighborhoods of regular singularities, whose
combination coefficients are determined by the Feynman integral on an ordinary
point or some regular singularities.
In the dimension regularization, the Feynman integral of the 1-loop self-energy
diagram is written as
\begin{eqnarray}
&&A_{_{1SE}}(p^2,m_{_1}^2,m_{_2}^2)=\Big(\Lambda_{_{\rm RE}}^2\Big)^{2-D/2}\int{d^Dq\over(2\pi)^D}
{1\over(q^2-m_{_1}^2)((q+p)^2-m_{_2}^2)}\;.
\label{1SE-0}
\end{eqnarray}
Adopting  Feynman parametric representation, we get the integral of zero virtual
masses as
\begin{eqnarray}
&&A_{_{1SE}}(p^2,0,0)=
{i(-1)^{D/2-2}\Gamma(2-{D\over2})\Big(\Lambda_{_{\rm RE}}^2\Big)^{2-D/2}
\over(4\pi)^{D/2}}
\nonumber\\
&&\hspace{2.9cm}\times
\int_0^1dt_{_1}t_{_1}^{D/2-2}(p^2t_{_1}-p^2)^{D/2-2}
\nonumber\\
&&\hspace{2.4cm}=
{i(-1)^{D/2-2}\Gamma(2-{D\over2})\Big(\Lambda_{_{\rm RE}}^2\Big)^{2-D/2}
\over(4\pi)^{D/2}}
\nonumber\\
&&\hspace{2.9cm}\times
\int_0^1\omega(t)t_{_1}^{D/2-2}t_{_2}^{2-D}(p^2t_{_1}+p^2t_{_2})^{D/2-2}\;.
\label{1SE-1}
\end{eqnarray}
In the last equation $t_{_2}=-1$, and $\omega(t)$
is the volume element of the projective line.
Eq.(\ref{1SE-1}) can be embedded in the subvariety of the Grassmannian $G_{_{2,3}}$
with splitting local coordinates as
\begin{eqnarray}
&&A^{1SE}=\left(\begin{array}{ccc}\;\;1\;\;&\;\;0\;\;&\;\;p^2\;\;\\
\;\;0\;\;&\;\;1\;\;&\;\;p^2\;\;\end{array}\right)\;.
\label{1SE-2}
\end{eqnarray}
In the matrix above, the first row corresponds to the integration variable $t_{_1}$,
the second row  corresponds to the integration variable $t_{_2}$, the first column represents
the power function $t_{_1}^{D/2-2}$, the second column represents the power function $t_{_2}^{2-D}$,
and the third column represents the power function $(z_{_{1,3}}t_{_1}+z_{_{2,3}}t_{_2})^{D/2-2}
=(t_{_1}p^2+t_{_2}p^2)^{D/2-2}$, respectively.
The Feynman integral satisfies the following GKZ-system
\begin{eqnarray}
&&\Big\{\vartheta_{_{1,1}}+\vartheta_{_{1,3}}\Big\}A_{_{1SE}}=-A_{_{1SE}}
\;,\nonumber\\
&&\Big\{\vartheta_{_{2,2}}+\vartheta_{_{2,3}}\Big\}A_{_{1SE}}=-A_{_{1SE}}
\;,\nonumber\\
&&\vartheta_{_{1,1}}A_{_{1SE}}=({D\over2}-2)A_{_{1SE}}
\;,\nonumber\\
&&\vartheta_{_{2,2}}A_{_{1SE}}=(2-D)A_{_{1SE}}
\;,\nonumber\\
&&\Big\{\vartheta_{_{1,3}}+\vartheta_{_{2,3}}\Big\}A_{_{1SE}}=({D\over2}-2)A_{_{1SE}}\;.
\label{1SE-2a}
\end{eqnarray}
Obviously the dual space of the GKZ-system of Eq.(\ref{1SE-2a}) is the null
space $\{0\}$ which is consistent with the exponent matrix
\begin{eqnarray}
&&\left(\begin{array}{ccc}\;\;{D\over2}-2\;\;&\;\;0\;\;&\;\;1-{D\over2}\;\;\\
\;\;0\;\;&\;\;2-D\;\;&\;\;D-3\;\;\end{array}\right)\;.
\label{1SE-3}
\end{eqnarray}
Using the matrix of exponents, one obtains the solution of
the GKZ-system of Eq.(\ref{1SE-2a})
\begin{eqnarray}
&&A_{_{1SE}}(p^2,0,0)={C_{_{1SE}}^{(0)}\over\Gamma(2-{D\over2})\Gamma(D-2)}
\Big(p^2\Big)^{1-D/2}\Big(p^2\Big)^{D-3}
\nonumber\\
&&\hspace{2.4cm}=
{C_{_{1SE}}^{(0)}\over\Gamma(2-{D\over2})\Gamma(D-2)}\Big(p^2\Big)^{D/2-2}\;,
\label{1SE-4}
\end{eqnarray}
with the integration constant
\begin{eqnarray}
&&C_{_{1SE}}^{(0)}={i\Gamma^2(2-{D\over2})\Gamma^2(1-{D\over2})
\Big(\Lambda_{_{\rm RE}}^2\Big)^{2-D/2}\over(4\pi)^{D/2}}\;.
\label{1SE-5}
\end{eqnarray}
The solution Eq.(\ref{1SE-4}) is obtained where we take the first and second column vectors of Eq.(\ref{1SE-2})
as the basis of the projective line. Because
\begin{eqnarray}
&&\det A_{_{13}}^{1SE}=\det\left(\begin{array}{cc}\;\;1\;\;&\;\;p^2\;\;\\\;\;0\;\;&\;\;p^2\;\;\end{array}\right)=p^2\neq0\;,
\nonumber\\
&&\det A_{_{23}}^{1SE}=\det\left(\begin{array}{cc}\;\;0\;\;&\;\;p^2\;\;\\\;\;1\;\;&\;\;p^2\;\;\end{array}\right)=-p^2\neq0\;,
\label{1SE-6}
\end{eqnarray}
we can choose $\{(1,\;0)^T,\;(p^2,\;p^2)^T\}$ or $\{(0,\;1)^T,\;(p^2,\;p^2)^T\}$ as the basis of
the projective line respectively. Using two different choices of the basis of the projective line,
the solutions of the GKZ-system are all proportional to $(p^2)^{D/2-2}$.

Along with increasing of the number of independent arguments of the Feynman integral,
the dimension of dual space of the GKZ-system increases. The bases of the dual space
determine the PDEs satisfied by the Feynman integral in the combined local coordinates.

\subsection{The fundamental solution systems as $m_{_1}^2=0$}
\begin{figure}[ht]
\setlength{\unitlength}{1cm}
\centering
\vspace{0.0cm}\hspace{-1.5cm}
\includegraphics[height=8cm,width=8.0cm]{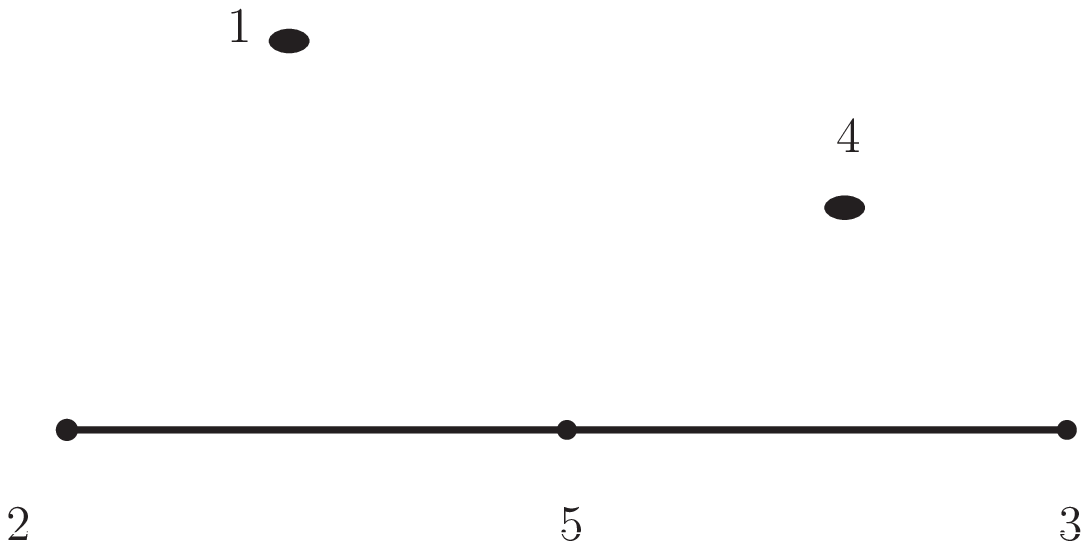}
\vspace{0cm}
\caption[]{The geometric description of the matrix in
Eq.(\ref{1SE-9}) on the projective plane $CP^{2}$, where the points $1,\cdots,5$
denote the indices of columns of the $3\times5$ matrix.}
\label{fig1}
\end{figure}

\begin{figure}[ht]
\setlength{\unitlength}{1cm}
\centering
\vspace{0.0cm}\hspace{-1.5cm}
\includegraphics[height=6cm,width=10.0cm]{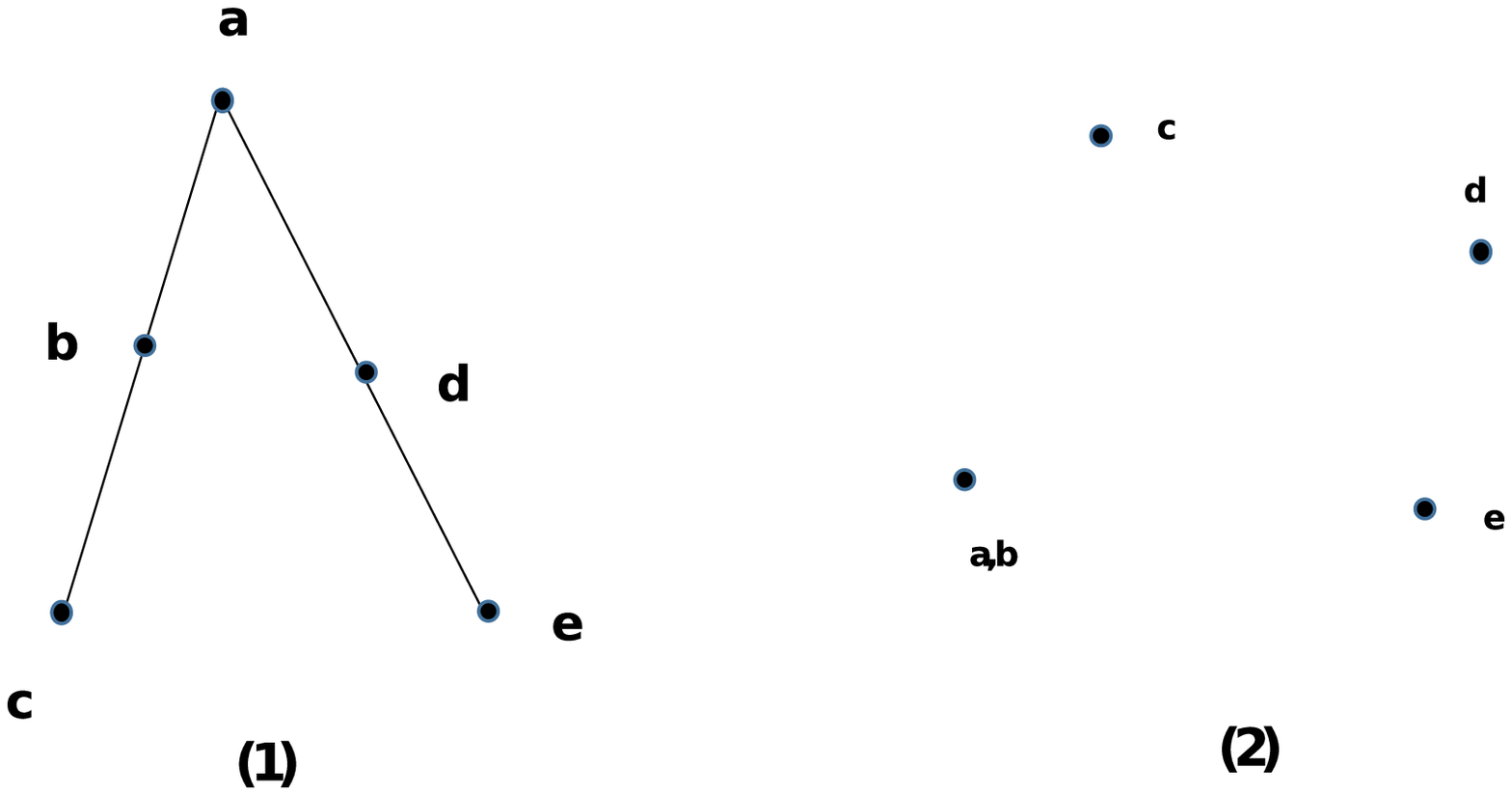}
\vspace{0cm}
\caption[]{The geometric description of the single orbits in
Eq.(\ref{1SE-13}),Eq.(\ref{1SE-14}),Eq.(\ref{1SE-16}), and Eq.(\ref{1SE-17})
on the projective plane $CP^{2}$, where the points $1,\cdots,5$
denote the indices of columns of the $3\times5$ matrix.}
\label{fig2}
\end{figure}

\indent\indent
As $m_{_1}^2=0$, $p^2\neq0$, and $m_{_2}^2\neq0$,
Feynman parametric representation of the 1-loop self energy is given as
\begin{eqnarray}
&&A_{_{1SE}}(p^2,0,m_{_2}^2)=
{i\Gamma(2-{D\over2})\Big(\Lambda_{_{\rm RE}}^2\Big)^{2-D/2}\over(-)^{2-D/2}
(4\pi)^{D/2}}\int_0^1dt_{_1}dt_{_3}{t_{_1}^{D/2-2}\delta(t_{_1}+t_{_3}-1)\over
(t_{_3}p^2-m_{_2}^2)^{2-D/2}}
\nonumber\\
&&\hspace{2.8cm}=
{i\Gamma(2-{D\over2})\Big(\Lambda_{_{\rm RE}}^2\Big)^{2-D/2}\over(-)^{1-D/2}
(4\pi)^{D/2}}\int\omega_{_3}(t){t_{_1}^{D/2-2}t_{_2}^{2-D}\delta(t_{_1}+t_{_2}+t_{_3})\over
(t_{_3}p^2+t_{_2}m_{_2}^2)^{2-D/2}}\;,
\label{1SE-7}
\end{eqnarray}
where
\begin{eqnarray}
&&\omega_{_k}(t)=\sum\limits_{i=1}^k(-)^{i-1}t_{_i}dt_{_1}\wedge\cdots\widehat{dt_{_i}}\wedge\cdots dt_{_k}
\label{1SE-8}
\end{eqnarray}
denotes the volume element of $(k-1)$-dimension projective space, and the
hat symbol implies that the wedge product does not involve $dt_{_i}$.
Eq.(\ref{1SE-7}) can be taken as an integral on the subvariety of the
Grassmannian $G_{_{3,5}}$ with splitting local coordinates as
\begin{eqnarray}
&&A^{(1)}=\left(\begin{array}{ccccc}\;\;1\;\;&\;\;0\;\;&\;\;0\;\;&\;\;1\;\;&\;\;0\;\;\\
\;\;0\;\;&\;\;1\;\;&\;\;0\;\;&\;\;1\;\;&\;\;m_{_2}^2\;\;\\
\;\;0\;\;&\;\;0\;\;&\;\;1\;\;&\;\;1\;\;&\;\;p^2\;\;\end{array}\right)\;.
\label{1SE-9}
\end{eqnarray}
Here the first row corresponds to the integration variable $t_{_1}$,
the second row  corresponds to the integration variable $t_{_2}$,
and the third row  corresponds to the integration variable $t_{_3}$, respectively.
Meanwhile the first column represents the power function $t_{_1}^{D/2-2}$,
the second column represents the power function $t_{_2}^{2-D}$, the third column
represents the power function $t_{_3}^0=1$,
the fourth column represents the function $\delta(t_{_1}+t_{_2}+t_{_3})$,
and the fifth column represents $(z_{_{1,5}}t_{_1}+z_{_{2,5}}t_{_2}+z_{_{3,5}}t_{_3})^{D/2-2}
=(t_{_2}m_{_2}^2+t_{_3}p^2)^{D/2-2}$, respectively.
The geometric description of the matrix of Eq.(\ref{1SE-9}) can be drawn in
2-dimension projective plane~\cite{Oxley2011} as Fig.\ref{fig1}.
Accordingly the Feynman integral satisfies the following GKZ-system in splitting local coordinates
\begin{eqnarray}
&&\Big\{\vartheta_{_{1,1}}+\vartheta_{_{1,4}}\Big\}A_{_{1SE}}(p^2,0,m_{_2}^2)=
-A_{_{1SE}}(p^2,0,m_{_2}^2)
\;,\nonumber\\
&&\Big\{\vartheta_{_{2,2}}+\vartheta_{_{2,4}}+\vartheta_{_{2,5}}\Big\}A_{_{1SE}}(p^2,0,m_{_2}^2)=
-A_{_{1SE}}(p^2,0,m_{_2}^2)
\;,\nonumber\\
&&\Big\{\vartheta_{_{3,3}}+\vartheta_{_{3,4}}+\vartheta_{_{3,5}}\Big\}A_{_{1SE}}(p^2,0,m_{_2}^2)=
-A_{_{1SE}}(p^2,0,m_{_2}^2)
\;,\nonumber\\
&&\vartheta_{_{1,1}}A_{_{1SE}}(p^2,0,m_{_2}^2)=({D\over2}-2)A_{_{1SE}}(p^2,0,m_{_2}^2)
\;,\nonumber\\
&&\vartheta_{_{2,2}}A_{_{1SE}}(p^2,0,m_{_2}^2)=(2-D)A_{_{1SE}}(p^2,0,m_{_2}^2)
\;,\nonumber\\
&&\vartheta_{_{3,3}}A_{_{1SE}}(p^2,0,m_{_2}^2)=0
\;,\nonumber\\
&&\Big\{\vartheta_{_{1,4}}+\vartheta_{_{2,4}}+\vartheta_{_{3,4}}\Big\}A_{_{1SE}}(p^2,0,m_{_2}^2)=
-A_{_{1SE}}(p^2,0,m_{_2}^2)
\;,\nonumber\\
&&\Big\{\vartheta_{_{2,5}}+\vartheta_{_{3,5}}\Big\}A_{_{1SE}}(p^2,0,m_{_2}^2)=
({D\over2}-2)A_{_{1SE}}(p^2,0,m_{_2}^2)\;,
\label{1SE-9a}
\end{eqnarray}
which induces the exponent matrix as
\begin{eqnarray}
&&\left(\begin{array}{ccccc}\;\;{D\over2}-2\;\;&\;\;0\;\;&\;\;0\;\;&\;\;1-{D\over2}\;\;&\;\;0\;\;\\
\;\;0\;\;&\;\;2-D\;\;&\;\;0\;\;&\;\alpha_{_{2,4}}\;\;&\;\alpha_{_{2,5}}\;\;\\
\;\;0\;\;&\;\;0\;\;&\;\;0\;\;&\;\alpha_{_{3,4}}\;\;&\;\alpha_{_{3,5}}\;\;\end{array}\right)\;.
\label{1SE-10}
\end{eqnarray}
Where the exponents $\alpha_{_{i,j}}$ satisfy the following relations
\begin{eqnarray}
&&\alpha_{_{2,4}}+\alpha_{_{2,5}}=D-3,\;\alpha_{_{3,4}}+\alpha_{_{3,5}}=-1,
\nonumber\\
&&\alpha_{_{2,4}}+\alpha_{_{3,4}}={D\over2}-2,\;\alpha_{_{2,5}}+\alpha_{_{3,5}}={D\over2}-2.
\label{1SE-11}
\end{eqnarray}
Furthermore, the dual space of the GKZ-system of Eq.(\ref{1SE-9a}) is spanned by
a $3\times5$ matrix $(0_{_{3\times3}}\Big|E_{_3}^{(1)})$ with
\begin{eqnarray}
&&E_{_3}^{(1)}=\left(\begin{array}{cc}\;\;0\;\;&\;\;0\;\;\\\;\;1\;\;&\;\;-1\;\;\\
\;\;-1\;\;&\;\;1\;\;\end{array}\right)\;.
\label{1SE-12}
\end{eqnarray}
The base of the dual space implies the Feynman integral also satisfying the PDE as
\begin{eqnarray}
&&{\partial^2A_{_{1SE}}\over\partial z_{_{2,4}}\partial z_{_{3,5}}}(p^2,0,m_{_2}^2)
={\partial^2A_{_{1SE}}\over\partial z_{_{2,5}}\partial z_{_{3,4}}}(p^2,0,m_{_2}^2)\;.
\label{1SE-12a}
\end{eqnarray}
When the exponent matrix is given, the Feynman integral can be formally expressed as
\begin{eqnarray}
&&A_{_{1SE}}(p^2,0,m_{_2}^2)=\prod\limits_{i,j}z_{_{i,j}}^{\alpha_{_{i,j}}}
\varphi({m_{_2}^2\over p^2})\;,
\label{1SE-12b}
\end{eqnarray}
where $\varphi(x)$ satisfies the PDE in the combined coordinate $x$ as
\begin{eqnarray}
&&x\Big[\alpha_{_{2,4}}\alpha_{_{3,5}}+(1-\alpha_{_{2,4}}-\alpha_{_{3,5}})x{d\over dx}
-x^2{d^2\over dx^2}\Big]\varphi(x)
\nonumber\\
&&\hspace{-0.5cm}=
\Big[\alpha_{_{2,5}}\alpha_{_{3,4}}+(1+\alpha_{_{2,5}}+\alpha_{_{3,4}})x{d\over dx}
-x^2{d^2\over dx^2}\Big]\varphi(x)\;.
\label{1SE-12c}
\end{eqnarray}
As $\alpha_{_{2,5}}\alpha_{_{3,4}}=0$ this equation recovers the well-known Gauss equation.
Note that the combined local coordinate $x=m_{_2}^2/p^2=z_{_{2,5}}z_{_{3,4}}/z_{_{2,4}}z_{_{3,5}}$
is determined by the base of the dual space.
The integer lattice $(0_{_{3\times3}}\Big|nE_{_3}^{(1)})$ ($n\ge0$) is compatible with two choices
of the exponents. The first choice is written as
\begin{eqnarray}
&&\alpha_{_{2,4}}=0,\;\alpha_{_{2,5}}=D-3,\;\alpha_{_{3,4}}={D\over2}-2,\;\alpha_{_{3,5}}=1-{D\over2}\;,
\label{1SE-13}
\end{eqnarray}
whose geometric description is presented as Fig.\ref{fig2}(1)
with $a=3$, $\{(b,c),\;(d,e)\}=\{(1,4),\;(2,5)\}$. The second choice is
\begin{eqnarray}
&&\alpha_{_{2,4}}={D\over2}-1,\;\alpha_{_{2,5}}={D\over2}-2,\;\alpha_{_{3,4}}=-1,\;\alpha_{_{3,5}}=0\;,
\label{1SE-14}
\end{eqnarray}
whose geometric description is presented as Fig.\ref{fig2}(2)
with $(a,b)=(2,5)$, $(c,d,e)=(1,3,4)$.
Adopting the matrix of integer lattice $(0_{_{3\times3}}\Big|nE_{_3}^{(1)})$
and the corresponding matrices of exponents, we obtain
two linear independent hypergeometric functions as
\begin{eqnarray}
&&\psi_{_{\{1,2,3\}}}^{(1)}(p^2,\;0,\;m_{_2}^2)\sim
(p^2)^{1-D/2}(m_{_2}^2)^{D-3}\sum\limits_{n=0}^\infty
{\Gamma(3-D+n)\over n!}\Big({p^2\over m_{_2}^2}\Big)^n
\nonumber\\
&&\hspace{3.2cm}\sim
(p^2)^{1-D/2}(m_{_2}^2-p^2)^{D-3}\;,
\nonumber\\
&&\psi_{_{\{1,2,3\}}}^{(2)}(p^2,\;0,\;m_{_2}^2)\sim
(m_{_2}^2)^{D/2-2}\sum\limits_{n=0}^\infty
{\Gamma(2-{D\over2}+n)\over\Gamma({D\over2}+n)}\Big({p^2\over m_{_2}^2}\Big)^n\;,
\label{1SE-15}
\end{eqnarray}
where the intersection of their convergent regions is $|p^2/m_{_2}^2|\le1$.
Two linearly independent Gaussian functions constitute the fundamental
solution system of the GKZ-system of Eq.(\ref{1SE-9a}) in the region $|p^2/m_{_2}^2|\le1$
which is consistent with the results in the literature. As $y_{_i}=0$ in Eq.(46)
of Ref.~\cite{Feng2020}, the fourth type Appell functions of the last two terms
are transformed into the Gauss functions above. In the first equation of Eq.(\ref{1SE-15}),
we apply the well-known reduction~\cite{L.J.Slater66}
\begin{eqnarray}
&&_{_2}F_{_1}\left(\left.\begin{array}{c}\;a,\;b\;\\b\end{array}\right|\;x\right)=(1-x)^{-a}\;.
\label{1SE-15a}
\end{eqnarray}
Actually the last expression can be taken as the analytical continuation
of the hypergeometric series in the whole complex plane except the regular singularities
$p^2=0$ and $p^2-m_{_2}^2=0$.

Similarly the matrix of integer lattice $(0_{_{3\times3}}\Big|-nE_{_3}^{(1)})$ ($n\ge0$)
permits to two possibilities of the exponents. The first possibility is
\begin{eqnarray}
&&\alpha_{_{2,4}}=D-3,\;\alpha_{_{2,5}}=0,\;\alpha_{_{3,4}}=1-{D\over2},\;\alpha_{_{3,5}}={D\over2}-2\;,
\label{1SE-16}
\end{eqnarray}
whose geometric description is presented as Fig.\ref{fig2}(2) with $(a,b)=(3,5)$, $(c,d,e)=(1,2,4)$.
The second possibility of the exponents is
\begin{eqnarray}
&&\alpha_{_{2,4}}={D\over2}-2,\;\alpha_{_{2,5}}={D\over2}-1,\;\alpha_{_{3,4}}=0,\;\alpha_{_{3,5}}=-1\;,
\label{1SE-17}
\end{eqnarray}
whose geometric description is drawn as Fig.\ref{fig2}(1) with $a=2$,
$\{(b,c),\;(d,e)\}=\{(1,4),\;(3,5)\}$, respectively.
Similarly using the matrix of integer lattice
and the matrices of exponents, one obtains
two linear independent hypergeometric functions as
\begin{eqnarray}
&&\psi_{_{\{1,2,3\}}}^{(3)}(p^2,\;0,\;m_{_2}^2)\sim
(p^2)^{D/2-2}\sum\limits_{n=0}^\infty
{\Gamma(3-D+n)\over n!}\Big({m_{_2}^2\over p^2}\Big)^n
\nonumber\\
&&\hspace{3.2cm}\sim
(p^2)^{1-D/2}(p^2-m_{_2}^2)^{D-3}\;,
\nonumber\\
&&\psi_{_{\{1,2,3\}}}^{(4)}(p^2,\;0,\;m_{_2}^2)\sim
{(m_{_2}^2)^{D/2-1}\over p^2}\sum\limits_{n=0}^\infty
{\Gamma(2-{D\over2}+n)\over\Gamma({D\over2}+n)}\Big({m_{_2}^2\over p^2}\Big)^n\;,
\label{1SE-18}
\end{eqnarray}
where the intersection of their convergent region is $|m_{_2}^2/p^2|\le1$.
These two linearly independent Gaussian functions constitute the fundamental
solution system of the GKZ-system of Eq.(\ref{1SE-9a}) in the region $|m_{_2}^2/p^2|\le1$
which is consistent with the results in the literature. Choosing $y_{_i}=0$ in Eq.(45)
of Ref.~\cite{Feng2020}, the fourth type Appell functions of the second and the fourth terms
are reduced as the Gauss functions above.

If the GKZ-system of Eq.(\ref{1SE-9a}) is embedded in a compact $n$-dimensional manifold,
we can only construct the hypergeometric series in neighborhoods of the regular
singularities $m_{_2}^2/p^2=0,\infty$. In order to construct the hypergeometric series
in neighborhoods of the regular singularity $m_{_2}^2/p^2=1$, we embed the GKZ-system
of Eq.(\ref{1SE-9a}) into the subvariety of the Grassmannian $G_{_{3,5}}$ where the
splitting local coordinates are presented in Eq.(\ref{1SE-9}).

Because $\det(A_{_{\{1,2,5\}}}^{(1)})=p^2$ where the matrix $A_{_{\{1,2,5\}}}^{(1)}$ denotes
the submatrix of $A^{(1)}$ composed of the  first-, second-, and fifth columns, one can choose the
first, the second, and the fifth column vectors spanning the projective plane. Obviously we have
\begin{eqnarray}
&&\Big(A_{_{\{1,2,5\}}}^{(1)}\Big)^{-1}\cdot A^{(1)}=\left(\begin{array}{ccccc}
\;\;1\;\;&\;\;0\;\;&\;\;0\;\;&\;\;1\;\;&\;\;0\;\;\\
\;\;0\;\;&\;\;1\;\;&\;\;-{m_{_2}^2\over p^2}\;\;&\;\;1-{m_{_2}^2\over p^2}\;\;&\;\;0\;\;\\
\;\;0\;\;&\;\;0\;\;&\;\;{1\over p^2}\;\;&\;\;{1\over p^2}\;\;&\;\;1\;\;\end{array}\right)\;,
\label{1SE-19}
\end{eqnarray}
and the matrix of exponents
\begin{eqnarray}
&&\left(\begin{array}{ccccc}
\;\;{D\over2}-2\;\;&\;\;0\;\;&\;\;0\;\;&\;\;1-{D\over2}\;\;&\;\;0\;\;\\
\;\;0\;\;&\;\;2-D\;\;&\;\alpha_{_{2,3}}\;\;&\;\alpha_{_{2,4}}\;\;&\;\;0\;\;\\
\;\;0\;\;&\;\;0\;\;&\;\alpha_{_{3,3}}\;\;&\;\alpha_{_{3,4}}\;\;&\;\;{D\over2}-2\;\;\end{array}\right)\;.
\label{1SE-20}
\end{eqnarray}
Where the matrix elements satisfy the relations
\begin{eqnarray}
&&\alpha_{_{2,3}}+\alpha_{_{2,4}}=D-3\;,\;\alpha_{_{3,3}}+\alpha_{_{3,4}}=1-{D\over2}\;,
\nonumber\\
&&\alpha_{_{2,3}}+\alpha_{_{3,3}}=0\;,\;\alpha_{_{2,4}}+\alpha_{_{3,4}}={D\over2}-2\;.
\label{1SE-21}
\end{eqnarray}
Correspondingly there are two choices on the matrix of integer lattice whose submatrix composed of the third and
fourth columns is formulated as $\pm nE_{_3}^{(1)}$ with integer $n\ge0$, and
other elements are all zero. The matrix of integer lattice $nE_{_3}^{(1)}$ is compatible with two
choices of the exponents. The first choice is
\begin{eqnarray}
&&\alpha_{_{2,3}}=0,\;\alpha_{_{2,4}}=D-3,\;\alpha_{_{3,3}}=0,\;\alpha_{_{3,4}}=1-{D\over2}\;,
\label{1SE-22}
\end{eqnarray}
whose geometric representation is presented by Fig.\ref{fig2}(2)
with $(a,b)=(3,5)$, $(c,d,e)=(1,2,4)$. The second choice is
\begin{eqnarray}
&&\alpha_{_{2,3}}={D\over2}-1,\;\alpha_{_{2,4}}={D\over2}-2,\;\alpha_{_{3,3}}=1-{D\over2},\;\alpha_{_{3,4}}=0\;,
\label{1SE-23}
\end{eqnarray}
whose geometric representation is drawn as Fig.\ref{fig2}(1) with $a=2$,
$\{(b,c),\;(d,e)\}=\{(1,4),\;(3,5)\}$, respectively.
Adopting the matrix of integer lattice and the matrices of exponents,
we obtain a fundamental solution system composed of two linear independent
hypergeometric functions as
\begin{eqnarray}
&&\psi_{_{\{1,2,5\}}}^{(1)}(p^2,\;0,\;m_{_2}^2)\sim(p^2)^{1-D/2}(p^2-m_{_2}^2)^{D-3}
\nonumber\\
&&\hspace{3.7cm}\times
\sum\limits_{n=0}^\infty
{\Gamma(3-D+n)\over n!\Gamma(1-n)\Gamma(2-{D\over2}+n)}\Big({m_{_2}^2\over m_{_2}^2-p^2}\Big)^n
\nonumber\\
&&\hspace{3.2cm}\sim
(p^2)^{1-D/2}(m_{_2}^2-p^2)^{D-3}\;,
\nonumber\\
&&\psi_{_{\{1,2,5\}}}^{(2)}(p^2,\;0,\;m_{_2}^2)\sim
\Big({m_{_2}^2\over p^2}\Big)^{D/2-1}(m_{_2}^2-p^2)^{D/2-2}
\nonumber\\
&&\hspace{3.9cm}\times
\sum\limits_{n=0}^\infty{\Gamma(2-{D\over2}+n)
\Gamma({D\over2}-1+n)\over n!\Gamma({D\over2}+n)}
\Big({m_{_2}^2\over m_{_2}^2-p^2}\Big)^n\;,
\label{1SE-24}
\end{eqnarray}
where the intersection of convergent regions of two hypergeometric functions is
$|m_{_2}^2/(m_{_2}^2-p^2)|\le1$. In fact the basis $\{1,2,5\}$ of the project
plane composed of the first, second and fifth columns of the matrix Eq.(\ref{1SE-9})
is obtained from the basis $\{1,2,3\}$ through the permutation $\widehat{(35)}$,
accordingly the geometric description of the exponents in Eq.(\ref{1SE-22})
is derived from that of the exponents in Eq.(\ref{1SE-16}) through the permutation $\widehat{(35)}$.
Because the geometric description of Eq.(\ref{1SE-22}) is same as that of Eq.(\ref{1SE-16}),
the corresponding function $\psi_{_{\{1,2,5\}}}^{(1)}$ is proportionate to
$\psi_{_{\{1,2,3\}}}^{(3)}$ in the nonempty intersection of their convergent regions.
Similarly the geometric description of the exponents in Eq.(\ref{1SE-23})
is obtained through the permutation $\widehat{(35)}$ from that of Eq.(\ref{1SE-17}).
Because the geometric description of Eq.(\ref{1SE-23}) is same as that of Eq.(\ref{1SE-17}),
the function $\psi_{_{\{1,2,5\}}}^{(2)}$ is proportionate to
$\psi_{_{\{1,2,3\}}}^{(4)}$ in the nonempty intersection of their convergent regions.
Actually $\psi_{_{\{1,2,5\}}}^{(2)}\sim \psi_{_{\{1,2,3\}}}^{(4)}$ is consistent
with the well-known relation~\cite{L.J.Slater66}
\begin{eqnarray}
&&_{_2}F_{_1}\left(\left.\begin{array}{c}\;a,\;b\;\\c\end{array}\right|\;x\right)=
(1-x)^{-a}\;_{_2}F_{_1}\left(\left.\begin{array}{c}\;a,\;c-b\;\\c\end{array}\right|\;{x\over x-1}\right)\;.
\label{1SE-24a}
\end{eqnarray}

Similarly the matrix of integer lattice $-nE_{_3}^{(1)}$ ($n\ge0$) permits two
possibilities of the exponents. The first choice is
\begin{eqnarray}
&&\alpha_{_{2,3}}=D-3,\;\alpha_{_{2,4}}=0,\;\alpha_{_{3,3}}=3-D,\;\alpha_{_{3,4}}={D\over2}-2\;,
\label{1SE-25}
\end{eqnarray}
whose geometric representation is presented in Fig.\ref{fig2}(1)
with $a=5$, $\{(b,c),\;(d,e)\}=\{(1,4),\;(2,3)\}$. And the second possibility is
\begin{eqnarray}
&&\alpha_{_{2,3}}=0,\;\alpha_{_{2,4}}=D-3,\;\alpha_{_{3,3}}=0,\;\alpha_{_{3,4}}=1-{D\over2}\;,
\label{1SE-26}
\end{eqnarray}
whose geometric representation is presented in Fig.\ref{fig2}(2) with $(a,b)=(2,3)$,
$(c,d,e)=(1,4,5)$, respectively. Adopting the matrix of integer lattice
and the matrices of exponents, we obtain a fundamental solution
system composed of two linear independent hypergeometric functions as
\begin{eqnarray}
&&\psi_{_{\{1,2,5\}}}^{(3)}(p^2,\;0,\;m_{_2}^2)\sim
(m_{_2}^2)^{D-3}(p^2)^{1-D/2}
\nonumber\\
&&\hspace{3.9cm}\times
\sum\limits_{n=0}^\infty{\Gamma(3-D+n)
\Gamma(2-{D\over2}+n)\over n!\Gamma(4-D+n)}
\Big({m_{_2}^2-p^2\over m_{_2}^2}\Big)^n\;,
\nonumber\\
&&\psi_{_{\{1,2,5\}}}^{(4)}(p^2,\;0,\;m_{_2}^2)\sim
(p^2)^{1-D/2}(m_{_2}^2-p^2)^{D-3}\;,
\label{1SE-27}
\end{eqnarray}
whose convergent region is $|(m_{_2}^2-p^2)/m_{_2}^2|\le1$.

We take the first, third, and fifth columns as a basis spanning the projective plane
because $\det(A_{_{\{1,3,5\}}}^{(1)})=-m_{_2}^2$ where the matrix $A_{_{\{1,3,5\}}}^{(1)}$ denotes
the submatrix of $A^{(1)}$ composed of the first, third, and fifth columns. It is easily derived
\begin{eqnarray}
&&\Big(A_{_{\{1,3,5\}}}^{(1)}\Big)^{-1}\cdot A^{(1)}=\left(\begin{array}{ccccc}
\;\;1\;\;&\;\;0\;\;&\;\;0\;\;&\;\;1\;\;&\;\;0\;\;\\
\;\;0\;\;&\;\;-{p^2\over m_{_2}^2}\;\;&\;\;1\;\;&\;\;1-{p^2\over m_{_2}^2}\;\;&\;\;0\;\;\\
\;\;0\;\;&\;\;{1\over m_{_2}^2}\;\;&\;\;0\;\;&\;\;{1\over m_{_2}^2}\;\;&\;\;1\;\;\end{array}\right)\;.
\label{1SE-28}
\end{eqnarray}
Accordingly the matrix of exponents is written as
\begin{eqnarray}
&&\left(\begin{array}{ccccc}
\;\;{D\over2}-2\;\;&\;\;0\;\;&\;\;0\;\;&\;\;1-{D\over2}\;\;&\;\;0\;\;\\
\;\;0\;\;&\;\alpha_{_{2,2}}\;\;&\;\;0\;\;&\;\alpha_{_{2,4}}\;\;&\;\;0\;\;\\
\;\;0\;\;&\;\alpha_{_{3,2}}\;\;&\;\;0\;\;&\;\alpha_{_{3,4}}\;\;&\;\;{D\over2}-2\;\;\end{array}\right)\;,
\label{1SE-29}
\end{eqnarray}
where the matrix elements satisfy the relations
\begin{eqnarray}
&&\alpha_{_{2,2}}+\alpha_{_{2,4}}=-1\;,\;\alpha_{_{3,2}}+\alpha_{_{3,4}}=1-{D\over2}\;,
\nonumber\\
&&\alpha_{_{2,2}}+\alpha_{_{3,2}}=2-D\;,\;\alpha_{_{2,4}}+\alpha_{_{3,4}}={D\over2}-2\;.
\label{1SE-30}
\end{eqnarray}
There are two choices on the matrix of integer lattice whose submatrix composed of the second and
fourth columns is formulated as $\pm nE_{_3}^{(1)}$ ($n\ge0$), and other elements are all zero.
The matrix of integer lattice $nE_{_3}^{(1)}$ permits two possible choices of the exponents.
The first choice is
\begin{eqnarray}
&&\alpha_{_{2,2}}=0,\;\alpha_{_{2,4}}=-1,\;\alpha_{_{3,2}}=2-D,\;\alpha_{_{3,4}}={D\over2}-1\;,
\label{1SE-31}
\end{eqnarray}
whose geometric representation is presented in Fig.\ref{fig2}(2)
with $(a,b)=(2,5)$, $(c,d,e)=(1,3,4)$. The second choice of the exponents is
\begin{eqnarray}
&&\alpha_{_{2,2}}=1-{D\over2},\;\alpha_{_{2,4}}={D\over2}-2,\;\alpha_{_{3,2}}=1-{D\over2},\;\alpha_{_{3,4}}=0\;,
\label{1SE-32}
\end{eqnarray}
whose geometric representation is presented in Fig.\ref{fig2}(1) with $a=3$,
$\{(b,c),\;(d,e)\}=\{(1,4),\;(2,5)\}$, respectively.
Adopting the matrix of integer lattice $nE_{_3}^{(1)}$
and the matrices of exponents, we obtain a fundamental solution
system composed of two hypergeometric solutions as
\begin{eqnarray}
&&\psi_{_{\{1,3,5\}}}^{(1)}(p^2,\;0,\;m_{_2}^2)\sim{(m_{_2}^2)^{D/2-1}\over p^2-m_{_2}^2}
\sum\limits_{n=0}^\infty{\Gamma(D-2+n)\over\Gamma({D\over2}+n)}\Big({p^2\over p^2-m_{_2}^2}\Big)^n\;,
\nonumber\\
&&\psi_{_{\{1,3,5\}}}^{(2)}(p^2,\;0,\;m_{_2}^2)\sim
\Big({m_{_2}^2\over p^2}\Big)^{D/2-1}(p^2-m_{_2}^2)^{D/2-2}
\nonumber\\
&&\hspace{3.7cm}\times
\sum\limits_{n=0}^\infty{\Gamma(2-{D\over2}+n)
\Gamma({D\over2}-1+n)\over n!\Gamma(2-{D\over2}+n)}
\Big({p^2\over p^2-m_{_2}^2}\Big)^n
\nonumber\\
&&\hspace{3.2cm}\sim
(p^2)^{1-D/2}(p^2-m_{_2}^2)^{D-3}\;,
\label{1SE-33}
\end{eqnarray}
which are convergent in the region $|p^2/(p^2-m_{_2}^2)|\le1$.
In fact the basis $\{1,3,5\}$ of the projective plane
is obtained from the basis $\{1,2,3\}$ through the permutation $\widehat{(25)}$,
accordingly the geometric description of the exponents Eq.(\ref{1SE-31})
is derived from that of the exponents Eq.(\ref{1SE-14}) through the permutation $\widehat{(25)}$.
Because the geometric description of Eq.(\ref{1SE-31}) is same as that of Eq.(\ref{1SE-14}),
the corresponding function $\psi_{_{\{1,2,5\}}}^{(1)}$ is proportionate to
$\psi_{_{\{1,2,3\}}}^{(2)}$ in the nonempty intersection of their convergent regions.
Similarly the geometric description of the exponents Eq.(\ref{1SE-32})
is obtained through the permutation $\widehat{(25)}$ from that of Eq.(\ref{1SE-13}).
Because the geometric description of Eq.(\ref{1SE-32}) is same as that of Eq.(\ref{1SE-13}),
the function $\psi_{_{\{1,2,5\}}}^{(2)}$ is proportionate to
$\psi_{_{\{1,2,3\}}}^{(1)}$ in the nonempty intersection of their convergent regions.
Actually $\psi_{_{\{1,3,5\}}}^{(2)}\sim \psi_{_{\{1,2,3\}}}^{(1)}$ is consistent
with the well-known relation Eq.(\ref{1SE-24a}).

Similarly the matrix of integer lattice $-nE_{_3}^{(1)}$ permits two possibilities of the exponents.
The first one is
\begin{eqnarray}
&&\alpha_{_{2,2}}=-1,\;\alpha_{_{2,4}}=0,\;\alpha_{_{3,2}}=3-D,\;\alpha_{_{3,4}}={D\over2}-2\;,
\label{1SE-34}
\end{eqnarray}
where the geometric representation is presented in Fig.\ref{fig2}(1)
with $a=5$, $\{(b,c),\;(d,e)\}=\{(1,4),\;(2,3)\}$. The second possibility is
\begin{eqnarray}
&&\alpha_{_{2,2}}=2-D,\;\alpha_{_{2,4}}=D-3,\;\alpha_{_{3,2}}=0,\;\alpha_{_{3,4}}=1-{D\over2}\;,
\label{1SE-35}
\end{eqnarray}
where the geometric representation is presented in Fig.\ref{fig2}(2) with $(a,b)=(2,3)$,
$(c,d,e)=(1,4,5)$, respectively. Adopting the matrix of integer lattice
and the matrices of exponents, we obtain a fundamental solution
system composed of two linear independent hypergeometric solutions as
\begin{eqnarray}
&&\psi_{_{\{1,3,5\}}}^{(3)}(p^2,\;0,\;m_{_2}^2)\sim
{(m_{_2}^2)^{D/2-1}\over p^2}
\sum\limits_{n=0}^\infty{\Gamma(2-{D\over2}+n)\over\Gamma(4-D+n)}
\Big({p^2-m_{_2}^2\over p^2}\Big)^n\;,
\nonumber\\
&&\psi_{_{\{1,3,5\}}}^{(4)}(p^2,\;0,\;m_{_2}^2)\sim
{(p^2-m_{_2}^2)^{D-3}(m_{_2}^2)^{D/2-1}\over(p^2)^{D-2}}
\sum\limits_{n=0}^\infty{\Gamma({D\over2}-1+n)\over n!}
\Big({p^2-m_{_2}^2\over p^2}\Big)^n
\nonumber\\
&&\hspace{3.3cm}\sim(p^2-m_{_2}^2)^{D-3}(p^2)^{1-D/2}\;,
\label{1SE-36}
\end{eqnarray}
whose convergent region is $|1-m_{_2}^2/p^2|\le1$.
Certainly the basis $\{1,3,5\}$ of the projective
plane composed of the first, third and fifth columns of the matrix Eq.(\ref{1SE-9})
is obtained from the basis $\{1,2,5\}$ through the permutation $\widehat{(23)}$,
accordingly the geometric description of the exponents Eq.(\ref{1SE-34})
is derived from that of the exponents Eq.(\ref{1SE-25}) through the permutation $\widehat{(23)}$.
Because the geometric description of Eq.(\ref{1SE-34}) is same as that of Eq.(\ref{1SE-25}),
the function $\psi_{_{\{1,3,5\}}}^{(3)}$ is proportionate to
$\psi_{_{\{1,2,5\}}}^{(3)}$ in the nonempty intersection of their convergent regions.
Similarly the geometric description of the exponents Eq.(\ref{1SE-35})
is obtained through the permutation $\widehat{(23)}$ from that of Eq.(\ref{1SE-26}).
Because the geometric description of Eq.(\ref{1SE-35}) is same as that of Eq.(\ref{1SE-26}),
the function $\psi_{_{\{1,3,5\}}}^{(4)}$ is proportionate to
$\psi_{_{\{1,2,5\}}}^{(4)}$ in the nonempty intersection of their convergent regions.
Actually $\psi_{_{\{1,3,5\}}}^{(3)}\sim \psi_{_{\{1,2,5\}}}^{(3)}$ is consistent
with the well-known relation Eq.(\ref{1SE-24a}).

Because $\det(A_{_{\{2,3,4\}}}^{(1)})=1$ where the matrix $A_{_{\{2,3,4\}}}^{(1)}$ denotes
the submatrix of $A^{(1)}$ composed of the second, third, and fourth columns, one derives
\begin{eqnarray}
&&\Big(A_{_{\{2,3,4\}}}^{(1)}\Big)^{-1}\cdot A^{(1)}=\left(\begin{array}{ccccc}
\;\;-1\;\;&\;\;1\;\;&\;\;0\;\;&\;\;0\;\;&\;\;m_{_2}^2\;\;\\
\;\;-1\;\;&\;\;0\;\;&\;\;1\;\;&\;\;0\;\;&\;\;p^2\;\;\\
\;\;1\;\;&\;\;0\;\;&\;\;0\;\;&\;\;1\;\;&\;\;0\;\;\end{array}\right)\;.
\label{1SE-37}
\end{eqnarray}
Correspondingly the matrix of exponents is written as
\begin{eqnarray}
&&\left(\begin{array}{ccccc}
\;\alpha_{_{1,1}}\;\;&\;\;2-D\;\;&\;\;0\;\;&\;\;0\;\;&\;\alpha_{_{1,5}}\;\;\\
\;\alpha_{_{2,1}}\;\;&\;\;0\;\;&\;\;0\;\;&\;\;0\;\;&\;\alpha_{_{2,5}}\;\;\\
\;\;0\;\;&\;\;0\;\;&\;\;0\;\;&\;\;-1\;\;&\;\;0\;\;\end{array}\right)\;,
\label{1SE-38}
\end{eqnarray}
where the matrix elements satisfy the relations
\begin{eqnarray}
&&\alpha_{_{1,1}}+\alpha_{_{1,5}}=D-3\;,\;\alpha_{_{2,1}}+\alpha_{_{2,5}}=-1\;,
\nonumber\\
&&\alpha_{_{1,1}}+\alpha_{_{2,1}}={D\over2}-2\;,\;\alpha_{_{1,5}}+\alpha_{_{2,5}}={D\over2}-2\;.
\label{1SE-39}
\end{eqnarray}
There are two choices on the matrix of integer lattice whose submatrix composed of the first and
fifth columns is formulated as $\pm nE_{_3}^{(3)}$ ($n\ge0$) with
\begin{eqnarray}
&&E_{_3}^{(3)}=\left(\begin{array}{cc}\;\;1\;\;&\;\;-1\;\;\\\;\;-1\;\;&\;\;1\;\;\\
\;\;0\;\;&\;\;0\;\;\end{array}\right)\;,
\label{1SE-40}
\end{eqnarray}
and other elements are all zero. The matrix of integer lattice $nE_{_3}^{(3)}$
is compatible with two choices of the exponents. The first choice is
\begin{eqnarray}
&&\alpha_{_{1,1}}=0\;,\;\alpha_{_{1,5}}=D-3\;,\;\alpha_{_{2,1}}={D\over2}-2\;,\;
\alpha_{_{2,5}}=1-{D\over2}\;,
\label{1SE-41}
\end{eqnarray}
whose geometric representation is presented in Fig.\ref{fig2}(1)
with $a=3$, $\{(b,c),\;(d,e)\}=\{(1,4),\;(2,5)\}$. The second choice is
\begin{eqnarray}
&&\alpha_{_{1,1}}={D\over2}-1\;,\;\alpha_{_{1,5}}={D\over2}-2\;,\;\alpha_{_{2,1}}=-1\;,
\alpha_{_{2,5}}=0\;.
\label{1SE-42}
\end{eqnarray}
whose geometric representation is presented in Fig.\ref{fig2}(2) with $(a,b)=(2,5)$,
$(c,d,e)=(1,3,4)$, respectively. Adopting the matrix of integer lattice $nE_{_3}^{(3)}$
and the matrices of exponents, we obtain a fundamental solution
system composed by two linear independent hypergeometric functions as
\begin{eqnarray}
&&\psi_{_{\{2,3,4\}}}^{(1)}(p^2,\;0,\;m_{_2}^2)\sim
(p^2)^{1-D/2}(m_{_2}^2)^{D-3}\sum\limits_{n=0}^\infty
{\Gamma(3-D+n)\over n!}\Big({p^2\over m_{_2}^2}\Big)^n
\nonumber\\
&&\hspace{3.2cm}\sim
(p^2)^{1-D/2}(m_{_2}^2-p^2)^{D-3}\;,
\nonumber\\
&&\psi_{_{\{2,3,4\}}}^{(2)}(p^2,\;0,\;m_{_2}^2)\sim
(m_{_2}^2)^{D/2-2}\sum\limits_{n=0}^\infty
{\Gamma(2-{D\over2}+n)\over\Gamma({D\over2}+n)}\Big({p^2\over m_{_2}^2}\Big)^n\;,
\label{1SE-43}
\end{eqnarray}
which are convergent in the region is $|p^2/m_{_2}^2|\le1$.
Similarly the basis $\{2,3,4\}$ of the projective
plane composed of the second, third and fourth columns of the matrix Eq.(\ref{1SE-9})
is obtained from the basis $\{1,2,3\}$ through the permutation $\widehat{(14)}$,
or equivalently from the basis $\{1,3,5\}$ through the permutation $\widehat{(14)(25)}$.
Accordingly the geometric description of Eq.(\ref{1SE-41})
is taken from that of Eq.(\ref{1SE-13}) through the permutation $\widehat{(14)}$,
or from that of Eq.(\ref{1SE-32}) through the permutation $\widehat{(14)(25)}$.
Because the geometric description of Eq.(\ref{1SE-41}) is same as that of Eq.(\ref{1SE-13})
and Eq.(\ref{1SE-32}), the functions
$\psi_{_{\{1,2,3\}}}^{(1)}\sim\psi_{_{\{1,2,5\}}}^{(2)}\sim\psi_{_{\{2,3,4\}}}^{(1)}$
in nonempty intersection of their convergent regions.
Similarly the geometric description of Eq.(\ref{1SE-42})
is taken from that of Eq.(\ref{1SE-14}) through the permutation $\widehat{(14)}$,
or from that of Eq.(\ref{1SE-31}) through the permutation $\widehat{(14)(25)}$.
Because the geometric description of Eq.(\ref{1SE-42}) is same as that of Eq.(\ref{1SE-14})
and Eq.(\ref{1SE-31}), the functions
$\psi_{_{\{1,2,3\}}}^{(2)}\sim\psi_{_{\{1,2,5\}}}^{(1)}\sim\psi_{_{\{2,3,4\}}}^{(2)}$
in nonempty intersection of their convergent regions.

The matrix of integer lattice $-nE_{_3}^{(3)}$ permits two possibilities of the exponents.
The first possibility is given as
\begin{eqnarray}
&&\alpha_{_{1,1}}=D-3\;,\;\alpha_{_{1,5}}=0\;,\;\alpha_{_{2,1}}=1-{D\over2}\;,\;
\alpha_{_{2,5}}={D\over2}-2\;,
\label{1SE-44}
\end{eqnarray}
whose geometric representation is presented in Fig.\ref{fig2}(2)
with $(a,b)=(3,5)$, $(c,d,e)=(1,2,4)$. Meanwhile the second possibility is
\begin{eqnarray}
&&\alpha_{_{1,1}}={D\over2}-2\;,\;\alpha_{_{1,5}}={D\over2}-1\;,\;\alpha_{_{2,1}}=0\;,
\alpha_{_{2,5}}=-1\;,
\label{1SE-45}
\end{eqnarray}
whose geometric representation is drawn in Fig.\ref{fig2}(1) with $a=2$,
$\{(b,c),\;(d,e)\}=\{(1,4),\;(3,5)\}$, respectively.
Adopting the matrix of integer lattice $-nE_{_3}^{(3)}$
and the matrices of exponents, we construct a fundamental solution
system composed of two linear independent hypergeometric functions as
\begin{eqnarray}
&&\psi_{_{\{2,3,4\}}}^{(3)}(p^2,\;0,\;m_{_2}^2)\sim
(p^2)^{D/2-2}
\sum\limits_{n=0}^\infty{\Gamma(3-D+n)\over n!}
\Big({m_{_2}^2\over p^2}\Big)^n\;,
\nonumber\\
&&\hspace{3.2cm}\sim
(p^2)^{1-D/2}(p^2-m_{_2}^2)^{D-3}\;,
\nonumber\\
&&\psi_{_{\{2,3,4\}}}^{(4)}(p^2,\;0,\;m_{_2}^2)\sim
{(m_{_2}^2)^{D/2-1}\over p^2}\sum\limits_{n=0}^\infty
{\Gamma(2-{D\over2}+n)\over\Gamma({D\over2}+n)}\Big({m_{_2}^2\over p^2}\Big)^n\;,
\label{1SE-46}
\end{eqnarray}
which are convergent in the region is $|m_{_2}^2/p^2|\le1$.
Similarly the basis $\{2,3,4\}$ of the projective
plane composed of the second, third and fourth columns of the matrix Eq.(\ref{1SE-9})
is obtained from the basis $\{1,2,3\}$ through the permutation $\widehat{(14)}$,
or equivalently from the basis $\{1,2,5\}$ through the permutation $\widehat{(14)(35)}$.
Similarly the geometric description of Eq.(\ref{1SE-44})
is taken from that of Eq.(\ref{1SE-16}) through the permutation $\widehat{(14)}$,
or from that of Eq.(\ref{1SE-22}) through the permutation $\widehat{(14)(25)}$.
Because the geometric description of Eq.(\ref{1SE-44}) is same as that of Eq.(\ref{1SE-16})
and Eq.(\ref{1SE-22}), the functions
$\psi_{_{\{1,2,3\}}}^{(3)}\sim\psi_{_{\{1,2,5\}}}^{(1)}\sim\psi_{_{\{2,3,4\}}}^{(3)}$
in nonempty intersection of their convergent regions.
Similarly the geometric description of Eq.(\ref{1SE-45})
is derived from that of Eq.(\ref{1SE-17}) through the permutation $\widehat{(14)}$,
or from that of Eq.(\ref{1SE-23}) through the permutation $\widehat{(14)(25)}$.
Because the geometric description of Eq.(\ref{1SE-45}) is same as that of Eq.(\ref{1SE-17})
and Eq.(\ref{1SE-23}), the functions
$\psi_{_{\{1,2,3\}}}^{(4)}\sim\psi_{_{\{1,2,5\}}}^{(2)}\sim\psi_{_{\{2,3,4\}}}^{(4)}$
in nonempty intersection of their convergent regions.

Because $\det(A_{_{\{2,4,5\}}}^{(1)})=-p^2$ where the matrix $A_{_{\{2,4,5\}}}^{(1)}$ denotes
the submatrix of $A^{(1)}$ composed of the  second, fourth, and fifth columns, one derives
\begin{eqnarray}
&&\Big(A_{_{\{2,4,5\}}}^{(1)}\Big)^{-1}\cdot A^{(1)}=\left(\begin{array}{ccccc}
\;\;-1+{m_{_2}^2\over p^2}\;\;&\;\;1\;\;&\;\;-{m_{_2}^2\over p^2}\;\;&\;\;0\;\;&\;\;0\;\;\\
\;\;1\;\;&\;\;0\;\;&\;\;0\;\;&\;\;1\;\;&\;\;0\;\;\\
\;\;-{1\over p^2}\;\;&\;\;0\;\;&\;\;{1\over p^2}\;\;&\;\;0\;\;&\;\;1\;\;\end{array}\right)\;.
\label{1SE-47}
\end{eqnarray}
Obviously the matrix of exponents is written as
\begin{eqnarray}
&&\left(\begin{array}{ccccc}
\;\alpha_{_{1,1}}\;\;&\;\;2-D\;\;&\;\alpha_{_{1,3}}\;\;&\;\;0\;\;&\;\;0\;\;\\
\;\;0\;\;&\;\;0\;\;&\;\;0\;\;&\;\;-1\;\;&\;\;0\;\;\\
\;\alpha_{_{3,1}}\;\;&\;\;0\;\;&\;\alpha_{_{3,3}}\;\;&\;\;0\;\;&\;\;{D\over2}-2\;\;\end{array}\right)\;,
\label{1SE-48}
\end{eqnarray}
where the matrix elements satisfy the relations
\begin{eqnarray}
&&\alpha_{_{1,1}}+\alpha_{_{1,3}}=D-3\;,\;\alpha_{_{3,1}}+\alpha_{_{3,3}}=1-{D\over2}\;,
\nonumber\\
&&\alpha_{_{1,1}}+\alpha_{_{3,1}}={D\over2}-2\;,\;\alpha_{_{1,3}}+\alpha_{_{3,3}}=0\;.
\label{1SE-49}
\end{eqnarray}
Correspondingly there are two choices on the matrix of integer lattice
whose submatrix composed of the first and third columns is formulated
as $\pm nE_{_3}^{(2)}$ $(n\ge0)$, with
\begin{eqnarray}
&&E_{_3}^{(2)}=\left(\begin{array}{cc}\;\;1\;\;&\;\;-1\;\;\\\;\;0\;\;&\;\;0\;\;\\
\;\;-1\;\;&\;\;1\;\;\end{array}\right)\;,
\label{1SE-50}
\end{eqnarray}
and other elements are all zero. The matrix of integer lattice $nE_{_3}^{(2)}$ permits
two possibilities of the exponents. The first possibility is
\begin{eqnarray}
&&\alpha_{_{1,1}}=0\;,\;\alpha_{_{1,3}}=D-3\;,\;\alpha_{_{3,1}}={D\over2}-2\;,\;
\alpha_{_{3,3}}=3-D\;,
\label{1SE-51}
\end{eqnarray}
whose geometric representation is presented in Fig.\ref{fig2}(1)
with $a=5$, $\{(b,c),\;(d,e)\}=\{(1,4),\;(2,3)\}$. The second possibility is
\begin{eqnarray}
&&\alpha_{_{1,1}}=D-3\;,\;\alpha_{_{1,3}}=0\;,\;\alpha_{_{3,1}}=1-{D\over2}\;,\;
\alpha_{_{3,3}}=0\;,
\label{1SE-52}
\end{eqnarray}
whose geometric representation is presented in Fig.\ref{fig2}(2) with $(a,b)=(2,3)$,
$(c,d,e)=(1,4,5)$, respectively. Adopting the matrix of integer lattice
and the matrices of exponents, we obtain a fundamental solution
system composed of two linear independent hypergeometric solutions as
\begin{eqnarray}
&&\psi_{_{\{2,4,5\}}}^{(1)}(p^2,\;0,\;m_{_2}^2)\sim
(m_{_2}^2)^{D-3}(p^2)^{1-D/2}
\nonumber\\
&&\hspace{3.7cm}\times
\sum\limits_{n=0}^\infty{\Gamma(3-D+n)
\Gamma(2-{D\over2}+n)\over n!\Gamma(4-D+n)}
\Big({m_{_2}^2-p^2\over m_{_2}^2}\Big)^n\;,
\nonumber\\
&&\psi_{_{\{2,4,5\}}}^{(2)}(p^2,\;0,\;m_{_2}^2)\sim
(p^2)^{1-D/2}(m_{_2}^2-p^2)^{D-3}\;,
\label{1SE-53}
\end{eqnarray}
whose convergent region is $|(m_{_2}^2-p^2)/m_{_2}^2|\le1$.
Similarly the basis $\{2,4,5\}$ of the projective
plane composed of the second, fourth and fifth columns of the matrix Eq.(\ref{1SE-9})
is obtained from the basis $\{1,2,5\}$ through the permutation $\widehat{(14)}$,
or equivalently obtained from the basis $\{1,3,5\}$ through the permutation $\widehat{(14)(23)}$.
Accordingly the geometric description of Eq.(\ref{1SE-51})
is obtained from that of Eq.(\ref{1SE-25}) through the permutation $\widehat{(14)}$,
or obtained from that of Eq.(\ref{1SE-34}) through the permutation $\widehat{(14)(23)}$.
Because the geometric description of Eq.(\ref{1SE-51}) is same as that of Eq.(\ref{1SE-25})
and Eq.(\ref{1SE-34}), the functions
$\psi_{_{\{1,2,5\}}}^{(3)}\sim\psi_{_{\{1,3,5\}}}^{(3)}\sim\psi_{_{\{2,4,5\}}}^{(1)}$
in nonempty intersection of their convergent regions.
Similarly the geometric description of Eq.(\ref{1SE-52})
is obtained from that of Eq.(\ref{1SE-26}) through the permutation $\widehat{(14)}$,
or obtained from that of Eq.(\ref{1SE-35}) through the permutation $\widehat{(14)(23)}$.
Because the geometric description of Eq.(\ref{1SE-52}) is same as that of Eq.(\ref{1SE-26})
and Eq.(\ref{1SE-35}), the functions
$\psi_{_{\{1,2,5\}}}^{(4)}\sim\psi_{_{\{1,3,5\}}}^{(4)}\sim\psi_{_{\{2,4,5\}}}^{(2)}$
in nonempty intersection of their convergent regions.
Including appropriate scalar factors, each hypergeometric function is analytic continuation
of other two functions in the union of their convergent regions.

The matrix of integer lattice $-nE_{_3}^{(2)}$ permits two choices of the exponents.
The first choice is
\begin{eqnarray}
&&\alpha_{_{1,1}}=D-3\;,\;\alpha_{_{1,3}}=0\;,\;\alpha_{_{3,1}}=1-{D\over2}\;,\;
\alpha_{_{3,3}}=0\;,
\label{1SE-54}
\end{eqnarray}
whose geometric representation is presented in Fig.\ref{fig2}(2)
with $(a,b)=(3,5)$, $(c,d,e)=(1,2,4)$. Meanwhile the second choice is
\begin{eqnarray}
&&\alpha_{_{1,1}}={D\over2}-2\;,\;\alpha_{_{1,3}}={D\over2}-1\;,\;\alpha_{_{3,1}}=0\;,\;
\alpha_{_{3,3}}=1-{D\over2}\;,
\label{1SE-55}
\end{eqnarray}
whose geometric representation is presented in Fig.\ref{fig2}(1) with $a=2$,
$\{(b,c),\;(d,e)\}=\{(1,4),\;(3,5)\}$, respectively.
Adopting the matrix of integer lattice and the matrices of exponents,
we obtain a fundamental solution system composed of two linear independent
hypergeometric functions as
\begin{eqnarray}
&&\psi_{_{\{2,4,5\}}}^{(3)}(p^2,\;0,\;m_{_2}^2)\sim
(p^2)^{1-D/2}(m_{_2}^2-p^2)^{D-3}\;,
\nonumber\\
&&\psi_{_{\{2,4,5\}}}^{(4)}(p^2,\;0,\;m_{_2}^2)\sim
\Big({m_{_2}^2\over p^2}\Big)^{D/2-1}(m_{_2}^2-p^2)^{D/2-2}
\nonumber\\
&&\hspace{3.9cm}\times
\sum\limits_{n=0}^\infty{\Gamma(2-{D\over2}+n)
\Gamma({D\over2}-1+n)\over n!\Gamma({D\over2}+n)}
\Big({m_{_2}^2\over m_{_2}^2-p^2}\Big)^n\;,
\label{1SE-56}
\end{eqnarray}
which are convergent in the region $|m_{_2}^2/(m_{_2}^2-p^2)|\le1$.
Because the basis $\{2,4,5\}$ of the projective
plane composed of the second, fourth and fifth columns of the matrix Eq.(\ref{1SE-9})
is obtained from the basis $\{1,2,3\}$ through the permutation $\widehat{(14)(35)}$,
or equivalently obtained from the basis $\{1,2,5\}$ through the permutation $\widehat{(14)}$.
Accordingly the geometric description of Eq.(\ref{1SE-54})
is taken from that of Eq.(\ref{1SE-16}) through the permutation $\widehat{(14)(35)}$,
or taken from that of Eq.(\ref{1SE-22}) through the permutation $\widehat{(14)}$.
Since the geometric description of Eq.(\ref{1SE-54}) is same as that of Eq.(\ref{1SE-16})
and Eq.(\ref{1SE-22}), the functions
$\psi_{_{\{1,2,3\}}}^{(1)}\sim\psi_{_{\{1,2,5\}}}^{(1)}\sim\psi_{_{\{2,4,5\}}}^{(3)}$
in nonempty intersection of their convergent regions.
Similarly the geometric description of Eq.(\ref{1SE-55})
is taken from that of Eq.(\ref{1SE-17}) through the permutation $\widehat{(14)(35)}$,
or taken from that of Eq.(\ref{1SE-23}) through the permutation $\widehat{(14)}$.
Similarly the geometric description of Eq.(\ref{1SE-55}) is same as that of Eq.(\ref{1SE-17})
and Eq.(\ref{1SE-23}), the functions
$\psi_{_{\{1,2,3\}}}^{(2)}\sim\psi_{_{\{1,2,5\}}}^{(2)}\sim\psi_{_{\{2,4,5\}}}^{(4)}$
in nonempty intersection of their convergent regions.

Because $\det(A_{_{\{3,4,5\}}}^{(1)})=m_{_2}^2$ where the matrix $A_{_{\{3,4,5\}}}^{(1)}$ denotes
the submatrix of $A^{(1)}$ composed of the  third, fourth, and fifth columns, we get
\begin{eqnarray}
&&\Big(A_{_{\{3,4,5\}}}^{(1)}\Big)^{-1}\cdot A^{(1)}=\left(\begin{array}{ccccc}
\;\;-1+{p^2\over m_{_2}^2}\;\;&\;\;-{p^2\over m_{_2}^2}\;\;&\;\;1\;\;&\;\;0\;\;&\;\;0\;\;\\
\;\;1\;\;&\;\;0\;\;&\;\;0\;\;&\;\;1\;\;&\;\;0\;\;\\
\;\;-{1\over m_{_2}^2}\;\;&\;\;{1\over m_{_2}^2}\;\;&\;\;0\;\;&\;\;0\;\;&\;\;1\;\;\end{array}\right)\;.
\label{1SE-57}
\end{eqnarray}
Obviously the matrix of exponents is written as
\begin{eqnarray}
&&\left(\begin{array}{ccccc}
\;\alpha_{_{1,1}}\;\;&\;\alpha_{_{1,2}}\;\;&\;\;0\;\;&\;\;0\;\;&\;\;0\;\;\\
\;\;0\;\;&\;\;0\;\;&\;\;0\;\;&\;\;-1\;\;&\;\;0\;\;\\
\;\alpha_{_{3,1}}\;\;&\;\alpha_{_{3,2}}\;\;&\;\;0\;\;&\;\;0\;\;&\;\;{D\over2}-2\;\;\end{array}\right)\;,
\label{1SE-58}
\end{eqnarray}
where the matrix elements satisfy the relations
\begin{eqnarray}
&&\alpha_{_{1,1}}+\alpha_{_{1,2}}=-1\;,\;\alpha_{_{3,1}}+\alpha_{_{3,2}}=1-{D\over2}\;,
\nonumber\\
&&\alpha_{_{1,1}}+\alpha_{_{3,1}}={D\over2}-2\;,\;\alpha_{_{1,2}}+\alpha_{_{3,2}}=2-D\;.
\label{1SE-59}
\end{eqnarray}
Correspondingly there are two choices on the matrix of integer lattice whose
submatrix composed of the first and second columns is formulated as $\pm nE_{_3}^{(2)}$ ($n\ge0$),
and other elements are all zero. The matrix of integer lattice $nE_{_3}^{(2)}$ permits two
possibilities of the exponents. The first possibility is
\begin{eqnarray}
&&\alpha_{_{1,1}}=0\;,\;\alpha_{_{1,2}}=-1\;,\;\alpha_{_{3,1}}={D\over2}-2\;,\;
\alpha_{_{3,2}}=3-D\;,
\label{1SE-60}
\end{eqnarray}
whose geometric representation is presented in Fig.\ref{fig2}(1)
with $a=5$, $\{(b,c),\;(d,e)\}=\{(1,4),\;(2,3)\}$. The second possibility is
\begin{eqnarray}
&&\alpha_{_{1,1}}=D-3\;,\;\alpha_{_{1,2}}=2-D\;,\;\alpha_{_{3,1}}=1-{D\over2}\;,\;
\alpha_{_{3,2}}=0\;,
\label{1SE-61}
\end{eqnarray}
whose geometric representation is presented in Fig.\ref{fig2}(2) with $(a,b)=(2,3)$,
$(c,d,e)=(1,4,5)$, respectively. Adopting the matrix of integer lattice $nE_{_3}^{(2)}$
and the matrices of exponents, we obtain a fundamental solution
system composed of two linear independent hypergeometric solutions as
\begin{eqnarray}
&&\psi_{_{\{3,4,5\}}}^{(1)}(p^2,\;0,\;m_{_2}^2)\sim
{(m_{_2}^2)^{D/2-1}\over p^2}
\sum\limits_{n=0}^\infty{\Gamma(2-{D\over2}+n)\over\Gamma(4-D+n)}
\Big({p^2-m_{_2}^2\over p^2}\Big)^n\;,
\nonumber\\
&&\psi_{_{\{3,4,5\}}}^{(2)}(p^2,\;0,\;m_{_2}^2)\sim
{(p^2-m_{_2}^2)^{D-3}(m_{_2}^2)^{D/2-1}\over(p^2)^{D-2}}
\sum\limits_{n=0}^\infty{\Gamma({D\over2}-1+n)\over n!}
\Big({p^2-m_{_2}^2\over p^2}\Big)^n
\nonumber\\
&&\hspace{3.2cm}\sim(p^2-m_{_2}^2)^{D-3}(p^2)^{1-D/2}\;,
\label{1SE-62}
\end{eqnarray}
which are convergent in the region $|1-m_{_2}^2/p^2|\le1$.
Obviously the basis $\{3,4,5\}$ of the projective
plane composed of the third, fourth and fifth columns of the matrix Eq.(\ref{1SE-9})
is obtained from the basis $\{1,2,5\}$ through the permutation $\widehat{(14)(23)}$,
or equivalently obtained from the basis $\{1,3,5\}$ through the permutation $\widehat{(14)}$,
or equivalently obtained from the basis $\{2,3,5\}$ through the permutation $\widehat{(23)}$, respectively.
Accordingly the geometric description of Eq.(\ref{1SE-60})
is taken from that of Eq.(\ref{1SE-25}) through the permutation $\widehat{(14)(23)}$,
or taken from that of Eq.(\ref{1SE-34}) through the permutation $\widehat{(14)}$,
or taken from that of Eq.(\ref{1SE-51}) through the permutation $\widehat{(23)}$, respectively.
Because the geometric description of Eq.(\ref{1SE-60}) is same as that of Eq.(\ref{1SE-25}),
Eq.(\ref{1SE-34}), and Eq.(\ref{1SE-51}), the functions
$\psi_{_{\{1,2,5\}}}^{(3)}\sim\psi_{_{\{1,3,5\}}}^{(3)}\sim\psi_{_{\{2,4,5\}}}^{(1)}\sim\psi_{_{\{3,4,5\}}}^{(1)}$
in nonempty intersection of their convergent regions.
Similarly the geometric description of Eq.(\ref{1SE-61})
is obtained from that of Eq.(\ref{1SE-26}) through the permutation $\widehat{(14)(23)}$,
or obtained from that of Eq.(\ref{1SE-35}) through the permutation $\widehat{(14)}$,
or obtained from that of Eq.(\ref{1SE-52}) through the permutation $\widehat{(23)}$, respectively.
Because the geometric description of Eq.(\ref{1SE-61}) is same as that of Eq.(\ref{1SE-26}),
Eq.(\ref{1SE-35}), and Eq.(\ref{1SE-52}), the functions
$\psi_{_{\{1,2,5\}}}^{(4)}\sim\psi_{_{\{1,3,5\}}}^{(4)}\sim\psi_{_{\{2,4,5\}}}^{(2)}\sim\psi_{_{\{3,4,5\}}}^{(2)}$
in nonempty intersection of their convergent regions.
Including appropriate scalar factors, each hypergeometric function is analytic continuation
of other functions in the union of their convergent regions.

The matrix of integer lattice $-nE_{_3}^{(2)}$ permits two choices of the exponents.
The first choice is
\begin{eqnarray}
&&\alpha_{_{1,1}}=-1,\;\alpha_{_{1,2}}=0,\;\alpha_{_{3,1}}={D\over2}-1,\;\alpha_{_{3,2}}=2-D\;,
\label{1SE-63}
\end{eqnarray}
whose geometric representation is presented in Fig.\ref{fig2}(2)
with $(a,b)=(2,5)$, $(c,d,e)=(1,3,4)$. And the second choice is
\begin{eqnarray}
&&\alpha_{_{1,1}}={D\over2}-2,\;\alpha_{_{1,2}}=1-{D\over2},\;\alpha_{_{3,1}}=0,\;\alpha_{_{3,2}}=1-{D\over2}\;.
\label{1SE-64}
\end{eqnarray}
whose geometric representation is presented in Fig.\ref{fig2}(1) with $a=3$,
$\{(b,c),\;(d,e)\}=\{(1,4),\;(2,5)\}$, respectively.
Adopting the matrix of integer lattice $-nE_{_3}^{(2)}$
and the matrices of exponents, we obtain a fundamental solution
system composed of two hypergeometric solutions as
\begin{eqnarray}
&&\psi_{_{\{3,4,5\}}}^{(3)}(p^2,\;0,\;m_{_2}^2)\sim{(m_{_2}^2)^{D/2-1}\over p^2-m_{_2}^2}
\sum\limits_{n=0}^\infty{\Gamma(D-2+n)\over\Gamma({D\over2}+n)}\Big({p^2\over p^2-m_{_2}^2}\Big)^n\;,
\nonumber\\
&&\psi_{_{\{3,4,5\}}}^{(4)}(p^2,\;0,\;m_{_2}^2)\sim
\Big({m_{_2}^2\over p^2}\Big)^{D/2-1}(p^2-m_{_2}^2)^{D/2-2}
\nonumber\\
&&\hspace{3.9cm}\times
\sum\limits_{n=0}^\infty{\Gamma({D\over2}-1+n)\over n!}
\Big({p^2\over p^2-m_{_2}^2}\Big)^n
\nonumber\\
&&\hspace{3.5cm}=
(p^2)^{1-D/2}(p^2-m_{_2}^2)^{D-3}\;,
\label{1SE-65}
\end{eqnarray}
which are convergent in the region $|p^2/(p^2-m_{_2}^2)|\le1$.
Since the basis $\{3,4,5\}$ of the projective
plane composed of the third, fourth and fifth columns of the matrix Eq.(\ref{1SE-9})
is obtained from the basis $\{1,2,3\}$ through the permutation $\widehat{(14)(25)}$,
or equivalently obtained from the basis $\{1,3,5\}$ through the permutation $\widehat{(14)}$,
or equivalently obtained from the basis $\{2,3,4\}$ through the permutation $\widehat{(25)}$, respectively.
Accordingly the geometric description of Eq.(\ref{1SE-63})
is taken from that of Eq.(\ref{1SE-14}) through the permutation $\widehat{(14)(25)}$,
or taken from that of Eq.(\ref{1SE-31}) through the permutation $\widehat{(14)}$,
or taken from that of Eq.(\ref{1SE-42}) through the permutation $\widehat{(25)}$, respectively.
Because the geometric description of Eq.(\ref{1SE-63}) is same as that of Eq.(\ref{1SE-14}),
Eq.(\ref{1SE-31}), and Eq.(\ref{1SE-42}), the functions
$\psi_{_{\{1,2,3\}}}^{(2)}\sim\psi_{_{\{1,3,5\}}}^{(1)}\sim\psi_{_{\{2,3,4\}}}^{(2)}\sim\psi_{_{\{3,4,5\}}}^{(3)}$
in nonempty intersection of their convergent regions.
Similarly the geometric description of Eq.(\ref{1SE-64})
is taken from that of Eq.(\ref{1SE-13}) through the permutation $\widehat{(14)(25)}$,
or taken from that of Eq.(\ref{1SE-32}) through the permutation $\widehat{(14)}$,
or taken from that of Eq.(\ref{1SE-41}) through the permutation $\widehat{(25)}$, respectively.
Because the geometric description of Eq.(\ref{1SE-64}) is same as that of Eq.(\ref{1SE-13}),
Eq.(\ref{1SE-32}), and Eq.(\ref{1SE-41}), the functions
$\psi_{_{\{1,2,3\}}}^{(1)}\sim\psi_{_{\{1,3,5\}}}^{(2)}\sim\psi_{_{\{2,3,4\}}}^{(1)}\sim\psi_{_{\{3,4,5\}}}^{(4)}$
in nonempty intersection of their convergent regions.

\subsection{The fundamental solution system in general case}
\indent\indent
As the 1-loop self energy diagram involves 2 nonzero virtual masses, the integer
lattice is spanned by two linear independent vectors, and the number of
exponential matrices compatible with them also increases accordingly.

\begin{figure}[ht]
\setlength{\unitlength}{1cm}
\centering
\vspace{0.0cm}\hspace{-1.5cm}
\includegraphics[height=8cm,width=8.0cm]{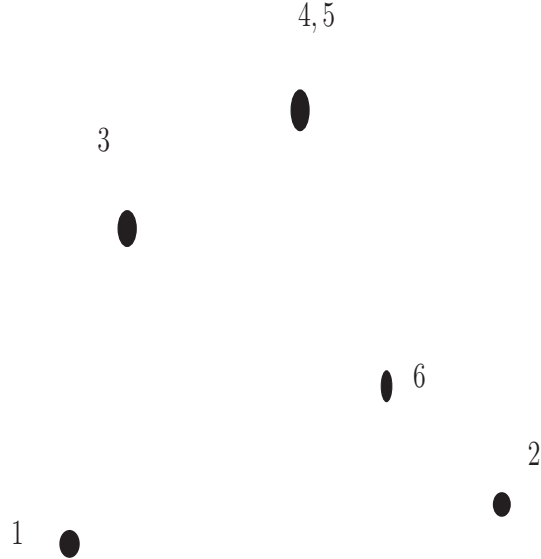}
\vspace{0cm}
\caption[]{The geometric description of the matrix in
Eq.(\ref{1SE-2-2}) on the projective plane $CP^{2}$, where the points $1,\cdots,6$
denote the indices of columns of the $3\times6$ matrix, and $4,\;5$ denote
a double point.}
\label{fig3}
\end{figure}

Using Feynman parametric representation, we write the 1-loop self energy in the general case as
\begin{eqnarray}
&&A^{1S}(p^2,m_{_1}^2,m_{_2}^2)=\Big(\Lambda_{_{\rm RE}}^2\Big)^{2-D/2}\int{d^Dq\over(2\pi)^D}
{1\over(q^2-m_{_1}^2)((q+p)^2-m_{_2}^2)}
\nonumber\\
&&\hspace{3.0cm}=
{i\Big(\Lambda_{_{\rm RE}}^2\Big)^{2-D/2}\Gamma(2-{D\over2})\over(-)^{2-D/2}(4\pi)^{D/2}}
\int_0^1 dt_{_1}dt_{_2}\delta(t_{_1}+t_{_2}-1)
\nonumber\\
&&\hspace{3.5cm}\times
(t_{_1}t_{_2}p^2-t_{_1}m_{_1}^2-t_{_2}m_{_2}^2)^{D/2-2}
\nonumber\\
&&\hspace{3.0cm}=
-{i\Big(\Lambda_{_{\rm RE}}^2\Big)^{2-D/2}\Gamma(2-{D\over2})\over(-)^{2-D/2}(4\pi)^{D/2}}
\int\omega_{_3}(t)\delta(t_{_1}+t_{_2}+t_{_3})
\nonumber\\
&&\hspace{3.5cm}\times
t_{_3}^{2-D}(t_{_1}t_{_2}p^2+t_{_1}t_{_3}m_{_1}^2+t_{_2}t_{_3}m_{_2}^2)^{D/2-2}\;.
\label{1SE-2-1}
\end{eqnarray}
The integral can be embedded in the subvariety of the Grassmannian $G_{_{3,6}}$
where the first row corresponds to the integration variable $t_{_1}$,
the second row  corresponds to the integration variable $t_{_2}$,
and the third row  corresponds to the integration variable $t_{_3}$, respectively.
Meanwhile the first column represents the power function $t_{_1}^0=1$,
the second column represents the power function $t_{_2}^0=1$, the third column
represents the power function $t_{_3}^{2-D}$,
the fourth column represents the function $\delta(t_{_1}+t_{_2}+t_{_3})$. In order to
embed the homogeneous polynomial $(t_{_1}t_{_2}p^2+t_{_1}t_{_3}m_{_1}^2+t_{_2}t_{_3}m_{_2}^2)$
as the fifth and sixth columns of the Grassmannian  $G_{_{3,6}}$, we rewrite
\begin{eqnarray}
&&t_{_1}t_{_2}p^2+t_{_1}t_{_3}m_{_1}^2+t_{_2}t_{_3}m_{_2}^2
\nonumber\\
&&\hspace{-0.5cm}=
z_{_{1,\widehat{\sigma}_{_1}(5)}}z_{_{2,\widehat{\sigma}_{_1}(6)}}t_{_1}t_{_2}
+z_{_{1,\widehat{\sigma}_{_2}(5)}}z_{_{3,\widehat{\sigma}_{_2}(6)}}t_{_1}t_{_3}
+z_{_{2,\widehat{\sigma}_{_3}(5)}}z_{_{3,\widehat{\sigma}_{_3}(6)}}t_{_2}t_{_3}\;,
\label{1SE-2-1+a}
\end{eqnarray}
where $\widehat{\sigma}_{_i},\;(i=1,\;2,\;3)$ are elements of the permutation
group $S_{_2}=\{\widehat{e},\;\widehat{(56)}\}$ on the column indices $5,\;6$.
Taking $\widehat{\sigma}_{_1}=\widehat{\sigma}_{_3}=\widehat{(56)}$, $\widehat{\sigma}_{_2}=\widehat{e}$,
we have
\begin{eqnarray}
&&z_{_{1,6}}z_{_{2,5}}=p^2\;,\;\;z_{_{1,5}}z_{_{3,6}}=m_{_1}^2\;,\;\;
z_{_{2,6}}z_{_{3,5}}=m_{_2}^2\;.
\label{1SE-2-1+b}
\end{eqnarray}
The solution of Eq.(\ref{1SE-2-1+b})
\begin{eqnarray}
&&z_{_{1,5}}=z_{_{2,5}}=z_{_{3,5}}=1\;,\;\;z_{_{1,6}}=p^2\;,\;\;z_{_{3,6}}=m_{_1}^2\;,\;\;
z_{_{2,6}}=m_{_2}^2\;,
\label{1SE-2-1+c}
\end{eqnarray}
indicates that the splitting local coordinates are written as
\begin{eqnarray}
&&A^{(1S)}=\left(\begin{array}{cccccc}\;1\;&\;0\;&\;0\;&\;1\;&\;1\;&\;p^2\;\\
\;0\;&\;1\;&\;0\;&\;1\;&\;1\;&\;m_{_2}^2\;\\
\;0\;&\;0\;&\;1\;&\;1\;&\;1\;&\;m_{_1}^2\;\end{array}\right)\;,
\label{1SE-2-2}
\end{eqnarray}
whose geometric description is drawn in Fig.\ref{fig3}.
Taking Feynman parametric representation of the 1-loop self energy
as a function of the splitting local coordinates,
\begin{eqnarray}
&&A^{1S}(p^2,m_{_1}^2,m_{_2}^2)=\Psi^{1S}(z_{_1},\;\cdots,\;z_{_6})\;,
\label{1SE-2-2+a}
\end{eqnarray}
we obviously have
\begin{eqnarray}
&&\Psi^{1S}(\chi_{_1}z_{_1},\;\cdots,\;\chi_{_6}z_{_6})=
\chi_{_3}^{2-D}\chi_{_4}^{-1}\chi_{_5}^{D/2-2}\chi_{_6}^{D/2-2}\Psi^{1S}(z_{_1},\;\cdots,\;z_{_6})\;.
\label{1SE-2-2+b}
\end{eqnarray}
Where $z_{_i}\;(i=1,\cdots,6)$ denotes the i-th column vector in the matrix Eq.(\ref{1SE-2-2}),
and $\chi_{_i}$ is a nonzero constant, respectively.
If we choose the first, second and third column vectors of $A^{(1S)}$ as the basis of the projective plane,
constructed hypergeometric functions contain redundant dependent variables.
However $\det(A_{_{\{1,2,4\}}}^{(1S)})=1$ where the matrix $A_{_{\{1,2,4\}}}^{(1S)}$ denotes
the submatrix of $A^{(1S)}$ composed of the first, second, and fourth columns, one derives
\begin{eqnarray}
&&Z_{_{3\times6}}=\Big(A_{_{\{1,2,4\}}}^{(1S)}\Big)^{-1}\cdot A^{(1S)}=\left(\begin{array}{cccccc}
\;\;1\;\;&\;\;0\;\;&\;\;-1\;\;&\;\;0\;\;&\;\;0\;\;&\;\;p^2-m_{_1}^2\;\;\\
\;\;0\;\;&\;\;1\;\;&\;\;-1\;\;&\;\;0\;\;&\;\;0\;\;&\;\;m_{_2}^2-m_{_1}^2\;\;\\
\;\;0\;\;&\;\;0\;\;&\;\;1\;\;&\;\;1\;\;&\;\;1\;\;&\;\;m_{_1}^2\;\;\end{array}\right)\;.
\label{1SE-2-3}
\end{eqnarray}
Accordingly, the Feynman integral satisfies the following GKZ-system in splitting local coordinates
\begin{eqnarray}
&&\Big\{\vartheta_{_{1,1}}+\vartheta_{_{1,3}}+\vartheta_{_{1,6}}\Big\}A_{_{1SE}}(p^2,m_{_1}^2,m_{_2}^2)=
-A_{_{1SE}}(p^2,m_{_1}^2,m_{_2}^2)
\;,\nonumber\\
&&\Big\{\vartheta_{_{2,2}}+\vartheta_{_{2,3}}+\vartheta_{_{2,6}}\Big\}A_{_{1SE}}(p^2,m_{_1}^2,m_{_2}^2)=
-A_{_{1SE}}(p^2,m_{_1}^2,m_{_2}^2)
\;,\nonumber\\
&&\Big\{\vartheta_{_{3,3}}+\vartheta_{_{3,4}}+\vartheta_{_{3,5}}
+\vartheta_{_{3,6}}\Big\}A_{_{1SE}}(p^2,m_{_1}^2,m_{_2}^2)=
-A_{_{1SE}}(p^2,m_{_1}^2,m_{_2}^2)
\;,\nonumber\\
&&\vartheta_{_{1,1}}A_{_{1SE}}(p^2,m_{_1}^2,m_{_2}^2)=0
\;,\nonumber\\
&&\vartheta_{_{2,2}}A_{_{1SE}}(p^2,m_{_1}^2,m_{_2}^2)=0
\;,\nonumber\\
&&\Big\{\vartheta_{_{1,3}}+\vartheta_{_{2,3}}+\vartheta_{_{3,3}}\Big\}
A_{_{1SE}}(p^2,m_{_1}^2,m_{_2}^2)=(2-D)A_{_{1SE}}(p^2,m_{_1}^2,m_{_2}^2)
\;,\nonumber\\
&&\vartheta_{_{3,4}}A_{_{1SE}}(p^2,m_{_1}^2,m_{_2}^2)=-A_{_{1SE}}(p^2,m_{_1}^2,m_{_2}^2)
\;,\nonumber\\
&&\vartheta_{_{3,5}}A_{_{1SE}}(p^2,m_{_1}^2,m_{_2}^2)=({D\over2}-2)A_{_{1SE}}(p^2,m_{_1}^2,m_{_2}^2)
\;,\nonumber\\
&&\Big\{\vartheta_{_{1,6}}+\vartheta_{_{2,6}}+\vartheta_{_{3,6}}\Big\}A_{_{1SE}}(p^2,m_{_1}^2,m_{_2}^2)=
({D\over2}-2)A_{_{1SE}}(p^2,m_{_1}^2,m_{_2}^2)\;.
\label{1SE-2-3a}
\end{eqnarray}
Obviously the matrix of exponents is written as
\begin{eqnarray}
&&\left(\begin{array}{cccccc}
\;\;0\;\;&\;\;0\;\;&\;\alpha_{_{1,3}}\;\;&\;\;0\;\;&\;\;0\;\;&\;\alpha_{_{1,6}}\;\;\\
\;\;0\;\;&\;\;0\;\;&\;\alpha_{_{2,3}}\;\;&\;\;0\;\;&\;\;0\;\;&\;\alpha_{_{2,6}}\;\;\\
\;\;0\;\;&\;\;0\;\;&\;\alpha_{_{3,3}}\;\;&\;\;-1\;\;&\;\;{D\over2}-2\;\;&\;\alpha_{_{3,6}}\;\;
\end{array}\right)\;,
\label{1SE-2-105}
\end{eqnarray}
where the matrix elements satisfy the relations
\begin{eqnarray}
&&\alpha_{_{1,3}}+\alpha_{_{1,6}}=-1\;,\;\alpha_{_{2,3}}+\alpha_{_{2,6}}=-1\;,\;
\alpha_{_{3,3}}+\alpha_{_{3,6}}=2-{D\over2}\;,
\nonumber\\
&&\alpha_{_{1,3}}+\alpha_{_{2,3}}+\alpha_{_{3,3}}=2-D\;,\;
\alpha_{_{1,6}}+\alpha_{_{2,6}}+\alpha_{_{3,6}}={D\over2}-2\;.
\label{1SE-2-106}
\end{eqnarray}
Correspondingly there are 12 choices on the matrix of integer lattice whose
submatrix composed of the third and
sixth columns is formulated as $\pm n_{_1}E_{_{3}}^{(i)}\pm n_{_2}E_{_{3}}^{(j)}$,
where $n_{_{1,2}}\ge0$, $(i,\;j)\in\{(1,\;2),\;(1,\;3),\;(2,\;3)\}$,
and other elements are all zero. Furthermore, the matrix
\begin{eqnarray}
&&E_{_{3}}^{(1)}=\left(\begin{array}{cc}\;0\;&\;0\;\\\;1\;&\;-1\;\\
\;-1\;&\;1\;\end{array}\right)\;,
\label{1SE-2-107}
\end{eqnarray}
and the matrices $E_{_{3}}^{(2)},\;E_{_{3}}^{(3)}$ are presented in Eq.(\ref{1SE-50})
and Eq.(\ref{1SE-40}), respectively.
Note that $E_{_{3}}^{(3)}=E_{_{3}}^{(2)}-E_{_{3}}^{(1)}$, two linear independent bases of
integer lattice matrices $E_{_{3}}^{(1)}$, $E_{_{3}}^{(2)}$ induce two independent PDEs as:
\begin{eqnarray}
&&{\partial^2A^{(1S)}\over\partial z_{_{1,3}}\partial z_{_{3,6}}}(p^2,m_{_1}^2,m_{_2}^2)
={\partial^2A^{(1S)}\over\partial z_{_{1,6}}\partial z_{_{3,3}}}(p^2,m_{_1}^2,m_{_2}^2)
\;,\nonumber\\
&&{\partial^2A^{(1S)}\over\partial z_{_{2,3}}\partial z_{_{3,6}}}(p^2,m_{_1}^2,m_{_2}^2)
={\partial^2A^{(1S)}\over\partial z_{_{2,6}}\partial z_{_{3,3}}}(p^2,m_{_1}^2,m_{_2}^2)\;.
\label{1SE-2-107a}
\end{eqnarray}
When the exponential matrix is given, the Feynman integral can be formally expressed as
\begin{eqnarray}
&&A_{_{1SE}}(p^2,m_{_1}^2,m_{_2}^2)=\prod\limits_{i,j}z_{_{i,j}}^{\alpha_{_{i,j}}}
\varphi(1-{p^2\over m_{_1}^2},1-{m_{_2}^2\over m_{_1}^2})\;,
\label{1SE-2-107b}
\end{eqnarray}
where $(i,j)\in\{(1,3),(1,6),(2,3),(2,6),(3,3),(3,6)\}$, and $\varphi(x_{_1},x_{_2})$
satisfies the PDEs in the combined local coordinates $x_{_1}=z_{_{1,6}}z_{_{3,3}}/(z_{_{1,3}}z_{_{3,6}})$
$=1-p^2/m_{_1}^2$, $x_{_2}=z_{_{2,6}}z_{_{3,3}}/(z_{_{2,3}}z_{_{3,6}})$$=1-m_{_2}^2/m_{_1}^2$ as
\begin{eqnarray}
&&x_{_1}\Big[\alpha_{_{1,3}}\alpha_{_{3,6}}+(1-\alpha_{_{1,3}}
-\alpha_{_{3,6}})x_{_1}{\partial\over\partial x_{_1}}
-\alpha_{_{1,3}}x_{_2}{\partial\over\partial x_{_2}}
+x_{_1}x_{_2}{\partial^2\over\partial x_{_1}\partial x_{_2}}
+x_{_1}^2{\partial^2\over\partial x_{_1}^2}\Big]
\varphi(x_{_1},x_{_2})
\nonumber\\
&&\hspace{-0.5cm}=
\Big[\alpha_{_{1,6}}\alpha_{_{3,3}}+(1+\alpha_{_{1,6}}
+\alpha_{_{3,3}})x_{_1}{\partial\over\partial x_{_1}}
+\alpha_{_{1,6}}x_{_2}{\partial\over\partial x_{_2}}
+x_{_1}x_{_2}{\partial^2\over\partial x_{_1}\partial x_{_2}}
+x_{_1}^2{\partial^2\over\partial x_{_1}^2}\Big]
\varphi(x_{_1},x_{_2})
\;,\nonumber\\
&&x_{_2}\Big[\alpha_{_{2,3}}\alpha_{_{3,6}}
-\alpha_{_{2,3}}x_{_1}{\partial\over\partial x_{_1}}
+(1-\alpha_{_{2,3}}-\alpha_{_{3,6}})x_{_2}{\partial\over\partial x_{_2}}
+x_{_1}x_{_2}{\partial^2\over\partial x_{_1}\partial x_{_2}}
+x_{_2}^2{\partial^2\over\partial x_{_2}^2}\Big]
\varphi(x_{_1},x_{_2})
\nonumber\\
&&\hspace{-0.5cm}=
\Big[\alpha_{_{2,6}}\alpha_{_{3,3}}
+\alpha_{_{2,6}}x_{_1}{\partial\over\partial x_{_1}}
+(1+\alpha_{_{2,6}}+\alpha_{_{3,3}})x_{_2}{\partial\over\partial x_{_2}}
+x_{_1}x_{_2}{\partial^2\over\partial x_{_1}\partial x_{_2}}
+x_{_2}^2{\partial^2\over\partial x_{_2}^2}\Big]
\varphi(x_{_1},x_{_2})\;.
\label{1SE-2-107c}
\end{eqnarray}
As $\alpha_{_{1,6}}=\alpha_{_{2,6}}=0$, the PDEs are changed into
the first type Appell PDEs. The holonomic rank of the ideal composed of the
PDEs is 3, which indicates the fundamental
solution system constituted by three linear independent hypergeometric
functions in neighborhoods of regular singularities.
It is known that the fourth type Appell function cannot be embedded into
Grassmannians~\cite{Gelfand1989}. Nevertheless, including the appropriate power factor
and the transformation of the variables, one proves that the Feynman integral presented by
the fourth Appell functions~\cite{Feng2018,Feng2020} satisfies the above PDEs.
Those matrices of integer lattice permit 12 possibilities of the exponents.
Corresponding to the integer lattice $n_{_1}E_{_3}^{(3)}+n_{_2}E_{_3}^{(2)}$
the choice of exponents is written as
\begin{eqnarray}
&&\alpha_{_{1,3}}=1-{D\over2}\;,\;\alpha_{_{1,6}}={D\over2}-2\;,\;\alpha_{_{2,3}}=-1\;,\;
\alpha_{_{2,6}}=0\;,\;\alpha_{_{3,3}}=2-{D\over2},\;\alpha_{_{3,6}}=0\;,
\label{1SE-2-108}
\end{eqnarray}
\begin{figure}[ht]
\setlength{\unitlength}{1cm}
\centering
\vspace{0.0cm}\hspace{-1.5cm}
\includegraphics[height=8cm,width=8.0cm]{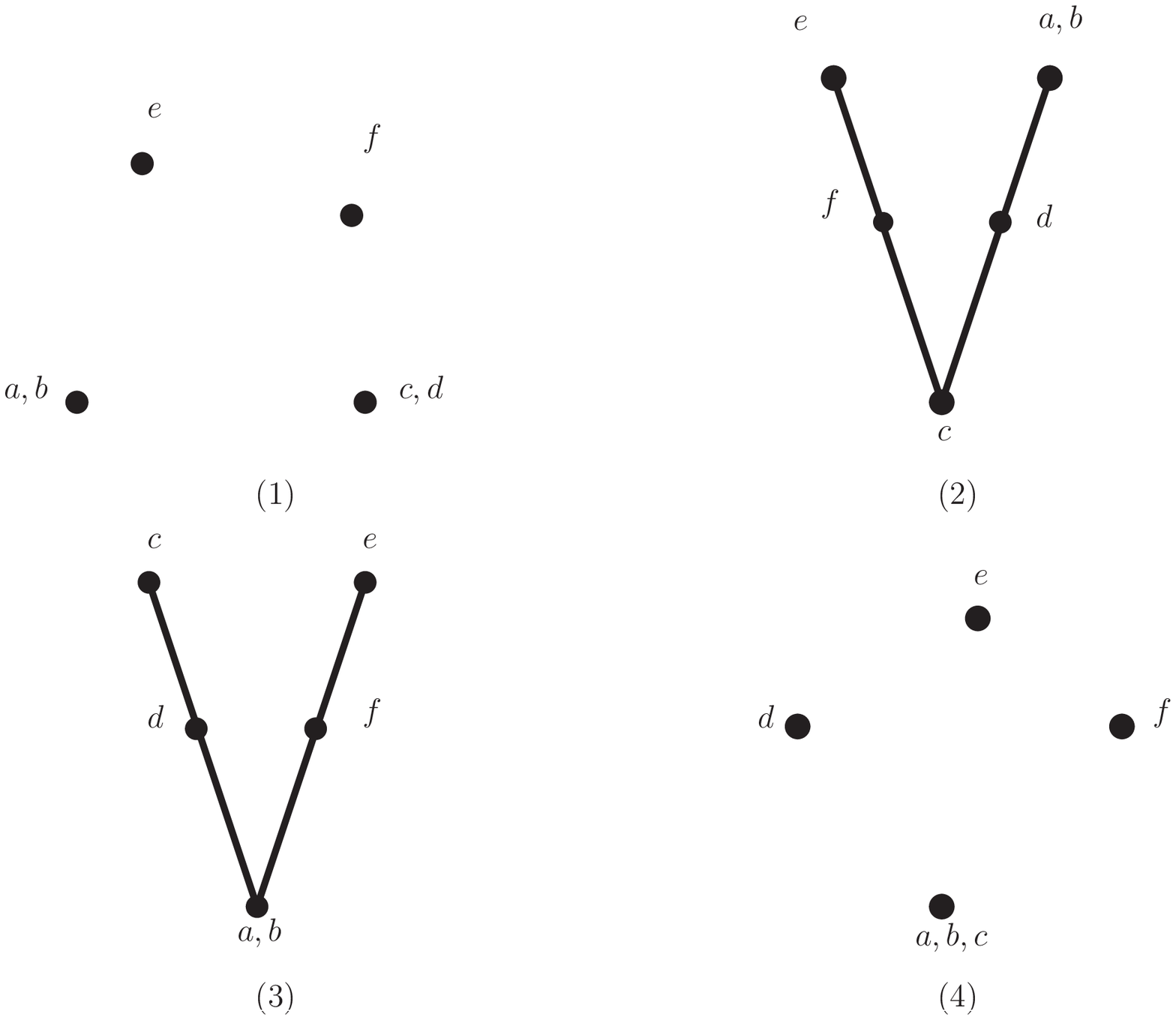}
\vspace{0cm}
\caption[]{The geometric descriptions of the single orbits in
Eq.(\ref{1SE-2-3}) on the projective plane $CP^{2}$, where the points $1,\cdots,6$
denote the indices of columns of the $3\times6$ matrix.}
\label{fig4}
\end{figure}
whose geometric representation is presented in Fig.\ref{fig4}(1)
with $\{(a,b)$, $(c,d)\}=$$\{(1,6)$, $(4,5)\}$, $\{e,\;f\}=$$\{2,\;3\}$.
As the integer lattice is chosen as $n_{_1}E_{_3}^{(3)}+n_{_2}E_{_3}^{(1)}$,
the compatible matrix of exponents is
\begin{eqnarray}
&&\alpha_{_{1,3}}=0\;,\;\alpha_{_{1,6}}=-1\;,\;\alpha_{_{2,3}}=-{D\over2}\;,\;
\alpha_{_{2,6}}={D\over2}-1\;,\;\alpha_{_{3,3}}=2-{D\over2},\;\alpha_{_{3,6}}=0\;,
\label{1SE-2-112}
\end{eqnarray}
whose geometric representation is presented in Fig.\ref{fig4}(2)
with $(a,b)=$ $(4,5)$, $c=2$, $d=3$, $\{e,\;f\}=$ $\{1,\;6\}$.
The exponents of the integer lattice $n_{_1}E_{_3}^{(2)}$ $+n_{_2}E_{_3}^{(1)}$ are
\begin{eqnarray}
&&\alpha_{_{1,3}}=0\;,\;\alpha_{_{1,6}}=-1\;,\;\alpha_{_{2,3}}=0\;,\;
\alpha_{_{2,6}}=-1\;,\;\alpha_{_{3,3}}=2-D,\;\alpha_{_{3,6}}={D\over2}\;,
\label{1SE-2-116}
\end{eqnarray}
whose geometric representation is presented in Fig.\ref{fig4}(4)
with $(a,b,c)=$ $(3,4,5)$, $\{d,e,f\}=$ $\{1,2,6\}$.
Accordingly three linear independent hypergeometric functions are obtained
\begin{eqnarray}
&&\psi_{_{\{1,2,4\}}}^{(1)}(p^2,m_{_1}^2,m_{_2}^2)
\sim(m_{_1}^2-p^2)^{D/2-2}\sum\limits_{n_{_1},n_{_2}}^\infty
{\Gamma({D\over2}-2+n_{_2})\over n_{_2}!}
\nonumber\\
&&\hspace{3.9cm}\times
\Big({m_{_1}^2-m_{_2}^2\over m_{_1}^2-p^2}\Big)^{n_{_1}}
\Big({m_{_1}^2\over m_{_1}^2-p^2}\Big)^{n_{_2}}
\nonumber\\
&&\hspace{3.4cm}\sim
{(p^2-m_{_1}^2)^{D-3}(p^2)^{2-D/2}\over p^2-m_{_2}^2}
\;,\nonumber\\
&&\psi_{_{\{1,2,4\}}}^{(2)}(p^2,m_{_1}^2,m_{_2}^2)\sim
{(m_{_2}^2-m_{_1}^2)^{D/2-1}\over p^2-m_{_1}^2}
\sum\limits_{n_{_1},n_{_2}}^\infty
{\Gamma({D\over2}-2+n_{_2})\over n_{_2}!}
\nonumber\\
&&\hspace{3.9cm}\times
\Big({m_{_1}^2-m_{_2}^2\over m_{_1}^2-p^2}\Big)^{n_{_1}}
\Big({m_{_1}^2\over m_{_1}^2-m_{_2}^2}\Big)^{n_{_2}}
\nonumber\\
&&\hspace{3.4cm}\sim
{(m_{_2}^2-m_{_1}^2)^{D-3}(m_{_2}^2)^{2-D/2}\over p^2-m_{_2}^2}
\;,\nonumber\\
&&\psi_{_{\{1,2,4\}}}^{(3)}(p^2,m_{_1}^2,m_{_2}^2)\sim
{(m_{_1}^2)^{D/2}\over(p^2-m_{_1}^2)(m_{_2}^2-m_{_1}^2)}
\sum\limits_{n_{_1},n_{_2}}^\infty
{\Gamma(D-2+n_{_1}+n_{_2})\over\Gamma({D\over2}+1+n_{_1}+n_{_2})}
\nonumber\\
&&\hspace{3.9cm}\times
\Big({m_{_1}^2\over m_{_1}^2-p^2}\Big)^{n_{_1}}
\Big({m_{_1}^2\over m_{_1}^2-m_{_2}^2}\Big)^{n_{_2}}\;,
\label{1SE-2-116a}
\end{eqnarray}
whose intersection of their convergent regions is nonempty proper subset
of the whole parameter space. Three hypergeometric functions constitute a
fundamental solution system of the GKZ-system in the nonempty intersection.

Similarly the exponents of the integer lattice $-n_{_1}E_{_3}^{(3)}-n_{_2}E_{_3}^{(2)}$ are
\begin{eqnarray}
&&\alpha_{_{1,3}}=2-D\;,\;\alpha_{_{1,6}}=D-3\;,\;\alpha_{_{2,3}}=0\;,\;
\alpha_{_{2,6}}=-1\;,\;\alpha_{_{3,3}}=0,\;\alpha_{_{3,6}}=2-{D\over2}\;,
\label{1SE-2-111}
\end{eqnarray}
whose geometric representation is drawn in Fig.\ref{fig4}(1)
with $\{(a,b)$, $(c,d)\}=$ $\{(1,3)$, $(4,5)\}$, $\{e, f\}=$ $\{2, 6\}$.
Corresponding to the integer lattice $-n_{_1}E_{_3}^{(3)}-n_{_2}E_{_3}^{(1)}$
the matrix of exponents is
\begin{eqnarray}
&&\alpha_{_{1,3}}=-1\;,\;\alpha_{_{1,6}}=0\;,\;\alpha_{_{2,3}}=3-D\;,\;
\alpha_{_{2,6}}=D-4\;,\;\alpha_{_{3,3}}=0,\;\alpha_{_{3,6}}=2-{D\over2}\;,
\label{1SE-2-115}
\end{eqnarray}
whose geometric representation is given in Fig.\ref{fig4}(2)
with $(a,b)=$ $(4,5)$, $c=2$, $d=6$, $\{e, f\}=$ $\{1, 3\}$.
Meanwhile the exponents of the integer lattice
$-n_{_1}E_{_3}^{(2)}-n_{_2}E_{_3}^{(1)}$ are
\begin{eqnarray}
&&\alpha_{_{1,3}}=-1\;,\;\alpha_{_{1,6}}=0\;,\;\alpha_{_{2,3}}=-1\;,\;
\alpha_{_{2,6}}=0\;,\;\alpha_{_{3,3}}=4-D,\;\alpha_{_{3,6}}={D\over2}-2\;,
\label{1SE-2-119}
\end{eqnarray}
whose geometric representation is presented in Fig.\ref{fig4}(4)
with $(a, b, c)=$ $(4, 5, 6)$, $\{d, e, f\}=$ $\{1, 2, 3\}$.
Correspondingly the linear independent hypergeometric functions are
\begin{eqnarray}
&&\psi_{_{\{1,2,4\}}}^{(4)}(p^2,m_{_1}^2,m_{_2}^2)\sim
{(m_{_1}^2)^{2-D/2}(m_{_1}^2-p^2)^{D-3}\over m_{_2}^2-m_{_1}^2}
\sum\limits_{n_{_1},n_{_2}}^\infty
{\Gamma({D\over2}-2+n_{_2})\over n_{_2}!}
\nonumber\\
&&\hspace{3.9cm}\times
\Big({m_{_1}^2-p^2\over m_{_1}^2-m_{_2}^2}\Big)^{n_{_1}}
\Big({m_{_1}^2-p^2\over m_{_1}^2}\Big)^{n_{_2}}
\nonumber\\
&&\hspace{3.4cm}\sim
{(m_{_1}^2-p^2)^{D-3}(p^2)^{2-D/2}\over p^2-m_{_2}^2}
\;,\nonumber\\
&&\psi_{_{\{1,2,4\}}}^{(5)}(p^2,m_{_1}^2,m_{_2}^2)\sim
(m_{_2}^2-m_{_1}^2)^{D-4}(m_{_1}^2)^{2-D/2}
\sum\limits_{n_{_1},n_{_2}}^\infty
{\Gamma({D\over2}-2+n_{_2})\over n_{_2}!}
\nonumber\\
&&\hspace{3.9cm}\times
\Big({m_{_1}^2-p^2\over m_{_1}^2-m_{_2}^2}\Big)^{n_{_1}}
\Big({m_{_1}^2-m_{_2}^2\over m_{_1}^2}\Big)^{n_{_2}}
\nonumber\\
&&\hspace{3.4cm}\sim
{(m_{_2}^2-m_{_1}^2)^{D-3}(m_{_2}^2)^{2-D/2}\over p^2-m_{_2}^2}
\;,\nonumber\\
&&\psi_{_{\{1,2,4\}}}^{(6)}(p^2,m_{_1}^2,m_{_2}^2)\sim
(m_{_1}^2)^{D/2-2}\sum\limits_{n_{_1},n_{_2}}^\infty
{\Gamma(2-{D\over2}+n_{_1}+n_{_2})\over\Gamma(5-D+n_{_1}+n_{_2})}
\nonumber\\
&&\hspace{3.9cm}\times
\Big({m_{_1}^2-p^2\over p^2}\Big)^{n_{_1}}
\Big({m_{_1}^2-m_{_2}^2\over p^2}\Big)^{n_{_2}}
\;,\nonumber\\
\label{1SE-2-119a}
\end{eqnarray}
whose intersection of their convergent regions is a nonempty proper subset
of the parameter space. The hypergeometric functions constitute a
fundamental solution system of the GKZ-system in a nonempty proper
subset of the whole parameter space.

The exponents of the integer lattice
$n_{_1}E_{_3}^{(3)}-n_{_2}E_{_3}^{(2)}$ are
\begin{eqnarray}
&&\alpha_{_{1,3}}=3-D\;,\;\alpha_{_{1,6}}=D-4\;,\;\alpha_{_{2,3}}=-1\;,\;
\alpha_{_{2,6}}=0\;,\;\alpha_{_{3,3}}=0,\;\alpha_{_{3,6}}=2-{D\over2}\;,
\label{1SE-2-109}
\end{eqnarray}
whose geometric representation is plotted in Fig.\ref{fig4}(2)
with $(a,b)=$ $(4,5)$, $c=1$, $d=6$, $\{e, f\}=$ $\{2, 3\}$.
In a similar way the exponent matrix of the integer lattice
$n_{_1}E_{_3}^{(3)}-n_{_2}E_{_3}^{(1)}$ is
\begin{eqnarray}
&&\alpha_{_{1,3}}=0\;,\;\alpha_{_{1,6}}=-1\;,\;\alpha_{_{2,3}}=2-D\;,\;
\alpha_{_{2,6}}=D-3\;,\;\alpha_{_{3,3}}=0,\;\alpha_{_{3,6}}=2-{D\over2}\;,
\label{1SE-2-113}
\end{eqnarray}
whose geometric representation is shown in Fig.\ref{fig4}(1)
with $\{(a,b)$, $(c,d)\}=$ $\{(2,3)$, $(4,5)\}$, $\{e, f\}=$ $\{1, 6\}$.
Meanwhile the exponent matrix of $n_{_1}E_{_3}^{(2)}-n_{_2}E_{_3}^{(1)}$
is given by
\begin{eqnarray}
&&\alpha_{_{1,3}}=0\;,\;\alpha_{_{1,6}}=-1\;,\;\alpha_{_{2,3}}=-1\;,\;
\alpha_{_{2,6}}=0\;,\;\alpha_{_{3,3}}=3-D,\;\alpha_{_{3,6}}={D\over2}-1\;,
\label{1SE-2-117}
\end{eqnarray}
whose geometric representation is presented in Fig.\ref{fig4}(3)
with $(a,b)=$ $(4,5)$, $\{(c, d)$, $(e, f)\}=$ $\{(1, 6)$, $(2, 3)\}$.
Accordingly the hypergeometric functions are
\begin{eqnarray}
&&\psi_{_{\{1,2,4\}}}^{(7)}(p^2,m_{_1}^2,m_{_2}^2)
\sim(m_{_1}^2)^{2-{D/2}}(p^2-m_{_1}^2)^{D-4}\sum\limits_{n_{_1},n_{_2}}^\infty
{\Gamma({D\over2}-2+n_{_2})\over n_{_2}!}
\nonumber\\
&&\hspace{3.9cm}\times
\Big({m_{_1}^2-m_{_2}^2\over m_{_1}^2-p^2}\Big)^{n_{_1}}
\Big({m_{_1}^2-p^2\over m_{_1}^2}\Big)^{n_{_2}}
\nonumber\\
&&\hspace{3.4cm}\sim
{(p^2-m_{_1}^2)^{D-3}(p^2)^{2-D/2}\over p^2-m_{_2}^2}
\;,\nonumber\\
&&\psi_{_{\{1,2,4\}}}^{(8)}(p^2,m_{_1}^2,m_{_2}^2)\sim
{(m_{_2}^2-m_{_1}^2)^{D-3}(m_{_1}^2)^{2-D/2}\over p^2-m_{_1}^2}
\sum\limits_{n_{_1},n_{_2}}^\infty
{\Gamma({D\over2}-2+n_{_2})\over n_{_2}!}
\nonumber\\
&&\hspace{3.9cm}\times
\Big({m_{_1}^2-m_{_2}^2\over m_{_1}^2-p^2}\Big)^{n_{_1}}
\Big({m_{_1}^2-m_{_2}^2\over m_{_1}^2}\Big)^{n_{_2}}
\nonumber\\
&&\hspace{3.4cm}\sim
{(m_{_2}^2-m_{_1}^2)^{D-3}(m_{_2}^2)^{2-D/2}\over p^2-m_{_2}^2}
\;,\nonumber\\
&&\psi_{_{\{1,2,4\}}}^{(9)}(p^2,m_{_1}^2,m_{_2}^2)\sim
{(m_{_1}^2)^{D/2-1}\over p^2-m_{_1}^2}\sum\limits_{n_{_1},n_{_2}}^\infty
{\Gamma(D-3+n_{_1}-n_{_2})\over\Gamma({D\over2}+n_{_1}-n_{_2})}
\nonumber\\
&&\hspace{3.9cm}\times
\Big({m_{_1}^2\over m_{_1}^2-p^2}\Big)^{n_{_1}}
\Big({m_{_1}^2-m_{_2}^2\over m_{_1}^2}\Big)^{n_{_2}}\;.
\label{1SE-2-117a}
\end{eqnarray}
Although the intersection of the convergent regions of these series is
the empty set, the sets of the hypergeometric functions
$\{\psi_{_{\{1,2,4\}}}^{(1)},\psi_{_{\{1,2,4\}}}^{(8)},\psi_{_{\{1,2,4\}}}^{(9)}\}$
and $\{\psi_{_{\{1,2,4\}}}^{(6)},\psi_{_{\{1,2,4\}}}^{(7)},\psi_{_{\{1,2,4\}}}^{(8)}\}$
constitute the fundamental solution systems of the nonempty proper subsets
of the whole parameter space, respectively.

Corresponding to the integer lattice $-n_{_1}E_{_3}^{(3)}+n_{_2}E_{_3}^{(2)}$,
the matrix of exponents is
\begin{eqnarray}
&&\alpha_{_{1,3}}=-{D\over2}\;,\;\alpha_{_{1,6}}={D\over2}-1\;,\;\alpha_{_{2,3}}=0\;,\;
\alpha_{_{2,6}}=-1\;,\;\alpha_{_{3,3}}=2-{D\over2},\;\alpha_{_{3,6}}=0\;,
\label{1SE-2-110}
\end{eqnarray}
whose geometric representation is presented in Fig.\ref{fig4}(2)
with $(a,b)=$ $(4,5)$, $c=1$, $d=3$, $\{e,f\}=$ $\{2,6\}$.
Similarly the exponents of the integer lattice
$-n_{_1}E_{_3}^{(3)}+n_{_2}E_{_3}^{(1)}$ are
\begin{eqnarray}
&&\alpha_{_{1,3}}=-1\;,\;\alpha_{_{1,6}}=0\;,\;\alpha_{_{2,3}}=1-{D\over2}\;,\;
\alpha_{_{2,6}}={D\over2}-2\;,\;\alpha_{_{3,3}}=2-{D\over2},\;\alpha_{_{3,6}}=0\;,
\label{1SE-2-114}
\end{eqnarray}
whose geometric representation is plotted in Fig.\ref{fig4}(1)
with $\{(a,b)$, $(c,d)\}=$ $\{(2,6)$, $(4,5)\}$, $\{e,f\}=$ $\{1,3\}$.
In addition the exponents of the integer
lattice $-n_{_1}E_{_3}^{(2)}+n_{_2}E_{_3}^{(1)}$ are
\begin{eqnarray}
&&\alpha_{_{1,3}}=-1\;,\;\alpha_{_{1,6}}=0\;,\;\alpha_{_{2,3}}=0\;,\;
\alpha_{_{2,6}}=-1\;,\;\alpha_{_{3,3}}=3-D,\;\alpha_{_{3,6}}={D\over2}-1\;,
\label{1SE-2-118}
\end{eqnarray}
whose geometric representation is shown in Fig.\ref{fig4}(3)
with $(a,b)=$ $(4,5)$, $c=2$, $d=3$, $\{(c,d)$, $(e,f)\}=$ $\{(1,3)$, $(2,6)\}$.
We obtain the hypergeometric functions according to the exponents and
integer lattices as
\begin{eqnarray}
&&\psi_{_{\{1,2,4\}}}^{(10)}(p^2,m_{_1}^2,m_{_2}^2)\sim
{(p^2-m_{_1}^2)^{D/2-1}\over m_{_2}^2-m_{_1}^2}
\sum\limits_{n_{_1},n_{_2}}^\infty{\Gamma({D\over2}-2+n_{_2})\over n_{_2}!}
\nonumber\\
&&\hspace{3.9cm}\times
\Big({m_{_1}^2-p^2\over m_{_1}^2-m_{_2}^2}\Big)^{n_{_1}}
\Big({m_{_1}^2\over m_{_1}^2-p^2}\Big)^{n_{_2}}
\nonumber\\
&&\hspace{3.4cm}\sim
{(p^2-m_{_1}^2)^{D-3}(p^2)^{2-D/2}\over p^2-m_{_2}^2}
\;,\nonumber\\
&&\psi_{_{\{1,2,4\}}}^{(11)}(p^2,m_{_1}^2,m_{_2}^2)\sim
(m_{_2}^2-m_{_1}^2)^{D/2-2}\sum\limits_{n_{_1},n_{_2}}^\infty
{\Gamma({D\over2}-2+n_{_2})\over n_{_2}!}
\nonumber\\
&&\hspace{3.9cm}\times
\Big({m_{_1}^2-p^2\over m_{_1}^2-m_{_2}^2}\Big)^{n_{_1}}
\Big({m_{_1}^2\over m_{_1}^2-m_{_2}^2}\Big)^{n_{_2}}
\nonumber\\
&&\hspace{3.4cm}\sim
{(m_{_2}^2-m_{_1}^2)^{D-3}(m_{_2}^2)^{2-D/2}\over p^2-m_{_2}^2}
\;,\nonumber\\
&&\psi_{_{\{1,2,4\}}}^{(12)}(p^2,m_{_1}^2,m_{_2}^2)\sim
{(m_{_1}^2)^{D/2-1}\over m_{_2}^2-m_{_1}^2}\sum\limits_{n_{_1},n_{_2}}^\infty
{\Gamma(1-{D\over2}+n_{_1}-n_{_2})\over\Gamma(4-D+n_{_1}-n_{_2})}
\nonumber\\
&&\hspace{3.9cm}\times
\Big({m_{_1}^2-p^2\over m_{_1}^2}\Big)^{n_{_1}}
\Big({m_{_1}^2\over m_{_1}^2-m_{_2}^2}\Big)^{n_{_2}}\;.
\label{1SE-2-120}
\end{eqnarray}
Certainly the intersection of the convergent regions of these series is
the empty set. However, the sets of the hypergeometric functions
$\{\psi_{_{\{1,2,4\}}}^{(3)},\psi_{_{\{1,2,4\}}}^{(10)},\psi_{_{\{1,2,4\}}}^{(11)}\}$
and $\{\psi_{_{\{1,2,4\}}}^{(4)},\psi_{_{\{1,2,4\}}}^{(11)},\psi_{_{\{1,2,4\}}}^{(12)}\}$
constitute the fundamental solution systems of the corresponding nonempty proper subsets
of the whole parameter space, respectively.

Because $\det(A_{_{\{1,3,4\}}}^{(1S)})=-1$ where the matrix $A_{_{\{1,3,4\}}}^{(1S)}$ denotes
the submatrix of $A^{(1S)}$ composed of the  first, third, and fourth columns, one derives
\begin{eqnarray}
&&\Big(A_{_{\{1,3,4\}}}^{(1S)}\Big)^{-1}\cdot A^{(1S)}=\left(\begin{array}{cccccc}
\;\;1\;\;&\;\;-1\;\;&\;\;0\;\;&\;\;0\;\;&\;\;0\;\;&\;\;m_{_1}^2-m_{_2}^2\;\;\\
\;\;0\;\;&\;\;-1\;\;&\;\;1\;\;&\;\;0\;\;&\;\;0\;\;&\;\;p^2-m_{_2}^2\;\;\\
\;\;0\;\;&\;\;1\;\;&\;\;0\;\;&\;\;1\;\;&\;\;1\;\;&\;\;m_{_2}^2\;\;\end{array}\right)\;.
\label{1SE-2-122}
\end{eqnarray}
Obviously the matrix of exponents is written as
\begin{eqnarray}
&&\left(\begin{array}{cccccc}
\;\;0\;\;&\;\alpha_{_{1,2}}\;\;&\;\;0\;\;&\;\;0\;\;&\;\;0\;\;&\;\alpha_{_{1,6}}\;\;\\
\;\;0\;\;&\;\alpha_{_{2,2}}\;\;&\;\;2-D\;\;&\;\;0\;\;&\;\;0\;\;&\;\alpha_{_{2,6}}\;\;\\
\;\;0\;\;&\;\alpha_{_{3,2}}\;\;&\;\;0\;\;&\;\;-1\;\;&\;\;{D\over2}-2\;\;&\;\alpha_{_{3,6}}\;\;
\end{array}\right)\;,
\label{1SE-2-123}
\end{eqnarray}
where the matrix elements satisfy the relations
\begin{eqnarray}
&&\alpha_{_{1,2}}+\alpha_{_{1,6}}=-1\;,\;\alpha_{_{2,2}}+\alpha_{_{2,6}}=D-3\;,\;
\alpha_{_{3,2}}+\alpha_{_{3,6}}=2-{D\over2}\;,
\nonumber\\
&&\alpha_{_{1,2}}+\alpha_{_{2,2}}+\alpha_{_{3,2}}=0\;,\;\alpha_{_{1,6}}+\alpha_{_{2,6}}+\alpha_{_{3,6}}={D\over2}-2\;.
\label{1SE-2-124}
\end{eqnarray}
There are 12 choices on the matrix of integer lattice whose the
submatrix composed of the second and sixth columns is formulated
as $\pm n_{_1}E_{_{3}}^{(i)}\pm n_{_2}E_{_{3}}^{(j)}$,
where $n_{_{1,2}}\ge0$, and $(i,j)=$ $\{(1,2)$, $(1,3)$, $(2,3)\}$,
and other elements are all zero.

Basing on the matrices of integer lattice, we have 12 choices on the
matrix of exponents. Corresponding to the integer lattice
$n_{_1}E_{_{3}}^{(3)}+n_{_2}E_{_{3}}^{(2)}$, the exponents are
\begin{eqnarray}
&&\alpha_{_{1,2}}=1-{D\over2}\;,\;\alpha_{_{1,6}}={D\over2}-2\;,\;\alpha_{_{2,2}}=D-3\;,\;
\alpha_{_{2,6}}=0\;,\;\alpha_{_{3,2}}=2-{D\over2},\;\alpha_{_{3,6}}=0\;.
\label{1SE-2-125}
\end{eqnarray}
whose geometric representation is drawn in Fig.\ref{fig4}(1)
with $\{(a,b)$, $(c,d)\}=$ $\{(1,6)$, $(4,5)\}$, $\{e,f\}=$ $\{2,3\}$.
The exponents of the integer lattice
$n_{_1}E_{_{3}}^{(3)}+n_{_2}E_{_{3}}^{(1)}$ are
\begin{eqnarray}
&&\alpha_{_{1,2}}=0\;,\;\alpha_{_{1,6}}=-1\;,\;\alpha_{_{2,2}}={D\over2}-2\;,\;
\alpha_{_{2,6}}={D\over2}-1\;,\;\alpha_{_{3,2}}=2-{D\over2},\;\alpha_{_{3,6}}=0\;,
\label{1SE-2-129}
\end{eqnarray}
whose geometric representation is presented in Fig.\ref{fig4}(2)
with $(a,b)=$ $(4,5)$, $c=3$, $d=2,$, $\{e,f\}=$ $\{1,6\}$.
In a similar way the exponents of the integer lattice
$n_{_1}E_{_{3}}^{(2)}+n_{_2}E_{_{3}}^{(1)}$ are
\begin{eqnarray}
&&\alpha_{_{1,2}}=0\;,\;\alpha_{_{1,6}}=-1\;,\;\alpha_{_{2,2}}=0\;,\;
\alpha_{_{2,6}}=D-3\;,\;\alpha_{_{3,2}}=0,\;\alpha_{_{3,6}}=2-{D\over2}\;,
\label{1SE-2-133}
\end{eqnarray}
whose geometric representation is drawn in Fig.\ref{fig4}(4)
with $(a,b,c)=$ $(2,4,5)$, $\{d,e,f\}=$ $\{1,3,6\}$.
Accordingly the hypergeometric functions are
\begin{eqnarray}
&&\psi_{_{\{1,3,4\}}}^{(1)}(p^2,m_{_1}^2,m_{_2}^2)\sim
(p^2-m_{_2}^2)^{D/2-2}\sum\limits_{n_{_1},n_{_2}}^\infty
{\Gamma(3-D+n_{_1})\Gamma({D\over2}-2+n_{_2})\over n_{_1}!n_{_2}!}
\nonumber\\
&&\hspace{3.9cm}\times
\Big({m_{_2}^2-m_{_1}^2\over m_{_2}^2-p^2}\Big)^{n_{_1}}
\Big({m_{_2}^2\over m_{_2}^2-p^2}\Big)^{n_{_2}}
\nonumber\\
&&\hspace{3.5cm}\sim
{(p^2-m_{_1}^2)^{D-3}(p^2)^{2-D/2}\over p^2-m_{_2}^2}
\;,\nonumber\\
&&\psi_{_{\{1,3,4\}}}^{(2)}(p^2,m_{_1}^2,m_{_2}^2)\sim
{(m_{_2}^2-m_{_1}^2)^{D/2-1}\over p^2-m_{_2}^2}
\sum\limits_{n_{_1},n_{_2}}^\infty
{\Gamma(2-{D\over2}+n_{_1}-n_{_2})\over n_{_2}!\Gamma({D\over2}+n_{_1}-n_{_2})}
\nonumber\\
&&\hspace{3.9cm}\times
\Gamma({D\over2}-2+n_{_2})
\Big({m_{_2}^2-m_{_1}^2\over m_{_2}^2-p^2}\Big)^{n_{_1}}
\Big({m_{_2}^2\over m_{_2}^2-m_{_1}^2}\Big)^{n_{_2}}
\;,\nonumber\\
&&\psi_{_{\{1,3,4\}}}^{(3)}(p^2,m_{_1}^2,m_{_2}^2)\sim
{(m_{_2}^2)^{2-D/2}(m_{_2}^2-m_{_1}^2)^{D-3}\over m_{_2}^2-p^2}
\sum\limits_{n_{_1},n_{_2}}^\infty
{(-)^{n_{_1}+n_{_2}}\Gamma(4-D+n_{_2})\over n_{_2}!
\Gamma(1-n_{_1}-n_{_2})}
\nonumber\\
&&\hspace{3.9cm}\times
{1\over\Gamma(3-{D\over2}+n_{_1}+n_{_2})}
\Big({m_{_2}^2\over m_{_2}^2-p^2}\Big)^{n_{_1}}
\Big({m_{_2}^2\over m_{_2}^2-m_{_1}^2}\Big)^{n_{_2}}
\nonumber\\
&&\hspace{3.5cm}\sim
{(m_{_2}^2)^{2-D/2}(m_{_2}^2-m_{_1}^2)^{D-3}\over m_{_2}^2-p^2}\;,
\label{1SE-2-133a}
\end{eqnarray}
whose intersection is a nonempty proper subset of the whole parameter space.
In other words, three hypergeometric functions constitute a fundamental solution system of
the GKZ-system in the nonempty intersection.

The matrix of exponents of the integer lattice
$-n_{_1}E_{_{3}}^{(3)}-n_{_2}E_{_{3}}^{(2)}$ is
\begin{eqnarray}
&&\alpha_{_{1,2}}=0\;,\;\alpha_{_{1,6}}=-1\;,\;\alpha_{_{2,2}}=0\;,\;
\alpha_{_{2,6}}=D-3\;,\;\alpha_{_{3,2}}=0,\;\alpha_{_{3,6}}=2-{D\over2}\;,
\label{1SE-2-128}
\end{eqnarray}
whose geometric representation is shown in Fig.\ref{fig4}(1)
with $\{(a,b)$, $(c,d)\}=$ $\{(1,2)$, $(4,5)\}$, $\{e,f\}=$ $\{3,6\}$.
Similarly the exponents of the integer lattice
$-n_{_1}E_{_{3}}^{(3)}-n_{_2}E_{_{3}}^{(1)}$ are
\begin{eqnarray}
&&\alpha_{_{1,2}}=-1\;,\;\alpha_{_{1,6}}=0\;,\;\alpha_{_{2,2}}=1\;,\;
\alpha_{_{2,6}}=D-4\;,\;\alpha_{_{3,2}}=0,\;\alpha_{_{3,6}}=2-{D\over2}\;,
\label{1SE-2-132}
\end{eqnarray}
whose geometric representation is presented in Fig.\ref{fig4}(2)
with $(a,b)=$ $(4,5)$, $c=3$, $d=6$, $\{e,f\}=$ $\{1,2\}$.
Furthermore the exponents of the integer lattice
$-n_{_1}E_{_{3}}^{(2)}-n_{_2}E_{_{3}}^{(1)}$ are
\begin{eqnarray}
&&\alpha_{_{1,2}}=-1\;,\;\alpha_{_{1,6}}=0\;,\;\alpha_{_{2,2}}=D-3\;,\;
\alpha_{_{2,6}}=0\;,\;\alpha_{_{3,2}}=4-D,\;\alpha_{_{3,6}}={D\over2}-2\;,
\label{1SE-2-136}
\end{eqnarray}
where the geometric representation is plotted in Fig.\ref{fig4}(4)
with $(a,b,c)=$ $(4,5,6)$, $\{d,e,f\}=$ $\{1,2,3\}$.
The corresponding hypergeometric functions are presented as
\begin{eqnarray}
&&\psi_{_{\{1,3,4\}}}^{(4)}(p^2,m_{_1}^2,m_{_2}^2)\sim
{(m_{_1}^2-m_{_2}^2)^{D-3}(m_{_2}^2)^{2-D/2}\over p^2-m_{_2}^2}
\sum\limits_{n_{_1},n_{_2}}^\infty
{\Gamma(3-D+n_{_1})\over n_{_1}!n_{_2}!}
\nonumber\\
&&\hspace{3.9cm}\times
\Gamma({D\over2}-2+n_{_2})
\Big({m_{_2}^2-p^2\over m_{_2}^2-m_{_1}^2}\Big)^{n_{_1}}
\Big({m_{_2}^2-p^2\over m_{_2}^2}\Big)^{n_{_2}}
\nonumber\\
&&\hspace{3.5cm}\sim
{(p^2-m_{_1}^2)^{D-3}(p^2)^{2-D/2}\over p^2-m_{_2}^2}
\;,\nonumber\\
&&\psi_{_{\{1,3,4\}}}^{(5)}(p^2,m_{_1}^2,m_{_2}^2)
\sim{(m_{_2}^2-m_{_1}^2)^{D}\over(m_{_2}^2)^{D/2}}
\sum\limits_{n_{_1},n_{_2}}^\infty
{\Gamma({D\over2}+n_{_1}+n_{_2})\Gamma(1+n_{_2})\over
\Gamma(3+n_{_1}+n_{_2})\Gamma(D-1+n_{_2})}
\nonumber\\
&&\hspace{3.9cm}\times
\Big({m_{_2}^2-p^2\over m_{_2}^2}\Big)^{n_{_1}}
\Big({m_{_2}^2-m_{_1}^2\over m_{_2}^2}\Big)^{n_{_2}}
\;,\nonumber\\
&&\psi_{_{\{1,3,4\}}}^{(6)}(p^2,m_{_1}^2,m_{_2}^2)
\sim(m_{_2}^2)^{D/2-2}\sum\limits_{n_{_1},n_{_2}}^\infty
{\Gamma(2-{D\over2}+n_{_1}+n_{_2})\over n_{_2}!\Gamma(5-D+n_{_1}+n_{_2})}
\nonumber\\
&&\hspace{3.9cm}\times
\Gamma(3-D+n_{_2})
\Big({m_{_2}^2-p^2\over m_{_2}^2}\Big)^{n_{_1}}
\Big({m_{_2}^2-m_{_1}^2\over m_{_2}^2}\Big)^{n_{_2}}\;,
\label{1SE-2-136a}
\end{eqnarray}
whose intersection is a nonempty proper subset of the whole parameter space.
Since the differences of the exponents between Eq.(\ref{1SE-2-128}) and
Eq.(\ref{1SE-2-132}) are all integers, the hypergeometric solution of
the exponents of Eq.(\ref{1SE-2-132}) is written as
\begin{eqnarray}
&&\psi_{_{\{1,3,4\}}}^{(5)\prime}(p^2,m_{_1}^2,m_{_2}^2)
\sim(m_{_2}^2-m_{_1}^2)^{D-4}(m_{_2}^2)^{2-D/2}\sum\limits_{n_{_1},n_{_2}}^\infty
{\Gamma(4-D+n_{_1}-n_{_2})\over n_{_2}!\Gamma(2+n_{_1}-n_{_2})}
\nonumber\\
&&\hspace{3.9cm}\times
\Gamma({D\over2}-2+n_{_2})
\Big({m_{_2}^2-p^2\over m_{_2}^2-m_{_1}^2}\Big)^{n_{_1}}
\Big({m_{_2}^2-m_{_1}^2\over m_{_2}^2}\Big)^{n_{_2}}
\nonumber\\
&&\hspace{3.5cm}\sim
-\psi_{_{\{1,3,4\}}}^{(4)}(p^2,m_{_1}^2,m_{_2}^2)
\nonumber\\
&&\hspace{3.9cm}
+\lim\limits_{\epsilon\rightarrow0}{\sin\pi\epsilon\over\sin\pi(2-D)}
\psi_{_{\{1,3,4\}}}^{(5)}(p^2,m_{_1}^2,m_{_2}^2)\;.
\label{1SE-2-136b}
\end{eqnarray}

For the integer lattice $n_{_1}E_{_{3}}^{(3)}-n_{_2}E_{_{3}}^{(2)}$,
those exponents are written as
\begin{eqnarray}
&&\alpha_{_{1,2}}=3-D\;,\;\alpha_{_{1,6}}=D-4\;,\;\alpha_{_{2,2}}=D-3\;,\;
\alpha_{_{2,6}}=0\;,\;\alpha_{_{3,2}}=0,\;\alpha_{_{3,6}}=2-{D\over2}\;,
\label{1SE-2-126}
\end{eqnarray}
whose geometric representation is drawn in Fig.\ref{fig4}(2)
with $(a,b)=$ $(4,5)$, $c=1$, $d=6$, $\{e,f\}=$ $\{2,3\}$.
Corresponding to the integer lattice $n_{_1}E_{_{3}}^{(3)}-n_{_2}E_{_{3}}^{(1)}$,
the exponents are
\begin{eqnarray}
&&\alpha_{_{1,2}}=0\;,\;\alpha_{_{1,6}}=-1\;,\;\alpha_{_{2,2}}=0\;,\;
\alpha_{_{2,6}}=D-3\;,\;\alpha_{_{3,2}}=0,\;\alpha_{_{3,6}}=2-{D\over2}\;,
\label{1SE-2-130}
\end{eqnarray}
where the geometric representation of the exponent matrix is presented in Fig.\ref{fig4}(1)
with $\{(a,b)$, $(c,d)\}=$ $\{(2,3)$, $(4,5)\}$, $\{e,f\}=$ $\{1,6\}$.
For the integer lattice $n_{_1}E_{_{3}}^{(2)}-n_{_2}E_{_{3}}^{(1)}$,
the matrix of exponents is given by
\begin{eqnarray}
&&\alpha_{_{1,2}}=0\;,\;\alpha_{_{1,6}}=-1\;,\;\alpha_{_{2,2}}=D-3\;,\;
\alpha_{_{2,6}}=0\;,\;\alpha_{_{3,2}}=3-D,\;\alpha_{_{3,6}}={D\over2}-1\;,
\label{1SE-2-134}
\end{eqnarray}
whose geometric representation is shown in Fig.\ref{fig4}(3)
with $(a,b)=$ $(4,5)$, $\{(c,d)$, $(e,f)\}=$ $\{(1,6)$, $(2,3)\}$.

Correspondingly the hypergeometric functions are
\begin{eqnarray}
&&\psi_{_{\{1,3,4\}}}^{(7)}(p^2,m_{_1}^2,m_{_2}^2)\sim
(p^2-m_{_2}^2)^{D-4}(m_{_2}^2)^{2-D/2}
\sum\limits_{n_{_1},n_{_2}}^\infty
{\Gamma(3-D+n_{_1})\Gamma({D\over2}-2+n_{_2})\over n_{_1}!n_{_2}!}
\nonumber\\
&&\hspace{3.9cm}\times
\Big({m_{_2}^2-m_{_1}^2\over m_{_2}^2-p^2}\Big)^{n_{_1}}
\Big({m_{_2}^2-p^2\over m_{_2}^2}\Big)^{n_{_2}}
\nonumber\\
&&\hspace{3.5cm}\sim
{(p^2-m_{_1}^2)^{D-3}(p^2)^{2-D/2}\over p^2-m_{_2}^2}
\;,\nonumber\\
&&\psi_{_{\{1,2,4\}}}^{(8)}(p^2,m_{_1}^2,m_{_2}^2)
\sim{(m_{_2}^2-m_{_1}^2)^{D-3}(m_{_2}^2)^{2-D/2}\over m_{_2}^2-p^2}
\sum\limits_{n_{_1},n_{_2}}^\infty
{(-)^{n_{_1}+n_{_2}}\Gamma({D\over2}-2+n_{_2})\over n_{_2}!
\Gamma(1-n_{_1}-n_{_2})}
\nonumber\\
&&\hspace{3.9cm}\times
{1\over\Gamma(D-2+n_{_1}+n_{_2})}
\Big({m_{_2}^2-m_{_1}^2\over m_{_2}^2-p^2}\Big)^{n_{_1}}
\Big({m_{_2}^2-m_{_1}^2\over m_{_2}^2}\Big)^{n_{_2}}
\nonumber\\
&&\hspace{3.5cm}\sim
{(m_{_2}^2-m_{_1}^2)^{D-3}(m_{_2}^2)^{2-D/2}\over m_{_2}^2-p^2}
\;,\nonumber\\
&&\psi_{_{\{1,3,4\}}}^{(9)}(p^2,m_{_1}^2,m_{_2}^2)
\sim{(m_{_2}^2)^{D/2-1}\over m_{_2}^2-p^2}
\sum\limits_{n_{_1},n_{_2}}^\infty
{\Gamma(D-3+n_{_1}-n_{_2})\Gamma(3-D+n_{_2})\over n_{_2}!\Gamma({D\over2}+n_{_1}-n_{_2})}
\nonumber\\
&&\hspace{3.9cm}\times
\Big({m_{_2}^2\over m_{_2}^2-p^2}\Big)^{n_{_1}}
\Big({m_{_2}^2-m_{_1}^2\over m_{_2}^2}\Big)^{n_{_2}}
\nonumber\\
&&\hspace{3.5cm}\sim
{(m_{_2}^2)^{D/2-1}(m_{_1}^2-p^2)^{D-3}\over(m_{_2}^2-p^2)^{D-2}}
\nonumber\\
&&\hspace{3.9cm}\times
\sum\limits_{n_{_1}=0}^\infty
{\Gamma(D-3+n_{_1})\over\Gamma({D\over2}+n_{_1})}
\Big({m_{_2}^2\over m_{_2}^2-p^2}\Big)^{n_{_1}}\;,
\label{1SE-2-130a}
\end{eqnarray}
Although the intersection of the convergent regions of these series is
the empty set, the sets of the hypergeometric functions
$\{\psi_{_{\{1,3,4\}}}^{(1)},\psi_{_{\{1,3,4\}}}^{(8)},\psi_{_{\{1,3,4\}}}^{(9)}\}$
and $\{\psi_{_{\{1,3,4\}}}^{(6)},\psi_{_{\{1,3,4\}}}^{(7)},\psi_{_{\{1,3,4\}}}^{(8)}\}$
constitute the fundamental solution systems of the nonempty proper subsets
of the whole parameter space, respectively.

In a similar way, the matrix of exponents of the integer lattice
$-n_{_1}E_{_{3}}^{(3)}+n_{_2}E_{_{3}}^{(2)}$ is
\begin{eqnarray}
&&\alpha_{_{1,2}}={D\over2}-2\;,\;\alpha_{_{1,6}}=1-{D\over2}\;,\;\alpha_{_{2,2}}=0\;,\;
\alpha_{_{2,6}}=D-3\;,\;\alpha_{_{3,2}}=2-{D\over2},\;\alpha_{_{3,6}}=0\;,
\label{1SE-2-127}
\end{eqnarray}
whose geometric representation is presented in Fig.\ref{fig4}(2)
with $(a,b)=$ $(4,5)$, $c=1$, $d=2$, $\{e,f\}=$ $\{3,6\}$.
The exponents of the integer lattice
$-n_{_1}E_{_{3}}^{(3)}+n_{_2}E_{_{3}}^{(1)}$ are
\begin{eqnarray}
&&\alpha_{_{1,2}}=-1\;,\;\alpha_{_{1,6}}=0\;,\;\alpha_{_{2,2}}={D\over2}-1\;,\;
\alpha_{_{2,6}}={D\over2}-2\;,\;\alpha_{_{3,2}}=2-{D\over2},\;\alpha_{_{3,6}}=0\;,
\label{1SE-2-131}
\end{eqnarray}
whose geometric representation is drawn in Fig.\ref{fig4}(1)
with $\{(a,b)$, $(c,d)\}=$ $\{(3,6)$, $(4,5)\}$, $\{e,f\}=$ $\{1,2\}$.
Finally the choice on exponents of the integer lattice
$-n_{_1}E_{_{3}}^{(2)}+n_{_2}E_{_{3}}^{(1)}$ is
\begin{eqnarray}
&&\alpha_{_{1,2}}=-1\;,\;\alpha_{_{1,6}}=0\;,\;\alpha_{_{2,2}}=0\;,\;
\alpha_{_{2,6}}=D-3\;,\;\alpha_{_{3,2}}=1,\;\alpha_{_{3,6}}=1-{D\over2}\;,
\label{1SE-2-135}
\end{eqnarray}
whose geometric representation is presented in Fig.\ref{fig4}(3)
with $(a,b)=$ $(4,5)$, $\{(c,d)$, $(e,f)\}=$ $\{(1,2)$, $(3,6)\}$.
Accordingly the hypergeometric functions are written as
\begin{eqnarray}
&&\psi_{_{\{1,3,4\}}}^{(10)}(p^2,m_{_1}^2,m_{_2}^2)\sim
(p^2-m_{_2}^2)^{1-D/2}(m_{_1}^2-m_{_2}^2)^{D-3}
\sum\limits_{n_{_1},n_{_2}}^\infty
{\Gamma(3-D+n_{_1})\over n_{_1}!n_{_2}!}
\nonumber\\
&&\hspace{3.9cm}\times
\Gamma({D\over2}-2+n_{_2})
\Big({m_{_2}^2-p^2\over m_{_2}^2-m_{_1}^2}\Big)^{n_{_1}}
\Big({m_{_2}^2\over m_{_2}^2-p^2}\Big)^{n_{_2}}
\nonumber\\
&&\hspace{3.5cm}\sim
{(p^2-m_{_1}^2)^{D-3}(p^2)^{2-D/2}\over p^2-m_{_2}^2}
\;,\nonumber\\
&&\psi_{_{\{1,3,4\}}}^{(11)}(p^2,m_{_1}^2,m_{_2}^2)
\sim(m_{_2}^2-m_{_1}^2)^{D/2-2}\sum\limits_{n_{_1},n_{_2}}^\infty
{\Gamma(2-{D\over2}+n_{_1}+n_{_2})\over n_{_2}!\Gamma({D\over2}+n_{_1}+n_{_2})}
\nonumber\\
&&\hspace{3.9cm}\times
\Gamma({D\over2}-2+n_{_2})
\Big({m_{_2}^2-p^2\over m_{_2}^2-m_{_1}^2}\Big)^{n_{_1}}
\Big({m_{_2}^2\over m_{_2}^2-m_{_1}^2}\Big)^{n_{_2}}
\;,\nonumber\\
&&\psi_{_{\{1,3,4\}}}^{(12)}(p^2,m_{_1}^2,m_{_2}^2)
\sim(m_{_2}^2)^{3-D/2}(m_{_2}^2-m_{_1}^2)^{D-5}\sum\limits_{n_{_1},n_{_2}}^\infty
{\Gamma(5-D+n_{_1}+n_{_2})\over\Gamma(3+n_{_1}+n_{_2})}
\nonumber\\
&&\hspace{3.9cm}\times
{\Gamma(1+n_{_2})\over\Gamma(4-{D\over2}+n_{_2})}
\Big({m_{_2}^2-p^2\over m_{_2}^2-m_{_1}^2}\Big)^{n_{_1}}
\Big({m_{_2}^2\over m_{_2}^2-m_{_1}^2}\Big)^{n_{_2}}\;,
\label{1SE-2-135a}
\end{eqnarray}
Certainly the intersection of the convergent regions of these series is
a nonempty proper subset of the whole parameter space. Furthermore the sets of the hypergeometric functions
$\{\psi_{_{\{1,3,4\}}}^{(3)},\psi_{_{\{1,3,4\}}}^{(10)},\psi_{_{\{1,3,4\}}}^{(11)}\}$
and $\{\psi_{_{\{1,3,4\}}}^{(4)},\psi_{_{\{1,3,4\}}}^{(11)},\psi_{_{\{1,3,4\}}}^{(12)}\}$
also constitute the fundamental solution systems of the corresponding nonempty proper subsets
of the whole parameter space, respectively. Similarly the hypergeometric solution
of the exponents of Eq.(\ref{1SE-2-135}) is
\begin{eqnarray}
&&\psi_{_{\{1,3,4\}}}^{(12)\prime}(p^2,m_{_1}^2,m_{_2}^2)
\sim(m_{_2}^2)^{1-D/2}(m_{_2}^2-p^2)^{D-3}\sum\limits_{n_{_1},n_{_2}}^\infty
{\Gamma({D\over2}-1+n_{_1}-n_{_2})\over n_{_2}!\Gamma(2+n_{_1}-n_{_2})}
\nonumber\\
&&\hspace{3.9cm}\times
\Gamma(3-D+n_{_2})\Big({m_{_2}^2-m_{_1}^2\over m_{_2}^2}\Big)^{n_{_1}}
\Big({m_{_2}^2\over m_{_2}^2-p^2}\Big)^{n_{_2}}
\nonumber\\
&&\hspace{3.5cm}\sim
-\psi_{_{\{1,3,4\}}}^{(10)}(p^2,m_{_1}^2,m_{_2}^2)
\nonumber\\
&&\hspace{3.9cm}
+\lim\limits_{\epsilon\rightarrow0}{\sin\pi\epsilon\over\sin\pi({D\over2}-3)}
\psi_{_{\{1,3,4\}}}^{(12)}(p^2,m_{_1}^2,m_{_2}^2)\;.
\label{1SE-2-135b}
\end{eqnarray}

Because $\det(A_{_{\{2,3,4\}}}^{(1S)})=1$, we choose the second, third, fourth
column vectors composing of the basis of the projective plane, and find
\begin{eqnarray}
&&\Big(A_{_{\{2,3,4\}}}^{(1S)}\Big)^{-1}\cdot A^{(1S)}=\left(\begin{array}{cccccc}
\;\;-1\;\;&\;\;1\;\;&\;\;0\;\;&\;\;0\;\;&\;\;0\;\;&\;\;m_{_2}^2-p^2\;\;\\
\;\;-1\;\;&\;\;0\;\;&\;\;1\;\;&\;\;0\;\;&\;\;0\;\;&\;\;m_{_1}^2-p^2\;\;\\
\;\;1\;\;&\;\;0\;\;&\;\;0\;\;&\;\;1\;\;&\;\;1\;\;&\;\;p^2\;\;\end{array}\right)\;.
\label{1SE-2-139}
\end{eqnarray}
Obviously the matrix of exponents is written as
\begin{eqnarray}
&&\left(\begin{array}{cccccc}
\;\alpha_{_{1,1}}\;\;&\;\;0\;\;&\;\;0\;\;&\;\;0\;\;&\;\;0\;\;&\;\alpha_{_{1,6}}\;\;\\
\;\alpha_{_{2,1}}\;\;&\;\;0\;\;&\;\;2-D\;\;&\;\;0\;\;&\;\;0\;\;&\;\alpha_{_{2,6}}\;\;\\
\;\alpha_{_{3,1}}\;\;&\;\;0\;\;&\;\;0\;\;&\;\;-1\;\;&\;\;{D\over2}-2\;\;&\;\alpha_{_{3,6}}\;\;\end{array}\right)\;,
\label{1SE-2-140}
\end{eqnarray}
where the matrix elements satisfy the relations
\begin{eqnarray}
&&\alpha_{_{1,1}}+\alpha_{_{1,6}}=-1\;,\;\alpha_{_{2,1}}+\alpha_{_{2,6}}=D-3\;,\;
\alpha_{_{3,1}}+\alpha_{_{3,6}}=2-{D\over2}\;,
\nonumber\\
&&\alpha_{_{1,1}}+\alpha_{_{2,1}}+\alpha_{_{3,1}}=0\;,\;\alpha_{_{1,6}}+\alpha_{_{2,6}}+\alpha_{_{3,6}}={D\over2}-2\;.
\label{1SE-2-141}
\end{eqnarray}
Correspondingly there are 12 choices on the matrix of integer lattice whose submatrix
composed of the first and sixth columns is formulated
as $\pm n_{_1}E_{_{3}}^{(i)}\pm n_{_2}E_{_{3}}^{(j)}$,
where $n_{_{1,2}}\ge0$, $(i,\;j)\in\{(1,\;2),\;(1,\;3),\;(2,\;3)\}$,
and other elements are all zero. The exponents are obtained from those
corresponding to $\psi_{_{\{1,3,4\}}}^{(i)}$
through the exchanging $\alpha_{_{i,2}}\rightarrow\alpha_{_{i,1}}$, $i=1,2,3$.
Certainly the geometric representations of those exponent matrices are
obtained from that of the hypergeometric functions by the permutation $\widehat{(12)}$.
The hypergeometric functions are accordingly written as
\begin{eqnarray}
&&\psi_{_{\{2,3,4\}}}^{(i)}(p^2,m_{_1}^2,m_{_2}^2)
=\psi_{_{\{1,3,4\}}}^{(i)}(p^2,m_{_2}^2,m_{_1}^2),\;\;(i=1,\cdots,12).
\label{1SE-2-142}
\end{eqnarray}

Since $\det(A_{_{\{1,4,6\}}}^{(1S)})=m_{_1}^2-m_{_2}^2$, we choose the first, fourth, and sixth
column vectors composing of the basis of the projective plane, and find
\begin{eqnarray}
&&\Big(A_{_{\{1,4,6\}}}^{(1S)}\Big)^{-1}\cdot A^{(1S)}=\left(\begin{array}{cccccc}
\;\;1\;\;&\;\;-{m_{_1}^2-p^2\over m_{_1}^2-m_{_2}^2}\;\;&\;\;-{p^2-m_{_2}^2\over m_{_1}^2-m_{_2}^2}
\;\;&\;\;0\;\;&\;\;0\;\;&\;\;0\;\;\\
\;\;0\;\;&\;\;{m_{_1}^2\over m_{_1}^2-m_{_2}^2}\;\;&\;\;-{m_{_2}^2\over m_{_1}^2-m_{_2}^2}
\;\;&\;\;1\;\;&\;\;1\;\;&\;\;0\;\;\\
\;\;0\;\;&\;\;-{1\over m_{_1}^2-m_{_2}^2}\;\;&\;\;{1\over m_{_1}^2-m_{_2}^2}
\;\;&\;\;0\;\;&\;\;0\;\;&\;\;1\;\;\end{array}\right)\;.
\label{1SE-2-143}
\end{eqnarray}
Accordingly the matrix of exponents is
\begin{eqnarray}
&&\left(\begin{array}{cccccc}
\;\;0\;\;&\;\alpha_{_{1,2}}\;\;&\;\alpha_{_{1,3}}\;\;&\;\;0\;\;&\;\;0\;\;&\;\;0\;\;\\
\;\;0\;\;&\;\alpha_{_{2,2}}\;\;&\;\alpha_{_{2,3}}\;\;&\;\;-1\;\;&\;\;{D\over2}-2\;\;&\;\;0\;\;\\
\;\;0\;\;&\;\alpha_{_{3,2}}\;\;&\;\alpha_{_{3,3}}\;\;&\;\;0\;\;&\;\;0\;\;&\;\;{D\over2}-2\;\;
\end{array}\right)\;,
\label{1SE-2-144}
\end{eqnarray}
where the matrix elements satisfy the relations
\begin{eqnarray}
&&\alpha_{_{1,2}}+\alpha_{_{1,3}}=-1\;,\;\alpha_{_{2,2}}+\alpha_{_{2,3}}=2-{D\over2}\;,\;
\alpha_{_{3,2}}+\alpha_{_{3,3}}=1-{D\over2}\;,
\nonumber\\
&&\alpha_{_{1,2}}+\alpha_{_{2,2}}+\alpha_{_{3,2}}=0\;,\;
\alpha_{_{1,3}}+\alpha_{_{2,3}}+\alpha_{_{3,3}}=2-D\;.
\label{1SE-2-145}
\end{eqnarray}
There are 12 choices on the matrix of integer lattice whose submatrix
composed of the second and third columns is formulated as
$\pm n_{_1}E_{_{3}}^{(i)}\pm n_{_2}E_{_{3}}^{(j)}$,
where $n_{_{1,2}}\ge0$, and $(i,j)=$ $\{(1,2)$, $(1,3)$, $(2,\;3)\}$,
and other elements are all zero. Corresponding to the integer lattice
$n_{_1}E_{_{3}}^{(3)}+n_{_2}E_{_{3}}^{(2)}$, the choice of exponents is written as
\begin{eqnarray}
&&\alpha_{_{1,2}}=D-3\;,\;\alpha_{_{1,3}}=2-D\;,\;\alpha_{_{2,2}}=2-{D\over2}\;,\;
\alpha_{_{2,3}}=0\;,\;\alpha_{_{3,2}}=1-{D\over2},\;\alpha_{_{3,3}}=0\;,
\label{1SE-2-146}
\end{eqnarray}
whose geometric representation is presented in Fig.\ref{fig4}(1)
with $\{(a,b)$, $(c,d)\}=$ $\{(1,3)$, $(4,5)\}$, $\{e,f\}=$ $\{2,6\}$.
In a similar way the exponents
of the integer lattice $n_{_1}E_{_{3}}^{(3)}+n_{_2}E_{_{3}}^{(1)}$ are
\begin{eqnarray}
&&\alpha_{_{1,2}}=0\;,\;\alpha_{_{1,3}}=-1\;,\;\alpha_{_{2,2}}={D\over2}-1\;,\;
\alpha_{_{2,3}}=3-D\;,\;\alpha_{_{3,2}}=1-{D\over2},\;\alpha_{_{3,3}}=0\;,
\label{1SE-2-150}
\end{eqnarray}
whose geometric representation is shown in Fig.\ref{fig4}(3)
with $(a,b)=$ $(4,5)$, $\{(c,d)$, $(e,f)\}=$ $\{(1,3)$, $(2,6)\}$.
The exponents of the integer lattice $n_{_1}E_{_{3}}^{(2)}+n_{_2}E_{_{3}}^{(1)}$
are
\begin{eqnarray}
&&\alpha_{_{1,2}}=0\;,\;\alpha_{_{1,3}}=-1\;,\;\alpha_{_{2,2}}=0\;,\;
\alpha_{_{2,3}}=2-{D\over2}\;,\;\alpha_{_{3,2}}=0,\;\alpha_{_{3,3}}=1-{D\over2}\;,
\label{1SE-2-154}
\end{eqnarray}
whose geometric representation is plotted in Fig.\ref{fig4}(1)
with $\{(a,b)$, $(c,d)\}=$ $\{(2,6)$, $(4,5)\}$, $\{e,f\}=$ $\{1,3\}$.
Three linear independent hypergeometric functions are
\begin{eqnarray}
&&\psi_{_{\{1,4,6\}}}^{(1)}(p^2,m_{_1}^2,m_{_2}^2)\sim
{[(m_{_1}^2-p^2)(m_{_1}^2-m_{_2}^2)]^{D-3}\over (p^2-m_{_2}^2)^{D-2}(m_{_1}^2)^{D/2-2}}
\sum\limits_{n_{_1},n_{_2}}^\infty
{\Gamma({D\over2}-2+n_{_1})\over n_{_1}!n_{_2}!}
\nonumber\\
&&\hspace{3.9cm}\times
\Gamma({D\over2}-1+n_{_2})
\Big({m_{_2}^2(p^2-m_{_1}^2)\over m_{_1}^2(p^2-m_{_2}^2)}\Big)^{n_{_1}}
\Big({p^2-m_{_1}^2\over p^2-m_{_2}^2}\Big)^{n_{_2}}
\nonumber\\
&&\hspace{3.5cm}\sim
{(p^2-m_{_1}^2)^{D-3}(p^2)^{2-D/2}\over p^2-m_{_2}^2}
\;,\nonumber\\
&&\psi_{_{\{1,4,6\}}}^{(2)}(p^2,m_{_1}^2,m_{_2}^2)\sim
{(m_{_2}^2-m_{_1}^2)^{D-3}(m_{_1}^2)^{D/2-1}\over(m_{_2}^2)^{D-3}(p^2-m_{_2}^2)}
\sum\limits_{n_{_1},n_{_2}}^\infty
{\Gamma(1-{D\over2}+n_{_1}-n_{_2})\over\Gamma(4-D+n_{_1}-n_{_2})n_{_2}!}
\nonumber\\
&&\hspace{3.9cm}\times
\Gamma({D\over2}-1+n_{_2})
\Big({m_{_2}^2(p^2-m_{_1}^2)\over m_{_1}^2(p^2-m_{_2}^2)}\Big)^{n_{_1}}
\Big({m_{_1}^2\over m_{_2}^2}\Big)^{n_{_2}}
\nonumber\\
&&\hspace{3.5cm}\sim
{(m_{_2}^2-m_{_1}^2)^{D-3}(m_{_1}^2)^{D/2-1}\over(m_{_2}^2)^{D-3}(p^2-m_{_2}^2)}
\sum\limits_{n_{_1},n_{_2}}^\infty
{\Gamma(1-{D\over2}+n_{_1})\over\Gamma(4-D+n_{_1})n_{_2}!}
\nonumber\\
&&\hspace{3.9cm}\times
\Gamma({D\over2}-1+n_{_2})
\Big({m_{_2}^2(p^2-m_{_1}^2)\over m_{_1}^2(p^2-m_{_2}^2)}\Big)^{n_{_1}}
\Big({p^2-m_{_1}^2\over p^2-m_{_2}^2}\Big)^{n_{_2}}
\nonumber\\
&&\hspace{3.5cm}\sim
{[(m_{_2}^2-m_{_1}^2)(p^2-m_{_2}^2)]^{D/2-2}(m_{_1}^2)^{D/2-1}\over(m_{_2}^2)^{D-3}}
\nonumber\\
&&\hspace{3.9cm}\times
\sum\limits_{n_{_1}=0}^\infty
{\Gamma(1-{D\over2}+n_{_1})\over\Gamma(4-D+n_{_1})}
\Big({m_{_2}^2(p^2-m_{_1}^2)\over m_{_1}^2(p^2-m_{_2}^2)}\Big)^{n_{_1}}
\;,\nonumber\\
&&\psi_{_{\{1,4,6\}}}^{(3)}(p^2,m_{_1}^2,m_{_2}^2)\sim
{(m_{_2}^2-m_{_1}^2)^{D-3}(m_{_2}^2)^{2-D/2}\over m_{_2}^2-p^2}
\sum\limits_{n_{_1},n_{_2}}^\infty
{(-)^{n_{_1}+n_{_2}}\Gamma({D\over2}-2+n_{_2})\over n_{_2}!\Gamma(1-n_{_1}-n_{_2})}
\nonumber\\
&&\hspace{3.9cm}\times
{1\over\Gamma(2-{D\over2}+n_{_1}+n_{_2})}
\Big({p^2-m_{_1}^2\over p^2-m_{_2}^2}\Big)^{n_{_1}}
\Big({m_{_1}^2\over m_{_2}^2}\Big)^{n_{_2}}
\nonumber\\
&&\hspace{3.5cm}\sim
{(m_{_2}^2-m_{_1}^2)^{D-3}(m_{_2}^2)^{2-D/2}\over m_{_2}^2-p^2}\;,
\label{1SE-2-154a}
\end{eqnarray}
whose intersection is a nonempty proper subset of the whole parameter space.
In other words, those hypergeometric functions constitute a fundamental solution
system of the GKZ-system of PDEs in a proper subset of the whole
parameter space.

Corresponding to the integer lattice $-n_{_1}E_{_{3}}^{(3)}-n_{_2}E_{_{3}}^{(2)}$,
the matrix of exponents is
\begin{eqnarray}
&&\alpha_{_{1,2}}=0\;,\;\alpha_{_{1,3}}=-1\;,\;\alpha_{_{2,2}}=0\;,\;
\alpha_{_{2,3}}=2-{D\over2}\;,\;\alpha_{_{3,2}}=0,\;\alpha_{_{3,3}}=1-{D\over2}\;,
\label{1SE-2-149}
\end{eqnarray}
whose geometric representation is presented in Fig.\ref{fig4}(1)
with $\{(a,b)$, $(c,d)\}=$ $\{(1,2)$, $(4,5)\}$, $\{e,f\}=$ $\{3,6\}$.
Similarly the exponents of the integer lattice
$-n_{_1}E_{_{3}}^{(3)}-n_{_2}E_{_{3}}^{(1)}$ are
\begin{eqnarray}
&&\alpha_{_{1,2}}=-1\;,\;\alpha_{_{1,3}}=0\;,\;\alpha_{_{2,2}}=1\;,\;
\alpha_{_{2,3}}=1-{D\over2}\;,\;\alpha_{_{3,2}}=0,\;\alpha_{_{3,3}}=1-{D\over2}\;,
\label{1SE-2-153}
\end{eqnarray}
whose geometric representation is drawn in Fig.\ref{fig4}(3)
with $(a,b)=(4,5)$, $\{(c,d)$, $(e,f)\}=$ $\{(1,2)$, $(3,6)\}$.
Meanwhile the exponents of the integer lattice
$-n_{_1}E_{_{3}}^{(2)}-n_{_2}E_{_{3}}^{(1)}$ are
\begin{eqnarray}
&&\alpha_{_{1,2}}=-1\;,\;\alpha_{_{1,3}}=0\;,\;\alpha_{_{2,2}}=2-{D\over2}\;,\;
\alpha_{_{2,3}}=0\;,\;\alpha_{_{3,2}}={D\over2}-1,\;\alpha_{_{3,3}}=2-D\;,
\label{1SE-2-157}
\end{eqnarray}
whose geometric representation is presented in Fig.\ref{fig4}(1)
with $\{(a,b)$, $(c,d)\}=$ $\{(3,6)$, $(4,5)\}$, $\{e,f\}=$ $\{1,2\}$.
The corresponding hypergeometric functions are
\begin{eqnarray}
&&\psi_{_{\{1,4,6\}}}^{(4)}(p^2,m_{_1}^2,m_{_2}^2)\sim
{(m_{_2}^2-m_{_1}^2)^{D-3}(m_{_2}^2)^{2-D/2}\over m_{_2}^2-p^2}
\sum\limits_{n_{_1},n_{_2}}^\infty
{\Gamma({D\over2}-2+n_{_1})\over n_{_1}!n_{_2}!}
\nonumber\\
&&\hspace{3.9cm}\times
\Gamma({D\over2}-1+n_{_2})
\Big({m_{_1}^2(p^2-m_{_2}^2)\over m_{_2}^2(p^2-m_{_1}^2)}\Big)^{n_{_1}}
\Big({p^2-m_{_2}^2\over p^2-m_{_1}^2}\Big)^{n_{_2}}
\nonumber\\
&&\hspace{3.5cm}\sim
{(p^2-m_{_1}^2)^{D-3}(p^2)^{2-D/2}\over p^2-m_{_2}^2}
\;,\nonumber\\
&&\psi_{_{\{1,4,6\}}}^{(5)}(p^2,m_{_1}^2,m_{_2}^2)\sim
{(m_{_2}^2-m_{_1}^2)^{D-3}(m_{_2}^2)^{3-D/2}\over m_{_1}^2(p^2-m_{_1}^2)}
\sum\limits_{n_{_1},n_{_2}}^\infty
{\Gamma({D\over2}+1+n_{_1}+n_{_2})\over \Gamma(3+n_{_1}+n_{_2})}
\nonumber\\
&&\hspace{3.9cm}\times
{\Gamma(1+n_{_2})\over\Gamma(4-{D\over2}+n_{_2})}
\Big({p^2-m_{_2}^2\over p^2-m_{_1}^2}\Big)^{n_{_1}}
\Big({m_{_2}^2\over m_{_1}^2}\Big)^{n_{_2}}
\;,\nonumber\\
&&\psi_{_{\{1,4,6\}}}^{(6)}(p^2,m_{_1}^2,m_{_2}^2)\sim
{(m_{_2}^2-m_{_1}^2)^{D/2-2}(m_{_2}^2)^{D/2-1}\over(p^2-m_{_1}^2)(m_{_1}^2)^{D/2-2}}
\sum\limits_{n_{_1},n_{_2}}^\infty
{\Gamma(D-2+n_{_1}+n_{_2})\over\Gamma({D\over2}+n_{_1}+n_{_2})n_{_2}!}
\nonumber\\
&&\hspace{3.9cm}\times
\Gamma({D\over2}-2+n_{_2})
\Big({p^2-m_{_2}^2\over p^2-m_{_1}^2}\Big)^{n_{_1}}
\Big({m_{_2}^2\over m_{_1}^2}\Big)^{n_{_2}}
\nonumber\\
&&\hspace{3.5cm}\sim
{(m_{_2}^2-m_{_1}^2)^{D/2-2}(m_{_2}^2)^{D/2-1}\over(p^2-m_{_1}^2)(m_{_1}^2)^{D/2-2}}
\sum\limits_{n_{_1},n_{_2}}^\infty
{\Gamma(D-2+n_{_1})\over\Gamma({D\over2}+n_{_1})n_{_2}!}
\nonumber\\
&&\hspace{3.9cm}\times
\Gamma({D\over2}-2+n_{_2})
\Big({p^2-m_{_2}^2\over p^2-m_{_1}^2}\Big)^{n_{_1}}
\Big({m_{_2}^2(p^2-m_{_1}^2)\over m_{_1}^2(p^2-m_{_2}^2)}\Big)^{n_{_2}}
\nonumber\\
&&\hspace{3.5cm}\sim
{(m_{_2}^2-p^2)^{D/2-2}(m_{_2}^2)^{D/2-1}\over(p^2-m_{_1}^2)(p^2)^{D/2-2}}
\nonumber\\
&&\hspace{3.9cm}\times
\sum\limits_{n_{_1},n_{_2}}^\infty
{\Gamma(D-2+n_{_1})\over\Gamma({D\over2}+n_{_1})}
\Big({p^2-m_{_2}^2\over p^2-m_{_1}^2}\Big)^{n_{_1}}\;,
\label{1SE-2-157a}
\end{eqnarray}
whose intersection of convergent regions is a nonempty proper subset of the whole parameter space.
Correspondingly the hypergeometric functions with the exponents of Eq.(\ref{1SE-2-153})
is written as
\begin{eqnarray}
&&\psi_{_{\{1,4,6\}}}^{(5)\prime}(p^2,m_{_1}^2,m_{_2}^2)\sim
{(m_{_2}^2-p^2)^{D-3}p^2\over (m_{_1}^2-p^2)(m_{_2}^2)^{D/2-1}}
\sum\limits_{n_{_1},n_{_2}}^\infty
{\Gamma({D\over2}-1+n_{_1}-n_{_2})\over \Gamma(2+n_{_1}-n_{_2})n_{_2}!}
\nonumber\\
&&\hspace{3.9cm}\times
\Gamma({D\over2}-1+n_{_2})
\Big({p^2(m_{_1}^2-m_{_2}^2)\over m_{_2}^2(m_{_1}^2-p^2)}\Big)^{n_{_1}}
\Big({m_{_2}^2\over p^2}\Big)^{n_{_2}}
\nonumber\\
&&\hspace{3.5cm}\sim
\psi_{_{\{1,4,6\}}}^{(4)}(p^2,m_{_1}^2,m_{_2}^2)
\nonumber\\
&&\hspace{3.9cm}
+\lim\limits_{\epsilon\rightarrow0}{\sin\pi\epsilon\over\sin\pi({D\over2}-3)}
\psi_{_{\{1,4,6\}}}^{(5)}(p^2,m_{_1}^2,m_{_2}^2)\;.
\label{1SE-2-157b}
\end{eqnarray}

As the integer lattice is chosen as $n_{_1}E_{_{3}}^{(3)}-n_{_2}E_{_{3}}^{(2)}$,
the exponents are written as
\begin{eqnarray}
&&\alpha_{_{1,2}}={D\over2}-2\;,\;\alpha_{_{1,3}}=1-{D\over2}\;,\;\alpha_{_{2,2}}=2-{D\over2}\;,\;
\alpha_{_{2,3}}=0\;,\;\alpha_{_{3,2}}=0,\;\alpha_{_{3,3}}=1-{D\over2}\;,
\label{1SE-2-147}
\end{eqnarray}
whose geometric representation is plotted in Fig.\ref{fig4}(2)
with $(a,b)=$ $(4,5)$, $c=1$, $d=2$, $\{e,f\}=$ $\{3,6\}$.
The exponents of the integer lattice
$n_{_1}E_{_{3}}^{(3)}-n_{_2}E_{_{3}}^{(1)}$ are
\begin{eqnarray}
&&\alpha_{_{1,2}}=0\;,\;\alpha_{_{1,3}}=-1\;,\;\alpha_{_{2,2}}=0\;,\;
\alpha_{_{2,3}}=2-{D\over2}\;,\;\alpha_{_{3,2}}=0,\;\alpha_{_{3,3}}=1-{D\over2}\;,
\label{1SE-2-151}
\end{eqnarray}
whose geometric representation is shown in Fig.\ref{fig4}(4)
with $(a,b,c)=$ $(2,4,5)\}$, $(d,e,f)=$ $(1,3,6)$. The
exponents of the integer lattice $n_{_1}E_{_{3}}^{(2)}-n_{_2}E_{_{3}}^{(1)}$
are given by
\begin{eqnarray}
&&\alpha_{_{1,2}}=0\;,\;\alpha_{_{1,3}}=-1\;,\;\alpha_{_{2,2}}=2-{D\over2}\;,\;
\alpha_{_{2,3}}=0\;,\;\alpha_{_{3,2}}={D\over2}-2,\;\alpha_{_{3,3}}=3-D\;,
\label{1SE-2-155}
\end{eqnarray}
where the geometric representation is presented in Fig.\ref{fig4}(2)
with $(a,b)=$ $(4,5)$, $c=6$, $d=2$, $(e,f)=$ $(1,3)$. Correspondingly,
the hypergeometric functions are
\begin{eqnarray}
&&\psi_{_{\{1,4,6\}}}^{(7)}(p^2,m_{_1}^2,m_{_2}^2)\sim
{(p^2-m_{_1}^2)^{D/2-2}(m_{_2}^2-m_{_1}^2)^{D-3}\over (p^2-m_{_2}^2)^{D/2-1}(m_{_1}^2)^{D/2-2}}
\sum\limits_{n_{_1},n_{_2}}^\infty{\Gamma({D\over2}-2+n_{_1})\over n_{_1}!n_{_2}!}
\nonumber\\
&&\hspace{3.9cm}\times
\Gamma({D\over2}-1+n_{_2})
\Big({m_{_2}^2(p^2-m_{_1}^2)\over m_{_1}^2(p^2-m_{_2}^2)}\Big)^{n_{_1}}
\Big({p^2-m_{_2}^2\over p^2-m_{_1}^2}\Big)^{n_{_2}}
\nonumber\\
&&\hspace{3.5cm}\sim
{(p^2-m_{_1}^2)^{D-3}(p^2)^{2-D/2}\over p^2-m_{_2}^2}
\;,\nonumber\\
&&\psi_{_{\{1,4,6\}}}^{(8)}(p^2,m_{_1}^2,m_{_2}^2)\sim
{(m_{_2}^2-m_{_1}^2)^{D-3}(m_{_2}^2)^{2-D/2}\over m_{_2}^2-p^2}
\sum\limits_{n_{_1},n_{_2}}^\infty
{(-)^{n_{_1}+n_{_2}}\Gamma({D\over2}-1+n_{_2})\over n_{_2}!\Gamma(1-n_{_1}-n_{_2})}
\nonumber\\
&&\hspace{3.9cm}\times
{1\over\Gamma(3-{D\over2}+n_{_1}+n_{_2})}
\Big({m_{_2}^2(p^2-m_{_1}^2)\over m_{_1}^2(p^2-m_{_2}^2)}\Big)^{n_{_1}}
\Big({m_{_2}^2\over m_{_1}^2}\Big)^{n_{_2}}
\nonumber\\
&&\hspace{3.5cm}\sim
{(m_{_2}^2-m_{_1}^2)^{D-3}(m_{_2}^2)^{2-D/2}\over m_{_2}^2-p^2}
\;,\nonumber\\
&&\psi_{_{\{1,4,6\}}}^{(9)}(p^2,m_{_1}^2,m_{_2}^2)\sim
{(m_{_2}^2-m_{_1}^2)^{D-3}(m_{_1}^2)^{2-D/2}\over m_{_2}^2-p^2}
\sum\limits_{n_{_1},n_{_2}}^\infty
{\Gamma(2-{D\over2}+n_{_1}-n_{_2})\over\Gamma(4-D+n_{_1}-n_{_2})n_{_2}!}
\nonumber\\
&&\hspace{3.9cm}\times
\Gamma({D\over2}-2+n_{_2})
\Big({p^2-m_{_1}^2\over p^2-m_{_2}^2}\Big)^{n_{_1}}
\Big({m_{_2}^2\over m_{_1}^2}\Big)^{n_{_2}}
\nonumber\\
&&\hspace{3.5cm}\sim
{(m_{_2}^2-m_{_1}^2)^{D-3}(m_{_1}^2)^{2-D/2}\over m_{_2}^2-p^2}
\sum\limits_{n_{_1},n_{_2}}^\infty
{\Gamma(2-{D\over2}+n_{_1})\over\Gamma(4-D+n_{_1})n_{_2}!}
\nonumber\\
&&\hspace{3.9cm}\times
\Gamma({D\over2}-2+n_{_2})
\Big({p^2-m_{_1}^2\over p^2-m_{_2}^2}\Big)^{n_{_1}}
\Big({m_{_2}^2(p^2-m_{_1}^2)\over m_{_1}^2(p^2-m_{_2}^2)}\Big)^{n_{_2}}
\nonumber\\
&&\hspace{3.5cm}\sim
{(m_{_2}^2-m_{_1}^2)^{D/2-1}(m_{_2}^2-p^2)^{D/2-3}\over(p^2)^{D/2-2}}
\nonumber\\
&&\hspace{3.9cm}\times
\sum\limits_{n_{_1}=0}^\infty
{\Gamma(2-{D\over2}+n_{_1})\over\Gamma(4-D+n_{_1})}
\Big({p^2-m_{_1}^2\over p^2-m_{_2}^2}\Big)^{n_{_1}}\;.
\label{1SE-2-155a}
\end{eqnarray}
Although the intersection of the convergent regions of these series is
the empty subset, the sets of the hypergeometric functions
$\{\psi_{_{\{1,4,6\}}}^{(1)},\psi_{_{\{1,4,6\}}}^{(8)},\psi_{_{\{1,4,6\}}}^{(9)}\}$
and $\{\psi_{_{\{1,4,6\}}}^{(6)},\psi_{_{\{1,4,6\}}}^{(7)},\psi_{_{\{1,4,6\}}}^{(8)}\}$
constitute the fundamental solution systems of the corresponding nonempty proper subsets
of the whole parameter space, respectively.

The matrix of exponents of the integer lattice
$-n_{_1}E_{_{3}}^{(3)}+n_{_2}E_{_{3}}^{(2)}$ is
\begin{eqnarray}
&&\alpha_{_{1,2}}={D\over2}-1\;,\;\alpha_{_{1,3}}=-{D\over2}\;,\;\alpha_{_{2,2}}=0\;,\;
\alpha_{_{2,3}}=2-{D\over2}\;,\;\alpha_{_{3,2}}=1-{D\over2},\;\alpha_{_{3,3}}=0\;,
\label{1SE-2-148}
\end{eqnarray}
whose geometric representation is drawn in Fig.\ref{fig4}(2)
with $(a,b)=$ $(4,5)$, $c=1$, $d=3$, $\{e,f\}=$ $\{2,6\}$.
The exponents of the integer lattice $-n_{_1}E_{_{3}}^{(3)}+n_{_2}E_{_{3}}^{(1)}$ are
\begin{eqnarray}
&&\alpha_{_{1,2}}=-1\;,\;\alpha_{_{1,3}}=0\;,\;\alpha_{_{2,2}}={D\over2}\;,\;
\alpha_{_{2,3}}=2-D\;,\;\alpha_{_{3,2}}=1-{D\over2},\;\alpha_{_{3,3}}=0\;,
\label{1SE-2-152}
\end{eqnarray}
whose geometric representation is presented in Fig.\ref{fig4}(4)
with $(a,b,c)=$ $(3,4,5)$, $(d,e,f\}=$ $(1,2,6)$.
Finally the choice on exponents of integer lattice
$-n_{_1}E_{_{3}}^{(2)}+n_{_2}E_{_{3}}^{(1)}$ is
\begin{eqnarray}
&&\alpha_{_{1,2}}=-1\;,\;\alpha_{_{1,3}}=0\;,\;\alpha_{_{2,2}}=0\;,\;
\alpha_{_{2,3}}=2-{D\over2}\;,\;\alpha_{_{3,2}}=1,\;\alpha_{_{3,3}}=-{D\over2}\;,
\label{1SE-2-156}
\end{eqnarray}
whose geometric representation is plotted in Fig.\ref{fig4}(2)
with $(a,b)=$ $(4,5)$, $c=6$, $d=3$, $(e,f)=$ $(1,2)$.
Three corresponding power series are presented as
\begin{eqnarray}
&&\psi_{_{\{1,4,6\}}}^{(10)}(p^2,m_{_1}^2,m_{_2}^2)\sim
{(p^2-m_{_1}^2)^{D/2-1}(m_{_2}^2-m_{_1}^2)^{D-3}\over(m_{_2}^2)^{D/2-2}
(p^2-m_{_2}^2)^{D/2}}\sum\limits_{n_{_1},n_{_2}}^\infty
{\Gamma({D\over2}-2+n_{_1})\over n_{_1}!n_{_2}!}
\nonumber\\
&&\hspace{3.9cm}\times
\Gamma({D\over2}-1+n_{_2})
\Big({m_{_1}^2(p^2-m_{_2}^2)\over m_{_2}^2(p^2-m_{_1}^2)}\Big)^{n_{_1}}
\Big({p^2-m_{_1}^2\over p^2-m_{_2}^2}\Big)^{n_{_2}}
\nonumber\\
&&\hspace{3.5cm}\sim
{(p^2-m_{_1}^2)^{D-3}(p^2)^{2-D/2}\over p^2-m_{_2}^2}
\;,\nonumber\\
&&\psi_{_{\{1,4,6\}}}^{(11)}(p^2,m_{_1}^2,m_{_2}^2)\sim
{(m_{_2}^2-m_{_1}^2)^{D-3}(m_{_1}^2)^{D/2}\over(p^2-m_{_1}^2)(m_{_2}^2)^{D-2}}
\sum\limits_{n_{_1},n_{_2}}^\infty
{\Gamma(D-2+n_{_1}+n_{_2})\over\Gamma(1+{D\over2}+n_{_1}+n_{_2})}
\nonumber\\
&&\hspace{3.9cm}\times
{\Gamma({D\over2}-1+n_{_2})\over n_{_2}!}
\Big({m_{_1}^2(p^2-m_{_2}^2)\over m_{_2}^2(p^2-m_{_1}^2)}\Big)^{n_{_1}}
\Big({m_{_1}^2\over m_{_2}^2}\Big)^{n_{_2}}
\nonumber\\
&&\hspace{3.5cm}\sim
{(m_{_2}^2-m_{_1}^2)^{D-3}(m_{_1}^2)^{D/2}\over(p^2-m_{_1}^2)(m_{_2}^2)^{D-2}}
\sum\limits_{n_{_1},n_{_2}}^\infty
{\Gamma(D-2+n_{_1})\over\Gamma(1+{D\over2}+n_{_1})}
\nonumber\\
&&\hspace{3.9cm}\times
{\Gamma({D\over2}-1+n_{_2})\over n_{_2}!}
\Big({m_{_1}^2(p^2-m_{_2}^2)\over m_{_2}^2(p^2-m_{_1}^2)}\Big)^{n_{_1}}
\Big({p^2-m_{_1}^2\over p^2-m_{_2}^2}\Big)^{n_{_2}}
\nonumber\\
&&\hspace{3.5cm}\sim
{(m_{_2}^2-m_{_1}^2)^{D/2-2}(p^2-m_{_2}^2)^{D/2-1}(m_{_1}^2)^{D/2}
\over(p^2-m_{_1}^2)(m_{_2}^2)^{D-2}}
\nonumber\\
&&\hspace{3.9cm}\times
\sum\limits_{n_{_1},n_{_2}}^\infty
{\Gamma(D-2+n_{_1})\over\Gamma(1+{D\over2}+n_{_1})}
\Big({m_{_1}^2(p^2-m_{_2}^2)\over m_{_2}^2(p^2-m_{_1}^2)}\Big)^{n_{_1}}
\;,\nonumber\\
&&\psi_{_{\{1,4,6\}}}^{(12)}(p^2,m_{_1}^2,m_{_2}^2)\sim
{(m_{_1}^2)^2(m_{_2}^2-m_{_1}^2)^{D-3}\over (m_{_2}^2)^{2-D/2}(p^2-m_{_1}^2)}
\sum\limits_{n_{_1},n_{_2}}^\infty
{\Gamma({D\over2}+n_{_1}+n_{_2})\Gamma(1+n_{_2})
\over\Gamma(3+n_{_1}+n_{_2})\Gamma(-1-{D\over2}+n_{_2})}
\nonumber\\
&&\hspace{3.9cm}\times
\Big({m_{_1}^2(p^2-m_{_2}^2)\over m_{_2}^2(p^2-m_{_1}^2)}\Big)^{n_{_1}}
\Big({m_{_1}^2\over m_{_2}^2}\Big)^{n_{_2}}\;.
\label{1SE-2-156a}
\end{eqnarray}
Certainly intersection of convergent regions of these series is
the empty set, and the hypergeometric function with the exponents of
Eq.(\ref{1SE-2-156}) is written as
\begin{eqnarray}
&&\psi_{_{\{1,4,6\}}}^{(12)\prime}(p^2,m_{_1}^2,m_{_2}^2)\sim
{(m_{_2}^2-m_{_1}^2)^{D-3}(m_{_2}^2)^{2-D/2}\over p^2-m_{_1}^2}
\sum\limits_{n_{_1},n_{_2}}^\infty
{\Gamma({D\over2}+n_{_1}-n_{_2})\over\Gamma(2+n_{_1}-n_{_2})n_{_2}!}
\nonumber\\
&&\hspace{3.9cm}\times
\Gamma({D\over2}-2+n_{_2})
\Big({p^2-m_{_2}^2\over p^2-m_{_1}^2}\Big)^{n_{_1}}
\Big({m_{_1}^2\over m_{_2}^2}\Big)^{n_{_2}}
\nonumber\\
&&\hspace{3.5cm}\sim
\psi_{_{\{1,4,6\}}}^{(10)}(p^2,m_{_1}^2,m_{_2}^2)
\nonumber\\
&&\hspace{3.9cm}
+\lim\limits_{\epsilon\rightarrow0}{\sin\pi\epsilon\over\sin\pi({D\over2}+2)}
\psi_{_{\{1,4,6\}}}^{(12)}(p^2,m_{_1}^2,m_{_2}^2)\;.
\label{1SE-2-156b}
\end{eqnarray}
The sets of the hypergeometric functions
$\{\psi_{_{\{1,4,6\}}}^{(3)},\psi_{_{\{1,4,6\}}}^{(10)},\psi_{_{\{1,4,6\}}}^{(11)}\}$
and $\{\psi_{_{\{1,4,6\}}}^{(4)},\psi_{_{\{1,4,6\}}}^{(11)},\psi_{_{\{1,4,6\}}}^{(12)}\}$
evidently constitute the fundamental solution systems of the nonempty proper subsets
of the whole parameter space, respectively.

Because $\det(A_{_{\{2,4,6\}}}^{(1S)})=-p^2+m_{_1}^2$, and
\begin{eqnarray}
&&\Big(A_{_{\{2,4,6\}}}^{(1S)}\Big)^{-1}\cdot A^{(1S)}=\left(\begin{array}{cccccc}
\;\;-{m_{_1}^2-m_{_2}^2\over m_{_1}^2-p^2}\;\;&\;\;1\;\;&\;\;-{m_{_2}^2-p^2\over m_{_1}^2-p^2}
\;\;&\;\;0\;\;&\;\;0\;\;&\;\;0\;\;\\
\;\;{m_{_1}^2\over m_{_1}^2-p^2}\;\;&\;\;0\;\;&\;\;-{p^2\over m_{_1}^2-p^2}
\;\;&\;\;1\;\;&\;\;1\;\;&\;\;0\;\;\\
\;\;-{1\over m_{_1}^2-p^2}\;\;&\;\;0\;\;&\;\;{1\over m_{_1}^2-p^2}
\;\;&\;\;0\;\;&\;\;0\;\;&\;\;1\;\;\end{array}\right)\;,
\label{1SE-2-160}
\end{eqnarray}
the matrix of exponents is written as
\begin{eqnarray}
&&\left(\begin{array}{cccccc}
\;\;\alpha_{_{1,1}}\;\;&\;\;0\;\;&\;\alpha_{_{1,3}}\;\;&\;\;0\;\;&\;\;0\;\;&\;\;0\;\;\\
\;\;\alpha_{_{2,1}}\;\;&\;\;0\;\;&\;\alpha_{_{2,3}}\;\;&\;\;-1\;\;&\;\;{D\over2}-2\;\;&\;\;0\;\;\\
\;\;\alpha_{_{3,1}}\;\;&\;\;0\;\;&\;\alpha_{_{3,3}}\;\;&\;\;0\;\;&\;\;0\;\;&\;\;{D\over2}-2\;\;
\end{array}\right)\;,
\label{1SE-2-161}
\end{eqnarray}
where the matrix elements satisfy the relations
\begin{eqnarray}
&&\alpha_{_{1,1}}+\alpha_{_{1,3}}=-1\;,\;\alpha_{_{2,1}}+\alpha_{_{2,3}}=2-{D\over2}\;,\;
\alpha_{_{3,1}}+\alpha_{_{3,3}}=1-{D\over2}\;,
\nonumber\\
&&\alpha_{_{1,1}}+\alpha_{_{2,1}}+\alpha_{_{3,1}}=0\;,\;
\alpha_{_{1,3}}+\alpha_{_{2,3}}+\alpha_{_{3,3}}=2-D\;.
\label{1SE-2-162}
\end{eqnarray}
There are 12 choices on the matrix of integer lattice whose submatrix composed of the first and
third columns is formulated as $\pm n_{_1}E_{_{3}}^{(i)}\pm n_{_2}E_{_{3}}^{(j)}$,
where $n_{_{1,2}}\ge0$, $(i,\;j)\in\{(1,\;2),\;(1,\;3),\;(2,\;3)\}$,
and other elements are all zero. The exponents are obtained from those
corresponding to $\psi_{_{\{1,4,6\}}}^{(i)}$
through the exchanging $\alpha_{_{i,2}}\rightarrow\alpha_{_{i,1}}$, $i=1,2,3$.
Certainly the geometric representations of those exponent matrices are
obtained from that of the hypergeometric functions by the permutation $\widehat{(12)}$.
The hypergeometric functions are
\begin{eqnarray}
&&\psi_{_{\{2,4,6\}}}^{(i)}(p^2,m_{_1}^2,m_{_2}^2)
=\psi_{_{\{1,4,6\}}}^{(i)}(p^2,m_{_2}^2,m_{_1}^2),\;\;(i=1,\cdots,12).
\label{1SE-2-163}
\end{eqnarray}

Because $\det(A_{_{\{3,4,6\}}}^{(1S)})=m_{_2}^2-m_{_1}^2$, we choose the third,
fourth, and sixth columns as the basis of the projective plane, and get
\begin{eqnarray}
&&\Big(A_{_{\{3,4,6\}}}^{(1S)}\Big)^{-1}\cdot A^{(1S)}=\left(\begin{array}{cccccc}
\;\;{m_{_1}^2-m_{_2}^2\over m_{_2}^2-p^2}\;\;&\;\;-{m_{_1}^2-p^2\over m_{_2}^2-p^2}\;\;&\;\;1
\;\;&\;\;0\;\;&\;\;0\;\;&\;\;0\;\;\\
\;\;{m_{_2}^2\over m_{_2}^2-p^2}\;\;&\;\;-{p^2\over m_{_2}^2-p^2}\;\;&\;\;0
\;\;&\;\;1\;\;&\;\;1\;\;&\;\;0\;\;\\
\;\;-{1\over m_{_2}^2-p^2}\;\;&\;\;{1\over m_{_2}^2-p^2}\;\;&\;\;0
\;\;&\;\;0\;\;&\;\;0\;\;&\;\;1\;\;\end{array}\right)\;.
\label{1SE-2-164}
\end{eqnarray}
Certainly the matrix of exponents is written as
\begin{eqnarray}
&&\left(\begin{array}{cccccc}
\;\;\alpha_{_{1,1}}\;\;&\;\;\alpha_{_{1,2}}\;\;&\;\;2-D\;\;&\;\;0\;\;&\;\;0\;\;&\;\;0\;\;\\
\;\;\alpha_{_{2,1}}\;\;&\;\;\alpha_{_{2,2}}\;\;&\;\;0\;\;&\;\;-1\;\;&\;\;{D\over2}-2\;\;&\;\;0\;\;\\
\;\;\alpha_{_{3,1}}\;\;&\;\;\alpha_{_{3,2}}\;\;&\;\;0\;\;&\;\;0\;\;&\;\;0\;\;&\;\;{D\over2}-2\;\;
\end{array}\right)\;,
\label{1SE-2-165}
\end{eqnarray}
where the matrix elements satisfy the relations
\begin{eqnarray}
&&\alpha_{_{1,1}}+\alpha_{_{1,2}}=D-3\;,\;\alpha_{_{2,1}}+\alpha_{_{2,2}}=2-{D\over2}\;,\;
\alpha_{_{3,1}}+\alpha_{_{3,2}}=1-{D\over2}\;,
\nonumber\\
&&\alpha_{_{1,1}}+\alpha_{_{2,1}}+\alpha_{_{3,1}}=0\;,\;
\alpha_{_{1,2}}+\alpha_{_{2,2}}+\alpha_{_{3,2}}=0\;.
\label{1SE-2-166}
\end{eqnarray}
There are 12 choices on the matrix of integer lattice whose submatrix
composed of the first and second columns is formulated as
$\pm n_{_1}E_{_{3}}^{(i)}\pm n_{_2}E_{_{3}}^{(j)}$,
where $n_{_{1,2}}\ge0$, $(i,\;j)\in\{(1,\;2),\;(1,\;3),\;(2,\;3)\}$,
and other elements are all zero. As the integer lattice is chosen as
$n_{_1}E_{_{3}}^{(3)}+n_{_2}E_{_{3}}^{(2)}$,
the matrix of exponents is written as
\begin{eqnarray}
&&\alpha_{_{1,1}}=D-3\;,\;\alpha_{_{1,2}}=0\;,\;\alpha_{_{2,1}}=2-{D\over2}\;,\;
\alpha_{_{2,2}}=0\;,\;\alpha_{_{3,1}}=1-{D\over2},\;\alpha_{_{3,2}}=0\;,
\label{1SE-2-167}
\end{eqnarray}
whose geometric representation is plotted in Fig.\ref{fig4}(1)
with $\{(a,b)$, $(c,d)\}=$ $\{(2,3)$, $(4,5)\}$, $\{e,f\}=$ $\{1,6\}$.
In a similar way the exponents
of the integer lattice $n_{_1}E_{_{3}}^{(3)}+n_{_2}E_{_{3}}^{(1)}$ are
\begin{eqnarray}
&&\alpha_{_{1,1}}=0\;,\;\alpha_{_{1,2}}=D-3\;,\;\alpha_{_{2,1}}={D\over2}-1\;,\;
\alpha_{_{2,2}}=3-D\;,\;\alpha_{_{3,1}}=1-{D\over2},\;\alpha_{_{3,2}}=0\;,
\label{1SE-2-171}
\end{eqnarray}
where the geometric representation is presented in Fig.\ref{fig4}(3)
with $(a,b)=$ $(4,5)$, $\{(c,d)$, $(e,f)\}=$ $\{(1,6)$, $(2,3)\}$.
The exponents of the integer lattice $n_{_1}E_{_{3}}^{(2)}+n_{_2}E_{_{3}}^{(1)}$ are
\begin{eqnarray}
&&\alpha_{_{1,1}}=0\;,\;\alpha_{_{1,2}}=D-3\;,\;\alpha_{_{2,1}}=0\;,\;
\alpha_{_{2,2}}=2-{D\over2}\;,\;\alpha_{_{3,1}}=0,\;\alpha_{_{3,2}}=1-{D\over2}\;,
\label{1SE-2-175}
\end{eqnarray}
whose geometric representation is drawn in Fig.\ref{fig4}(1)
with $\{(a,b)$, $(c,d)\}=$ $\{(1,6)$, $(4,5)\}$, $\{e,f\}=$ $\{2,3\}$.
Using the integer lattices and corresponding exponents, the hypergeometric
functions are
\begin{eqnarray}
&&\psi_{_{\{3,4,6\}}}^{(1)}(p^2,m_{_1}^2,m_{_2}^2)\sim
{(m_{_2}^2-m_{_1}^2)^{D-3}(m_{_2}^2)^{2-D/2}\over m_{_2}^2-p^2}
\sum\limits_{n_{_1},n_{_2}}^\infty
{(-)^{n_{_1}+n_{_2}}\Gamma({D\over2}-2+n_{_1})
\over n_{_1}!n_{_2}!\Gamma(1-n_{_1}-n_{_2})}
\nonumber\\
&&\hspace{3.9cm}\times
{\Gamma({D\over2}-1+n_{_2})\over\Gamma(D-2+n_{_1}+n_{_2})}
\Big({p^2(m_{_2}^2-m_{_1}^2)\over m_{_2}^2(p^2-m_{_1}^2)}\Big)^{n_{_1}}
\Big({m_{_2}^2-m_{_1}^2\over p^2-m_{_1}^2}\Big)^{n_{_2}}
\nonumber\\
&&\hspace{3.5cm}\sim
{(m_{_2}^2-m_{_1}^2)^{D-3}(m_{_2}^2)^{2-D/2}\over m_{_2}^2-p^2}
\;,\nonumber\\
&&\psi_{_{\{3,4,6\}}}^{(2)}(p^2,m_{_1}^2,m_{_2}^2)\sim
{(p^2-m_{_1}^2)^{D-3}(m_{_2}^2)^{D/2-1}\over(m_{_2}^2-p^2)(p^2)^{D-3}}
\nonumber\\
&&\hspace{3.9cm}\times
\sum\limits_{n_{_1},n_{_2}}^\infty
{\Gamma(1-{D\over2}+n_{_1}-n_{_2})\Gamma(3-D+n_{_1})\over n_{_1}!n_{_2}!
\Gamma(4-D+n_{_1}-n_{_2})}
\nonumber\\
&&\hspace{3.9cm}\times
\Gamma({D\over2}-1+n_{_2})
\Big({p^2(m_{_2}^2-m_{_1}^2)\over m_{_2}^2(p^2-m_{_1}^2)}\Big)^{n_{_1}}
\Big({m_{_2}^2\over p^2}\Big)^{n_{_2}}
\;,\nonumber\\
&&\psi_{_{\{3,4,6\}}}^{(3)}(p^2,m_{_1}^2,m_{_2}^2)\sim
{(p^2-m_{_1}^2)^{D-3}(p^2)^{2-D/2}\over p^2-m_{_2}^2}
\nonumber\\
&&\hspace{3.9cm}\times
\sum\limits_{n_{_1},n_{_2}}^\infty
{(-)^{n_{_1}+n_{_2}}\Gamma(3-D+n_{_1})\Gamma({D\over2}-2+n_{_2})
\over n_{_1}!n_{_2}!\Gamma(1-n_{_1}-n_{_2})}
\nonumber\\
&&\hspace{3.9cm}\times
{1\over\Gamma(2-{D\over2}+n_{_1}+n_{_2})}
\Big({m_{_2}^2-m_{_1}^2\over p^2-m_{_1}^2}\Big)^{n_{_1}}
\Big({m_{_2}^2\over p^2}\Big)^{n_{_2}}
\nonumber\\
&&\hspace{3.5cm}\sim
{(p^2-m_{_1}^2)^{D-3}(p^2)^{2-D/2}\over p^2-m_{_2}^2}\;,
\label{1SE-2-175a}
\end{eqnarray}
where the intersection of their convergent regions is a nonempty proper subset of the whole parameter space.
In other words, those hypergeometric functions constitute a fundamental solution
system of the GKZ-system in a proper subset of the whole parameter space.

Corresponding to the integer lattice $-n_{_1}E_{_{3}}^{(3)}-n_{_2}E_{_{3}}^{(2)}$,
the matrix of exponents is
\begin{eqnarray}
&&\alpha_{_{1,1}}=0\;,\;\alpha_{_{1,2}}=D-3\;,\;\alpha_{_{2,1}}=0\;,\;
\alpha_{_{2,2}}=2-{D\over2}\;,\;\alpha_{_{3,1}}=0,\;\alpha_{_{3,2}}=1-{D\over2}\;,
\label{1SE-2-170}
\end{eqnarray}
whose geometric representation is shown in Fig.\ref{fig4}(1)
with $\{(a,b)$, $(c,d)\}=$ $\{(1,3)$, $(4,5)\}$, $\{e,f\}=$ $\{2,6\}$.
Similarly the exponents of the integer lattice
$-n_{_1}E_{_{3}}^{(3)}-n_{_2}E_{_{3}}^{(1)}$ are
\begin{eqnarray}
&&\alpha_{_{1,1}}=D-3\;,\;\alpha_{_{1,2}}=0\;,\;\alpha_{_{2,1}}=3-D\;,\;
\alpha_{_{2,2}}={D\over2}-1\;,\;\alpha_{_{3,1}}=0,\;\alpha_{_{3,2}}=1-{D\over2}\;,
\label{1SE-2-174}
\end{eqnarray}
whose geometric representation is presented in Fig.\ref{fig4}(3)
with $(a,b)=$ $(4,5)$, $\{(c,d)$, $(e,f)\}=$ $\{(1,3)$, $(2,6)\}$.
The exponents of the integer lattice
$-n_{_1}E_{_{3}}^{(2)}-n_{_2}E_{_{3}}^{(1)}$ are
\begin{eqnarray}
&&\alpha_{_{1,1}}=D-3\;,\;\alpha_{_{1,2}}=0\;,\;\alpha_{_{2,1}}=2-{D\over2}\;,\;
\alpha_{_{2,2}}=0\;,\;\alpha_{_{3,1}}=1-{D\over2},\;\alpha_{_{3,2}}=0\;,
\label{1SE-2-178}
\end{eqnarray}
whose geometric representation is drawn in Fig.\ref{fig4}(1)
with $\{(a,b)$, $(c,d)\}=$ $\{(2,6)$, $(4,5)\}$, $\{e,f\}=$ $\{1,3\}$.
Accordingly the hypergeometric functions are written as
\begin{eqnarray}
&&\psi_{_{\{3,4,6\}}}^{(4)}(p^2,m_{_1}^2,m_{_2}^2)\sim
{(p^2-m_{_1}^2)^{D-3}(p^2)^{2-D/2}
\over p^2-m_{_2}^2}\sum\limits_{n_{_1},n_{_2}}^\infty
{(-)^{n_{_1}+n_{_2}}\Gamma({D\over2}-2+n_{_1})\over n_{_1}!n_{_2}!
\Gamma(1-n_{_1}-n_{_2})}
\nonumber\\
&&\hspace{3.9cm}\times
{\Gamma({D\over2}-1+n_{_2})\over\Gamma(D-2+n_{_1}+n_{_2})}
\Big({m_{_2}^2(p^2-m_{_1}^2)\over p^2(m_{_2}^2-m_{_1}^2)}\Big)^{n_{_1}}
\Big({p^2-m_{_1}^2\over m_{_2}^2-m_{_1}^2}\Big)^{n_{_2}}
\nonumber\\
&&\hspace{3.5cm}\sim
{(p^2-m_{_1}^2)^{D-3}(p^2)^{2-D/2}\over p^2-m_{_2}^2}
\;,\nonumber\\
&&\psi_{_{\{3,4,6\}}}^{(5)}(p^2,m_{_1}^2,m_{_2}^2)\sim
{(m_{_2}^2-m_{_1}^2)^{D-3}(p^2)^{D/2-1}\over(m_{_2}^2)^{D-3}(m_{_2}^2-p^2)}
\nonumber\\
&&\hspace{3.9cm}\times
\sum\limits_{n_{_1},n_{_2}}^\infty
{\Gamma(1-{D\over2}+n_{_1}-n_{_2})\Gamma(3-D+n_{_1})\over n_{_1}!n_{_2}!\Gamma(4-D+n_{_1}-n_{_2})}
\nonumber\\
&&\hspace{3.9cm}\times
\Gamma({D\over2}-1+n_{_2})
\Big({m_{_2}^2(p^2-m_{_1}^2)\over p^2(m_{_2}^2-m_{_1}^2)}\Big)^{n_{_1}}
\Big({m_{_2}^2\over p^2}\Big)^{n_{_2}}
\;,\nonumber\\
&&\psi_{_{\{3,4,6\}}}^{(6)}(p^2,m_{_1}^2,m_{_2}^2)\sim
{(m_{_2}^2-m_{_1}^2)^{D-3}(m_{_2}^2)^{2-D/2}\over m_{_2}^2-p^2}
\nonumber\\
&&\hspace{3.9cm}\times
\sum\limits_{n_{_1},n_{_2}}^\infty
{(-)^{n_{_1}+n_{_2}}\Gamma(3-D+n_{_1})\Gamma({D\over2}-2+n_{_2})
\over n_{_1}!n_{_2}!\Gamma(1-n_{_1}-n_{_2})}
\nonumber\\
&&\hspace{3.9cm}\times
{1\over\Gamma(2-{D\over2}+n_{_1}+n_{_2})}
\Big({p^2-m_{_1}^2\over m_{_2}^2-m_{_1}^2}\Big)^{n_{_1}}
\Big({p^2\over m_{_2}^2}\Big)^{n_{_2}}
\nonumber\\
&&\hspace{3.5cm}\sim
{(m_{_2}^2-m_{_1}^2)^{D-3}(m_{_2}^2)^{2-D/2}\over m_{_2}^2-p^2}
\;,\nonumber\\
\label{1SE-2-178a}
\end{eqnarray}
where the intersection of their convergent regions is a nonempty proper subset of the whole parameter space.
The fact indicates that those hypergeometric functions constitute a fundamental solution
system of the GKZ-system in a proper subset of the whole parameter space.

As the integer lattice is chosen as $n_{_1}E_{_{3}}^{(3)}-n_{_2}E_{_{3}}^{(2)}$
the exponents are
\begin{eqnarray}
&&\alpha_{_{1,1}}={D\over2}-2\;,\;\alpha_{_{1,2}}={D\over2}-1\;,\;\alpha_{_{2,1}}=2-{D\over2}\;,\;
\alpha_{_{2,2}}=0\;,\;\alpha_{_{3,1}}=0,\;\alpha_{_{3,2}}=1-{D\over2}\;,
\label{1SE-2-168}
\end{eqnarray}
whose geometric representation is plotted in Fig.\ref{fig4}(2)
with $(a,b)=$ $(4,5)$, $c=3$, $d=1$, $\{e,f\}=$ $\{2,6\}$.
Corresponding to the integer lattice $n_{_1}E_{_{3}}^{(3)}-n_{_2}E_{_{3}}^{(1)}$,
the exponents are
\begin{eqnarray}
&&\alpha_{_{1,1}}=0\;,\;\alpha_{_{1,2}}=D-3\;,\;\alpha_{_{2,1}}=0\;,\;
\alpha_{_{2,2}}=2-{D\over2}\;,\;\alpha_{_{3,1}}=0,\;\alpha_{_{3,2}}=1-{D\over2}\;,
\label{1SE-2-172}
\end{eqnarray}
whose geometric representation is presented in Fig.\ref{fig4}(4)
with $(a,b,c)=$ $(1,4,5)$, $(d,e,f)=$ $(2,3,6)$.
Similarly the exponents of the integer lattice
$n_{_1}E_{_{3}}^{(2)}-n_{_2}E_{_{3}}^{(1)}$ are
\begin{eqnarray}
&&\alpha_{_{1,1}}=0\;,\;\alpha_{_{1,2}}=D-3\;,\;\alpha_{_{2,1}}=2-{D\over2}\;,\;
\alpha_{_{2,2}}=0\;,\;\alpha_{_{3,1}}={D\over2}-2,\;\alpha_{_{3,2}}=3-D\;,
\label{1SE-2-176}
\end{eqnarray}
whose geometric representation is presented in Fig.\ref{fig4}(2)
with $(a,b)=$ $(4,5)$, $c=6$, $d=1$, $(e,f)=$ $(2,3)$.
Applying those matrices of exponents and integer lattices,
we obtain three linear independent hypergeometric functions
\begin{eqnarray}
&&\psi_{_{\{3,4,6\}}}^{(7)}(p^2,m_{_1}^2,m_{_2}^2)\sim
{(p^2-m_{_1}^2)^{D/2-1}(m_{_2}^2-m_{_1}^2)^{D/2-2}\over(m_{_2}^2-p^2)
(m_{_2}^2)^{D/2-2}}
\nonumber\\
&&\hspace{3.9cm}\times
\sum\limits_{n_{_1},n_{_2}}^\infty
{\Gamma(1-{D\over2}+n_{_1}-n_{_2})\Gamma({D\over2}-2+n_{_1})
\over n_{_1}!n_{_2}!\Gamma({D\over2}-1+n_{_1}-n_{_2})}
\nonumber\\
&&\hspace{3.9cm}\times
\Gamma({D\over2}-1+n_{_2})
\Big({p^2(m_{_2}^2-m_{_1}^2)\over m_{_2}^2(p^2-m_{_1}^2)}\Big)^{n_{_1}}
\Big({p^2-m_{_1}^2\over m_{_2}^2-m_{_1}^2}\Big)^{n_{_2}}
\;,\nonumber\\
&&\psi_{_{\{3,4,6\}}}^{(8)}(p^2,m_{_1}^2,m_{_2}^2)\sim
{(p^2-m_{_1}^2)^{D-3}(p^2)^{2-D/2}\over p^2-m_{_2}^2}
\nonumber\\
&&\hspace{3.9cm}\times
\sum\limits_{n_{_1},n_{_2}}^\infty{(-)^{n_{_1}+n_{_2}}
\Gamma(3-D+n_{_1})\Gamma({D\over2}-1+n_{_2})
\over n_{_1}!n_{_2}!\Gamma(1-n_{_1}-n_{_2})\Gamma(3-{D\over2}+n_{_1}+n_{_2})}
\nonumber\\
&&\hspace{3.9cm}\times
\Big({p^2(m_{_2}^2-m_{_1}^2)\over m_{_2}^2(p^2-m_{_1}^2)}\Big)^{n_{_1}}
\Big({p^2\over m_{_2}^2}\Big)^{n_{_2}}
\nonumber\\
&&\hspace{3.5cm}\sim
{(p^2-m_{_1}^2)^{D-3}(p^2)^{2-D/2}\over p^2-m_{_2}^2}
\;,\nonumber\\
&&\psi_{_{\{3,4,6\}}}^{(9)}(p^2,m_{_1}^2,m_{_2}^2)\sim
{(p^2-m_{_1}^2)^{D-3}(m_{_2}^2)^{2-D/2}\over m_{_2}^2-p^2}
\nonumber\\
&&\hspace{3.9cm}\times
\sum\limits_{n_{_1},n_{_2}}^\infty
{\Gamma(2-{D\over2}+n_{_1}-n_{_2})\Gamma(3-D+n_{_1})
\over n_{_1}!n_{_2}!\Gamma(4-D+n_{_1}-n_{_2})}
\nonumber\\
&&\hspace{3.9cm}\times
\Gamma({D\over2}-2+n_{_2})
\Big({m_{_2}^2-m_{_1}^2\over p^2-m_{_1}^2}\Big)^{n_{_1}}
\Big({p^2\over m_{_2}^2}\Big)^{n_{_2}}\;.
\label{1SE-2-176a}
\end{eqnarray}
Although the intersection of the convergent regions of these series is
the empty set, the sets of the hypergeometric functions
$\{\psi_{_{\{3,4,6\}}}^{(1)},\psi_{_{\{3,4,6\}}}^{(8)},\psi_{_{\{3,4,6\}}}^{(9)}\}$
and $\{\psi_{_{\{3,4,6\}}}^{(6)},\psi_{_{\{3,4,6\}}}^{(7)},\psi_{_{\{3,4,6\}}}^{(8)}\}$
constitute the fundamental solution systems of the nonempty proper subsets
of the whole parameter space, respectively.

As the integer lattice is chosen as $-n_{_1}E_{_{3}}^{(3)}+n_{_2}E_{_{3}}^{(2)}$,
the matrix of exponents is
\begin{eqnarray}
&&\alpha_{_{1,1}}={D\over2}-1\;,\;\alpha_{_{1,2}}={D\over2}-2\;,\;\alpha_{_{2,1}}=0\;,\;
\alpha_{_{2,2}}=2-{D\over2}\;,\;\alpha_{_{3,1}}=1-{D\over2},\;\alpha_{_{3,2}}=0\;,
\label{1SE-2-169}
\end{eqnarray}
whose geometric representation is plotted in Fig.\ref{fig4}(2)
with $(a,b)=$ $(4,5)$, $c=3$, $d=2$, $\{e,f\}=$ $\{1,6\}$. Corresponding to the integer lattice
$-n_{_1}E_{_{3}}^{(3)}+n_{_2}E_{_{3}}^{(1)}$, the matrix of exponents is
\begin{eqnarray}
&&\alpha_{_{1,1}}=D-3\;,\;\alpha_{_{1,2}}=0\;,\;\alpha_{_{2,1}}=2-{D\over2}\;,\;
\alpha_{_{2,2}}=0\;,\;\alpha_{_{3,1}}=1-{D\over2},\;\alpha_{_{3,2}}=0\;,
\label{1SE-2-173}
\end{eqnarray}
whose geometric representation is presented in Fig.\ref{fig4}(4)
with $(a,b,c)=$ $(2,4,5)$, $(d,e,f)=$ $(1,3,6)$.
The exponents of the integer lattice $-n_{_1}E_{_{3}}^{(2)}+n_{_2}E_{_{3}}^{(1)}$ are
\begin{eqnarray}
&&\alpha_{_{1,1}}=D-3\;,\;\alpha_{_{1,2}}=0\;,\;\alpha_{_{2,1}}=0\;,\;
\alpha_{_{2,2}}=2-{D\over2}\;,\;\alpha_{_{3,1}}=3-D,\;\alpha_{_{3,2}}={D\over2}-2\;,
\label{1SE-2-177}
\end{eqnarray}
whose geometric representation is shown in Fig.\ref{fig4}(2)
with $(a,b)=$ $(4,5)$, $c=6$, $d=2$, $(e,f)=$ $(1,3)$.
Accordingly the hypergeometric series are
\begin{eqnarray}
&&\psi_{_{\{3,4,6\}}}^{(10)}(p^2,m_{_1}^2,m_{_2}^2)\sim
{(p^2-m_{_1}^2)^{D/2-2}(m_{_2}^2-m_{_1}^2)^{D/2-1}\over(m_{_2}^2-p^2)
(p^2)^{D/2-2}}
\nonumber\\
&&\hspace{3.9cm}\times
\sum\limits_{n_{_1},n_{_2}}^\infty
{\Gamma(1-{D\over2}+n_{_1}-n_{_2})\Gamma({D\over2}-2+n_{_1})\over
n_{_1}!n_{_2}!\Gamma({D\over2}-1+n_{_1}-n_{_2})}
\nonumber\\
&&\hspace{3.9cm}\times
\Gamma({D\over2}-1+n_{_2})
\Big({m_{_2}^2(p^2-m_{_1}^2)\over p^2(m_{_2}^2-m_{_1}^2)}\Big)^{n_{_1}}
\Big({m_{_2}^2-m_{_1}^2\over p^2-m_{_1}^2}\Big)^{n_{_2}}
\;,\nonumber\\
&&\psi_{_{\{3,4,6\}}}^{(11)}(p^2,m_{_1}^2,m_{_2}^2)\sim
{(m_{_2}^2-m_{_1}^2)^{D-3}(m_{_2}^2)^{2-D/2}\over m_{_2}^2-p^2}
\nonumber\\
&&\hspace{3.9cm}\times
\sum\limits_{n_{_1},n_{_2}}^\infty{(-)^{n_{_1}+n_{_2}}
\Gamma(3-D+n_{_1})\Gamma({D\over2}-1+n_{_2})
\over n_{_1}!n_{_2}!\Gamma(1-n_{_1}-n_{_2})\Gamma(3-{D\over2}+n_{_1}+n_{_2})}
\nonumber\\
&&\hspace{3.9cm}\times
\Big({m_{_2}^2(p^2-m_{_1}^2)\over p^2(m_{_2}^2-m_{_1}^2)}\Big)^{n_{_1}}
\Big({m_{_2}^2\over p^2}\Big)^{n_{_2}}
\nonumber\\
&&\hspace{3.5cm}\sim
{(m_{_2}^2-m_{_1}^2)^{D-3}(m_{_2}^2)^{2-D/2}\over m_{_2}^2-p^2}
\;,\nonumber\\
&&\psi_{_{\{3,4,6\}}}^{(12)}(p^2,m_{_1}^2,m_{_2}^2)\sim
{(m_{_2}^2-m_{_1}^2)^{D-3}(p^2)^{2-D/2}\over m_{_2}^2-p^2}
\nonumber\\
&&\hspace{3.9cm}\times
\sum\limits_{n_{_1},n_{_2}}^\infty
{\Gamma(2-{D\over2}+n_{_1}-n_{_2})\Gamma(3-D+n_{_1})
\over n_{_1}!n_{_2}!\Gamma(4-D+n_{_1}-n_{_2})}
\nonumber\\
&&\hspace{3.9cm}\times
\Gamma({D\over2}-2+n_{_2})
\Big({p^2-m_{_1}^2\over m_{_2}^2-m_{_1}^2}\Big)^{n_{_1}}
\Big({m_{_2}^2\over p^2}\Big)^{n_{_2}}\;.
\label{1SE-2-177a}
\end{eqnarray}
Although the intersection of the convergent regions of these series is
the empty set, the sets of the hypergeometric functions
$\{\psi_{_{\{3,4,6\}}}^{(3)},\psi_{_{\{3,4,6\}}}^{(10)},\psi_{_{\{3,4,6\}}}^{(11)}\}$
and $\{\psi_{_{\{3,4,6\}}}^{(4)},\psi_{_{\{3,4,6\}}}^{(11)},\psi_{_{\{3,4,6\}}}^{(12)}\}$
also constitute the fundamental solution systems of the corresponding nonempty proper subsets
of the whole parameter space, respectively.

In principle, we can embed the Feynman parametric representation of the 1-loop
self energy into infinite subvarieties of the Grassmannian $G_{_{3,6}}$.
However, the dimension of dual space of the GKZ-systems of most subvarieties
is much larger than the number of independent dimensionless ratios between the
square of outgoing momentum and the square of virtual masses.
Correspondingly convergent regions of the obtained hypergeometric series solutions are empty.

\section{The 1-loop triangle with zero virtual masses\label{1Triangle}}
\indent\indent
The Feynman integral of the 1-loop triangle diagram with zero virtual masses is represented as
the linear combination of the generalized hypergeometric functions $F_{_4}$ in
literature~\cite{Davydychev1} through $\alpha$-parametric representation for the denominators.
Here, we construct the equivalent fundamental solution system through embedding
the Feynman integral on Grassmannians.

Denoting three incoming momenta as $p_{_i}$ $(i=1,2,3)$, we obtain the Feynman
parametric representation of the diagram as
\begin{eqnarray}
&&C(p_{_1}^2,\;p_{_2}^2,\;p_{_3}^2)=
\Big(\Lambda_{_{\rm RE}}^2\Big)^{4-D}\int{d^Dq\over(2\pi)^D}
{1\over q^2(q-p_{_1})^2(q+p_{_2})^2}
\nonumber\\
&&\hspace{2.4cm}=
-i\Big(\Lambda_{_{\rm RE}}^2\Big)^{4-D}{(-)^{D/2-3}\Gamma(3-{D\over2})\over(4\pi)^{D/2}}
\int_0^1dt_{_1}dt_{_2}dt_{_3}
\nonumber\\
&&\hspace{2.9cm}\times
{\delta(t_{_1}+t_{_2}+t_{_3}-1)\over
(t_{_1}t_{_2}p_{_3}^2+t_{_2}t_{_3}p_{_1}^2+t_{_1}t_{_3}p_{_2}^2)^{3-D/2}}\;,
\label{1TR2a}
\end{eqnarray}
where $p_{_1}+p_{_2}+p_{_3}=0$. As $p_{_2}\neq0$, $p_{_3}\neq0$, but $p_{_2}^2=p_{_3}^2=0$,
the Feynman parametric representation is simplified as
\begin{eqnarray}
&&C(p_{_1}^2,\;0,\;0)=
-i\Big(\Lambda_{_{\rm RE}}^2\Big)^{4-D}{(-)^{D/2-3}\Gamma(3-{D\over2})
\over(4\pi)^{D/2}}
\nonumber\\
&&\hspace{2.9cm}\times
\int_0^1dt_{_1}dt_{_2}t_{_2}^{D/2-3}(p_{_1}^2-p_{_1}^2t_{_1}-p_{_1}^2t_{_2})^{D/2-3}
\nonumber\\
&&\hspace{2.4cm}=
i\Big(\Lambda_{_{\rm RE}}^2\Big)^{4-D}{\Gamma(3-{D\over2})
\over(4\pi)^{D/2}}
\nonumber\\
&&\hspace{2.9cm}\times
\int\omega_{_3}(t)t_{_2}^{D/2-3}t_{_3}^{3-D}
(p_{_1}^2t_{_1}+p_{_1}^2t_{_2}+p_{_1}^2t_{_3})^{D/2-3}\;.
\label{1TR2a-1}
\end{eqnarray}
Eq.(\ref{1TR2a-1}) can be embedded in the subvariety of the Grassmannian $G_{_{3,4}}$
with splitting local coordinates as
\begin{eqnarray}
&&A^{(1)}=\left(\begin{array}{cccc}\;\;1\;\;&\;\;0\;\;&\;\;0\;\;&\;\;p_{_1}^2\;\;\\
\;\;0\;\;&\;\;1\;\;&\;\;0\;\;&\;\;p_{_1}^2\;\;\\\;\;0\;\;&\;\;0\;\;&\;\;1\;\;&\;\;p_{_1}^2\;\;
\end{array}\right)\;.
\label{1TR2a-2}
\end{eqnarray}
Here the first row corresponds to the integration variable $t_{_1}$,
the second row  corresponds to the integration variable $t_{_2}$,
and the third row  corresponds to the integration variable $t_{_3}$, respectively.
Meanwhile the first column represents the power function $t_{_1}^0=1$,
the second column represents the power function $t_{_2}^{D/2-3}$, and the third column
represents the power function $t_{_3}^{3-D}$,
the fourth column represents the function $p_{_1}^2(t_{_1}+t_{_2}+t_{_3})$, respectively.
Taking Feynman parametric representation of the 1-loop triangle diagram
as a function of the splitting local coordinates in Eq.(\ref{1TR2a-2}),
\begin{eqnarray}
&&C(p_{_1}^2,\;0,\;0)=\Psi^{1T}(z_{_1},\;\cdots,\;z_{_4})\;,
\label{1TR2a-2+a}
\end{eqnarray}
we obviously have
\begin{eqnarray}
&&\Psi^{1T}(\chi_{_1}z_{_1},\;\cdots,\;\chi_{_4}z_{_4})=
\chi_{_2}^{D/2-3}\chi_{_3}^{3-D}\chi_{_4}^{D/2-3}\Psi^{1T}(z_{_1},\;\cdots,\;z_{_4})\;.
\label{1TR2a-2+b}
\end{eqnarray}
Where $z_{_i}\;(i=1,\cdots,4)$ denotes the i-th column vector in the matrix Eq.(\ref{1TR2a-2}),
and $\chi_{_i}$ is a nonzero constant, respectively.
Eq.(\ref{1TR2a-2+b}) implies that the special Feynman integral
satisfies following GKZ-system
\begin{eqnarray}
&&\Big\{\vartheta_{_{1,1}}+\vartheta_{_{1,4}}\Big\}C(p_{_1}^2,\;0,\;0)
=-C(p_{_1}^2,\;0,\;0)
\;,\nonumber\\
&&\Big\{\vartheta_{_{2,2}}+\vartheta_{_{2,4}}\Big\}C(p_{_1}^2,\;0,\;0)
=-C(p_{_1}^2,\;0,\;0)
\;,\nonumber\\
&&\Big\{\vartheta_{_{3,3}}+\vartheta_{_{3,4}}\Big\}C(p_{_1}^2,\;0,\;0)
=-C(p_{_1}^2,\;0,\;0)
\;,\nonumber\\
&&\vartheta_{_{1,1}}C(p_{_1}^2,\;0,\;0)=0
\;,\nonumber\\
&&\vartheta_{_{2,2}}C(p_{_1}^2,\;0,\;0)=({D\over2}-3)C(p_{_1}^2,\;0,\;0)
\;,\nonumber\\
&&\vartheta_{_{3,3}}C(p_{_1}^2,\;0,\;0)=(3-D)C(p_{_1}^2,\;0,\;0)
\;,\nonumber\\
&&\Big\{\vartheta_{_{1,4}}+\vartheta_{_{2,4}}+\vartheta_{_{3,4}}\Big\}C(p_{_1}^2,\;0,\;0)
=({D\over2}-3)C(p_{_1}^2,\;0,\;0)\;.
\label{1TR2a-3}
\end{eqnarray}
Obviously the dual space of the GKZ-system of Eq.(\ref{1TR2a-3}) is the null
space $\{0\}$, and compatible with the exponent matrix
\begin{eqnarray}
&&\left(\begin{array}{cccc}\;\;0\;\;&\;\;0\;\;&\;\;0\;\;&\;\;-1\;\;\\
\;\;0\;\;&\;\;{D\over2}-3\;\;&\;\;0\;\;&\;\;2-{D\over2}\;\;\\
\;\;0\;\;&\;\;0\;\;&\;\;3-D\;\;&\;\;D-4\;\;\end{array}\right)\;.
\label{1TR2a-4}
\end{eqnarray}
Using the above matrices of exponent and integer lattice, one obtains the
unique solution of the GKZ-system as
\begin{eqnarray}
&&C(p_{_1}^2,\;0,\;0)={C_{_{1T}}\over\Gamma(3-{D\over2})\Gamma(D-3)}
\Big(p_{_1}^2\Big)^{D/2-3}\;,
\label{1TR2a-5}
\end{eqnarray}
where $C_{_{1T}}$ is the integration constant. In the case $p_{_1}=-p_{_2}\neq0$,
$p_{_3}=0$, the Feynman parametric representation is modified as
\begin{eqnarray}
&&C(p_{_1}^2,\;p_{_1}^2,\;0)=
-i\Big(\Lambda_{_{\rm RE}}^2\Big)^{4-D}{(-)^{D/2-3}\Gamma(3-{D\over2})\over(4\pi)^{D/2}}
\int_0^1dt_{_1}dt_{_2}(1-t_{_1}-t_{_2})^{D/2-3}
\nonumber\\
&&\hspace{2.9cm}\times
(t_{_1}p_{_1}^2+t_{_1}p_{_1}^2)^{D/2-3}
\nonumber\\
&&\hspace{2.4cm}=
i\Big(\Lambda_{_{\rm RE}}^2\Big)^{4-D}{\Gamma(3-{D\over2})\over(4\pi)^{D/2}}
\int\omega(t)t_{_3}^{3-D}(t_{_1}+t_{_2}+t_{_3})^{D/2-3}
\nonumber\\
&&\hspace{2.9cm}\times
(t_{_1}p_{_1}^2+t_{_2}p_{_1}^2)^{D/2-3}\;,
\label{1TR2a-6}
\end{eqnarray}
which can be embedded in the subvariety of the Grassmannian $G_{_{3,5}}$
with splitting local coordinates as
\begin{eqnarray}
&&A^{(1)\prime}=\left(\begin{array}{ccccc}
\;\;1\;\;&\;\;0\;\;&\;\;0\;\;&\;\;1\;\;&\;\;p_{_1}^2\;\;\\
\;\;0\;\;&\;\;1\;\;&\;\;0\;\;&\;\;1\;\;&\;\;p_{_1}^2\;\;\\
\;\;0\;\;&\;\;0\;\;&\;\;1\;\;&\;\;1\;\;&\;\;0\;\;
\end{array}\right)\;.
\label{1TR2a-7}
\end{eqnarray}
In the above splitting coordinates, the Feynman integral of 1-loop massless triangle
satisfies the following GKZ-system
\begin{eqnarray}
&&\Big\{\vartheta_{_{1,1}}+\vartheta_{_{1,4}}+\vartheta_{_{1,5}}\Big\}C(p_{_1}^2,\;p_{_1}^2,\;0)
=-C(p_{_1}^2,\;p_{_1}^2,\;0)
\;,\nonumber\\
&&\Big\{\vartheta_{_{2,2}}+\vartheta_{_{2,4}}+\vartheta_{_{2,5}}\Big\}C(p_{_1}^2,\;p_{_1}^2,\;0)
=-C(p_{_1}^2,\;p_{_1}^2,\;0)
\;,\nonumber\\
&&\Big\{\vartheta_{_{3,3}}+\vartheta_{_{3,4}}\Big\}C(p_{_1}^2,\;p_{_1}^2,\;0)
=-C(p_{_1}^2,\;p_{_1}^2,\;0)
\;,\nonumber\\
&&\vartheta_{_{1,1}}C(p_{_1}^2,\;p_{_1}^2,\;0)=0
\;,\nonumber\\
&&\vartheta_{_{2,2}}C(p_{_1}^2,\;p_{_1}^2,\;0)=0
\;,\nonumber\\
&&\vartheta_{_{3,3}}C(p_{_1}^2,\;p_{_1}^2,\;0)=(3-D)C(p_{_1}^2,\;p_{_1}^2,\;0)
\;,\nonumber\\
&&\Big\{\vartheta_{_{1,4}}+\vartheta_{_{2,4}}+\vartheta_{_{3,4}}\Big\}C(p_{_1}^2,\;p_{_1}^2,\;0)
=({D\over2}-3)C(p_{_1}^2,\;p_{_1}^2,\;0)
\;,\nonumber\\
&&\Big\{\vartheta_{_{1,5}}+\vartheta_{_{2,5}}\Big\}C(p_{_1}^2,\;p_{_1}^2,\;0)
=({D\over2}-3)C(p_{_1}^2,\;p_{_1}^2,\;0)\;.
\label{1TR2a-8}
\end{eqnarray}

Because the submatrix of Eq.(\ref{1TR2a-7}) composed of the first, fourth,
fifth columns and that composed of the second, fourth,
fifth columns satisfy
\begin{eqnarray}
&&\det A_{_{145}}^{(1)\prime}=\det\left(\begin{array}{ccc}\;\;1\;\;&\;\;1\;\;&\;\;p_{_1}^2\;\;\\
\;\;0\;\;&\;\;1\;\;&\;\;p_{_1}^2\;\;\\ \;\;0\;\;&\;\;1\;\;&\;\;0\;\;\\
\end{array}\right)=-p_{_1}^2\neq0\;,
\nonumber\\
&&\det A_{_{245}}^{{(1)\prime}}=\det\left(\begin{array}{ccc}\;\;0\;\;&\;\;1\;\;&\;\;p_{_1}^2\;\;\\
\;\;1\;\;&\;\;1\;\;&\;\;p_{_1}^2\;\;\\ \;\;0\;\;&\;\;1\;\;&\;\;0\;\;\\
\end{array}\right)=p_{_1}^2\neq0\;,
\label{1TR2a-13}
\end{eqnarray}
we can choose $\{(1,0,0)^T$, $(1,1,1)^T$, $(p_{_1}^2,p_{_1}^2,0)^T\}$
or $\{(0,1,0)^T$, $(1,1,1)^T$, $(p_{_1}^2,p_{_1}^2,0)^T\}$ spanning
the projective plane respectively. As we choose the vectors
$\{(1,0,0)^T$, $(1,1,1)^T$, $(p_{_1}^2,p_{_1}^2,0)^T\}$
as the basis of the projective plane, splitting local coordinates are written as
\begin{eqnarray}
&&\left(\begin{array}{ccccc}\;\;1\;\;&\;\;-1\;\;&\;\;0\;\;&\;\;0\;\;&\;\;0\;\;\\
\;\;0\;\;&\;\;0\;\;&\;\;1\;\;&\;\;1\;\;&\;\;0\;\;\\
\;\;0\;\;&\;\;{1\over p_{_1}^2}\;\;&\;\;-{1\over p_{_1}^2}\;\;&\;\;0\;\;&\;\;1\;\;
\end{array}\right)\;,
\label{1TR2a-14}
\end{eqnarray}
with the exponent matrix
\begin{eqnarray}
&&\left(\begin{array}{ccccc}\;\;0\;\;&\;\;-1\;\;&\;\;0\;\;&\;\;0\;\;&\;\;0\;\;\\
\;\;0\;\;&\;\;0\;\;&\;\;2-{D\over2}\;\;&\;\;{D\over2}-3\;\;&\;\;0\;\;\\
\;\;0\;\;&\;\;1\;\;&\;\;1-{D\over2}\;\;&\;\;0\;\;&\;\;{D\over2}-3\;\;
\end{array}\right)\;.
\label{1TR2a-14a}
\end{eqnarray}
Using the above matrix of exponents and the
integer lattice $0_{_{3\times5}}$, one obtains the unique solution as
\begin{eqnarray}
&&C(p_{_1}^2,\;p_{_1}^2,\;0)={C_{_{1T}}^{(145)}\over
\det(A_{_{145}}^{(1)\prime})\Gamma(2)\Gamma(2-{D\over2})}\Big(p_{_1}^2\Big)^{D/2-2}
\nonumber\\
&&\hspace{2.3cm}=
-{C_{_{1T}}^{(145)}\over\Gamma(2-{D\over2})}
\Big(p_{_1}^2\Big)^{D/2-3}\;.
\label{1TR2a-15}
\end{eqnarray}
Similarly taking the vectors $\{(0,1,0)^T$, $(1,1,1)^T$, $(p_{_1}^2,p_{_1}^2,0)^T\}$
as the basis of the projective plane, we write the splitting local coordinates as
\begin{eqnarray}
&&\left(\begin{array}{ccccc}\;\;-1\;\;&\;\;1\;\;&\;\;0\;\;&\;\;0\;\;&\;\;0\;\;\\
\;\;0\;\;&\;\;0\;\;&\;\;1\;\;&\;\;1\;\;&\;\;0\;\;\\
\;\;{1\over p_{_1}^2}\;\;&\;\;0\;\;&\;\;-{1\over p_{_1}^2}\;\;&\;\;0\;\;&\;\;1\;\;
\end{array}\right)\;,
\label{1TR2a-16}
\end{eqnarray}
with the exponent matrix
\begin{eqnarray}
&&\left(\begin{array}{ccccc}\;\;-1\;\;&\;\;0\;\;&\;\;0\;\;&\;\;0\;\;&\;\;0\;\;\\
\;\;0\;\;&\;\;0\;\;&\;\;2-{D\over2}\;\;&\;\;{D\over2}-3\;\;&\;\;0\;\;\\
\;\;1\;\;&\;\;0\;\;&\;\;1-{D\over2}\;\;&\;\;0\;\;&\;\;{D\over2}-3\;\;
\end{array}\right)\;.
\label{1TR2a-16a}
\end{eqnarray}
Using the above matrix of exponents and the
integer lattice $0_{_{3\times5}}$, we write the unique solution as
\begin{eqnarray}
&&C(p_{_1}^2,\;p_{_1}^2,\;0)={C_{_{1T}}^{(245)}\over
\det(A_{_{245}}^{(1)\prime})\Gamma(2)\Gamma(2-{D\over2})}\Big(p_{_1}^2\Big)^{D/2-2}
\nonumber\\
&&\hspace{2.3cm}=
{C_{_{1T}}^{(245)}\over\Gamma(2-{D\over2})}
\Big(p_{_1}^2\Big)^{D/2-3}\;.
\label{1TR2a-17}
\end{eqnarray}

\subsection{The fundamental solution systems as $p_{_3}\neq0$, $p_{_3}^2=0$}
\indent\indent
As $p_{_3}\neq0$, $p_{_3}^2=0$, the Feynman parametric representation is
\begin{eqnarray}
&&C(p_{_1}^2,\;p_{_2}^2,\;0)=
i\Big(\Lambda_{_{\rm RE}}^2\Big)^{4-D}{\Gamma(3-{D\over2})\over(4\pi)^{D/2}}
\int\omega(t)_{_3}{t_{_3}^{3-D}(t_{_1}+t_{_2}+t_{_3})^{D/2-3}\over
(t_{_2}p_{_1}^2+t_{_1}p_{_2}^2)^{3-D/2}}\;.
\label{1TR1-1}
\end{eqnarray}
The integral can be embedded in the subvariety of the Grassmannian $G_{_{3,5}}$
which local coordinates are written as
\begin{eqnarray}
&&A^{(2)}=\left(\begin{array}{ccccc}
\;\;1\;\;&\;\;0\;\;&\;\;0\;\;&\;\;p_{_2}^2\;\;&\;\;1\;\;\\
\;\;0\;\;&\;\;1\;\;&\;\;0\;\;&\;\;p_{_1}^2\;\;&\;\;1\;\;\\
\;\;0\;\;&\;\;0\;\;&\;\;1\;\;&\;\;0\;\;&\;\;1\;\;
\end{array}\right)\;.
\label{1TR1-2}
\end{eqnarray}
Here three rows correspond to the integration variables $t_{_i}\;(i=1,\;2,\;3)$,
the first column represents the power function $t_{_1}^0=1$,
the second column represents the power function $t_{_2}^0=1$, the third column
represents the power function $t_{_3}^{3-D}$,
the fourth column represents the function $(t_{_2}p_{_1}^2+t_{_1}p_{_2}^2)^{D/2-3}$,
and the fifth column represents $(t_{_1}+t_{_2}+t_{_3})^{D/2-3}$, respectively.
The geometric description of the matrix is similarly presented in Fig.\ref{fig1}
where $(2,3,5)$ is changed into $(1,2,4)$, $\{1,4\}$ is changed into $\{3,5\}$,
respectively. Accordingly, the Feynman integral satisfies the following
GKZ-system in splitting local coordinates
\begin{eqnarray}
&&\Big\{\vartheta_{_{1,1}}+\vartheta_{_{1,4}}+\vartheta_{_{1,5}}\Big\}C(p_{_1}^2,\;p_{_2}^2,\;0)=
-C(p_{_1}^2,\;p_{_2}^2,\;0)
\;,\nonumber\\
&&\Big\{\vartheta_{_{2,2}}+\vartheta_{_{2,4}}+\vartheta_{_{2,5}}\Big\}C(p_{_1}^2,\;p_{_2}^2,\;0)=
-C(p_{_1}^2,\;p_{_2}^2,\;0)
\;,\nonumber\\
&&\Big\{\vartheta_{_{3,3}}+\vartheta_{_{3,5}}\Big\}C(p_{_1}^2,\;p_{_2}^2,\;0)=
-C(p_{_1}^2,\;p_{_2}^2,\;0)
\;,\nonumber\\
&&\vartheta_{_{1,1}}C(p_{_1}^2,\;p_{_2}^2,\;0)=0
\;,\nonumber\\
&&\vartheta_{_{2,2}}C(p_{_1}^2,\;p_{_2}^2,\;0)=0
\;,\nonumber\\
&&\vartheta_{_{3,3}}C(p_{_1}^2,\;p_{_2}^2,\;0)=(3-D)C(p_{_1}^2,\;p_{_2}^2,\;0)
\;,\nonumber\\
&&\Big\{\vartheta_{_{1,4}}+\vartheta_{_{2,4}}\Big\}C(p_{_1}^2,\;p_{_2}^2,\;0)=
({D\over2}-3)C(p_{_1}^2,\;p_{_2}^2,\;0)
\;,\nonumber\\
&&\Big\{\vartheta_{_{1,5}}+\vartheta_{_{2,5}}+\vartheta_{_{3,5}}\Big\}C(p_{_1}^2,\;p_{_2}^2,\;0)=
({D\over2}-3)C(p_{_1}^2,\;p_{_2}^2,\;0)\;.
\label{1TR1-2a}
\end{eqnarray}
Obviously the matrix of exponents is written as
\begin{eqnarray}
&&\left(\begin{array}{ccccc}
\;\;0\;\;&\;\;0\;\;&\;\;0\;\;&\;\;\alpha_{_{1,4}}\;\;&\;\;\alpha_{_{1,5}}\;\;\\
\;\;0\;\;&\;\;0\;\;&\;\;0\;\;&\;\;\alpha_{_{2,4}}\;\;&\;\;\alpha_{_{2,5}}\;\;\\
\;\;0\;\;&\;\;0\;\;&\;\;3-D\;\;&\;\;0\;\;&\;\;D-4\;\;
\end{array}\right)\;,
\label{1TR1-3}
\end{eqnarray}
with
\begin{eqnarray}
&&\alpha_{_{1,4}}+\alpha_{_{1,5}}=-1,\;\alpha_{_{2,4}}+\alpha_{_{2,5}}=-1,
\;\nonumber\\
&&\alpha_{_{1,4}}+\alpha_{_{2,4}}={D\over2}-3,\;\alpha_{_{1,5}}+\alpha_{_{2,5}}=1-{D\over2}\;.
\label{1TR1-4}
\end{eqnarray}
The dual space of the GKZ-system is spanned by the integer matrix
whose submatrix composed of the fourth and fifth columns is formulated as $E_{_{3}}^{(3)}$,
and other elements are all zero. The basis of the dual space implies the Feynman
integral satisfying the PDE as
\begin{eqnarray}
&&{\partial^2C\over\partial z_{_{1,4}}\partial z_{_{2,5}}}(p_{_1}^2,\;p_{_2}^2,\;0)
={\partial^2C\over\partial z_{_{1,5}}\partial z_{_{2,4}}}(p_{_1}^2,\;p_{_2}^2,\;0)\;.
\label{1TR1-4a}
\end{eqnarray}
When the exponent matrix is given, the Feynman integral can be formally expressed as
\begin{eqnarray}
&&C(p_{_1}^2,\;p_{_2}^2,\;0)=\prod\limits_{i,j}z_{_{i,j}}^{\alpha_{_{i,j}}}
\varphi({p_{_1}^2\over p_{_2}^2})\;,
\label{1TR1-4a1}
\end{eqnarray}
where $\varphi(x)$ satisfies the PDE
\begin{eqnarray}
&&x\Big[\alpha_{_{1,4}}\alpha_{_{2,5}}+(1-\alpha_{_{1,4}}-\alpha_{_{2,5}})x{d\over dx}
-x^2{d^2\over dx^2}\Big]\varphi(x)
\nonumber\\
&&\hspace{-0.5cm}=
\Big[\alpha_{_{1,5}}\alpha_{_{2,4}}+(1+\alpha_{_{1,5}}+\alpha_{_{2,4}})x{d\over dx}
-x^2{d^2\over dx^2}\Big]\varphi(x)\;.
\label{1TR1-4a2}
\end{eqnarray}
As $\alpha_{_{1,5}}\alpha_{_{2,4}}=0$ this equation recovers the well-known Gauss equation.

There are two choices on the matrix of integer lattice whose
submatrix composed of the fourth and fifth columns is formulated as $\pm nE_{_{3}}^{(3)}$ with $n\ge0$.
The matrix of integer lattice $nE_{_{3}}^{(3)}$ permits two
choices on the exponents. The first one is
\begin{eqnarray}
&&\alpha_{_{1,4}}=0,\;\alpha_{_{1,5}}=-1,\;\alpha_{_{2,4}}={D\over2}-3,\;\alpha_{_{2,5}}=2-{D\over2},
\label{1TR1-5}
\end{eqnarray}
whose geometric representation is plotted in Fig.\ref{fig2}(2)
with $(a,b)=$ $(2,4)$, $\{c,d,e\}=$ $\{1,3,5\}$. The second choice
of exponents is
\begin{eqnarray}
&&\alpha_{_{1,4}}={D\over2}-2,\;\alpha_{_{1,5}}=1-{D\over2},\;\alpha_{_{2,4}}=-1,\;\alpha_{_{2,5}}=0,
\label{1TR1-6}
\end{eqnarray}
whose geometric description is presented in Fig.\ref{fig2}(1) with
$a=1$, $\{(b,c)$, $(d,e)\}=$ $\{(2,4)$, $(3,5)\}$, respectively.
The corresponding hypergeometric functions are evidently written as
\begin{eqnarray}
&&\psi_{_{\{1,2,3\}}}^{(1)}(p_{_1}^2,p_{_2}^2,0)\sim
(p_{_1}^2)^{D/2-3}\sum\limits_{n=0}^\infty\Big({p_{_2}^2\over p_{_1}^2}\Big)^n
\sim{(p_{_1}^2)^{D/2-2}\over p_{_1}^2-p_{_2}^2}
\;,\nonumber\\
&&\psi_{_{\{1,2,3\}}}^{(2)}(p_{_1}^2,p_{_2}^2,0)\sim{(p_{_2}^2)^{D/2-2}\over p_{_1}^2}
\sum\limits_{n=0}^\infty\Big({p_{_2}^2\over p_{_1}^2}\Big)^n
\sim{(p_{_2}^2)^{D/2-2}\over p_{_1}^2-p_{_2}^2}\;,
\label{1TR1-6a}
\end{eqnarray}
where the intersection of their convergent regions is $|p_{_2}^2/p_{_1}^2|\le1$.
The linearly independent Gauss functions constitute the fundamental
solution system of the GKZ-system of Eq.(\ref{1TR1-2a}) in the region $|p_{_2}^2/p_{_1}^2|\le1$
which is consistent with the results in the literature. As $\mu=\nu=\rho=1$
and $k^2=0$ in Eq.(7) of Ref.~\cite{Davydychev1}, the fourth type Appell functions of the second
and third terms are modified into the Gauss functions in Eq.(\ref{1TR1-6a}).

The integer lattice $-nE_{_{3}}^{(3)}$ permits two choices
of the exponents. The first choice is
\begin{eqnarray}
&&\alpha_{_{1,4}}=-1,\;\alpha_{_{1,5}}=0,\;\alpha_{_{2,4}}={D\over2}-2,\;\alpha_{_{2,5}}=1-{D\over2},
\label{1TR1-7}
\end{eqnarray}
whose geometric representation is drawn in Fig.\ref{fig2}(1) with $a=2$,
$\{(b,c)$, $(d,e)\}=$ $\{(1,4)$, $(3,5)\}$, and the second one is
\begin{eqnarray}
&&\alpha_{_{1,4}}={D\over2}-3,\;\alpha_{_{1,5}}=2-{D\over2},\;\alpha_{_{2,4}}=0,\;\alpha_{_{2,5}}=-1.
\label{1TR1-8}
\end{eqnarray}
whose geometric representation is presented in Fig.\ref{fig2}(2) with
$(a,b)=$ $(1,4)$, $\{c,d,e\}=$ $\{2,3,5\}$.
The hypergeometric functions are
\begin{eqnarray}
&&\psi_{_{\{1,2,3\}}}^{(3)}(p_{_1}^2,p_{_2}^2,0)\sim{(p_{_1}^2)^{D/2-2}\over p_{_2}^2}
\sum\limits_{n=0}^\infty\Big({p_{_1}^2\over p_{_2}^2}\Big)^n
\sim{(p_{_1}^2)^{D/2-2}\over p_{_1}^2-p_{_2}^2}
\;,\nonumber\\
&&\psi_{_{\{1,2,3\}}}^{(4)}(p_{_1}^2,p_{_2}^2,0)\sim
(p_{_2}^2)^{D/2-3}\sum\limits_{n=0}^\infty\Big({p_{_1}^2\over p_{_2}^2}\Big)^n
\sim{(p_{_2}^2)^{D/2-2}\over p_{_1}^2-p_{_2}^2}\;,
\label{1TR1-9}
\end{eqnarray}
whose intersection of their convergent regions is $|p_{_1}^2/p_{_2}^2|\le1$.
The projective plane can also spanned by the first, second, and fifth columns
because $\det(A_{_{\{1,2,5\}}}^{(2)})=1$,  the hypergeometric solutions based
on the matrix $(A_{_{\{1,2,5\}}}^{(2)})^{-1}\cdot A^{(2)}$ are same as
those presented above.

Because $\det(A_{_{\{1,3,4\}}}^{(2)})=-p_{_1}^2$, where the matrix $A_{_{\{1,3,4\}}}^{(2)}$ denotes
the submatrix of $A^{(2)}$ composed of the first, third, and fourth columns, one derives
\begin{eqnarray}
&&\Big(A_{_{\{1,3,4\}}}^{(2)}\Big)^{-1}\cdot A^{(2)}=\left(\begin{array}{ccccc}
\;\;1\;\;&\;\;-{p_{_2}^2\over p_{_1}^2}\;\;&\;\;0\;\;&\;\;0\;\;&\;\;1-{p_{_2}^2\over p_{_1}^2}\;\;\\
\;\;0\;\;&\;\;0\;\;&\;\;1\;\;&\;\;0\;\;&\;\;1\;\;\\
\;\;0\;\;&\;\;{1\over p_{_1}^2}\;\;&\;\;0\;\;&\;\;1\;\;&\;\;{1\over p_{_1}^2}\;\;\end{array}\right)\;.
\label{1TR1-10}
\end{eqnarray}
Obviously the matrix of exponents is written as
\begin{eqnarray}
&&\left(\begin{array}{ccccc}
\;0\;\;&\;\alpha_{_{1,2}}\;\;&\;\;0\;\;&\;\;0\;\;&\;\;\alpha_{_{1,5}}\;\;\\
\;\;0\;\;&\;\;0\;\;&\;\;3-D\;\;&\;\;0\;\;&\;\;D-4\;\;\\
\;0\;\;&\;\alpha_{_{3,2}}\;\;&\;\;0\;\;&\;\;{D\over2}-3\;\;&\;\;\alpha_{_{3,5}}\;\;\end{array}\right)\;,
\label{1TR1-11}
\end{eqnarray}
where the matrix elements satisfy the relations
\begin{eqnarray}
&&\alpha_{_{1,2}}+\alpha_{_{1,5}}=-1\;,\;\alpha_{_{3,2}}+\alpha_{_{3,5}}=2-{D\over2}\;,
\nonumber\\
&&\alpha_{_{1,2}}+\alpha_{_{3,2}}=0\;,\;\alpha_{_{1,5}}+\alpha_{_{3,5}}=1-{D\over2}\;.
\label{1TR1-12}
\end{eqnarray}
There are two choices on the matrix of integer lattice whose submatrix
composed of the second and fifth columns is formulated as $\pm nE_{_{3}}^{(2)}$ with $n\ge0$, and
other elements are all zero. The matrix of integer lattice $nE_{_{3}}^{(2)}$ permits two
choices on the exponents. The first choice is
\begin{eqnarray}
&&\alpha_{_{1,2}}=0,\;\alpha_{_{1,5}}=-1,\;\alpha_{_{3,2}}=0,\;\alpha_{_{3,5}}=2-{D\over2},
\label{1TR1-13}
\end{eqnarray}
whose geometric description is presented in Fig.\ref{fig2}(2)
with $(a,b)=(2,4)$, $\{c,d,e\}=\{1,3,5\}$. The second choice is
\begin{eqnarray}
&&\alpha_{_{1,2}}={D\over2}-2,\;\alpha_{_{1,5}}=1-{D\over2},\;\alpha_{_{3,2}}=2-{D\over2},\;\alpha_{_{3,5}}=0,
\label{1TR1-14}
\end{eqnarray}
whose geometric description is presented in Fig.\ref{fig2}(1) with $a=1$,
$\{(b,c),(d,e)\}=\{(2,4),(3,5)\}$, respectively.
Two hypergeometric functions are evidently written as
\begin{eqnarray}
&&\psi_{_{\{1,3,4\}}}^{(1)}(p_{_1}^2,p_{_2}^2,0)\sim
{(p_{_1}^2)^{D/2-2}\over p_{_1}^2-p_{_2}^2}
\sum\limits_{n=0}^\infty{1\over\Gamma(3-{D\over2}+n)\Gamma(1-n)}
\Big({p_{_2}^2\over p_{_1}^2-p_{_2}^2}\Big)^n
\nonumber\\
&&\hspace{2.8cm}\sim
{(p_{_1}^2)^{D/2-2}\over p_{_1}^2-p_{_2}^2}
\;,\nonumber\\
&&\psi_{_{\{1,3,4\}}}^{(2)}(p_{_1}^2,p_{_2}^2,0)\sim
{(p_{_1}^2p_{_2}^2)^{D/2-2}\over(p_{_1}^2-p_{_2}^2)^{D/2-1}}
\sum\limits_{n=0}^\infty
{\Gamma({D\over2}-2+n)\over n!}\Big({p_{_2}^2\over p_{_2}^2-p_{_1}^2}\Big)^n
\nonumber\\
&&\hspace{2.8cm}\sim
{(p_{_2}^2)^{D/2-2}\over p_{_1}^2-p_{_2}^2}\;,
\label{1TR1-14a}
\end{eqnarray}
whose intersection of their convergent regions is $|p_{_2}^2/(p_{_2}^2-p_{_1}^2)|\le1$.

Corresponding to the integer lattice $-nE_{_{3}}^{(2)}$, the first choice of
exponents is
\begin{eqnarray}
&&\alpha_{_{1,2}}=-1,\;\alpha_{_{1,5}}=0,\;\alpha_{_{3,2}}=1,\;\alpha_{_{3,5}}=1-{D\over2},
\label{1TR1-15}
\end{eqnarray}
whose geometric representation is drawn in Fig.\ref{fig2}(1) with $a=4$,
$\{(b,c)$, $(d,e)\}=$ $\{(1,2)$, $(3,5)\}$, and the second choice is
\begin{eqnarray}
&&\alpha_{_{1,2}}=0,\;\alpha_{_{1,5}}=-1,\;\alpha_{_{3,2}}=0,\;\alpha_{_{3,5}}=2-{D\over2}.
\label{1TR1-16}
\end{eqnarray}
where the geometric representation is drawn in Fig.\ref{fig2}(2) with
$(a,b)=$ $(1,2)$, $\{c,d,e\}=$ $\{3,4,5\}$.
The corresponding hypergeometric functions are written as
\begin{eqnarray}
&&\psi_{_{\{1,3,4\}}}^{(3)}(p_{_1}^2,p_{_2}^2,0)\sim
{(p_{_1}^2)^{D/2-2}\over p_{_2}^2}
\sum\limits_{n=0}^\infty
{\Gamma({D\over2}-1+n)\over\Gamma(2+n)}\Big({p_{_2}^2-p_{_1}^2\over p_{_2}^2}\Big)^n
\nonumber\\
&&\hspace{2.8cm}\sim
{(p_{_2}^2)^{D/2-2}-(p_{_1}^2)^{D/2-2}\over p_{_1}^2-p_{_2}^2}
\;,\nonumber\\
&&\psi_{_{\{1,3,4\}}}^{(4)}(p_{_1}^2,p_{_2}^2,0)\sim
{(p_{_1}^2)^{D/2-2}\over p_{_1}^2-p_{_2}^2}
\sum\limits_{n=1}^\infty
{\Gamma({D\over2}-2+n)\over \Gamma(n)}\Big({p_{_2}^2-p_{_1}^2\over p_{_2}^2}\Big)^n
\nonumber\\
&&\hspace{2.8cm}\sim
{(p_{_2}^2)^{D/2-2}\over p_{_1}^2}\;,
\label{1TR1-17}
\end{eqnarray}
whose intersection of their convergent regions is $|(p_{_2}^2-p_{_1}^2)/p_{_2}^2|\le1$.

Since $\det(A_{_{\{2,3,4\}}}^{(2)})=p_{_2}^2$, where the matrix $A_{_{\{2,3,4\}}}^{(2)}$ denotes
the submatrix of $A^{(2)}$ composed of the  second, third, and fourth columns, we have
\begin{eqnarray}
&&\Big(A_{_{\{2,3,4\}}}^{(2)}\Big)^{-1}\cdot A^{(2)}=\left(\begin{array}{ccccc}
\;\;-{p_{_1}^2\over p_{_2}^2}\;\;&\;\;1\;\;&\;\;0\;\;&\;\;0\;\;&\;\;1-{p_{_1}^2\over p_{_2}^2}\;\;\\
\;\;0\;\;&\;\;0\;\;&\;\;1\;\;&\;\;0\;\;&\;\;1\;\;\\
\;\;{1\over p_{_2}^2}\;\;&\;\;0\;\;&\;\;0\;\;&\;\;1\;\;&\;\;{1\over p_{_2}^2}\;\;\end{array}\right)\;.
\label{1TR1-18}
\end{eqnarray}
Obviously the matrix of exponents is written as
\begin{eqnarray}
&&\left(\begin{array}{ccccc}
\;\alpha_{_{1,1}}\;\;&\;0\;\;&\;\;0\;\;&\;\;0\;\;&\;\;\alpha_{_{1,5}}\;\;\\
\;\;0\;\;&\;\;0\;\;&\;\;3-D\;\;&\;\;0\;\;&\;\;D-4\;\;\\
\;\alpha_{_{3,1}}\;\;&\;0\;\;&\;\;0\;\;&\;\;{D\over2}-3\;\;&\;\;\alpha_{_{3,5}}\;\;\end{array}\right)\;,
\label{1TR1-19}
\end{eqnarray}
where the matrix elements satisfy the relations
\begin{eqnarray}
&&\alpha_{_{1,1}}+\alpha_{_{1,5}}=-1\;,\;\alpha_{_{3,1}}+\alpha_{_{3,5}}=2-{D\over2}\;,
\nonumber\\
&&\alpha_{_{1,1}}+\alpha_{_{3,1}}=0\;,\;\alpha_{_{1,5}}+\alpha_{_{3,5}}=1-{D\over2}\;.
\label{1TR1-20}
\end{eqnarray}
There are two choices on the matrix of integer lattice whose submatrix
composed of the first and fifth columns is formulated as $\pm nE_{_3}^{(2)}$ with $n\ge0$, and
other elements are all zero. The matrix of integer lattice $nE_{_3}^{(2)}$ permits two
choices on the exponents. The first choice is
\begin{eqnarray}
&&\alpha_{_{1,1}}=0,\;\alpha_{_{1,5}}=-1,\;\alpha_{_{3,1}}=0,\;\alpha_{_{3,5}}=2-{D\over2},
\label{1TR1-21}
\end{eqnarray}
whose geometric description is presented in Fig.\ref{fig2}(2)
with $(a,b)$ $=(1,4)$, $\{c,d,e\}$ $=\{2,3,5\}$. The second choice is
\begin{eqnarray}
&&\alpha_{_{1,1}}={D\over2}-2,\;\alpha_{_{1,5}}=1-{D\over2},\;\alpha_{_{3,1}}=2-{D\over2},\;\alpha_{_{3,5}}=0,
\label{1TR1-22}
\end{eqnarray}
whose geometric description is presented in Fig.\ref{fig2}(1) with $a=2$,
$\{(b,c)$, $(d,e)\}$ $=\{(1,4)$, $(3,5)\}$, respectively.

The first choice of exponents of the integer lattice $-nE_{_3}^{(2)}$ is
\begin{eqnarray}
&&\alpha_{_{1,1}}=-1,\;\alpha_{_{1,5}}=0,\;\alpha_{_{3,1}}=1,\;\alpha_{_{3,5}}=1-{D\over2},
\label{1TR1-23}
\end{eqnarray}
whose geometric representation is drawn in Fig.\ref{fig2}(1) with $a=4$,
$\{(b,c)$, $(d,e)\}$ $=\{(1,2)$, $(3,5)\}$. The second choice is
\begin{eqnarray}
&&\alpha_{_{1,1}}=0,\;\alpha_{_{1,5}}=-1,\;\alpha_{_{3,1}}=0,\;\alpha_{_{3,5}}=2-{D\over2}.
\label{1TR1-24}
\end{eqnarray}
whose geometric representation is drawn in Fig.\ref{fig2}(2) with
$(a,b)$ $=(1,2)$, $\{c,d,e\}$ $=\{3,4,5\}$.
The corresponding hypergeometric functions are evidently written as
\begin{eqnarray}
&&\psi_{_{\{2,3,4\}}}^{(i)}(p_{_1}^2,p_{_2}^2,0)\sim
\psi_{_{\{1,3,4\}}}^{(i)}(p_{_2}^2,p_{_1}^2,0),\;(i=1,2,3,4)\;.
\label{1TR1-25}
\end{eqnarray}

One obviously has $\det(A_{_{\{1,4,5\}}}^{(2)})=p_{_1}^2$, thus derives
\begin{eqnarray}
&&\Big(A_{_{\{1,4,5\}}}^{(2)}\Big)^{-1}\cdot A^{(2)}=\left(\begin{array}{ccccc}
\;\;1\;\;&\;\;-{p_{_2}^2\over p_{_1}^2}\;\;&\;\;-1+{p_{_2}^2\over p_{_1}^2}\;\;&\;\;0\;\;&\;\;0\;\;\\
\;\;0\;\;&\;\;{1\over p_{_1}^2}\;\;&\;\;-{1\over p_{_1}^2}\;\;&\;\;1\;\;&\;\;0\;\;\\
\;\;0\;\;&\;\;0\;\;&\;\;1\;\;&\;\;0\;\;&\;\;1\;\;\end{array}\right)\;,
\label{1TR1-26}
\end{eqnarray}
where the matrix $A_{_{\{1,4,5\}}}^{(2)}$ denotes
the submatrix of $A^{(2)}$ composed of the first, fourth, and fifth columns.
Obviously the matrix of exponents is written as
\begin{eqnarray}
&&\left(\begin{array}{ccccc}
\;0\;\;&\;\alpha_{_{1,2}}\;\;&\;\;\alpha_{_{1,3}}\;\;&\;\;0\;\;&\;\;0\;\;\\
\;\;0\;\;&\;\;\alpha_{_{2,2}}\;\;&\;\;\alpha_{_{2,3}}\;\;&\;\;{D\over2}-3\;\;&\;\;0\;\;\\
\;0\;\;&\;0\;\;&\;\;2-{D\over2}\;\;&\;\;0\;\;&\;\;{D\over2}-3\;\;\end{array}\right)\;,
\label{1TR1-27}
\end{eqnarray}
where the matrix elements satisfy the relations
\begin{eqnarray}
&&\alpha_{_{1,2}}+\alpha_{_{1,3}}=-1\;,\;\alpha_{_{2,2}}+\alpha_{_{2,3}}=2-{D\over2}\;,
\nonumber\\
&&\alpha_{_{1,2}}+\alpha_{_{2,2}}=0\;,\;\alpha_{_{1,3}}+\alpha_{_{2,3}}=1-{D\over2}\;.
\label{1TR1-28}
\end{eqnarray}
There are two choices on the matrix of integer lattice whose
submatrix composed of the second and third columns is formulated as $\pm nE_{_3}^{(1)}$ with $n\ge0$, and
other elements are all zero. The matrix of integer lattice $nE_{_3}^{(1)}$ permits two
choices on the exponents. The first choice is
\begin{eqnarray}
&&\alpha_{_{1,2}}=0,\;\alpha_{_{1,3}}=-1,\;\alpha_{_{2,2}}=0,\;\alpha_{_{2,3}}=2-{D\over2},
\label{1TR1-29}
\end{eqnarray}
whose geometric description is presented in Fig.\ref{fig2}(2)
with $(a,b)$ $=(2,4)$, $\{c,d,e\}$ $=\{1,3,5\}$. The second one is
\begin{eqnarray}
&&\alpha_{_{1,2}}={D\over2}-2,\;\alpha_{_{1,3}}=1-{D\over2},\;\alpha_{_{2,2}}=2-{D\over2},\;\alpha_{_{2,3}}=0,
\label{1TR1-30}
\end{eqnarray}
whose geometric description is presented in Fig.\ref{fig2}(1) with $a=1$,
$\{(b,c)$, $(d,e)\}$ $=\{(2,4)$, $(3,5)\}$, respectively.
The hypergeometric functions are
\begin{eqnarray}
&&\psi_{_{\{1,4,5\}}}^{(1)}(p_{_1}^2,p_{_2}^2,0)\sim
{(p_{_1}^2)^{D/2-2}\over p_{_1}^2-p_{_2}^2}
\sum\limits_{n=0}^\infty{1\over\Gamma(3-{D\over2}+n)\Gamma(1-n)}
\Big({p_{_2}^2\over p_{_1}^2-p_{_2}^2}\Big)^n
\nonumber\\
&&\hspace{2.8cm}\sim
{(p_{_1}^2)^{D/2-2}\over p_{_1}^2-p_{_2}^2}
\;,\nonumber\\
&&\psi_{_{\{1,4,5\}}}^{(2)}(p_{_1}^2,p_{_2}^2,0)\sim
{(p_{_1}^2p_{_2}^2)^{D/2-2}\over(p_{_1}^2-p_{_2}^2)^{D/2-1}}
\sum\limits_{n=0}^\infty
{\Gamma({D\over2}-2+n)\over n!}\Big({p_{_2}^2\over p_{_2}^2-p_{_1}^2}\Big)^n
\nonumber\\
&&\hspace{2.8cm}\sim
{(p_{_2}^2)^{D/2-2}\over p_{_1}^2-p_{_2}^2}\;,
\label{1TR1-30a}
\end{eqnarray}
whose intersection of their convergent regions is $|p_{_2}^2/(p_{_2}^2-p_{_1}^2)|\le1$.

Corresponding to the integer lattice $-nE_{_3}^{(2)}$, the first choice
of exponents is
\begin{eqnarray}
&&\alpha_{_{1,2}}=-1,\;\alpha_{_{1,3}}=0,\;\alpha_{_{2,2}}=1,\;\alpha_{_{2,3}}=1-{D\over2},
\label{1TR1-31}
\end{eqnarray}
whose geometric representation is drawn in Fig.\ref{fig2}(1) with $a=4$,
$\{(b,c)$, $(d,e)\}$ $=\{(1,2)$, $(3,5)\}$. The second choice is
\begin{eqnarray}
&&\alpha_{_{1,2}}=0,\;\alpha_{_{1,3}}=-1,\;\alpha_{_{2,2}}=0,\;\alpha_{_{2,3}}=2-{D\over2},
\label{1TR1-32}
\end{eqnarray}
whose geometric representation is drawn in Fig.\ref{fig2}(2) with
$(a,b)$ $=(1,2)$, $\{c,d,e\}$ $=\{3,4,5\}$.
The hypergeometric functions are presented as
\begin{eqnarray}
&&\psi_{_{\{1,4,5\}}}^{(3)}(p_{_1}^2,p_{_2}^2,0)\sim
{(p_{_1}^2)^{D/2-2}\over p_{_2}^2}\sum\limits_{n=0}^\infty
{\Gamma({D\over2}-2+n)\over\Gamma(2+n)}\Big({p_{_2}^2-p_{_1}^2\over p_{_2}^2}\Big)^n
\nonumber\\
&&\hspace{2.8cm}\sim
{(p_{_2}^2)^{D/2-2}-(p_{_1}^2)^{D/2-2}\over p_{_2}^2-p_{_1}^2}
\;,\nonumber\\
&&\psi_{_{\{1,4,5\}}}^{(4)}(p_{_1}^2,p_{_2}^2,0)\sim
{(p_{_1}^2)^{D/2-2}\over p_{_1}^2-p_{_2}^2}\;,
\label{1TR1-33}
\end{eqnarray}
whose intersection of their convergent regions is $|(p_{_2}^2-p_{_1}^2)/p_{_2}^2|\le1$.

Since $\det(A_{_{\{2,4,5\}}}^{(2)})=-p_{_2}^2$, one derives
\begin{eqnarray}
&&\Big(A_{_{\{2,4,5\}}}^{(2)}\Big)^{-1}\cdot A^{(2)}=\left(\begin{array}{ccccc}
\;\;-{p_{_1}^2\over p_{_2}^2}\;\;&\;\;1\;\;&\;\;-1+{p_{_1}^2\over p_{_2}^2}\;\;&\;\;0\;\;&\;\;0\;\;\\
\;\;{1\over p_{_2}^2}\;\;&\;\;0\;\;&\;\;-{1\over p_{_2}^2}\;\;&\;\;1\;\;&\;\;0\;\;\\
\;\;0\;\;&\;\;0\;\;&\;\;1\;\;&\;\;0\;\;&\;\;1\;\;\end{array}\right)\;,
\label{1TR1-34}
\end{eqnarray}
where the matrix $A_{_{\{2,4,5\}}}^{(2)}$ denotes
the submatrix of $A^{(2)}$ consisted in the second, fourth, and fifth columns.
Obviously the matrix of exponents is written as
\begin{eqnarray}
&&\left(\begin{array}{ccccc}
\;\;\alpha_{_{1,1}}\;\;&\;\;0\;\;&\;\;\alpha_{_{1,3}}\;\;&\;\;0\;\;&\;\;0\;\;\\
\;\;\alpha_{_{2,1}}\;\;&\;\;0\;\;&\;\;\alpha_{_{2,3}}\;\;&\;\;{D\over2}-3\;\;&\;\;0\;\;\\
\;0\;\;&\;0\;\;&\;\;2-{D\over2}\;\;&\;\;0\;\;&\;\;{D\over2}-3\;\;\end{array}\right)\;,
\label{1TR1-35}
\end{eqnarray}
where the matrix elements satisfy the relations
\begin{eqnarray}
&&\alpha_{_{1,1}}+\alpha_{_{1,3}}=-1\;,\;\alpha_{_{2,1}}+\alpha_{_{2,3}}=2-{D\over2}\;,
\nonumber\\
&&\alpha_{_{1,1}}+\alpha_{_{2,1}}=0\;,\;\alpha_{_{1,3}}+\alpha_{_{2,3}}=1-{D\over2}\;.
\label{1TR1-36}
\end{eqnarray}
There are two choices on the matrix of integer lattice whose
submatrix composed of the first and
third columns is formulated as $\pm nE_{_{3}}^{(3)}$ with $n\ge0$, and
other elements are all zero. The matrix of integer lattice $nE_{_{3}}^{(3)}$ permits two
choices on exponents. The first one is
\begin{eqnarray}
&&\alpha_{_{1,1}}=0,\;\alpha_{_{1,3}}=-1,\;\alpha_{_{2,1}}=0,\;\alpha_{_{2,3}}=2-{D\over2},
\label{1TR1-37}
\end{eqnarray}
whose geometric description is presented in Fig.\ref{fig2}(2)
with $(a,b)$ $=(1,4)$, $\{c,d,e\}$ $=\{2,3,5\}$. The second choice is
\begin{eqnarray}
&&\alpha_{_{1,1}}={D\over2}-2,\;\alpha_{_{1,3}}=1-{D\over2},\;\alpha_{_{2,1}}=2-{D\over2},\;\alpha_{_{2,3}}=0,
\label{1TR1-38}
\end{eqnarray}
whose geometric description is presented in Fig.\ref{fig2}(1) with $a=2$,
$\{(b,c)$, $(d,e)\}$ $=\{(1,4)$, $(3,5)\}$, respectively.

The first choice of exponents of the integer lattice $-nE_{_{3}}^{(3)}$ is
\begin{eqnarray}
&&\alpha_{_{1,1}}=-1,\;\alpha_{_{1,3}}=0,\;\alpha_{_{2,1}}=1,\;\alpha_{_{2,3}}=1-{D\over2},
\label{1TR1-39}
\end{eqnarray}
whose geometric representation is drawn in Fig.\ref{fig2}(1) with $a=4$,
$\{(b,c)$, $(d,e)\}$ $=\{(1,2)$, $(3,5)\}$. The second choice is
\begin{eqnarray}
&&\alpha_{_{1,1}}=0,\;\alpha_{_{1,3}}=-1,\;\alpha_{_{2,1}}=0,\;\alpha_{_{2,3}}=2-{D\over2}.
\label{1TR1-40}
\end{eqnarray}
whose geometric representation is drawn in Fig.\ref{fig2}(2) with
$(a,b)$ $=(1,2)$, $\{c,d,e\}$ $=\{3,4,5\}$.
The hypergeometric functions are
\begin{eqnarray}
&&\psi_{_{\{2,4,5\}}}^{(i)}(p_{_1}^2,p_{_2}^2,0)\sim
\psi_{_{\{1,4,5\}}}^{(i)}(p_{_2}^2,p_{_1}^2,0),\;(i=1,2,3,4)\;.
\label{1TR1-41}
\end{eqnarray}

\subsection{The fundamental solution systems in general case}
\begin{figure}[ht]
\setlength{\unitlength}{1cm}
\centering
\vspace{0.0cm}\hspace{-1.5cm}
\includegraphics[height=8cm,width=8.0cm]{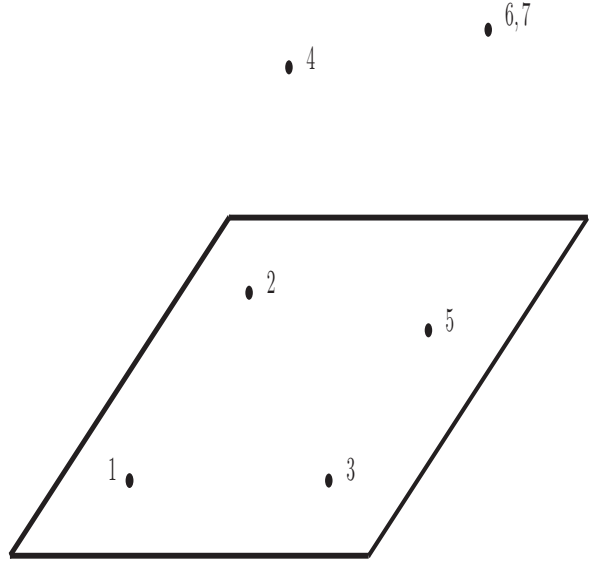}
\vspace{0cm}
\caption[]{The geometric description of the matroid $A^{(1T)}$ in
Eq.(\ref{1TR2-91}) on the projective plane $CP^{4}$, where the points $1,\cdots,7$
denote the indices of columns of the $3\times7$ matrix. In the representation,
point $4$ and double point $6,7$ lie out the hyperplane $\{1,2,3,5\}$.}
\label{fig9}
\end{figure}

\indent\indent
In the general case, the Feynman parametric representation is
\begin{eqnarray}
&&C(p_{_1}^2,\;p_{_2}^2,\;p_{_3}^2)=
i\Big(\Lambda_{_{\rm RE}}^2\Big)^{4-D}{(-)^{D/2-3}\Gamma(3-{D\over2})\over(4\pi)^{D/2}}
\int_0^1\omega_{_4}(t)t_{_4}^{3-D}
\nonumber\\
&&\hspace{2.9cm}\times
{\delta(t_{_1}+t_{_2}+t_{_3}+t_{_4})\over
(t_{_1}t_{_2}p_{_3}^2+t_{_2}t_{_3}p_{_1}^2+t_{_1}t_{_3}p_{_2}^2)^{3-D/2}}\;.
\label{1TR2-2}
\end{eqnarray}
The integral can be embedded in the subvariety of the Grassmannian $G_{_{4,7}}$
where the first row corresponds to the integration variable $t_{_1}$,
the second row corresponds to the integration variable $t_{_2}$,
the third row  corresponds to the integration variable $t_{_3}$,
and the fourth row  corresponds to the integration variable $t_{_4}$, respectively.
Meanwhile the first column represents the power function $t_{_1}^0=1$,
the second column represents the power function $t_{_2}^0=1$, the third column
represents the power function $t_{_3}^0=1$, the fourth column
represents the power function $t_{_4}^{3-D}$,
the seventh column represents the function $\delta(t_{_1}+t_{_2}+t_{_3})$. In order to
embed the homogeneous polynomial $t_{_1}t_{_2}p_{_3}^2+t_{_2}t_{_3}p_{_1}^2+t_{_1}t_{_3}p_{_2}^2$
as the fifth and sixth columns of the Grassmannian  $G_{_{4,7}}$, we rewrite
\begin{eqnarray}
&&t_{_1}t_{_2}p_{_3}^2+t_{_2}t_{_3}p_{_1}^2+t_{_1}t_{_3}p_{_2}^2
\nonumber\\
&&\hspace{-0.5cm}=
z_{_{1,\widehat{\sigma}_{_1}(5)}}z_{_{2,\widehat{\sigma}_{_1}(6)}}t_{_1}t_{_2}
+z_{_{1,\widehat{\sigma}_{_2}(5)}}z_{_{3,\widehat{\sigma}_{_2}(6)}}t_{_1}t_{_3}
+z_{_{2,\widehat{\sigma}_{_3}(5)}}z_{_{3,\widehat{\sigma}_{_3}(6)}}t_{_2}t_{_3}
\nonumber\\
&&\hspace{0.cm}
+z_{_{1,\widehat{\sigma}_{_4}(5)}}z_{_{4,\widehat{\sigma}_{_4}(6)}}t_{_1}t_{_4}
+z_{_{2,\widehat{\sigma}_{_5}(5)}}z_{_{4,\widehat{\sigma}_{_5}(6)}}t_{_2}t_{_4}
+z_{_{3,\widehat{\sigma}_{_6}(5)}}z_{_{4,\widehat{\sigma}_{_6}(6)}}t_{_3}t_{_4}\;,
\label{1TR2-2+a}
\end{eqnarray}
where $\widehat{\sigma}_{_i},\;(i=1,\cdots,\;6)$ are elements of the permutation
group $S_{_2}=\{\widehat{e},\;\widehat{(56)}\}$ on the column indices $5,\;6$.
Taking $\widehat{\sigma}_{_1}=\widehat{\sigma}_{_3}=\widehat{\sigma}_{_4}=\widehat{\sigma}_{_5}
=\widehat{\sigma}_{_6}=\widehat{(56)}$, $\widehat{\sigma}_{_2}=\widehat{e}$, we have
\begin{eqnarray}
&&z_{_{1,6}}z_{_{2,5}}=p_{_3}^2\;,\;\;z_{_{1,5}}z_{_{3,6}}=p_{_2}^2\;,\;\;
z_{_{2,6}}z_{_{3,5}}=p_{_1}^2\;,
\nonumber\\
&&z_{_{1,6}}z_{_{4,5}}=0\;,
\;\;z_{_{2,6}}z_{_{4,5}}=0\;,\;\;z_{_{3,6}}z_{_{4,5}}=0\;.
\label{1TR2-2+b}
\end{eqnarray}
The solution of Eq.(\ref{1TR2-2+b})
\begin{eqnarray}
&&z_{_{1,5}}=p_{_2}^2\;,z_{_{2,5}}=p_{_3}^2\;,z_{_{3,5}}=p_{_1}^2\;,z_{_{4,5}}=0\;,
\nonumber\\
&&z_{_{1,6}}=z_{_{2,6}}=z_{_{3,6}}=1\;,
\label{1TR2-2+c}
\end{eqnarray}
does not constrain the element $z_{_{4,6}}$. For convenience we choose
$z_{_{4,6}}=0$ or $z_{_{4,6}}=1$, respectively.
The integral can be embedded in the subvariety of the Grassmannian $G_{_{4,7}}$
where splitting local coordinates are written as
\begin{eqnarray}
&&A^{(1T)}=\left(\begin{array}{ccccccc}
\;\;1\;\;&\;\;0\;\;&\;\;0\;\;&\;\;0\;\;&\;\;p_{_2}^2\;\;&\;\;1\;\;&\;\;1\;\;\\
\;\;0\;\;&\;\;1\;\;&\;\;0\;\;&\;\;0\;\;&\;\;p_{_3}^2\;\;&\;\;1\;\;&\;\;1\;\;\\
\;\;0\;\;&\;\;0\;\;&\;\;1\;\;&\;\;0\;\;&\;\;p_{_1}^2\;\;&\;\;1\;\;&\;\;1\;\;\\
\;\;0\;\;&\;\;0\;\;&\;\;0\;\;&\;\;1\;\;&\;\;0\;\;&\;\;1\;\;&\;\;1\;\;\\
\end{array}\right)\;,
\label{1TR2-91}
\end{eqnarray}
as $z_{_{4,6}}=1$, whose geometric description is presented in Fig.\ref{fig9}.
Taking Feynman parametric representation of the 1-loop massless triangle diagram
as a function of the splitting local coordinates,
\begin{eqnarray}
&&C(p_{_1}^2,\;p_{_2}^2,\;p_{_3}^2)=\Psi^{1T}(z_{_1},\;\cdots,\;z_{_7})\;,
\label{1TR2-91+a}
\end{eqnarray}
we obviously have
\begin{eqnarray}
&&\Psi^{1T}(\chi_{_1}z_{_1},\;\cdots,\;\chi_{_7}z_{_7})=
\chi_{_4}^{3-D}\chi_{_5}^{D/2-3}\chi_{_6}^{D/2-3}\chi_{_7}^{-1}\Psi^{1T}(z_{_1},\;\cdots,\;z_{_7})\;.
\label{1TR2-91+b}
\end{eqnarray}
Where $z_{_i}\;(i=1,\cdots,7)$ denotes the i-th column vector in the matrix Eq.(\ref{1TR2-91}),
and $\chi_{_i}$ is a nonzero constant, respectively.

If we choose the first, second, third, and fourth column vectors
of $A^{(1T)}$ in Eq.(\ref{1TR2-91}) as the basis of the projective plane,
the convergent region of the fundamental solution system is the empty subset
of the whole parameter space because those constructed hypergeometric functions
contain redundant dependent variables.
However $\det(A_{_{\{1,2,3,6\}}}^{(1T)})=1$ where the matrix $A_{_{\{1,2,3,6\}}}^{(1T)}$ denotes
the submatrix of $A^{(1T)}$ composed of the first, second, third and sixth columns, one derives
the splitting local coordinates as
\begin{eqnarray}
&&Z_{_{4\times7}}=\Big(A_{_{\{1,2,3,6\}}}^{(1T)}\Big)^{-1}A^{(1T)}=\left(\begin{array}{ccccccc}
\;\;1\;\;&\;\;0\;\;&\;\;0\;\;&\;\;-1\;\;&\;\;p_{_2}^2\;\;&\;\;0\;\;&\;\;0\;\;\\
\;\;0\;\;&\;\;1\;\;&\;\;0\;\;&\;\;-1\;\;&\;\;p_{_3}^2\;\;&\;\;0\;\;&\;\;0\;\;\\
\;\;0\;\;&\;\;0\;\;&\;\;1\;\;&\;\;-1\;\;&\;\;p_{_1}^2\;\;&\;\;0\;\;&\;\;0\;\;\\
\;\;0\;\;&\;\;0\;\;&\;\;0\;\;&\;\;1\;\;&\;\;0\;\;&\;\;1\;\;&\;\;1\;\;
\end{array}\right)\;.
\label{1TR2-92}
\end{eqnarray}
Accordingly, the Feynman integral satisfies the following GKZ-system
\begin{eqnarray}
&&\Big\{\vartheta_{_{1,1}}+\vartheta_{_{1,4}}+\vartheta_{_{1,5}}\Big\}C(p_{_1}^2,\;p_{_2}^2,\;p_{_3}^2)=
-C(p_{_1}^2,\;p_{_2}^2,\;p_{_3}^2)
\;,\nonumber\\
&&\Big\{\vartheta_{_{2,2}}+\vartheta_{_{2,4}}+\vartheta_{_{2,5}}\Big\}C(p_{_1}^2,\;p_{_2}^2,\;p_{_3}^2)=
-C(p_{_1}^2,\;p_{_2}^2,\;p_{_3}^2)
\;,\nonumber\\
&&\Big\{\vartheta_{_{3,3}}+\vartheta_{_{3,4}}+\vartheta_{_{3,5}}\Big\}C(p_{_1}^2,\;p_{_2}^2,\;p_{_3}^2)=
-C(p_{_1}^2,\;p_{_2}^2,\;p_{_3}^2)
\;,\nonumber\\
&&\Big\{\vartheta_{_{4,4}}+\vartheta_{_{4,6}}+\vartheta_{_{4,7}}\Big\}C(p_{_1}^2,\;p_{_2}^2,\;p_{_3}^2)=
-C(p_{_1}^2,\;p_{_2}^2,\;p_{_3}^2)
\;,\nonumber\\
&&\vartheta_{_{1,1}}C(p_{_1}^2,\;p_{_2}^2,\;p_{_3}^2)=0
\;,\nonumber\\
&&\vartheta_{_{2,2}}C(p_{_1}^2,\;p_{_2}^2,\;p_{_3}^2)=0
\;,\nonumber\\
&&\vartheta_{_{3,3}}C(p_{_1}^2,\;p_{_2}^2,\;p_{_3}^2)=0
\;,\nonumber\\
&&\Big\{\vartheta_{_{1,4}}+\vartheta_{_{2,4}}+\vartheta_{_{3,4}}+\vartheta_{_{4,4}}\Big\}
C(p_{_1}^2,\;p_{_2}^2,\;p_{_3}^2)=(3-D)C(p_{_1}^2,\;p_{_2}^2,\;p_{_3}^2)
\;,\nonumber\\
&&\Big\{\vartheta_{_{1,5}}+\vartheta_{_{2,5}}+\vartheta_{_{3,5}}\Big\}C(p_{_1}^2,\;p_{_2}^2,\;p_{_3}^2)=
({D\over2}-3)C(p_{_1}^2,\;p_{_2}^2,\;p_{_3}^2)
\;,\nonumber\\
&&\vartheta_{_{4,6}}C(p_{_1}^2,\;p_{_2}^2,\;p_{_3}^2)=({D\over2}-3)C(p_{_1}^2,\;p_{_2}^2,\;p_{_3}^2)
\;,\nonumber\\
&&\vartheta_{_{4,7}}C(p_{_1}^2,\;p_{_2}^2,\;p_{_3}^2)=-C(p_{_1}^2,\;p_{_2}^2,\;p_{_3}^2)\;,
\label{1TR2-92a}
\end{eqnarray}
which induces the matrix of exponents as
\begin{eqnarray}
&&\left(\begin{array}{ccccccc}
\;\;0\;\;&\;\;0\;\;&\;\;0\;\;&\;\;\alpha_{_{1,4}}\;\;&\;\alpha_{_{1,5}}\;\;&\;\;0\;\;&\;\;0\;\;\\
\;\;0\;\;&\;\;0\;\;&\;\;0\;\;&\;\;\alpha_{_{2,4}}\;\;&\;\;\alpha_{_{2,5}}\;\;&\;0\;\;&\;\;0\;\;\\
\;\;0\;\;&\;\;0\;\;&\;\;0\;\;&\;\;\alpha_{_{3,4}}\;\;&\;\;\alpha_{_{3,5}}\;\;&\;0\;\;&\;\;0\;\;\\
\;\;0\;\;&\;\;0\;\;&\;\;0\;\;&\;\;3-{D\over2}\;\;&\;\;0\;\;&\;{D\over2}-3\;\;&\;\;-1\;\;
\end{array}\right)\;,
\label{1TR2-93}
\end{eqnarray}
with
\begin{eqnarray}
&&\alpha_{_{1,4}}+\alpha_{_{1,5}}=-1\;,\;\alpha_{_{2,4}}+\alpha_{_{2,5}}=-1\;,\;
\alpha_{_{3,4}}+\alpha_{_{3,5}}=-1\;,
\nonumber\\
&&\alpha_{_{1,4}}+\alpha_{_{2,4}}+\alpha_{_{3,4}}=-{D\over2}\;,\;\;
\alpha_{_{1,5}}+\alpha_{_{2,5}}+\alpha_{_{3,5}}={D\over2}-3\;.
\label{1TR2-94}
\end{eqnarray}
\begin{figure}[ht]
\setlength{\unitlength}{1cm}
\centering
\vspace{0.0cm}\hspace{-1.5cm}
\includegraphics[height=8cm,width=8.0cm]{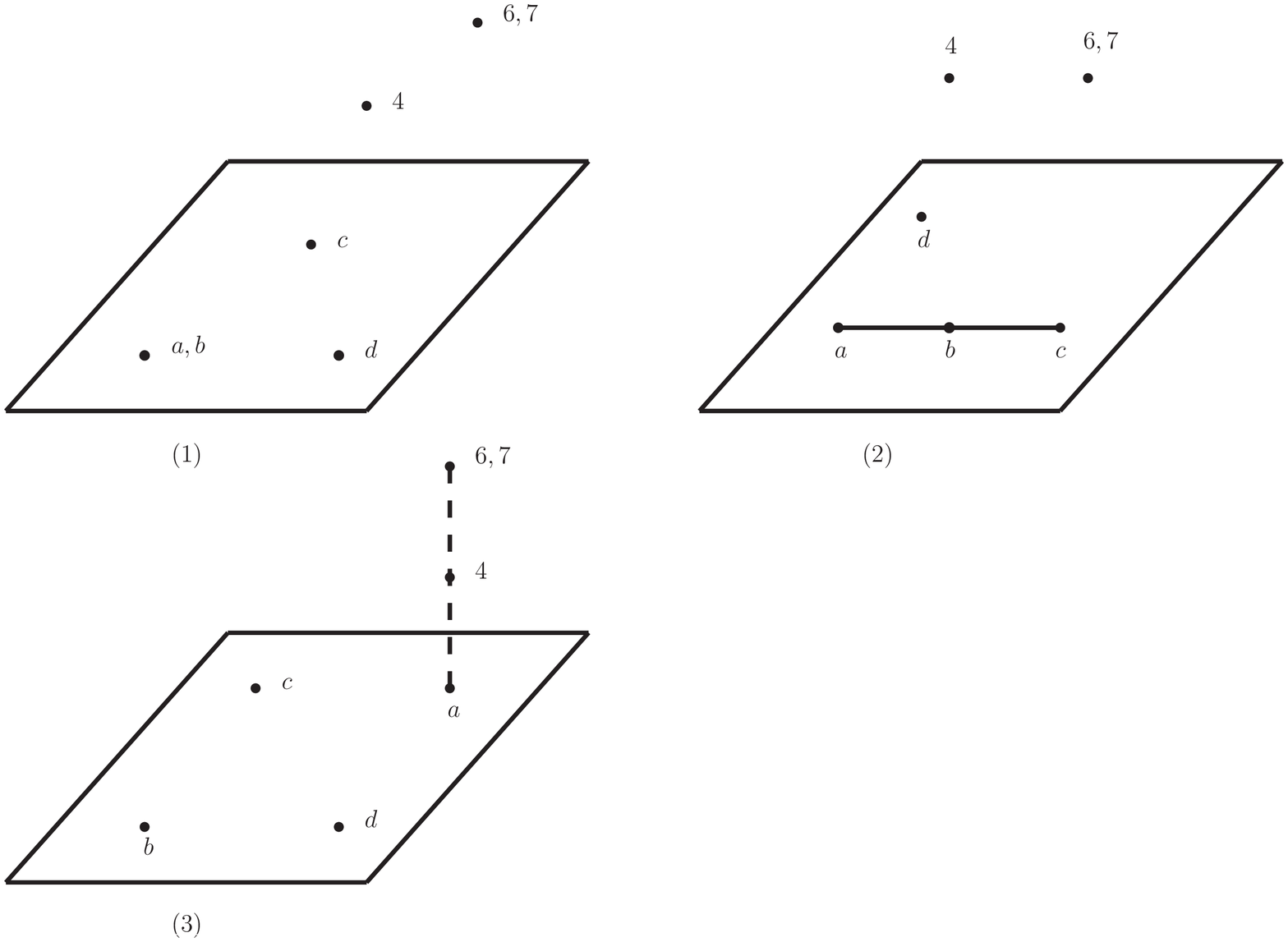}
\vspace{0cm}
\caption[]{The geometric descriptions of single orbits in
Eq.(\ref{1TR2-91}) on the projective plane $CP^{4}$, where point 4
and double point $(6,7)$ lie out the hyperplane $\{1,2,3,5\}$.}
\label{fig10}
\end{figure}
Corresponding to the matrix of local coordinates in Eq.(\ref{1TR2-92}),
we get 12 choices on the matrix of integer lattice whose submatrix composed of the fourth and
fifth columns is formulated as $\pm n_{_1}E_{_{4}}^{(i)}\pm n_{_2}E_{_{4}}^{(j)}$,
where $n_{_{1,2}}\ge0$, $(i,\;j)\in\{(1,\;2),\;(1,\;4),\;(2,\;4)\}$,
and other elements are all zero. Here the bases of integer lattice are written as
\begin{eqnarray}
&&E_{_{4}}^{(1)}=\left(\begin{array}{cc}\;1\;&\;-1\;\\\;-1\;&\;1\;\\
\;0\;&\;0\;\\\;0\;&\;0\;\end{array}\right),\;
E_{_{4}}^{(2)}=\left(\begin{array}{cc}\;1\;&\;-1\;\\\;0\;&\;0\;\\
\;-1\;&\;1\;\\ \;0\;&\;0\;\end{array}\right),\;
E_{_{4}}^{(3)}=\left(\begin{array}{cc}\;1\;&\;-1\;\\\;0\;&\;0\;\\
\;0\;&\;0\;\\ \;-1\;&\;1\;\end{array}\right),
\nonumber\\
&&E_{_{4}}^{(4)}=\left(\begin{array}{cc}\;0\;&\;0\;\\
\;1\;&\;-1\;\\\;-1\;&\;1\;\\ \;0\;&\;0\;\end{array}\right),\;
E_{_{4}}^{(5)}=\left(\begin{array}{cc}\;0\;&\;0\;\\
\;1\;&\;-1\;\\\;0\;&\;0\;\\ \;-1\;&\;1\;\end{array}\right),\;
E_{_{4}}^{(6)}=\left(\begin{array}{cc}\;0\;&\;0\;\\
\;0\;&\;0\;\\\;1\;&\;-1\;\\ \;-1\;&\;1\;\end{array}\right).
\label{1TR2-7}
\end{eqnarray}
Note that $E_{_{4}}^{(4)}=E_{_{4}}^{(2)}-E_{_{4}}^{(1)}$, two linear independent
integer lattice matrices $E_{_{4}}^{(1)}$ and $E_{_{4}}^{(4)}$ induce two independent
PDEs in splitting local coordinates as:
\begin{eqnarray}
&&{\partial^2C\over\partial z_{_{1,4}}\partial z_{_{2,5}}}(p_{_1}^2,\;p_{_2}^2,\;p_{_3}^2)
={\partial^2C\over\partial z_{_{1,5}}\partial z_{_{2,4}}}(p_{_1}^2,\;p_{_2}^2,\;p_{_3}^2)
\;,\nonumber\\
&&{\partial^2C\over\partial z_{_{2,4}}\partial z_{_{3,5}}}(p_{_1}^2,\;p_{_2}^2,\;p_{_3}^2)
={\partial^2C\over\partial z_{_{2,5}}\partial z_{_{3,4}}}(p_{_1}^2,\;p_{_2}^2,\;p_{_3}^2)\;.
\label{1TR2-94a2}
\end{eqnarray}
When the exponents are given, the Feynman integral can be formally expressed as
\begin{eqnarray}
&&C(p_{_1}^2,\;p_{_2}^2,\;p_{_3}^2)=\prod\limits_{i,j}z_{_{i,j}}^{\alpha_{_{i,j}}}
\varphi({p_{_1}^2\over p_{_3}^2},{p_{_2}^2\over p_{_3}^2})\;,
\label{1TR2-94a3}
\end{eqnarray}
where $(i,j)\in\{(1,4),(1,5),(2,4),(2,5),(3,4),(3,5)\}$, and $\varphi(x_{_1},x_{_2})$
satisfies the PDEs in the combined coordinate $x_{_1}$ $=z_{_{2,4}}z_{_{3,5}}/(z_{_{1,4}}z_{_{2,5}})$
$=p_{_1}^2/p_{_3}^2$, $x_{_2}$ $=z_{_{1,5}}z_{_{2,4}}/(z_{_{1,4}}z_{_{2,5}})$ $=p_{_2}^2/p_{_3}^2$ as
\begin{eqnarray}
&&x_{_1}\Big[\alpha_{_{1,4}}\alpha_{_{2,5}}+(1-\alpha_{_{1,4}}
-\alpha_{_{2,5}})x_{_1}{\partial\over\partial x_{_1}}
-\alpha_{_{1,4}}x_{_2}{\partial\over\partial x_{_2}}
+x_{_1}x_{_2}{\partial^2\over\partial x_{_1}\partial x_{_2}}
+x_{_1}^2{\partial^2\over\partial x_{_1}^2}\Big]
\varphi(x_{_1},x_{_2})
\nonumber\\
&&\hspace{-0.5cm}=
\Big[\alpha_{_{2,4}}\alpha_{_{3,5}}+(1+\alpha_{_{2,4}}
+\alpha_{_{3,5}})x_{_1}{\partial\over\partial x_{_1}}
+\alpha_{_{2,4}}x_{_2}{\partial\over\partial x_{_2}}
+x_{_1}x_{_2}{\partial^2\over\partial x_{_1}\partial x_{_2}}
+x_{_1}^2{\partial^2\over\partial x_{_1}^2}\Big]
\varphi(x_{_1},x_{_2})
\;,\nonumber\\
&&x_{_2}\Big[\alpha_{_{1,4}}\alpha_{_{2,5}}
-\alpha_{_{1,4}}x_{_1}{\partial\over\partial x_{_1}}
+(1-\alpha_{_{1,4}}-\alpha_{_{2,5}})x_{_2}{\partial\over\partial x_{_2}}
+x_{_1}x_{_2}{\partial^2\over\partial x_{_1}\partial x_{_2}}
+x_{_2}^2{\partial^2\over\partial x_{_2}^2}\Big]
\varphi(x_{_1},x_{_2})
\nonumber\\
&&\hspace{-0.5cm}=
\Big[\alpha_{_{1,5}}\alpha_{_{2,4}}
+\alpha_{_{1,5}}x_{_1}{\partial\over\partial x_{_1}}
+(1+\alpha_{_{1,5}}+\alpha_{_{2,4}})x_{_2}{\partial\over\partial x_{_2}}
+x_{_1}x_{_2}{\partial^2\over\partial x_{_1}\partial x_{_2}}
+x_{_2}^2{\partial^2\over\partial x_{_2}^2}\Big]
\varphi(x_{_1},x_{_2})\;.
\label{1TR2-94a4}
\end{eqnarray}
As $\alpha_{_{1,5}}=\alpha_{_{2,4}}=0$, the PDEs above are changed into
the first type Appell PDEs. The holonomic rank of the ideal composed of the
partial differential operators of above PDEs is 3, which indicates that
three linear independent hypergeometric functions constitute the fundamental
solution system in neighborhoods of regular singular points.
It is known that the fourth type Appell function cannot be embedded into the subvarieties of
Grassmannians~\cite{Gelfand1989}. Nevertheless, including the appropriate power factor,
the Feynman integral presented through the fourth Appell functions~\cite{Feng2018,Feng2020,Davydychev1}
satisfies the PDEs above. Those matrices of integer lattice permit 12 possibilities of the exponents.

The exponents of the integer lattice $n_{_1}E_{_{4}}^{(1)}+n_{_2}E_{_{4}}^{(2)}$ are
\begin{eqnarray}
&&\alpha_{_{1,4}}=2-{D\over2}\;,\;\alpha_{_{1,5}}={D\over2}-3\;,\;\alpha_{_{2,4}}=-1\;,\;
\alpha_{_{2,5}}=0\;,\;\alpha_{_{3,4}}=-1,\;\alpha_{_{3,5}}=0\;,
\label{1TR2-95}
\end{eqnarray}
whose geometric representation is presented in Fig.\ref{fig10}(1)
with $(a,b)$ $=(1,5)$, $\{c,d\}$ $=\{2,3\}$. As the integer lattice is chosen as
$n_{_1}E_{_{4}}^{(1)}$ $+n_{_2}E_{_{4}}^{(4)}$, the matrix of exponents is
\begin{eqnarray}
&&\alpha_{_{1,4}}=0\;,\;\alpha_{_{1,5}}=-1\;,\;\alpha_{_{2,4}}=1-{D\over2}\;,\;
\alpha_{_{2,5}}={D\over2}-2\;,\;\alpha_{_{3,4}}=-1,\;\alpha_{_{3,5}}=0\;,
\label{1TR2-99}
\end{eqnarray}
whose geometric representation is shown in Fig.\ref{fig10}(2)
with $\{a,b,c\}$ $=\{1,2,5\}$, $d=3$.
When the integer lattice is chosen as $n_{_1}E_{_{4}}^{(2)}$ $+n_{_2}E_{_{4}}^{(4)}$,
the exponents are
\begin{eqnarray}
&&\alpha_{_{1,4}}=0\;,\;\alpha_{_{1,5}}=-1\;,\;\alpha_{_{2,4}}=0\;,\;
\alpha_{_{2,5}}=-1\;,\;\alpha_{_{3,4}}=-{D\over2},\;\alpha_{_{3,5}}={D\over2}-1\;,
\label{1TR2-103}
\end{eqnarray}
whose geometric representation is drawn in Fig.\ref{fig10}(3)
with $a=3$, $(b,c,d)$ $=(1,2,5)$. Accordingly the constructed hypergeometric
functions are
\begin{eqnarray}
&&\psi_{_{\{1,2,3,6\}}}^{(1)}(p_{_1}^2,p_{_2}^2,p_{_3}^2)
\sim(p_{_2}^2)^{D/2-3}\sum\limits_{n_{_1},n_{_2}}^\infty
\Big({p_{_3}^2\over p_{_2}^2}\Big)^{n_{_1}}
\Big({p_{_1}^2\over p_{_2}^2}\Big)^{n_{_2}}
\nonumber\\
&&\hspace{3.2cm}\sim
{(p_{_2}^2)^{D/2-1}\over(p_{_2}^2-p_{_1}^2)(p_{_2}^2-p_{_3}^2)}
\;,\nonumber\\
&&\psi_{_{\{1,2,3,6\}}}^{(2)}(p_{_1}^2,p_{_2}^2,p_{_3}^2)
\sim{(p_{_3}^2)^{D/2-2}\over p_{_2}^2}\sum\limits_{n_{_1},n_{_2}}^\infty
\Big({p_{_3}^2\over p_{_2}^2}\Big)^{n_{_1}}
\Big({p_{_1}^2\over p_{_3}^2}\Big)^{n_{_2}}
\nonumber\\
&&\hspace{3.2cm}\sim
{(p_{_3}^2)^{D/2-1}\over(p_{_3}^2-p_{_1}^2)(p_{_3}^2-p_{_2}^2)}
\;,\nonumber\\
&&\psi_{_{\{1,2,3,6\}}}^{(3)}(p_{_1}^2,p_{_2}^2,p_{_3}^2)
\sim{(p_{_1}^2)^{D/2-1}\over p_{_2}^2p_{_3}^2}
\sum\limits_{n_{_1},n_{_2}}^\infty\Big({p_{_1}^2\over p_{_2}^2}\Big)^{n_{_1}}
\Big({p_{_1}^2\over p_{_3}^2}\Big)^{n_{_2}}
\nonumber\\
&&\hspace{3.2cm}\sim
{(p_{_1}^2)^{D/2-1}\over(p_{_1}^2-p_{_2}^2)(p_{_1}^2-p_{_3}^2)}\;,
\label{1TR2-103a}
\end{eqnarray}
where the intersection of their convergent regions is a nonempty proper subset of the whole parameter space.

The exponents corresponding to the integer lattice
$-n_{_1}E_{_{4}}^{(1)}$ $-n_{_2}E_{_{4}}^{(2)}$ are
\begin{eqnarray}
&&\alpha_{_{1,4}}=-{D\over2}\;,\;\alpha_{_{1,5}}={D\over2}-1\;,\;\alpha_{_{2,4}}=0\;,\;
\alpha_{_{2,5}}=-1\;,\;\alpha_{_{3,4}}=0,\;\alpha_{_{3,5}}=-1\;,
\label{1TR2-98}
\end{eqnarray}
whose geometric representation is plotted in Fig.\ref{fig10}(3)
with $a=1$, $\{b,c,d\}$ $=\{2,3,5\}$.
The exponents of the inter lattice $-n_{_1}E_{_{4}}^{(1)}$ $-n_{_2}E_{_{4}}^{(4)}$ are
\begin{eqnarray}
&&\alpha_{_{1,4}}=-1\;,\;\alpha_{_{1,5}}=0\;,\;\alpha_{_{2,4}}=1-{D\over2}\;,\;
\alpha_{_{2,5}}={D\over2}-2\;,\;\alpha_{_{3,4}}=0,\;\alpha_{_{3,5}}=-1\;,
\label{1TR2-102}
\end{eqnarray}
whose geometric representation is plotted in Fig.\ref{fig10}(2)
with $\{a,b,c\}$ $=\{2,3,5\}$, $d=1$. Similarly the exponents
of the integer lattice $-n_{_1}E_{_{4}}^{(2)}$ $-n_{_2}E_{_{4}}^{(4)}$ are
\begin{eqnarray}
&&\alpha_{_{1,4}}=-1\;,\;\alpha_{_{1,5}}=0\;,\;\alpha_{_{2,4}}=-1\;,\;
\alpha_{_{2,5}}=0\;,\;\alpha_{_{3,4}}=2-{D\over2},\;\alpha_{_{3,5}}={D\over2}-3\;,
\label{1TR2-106}
\end{eqnarray}
whose geometric representation is presented in Fig.\ref{fig10}(1)
with $(a,b)$ $=(3,5)$, $\{c,d\}$ $=\{1,2\}$.
The corresponding hypergeometric functions are
\begin{eqnarray}
&&\psi_{_{\{1,2,3,6\}}}^{(4)}(p_{_1}^2,p_{_2}^2,p_{_3}^2)
\sim{(p_{_2}^2)^{D/2-1}\over p_{_1}^2p_{_3}^2}
\sum\limits_{n_{_1},n_{_2}}^\infty\Big({p_{_2}^2\over p_{_3}^2}\Big)^{n_{_1}}
\Big({p_{_2}^2\over p_{_1}^2}\Big)^{n_{_2}}
\nonumber\\
&&\hspace{3.2cm}\sim
{(p_{_2}^2)^{D/2-1}\over(p_{_2}^2-p_{_1}^2)(p_{_2}^2-p_{_3}^2)}
\;,\nonumber\\
&&\psi_{_{\{1,2,3,6\}}}^{(5)}(p_{_1}^2,p_{_2}^2,p_{_3}^2)
\sim{(p_{_3}^2)^{D/2-2}\over p_{_1}^2}
\sum\limits_{n_{_1},n_{_2}}^\infty\Big({p_{_2}^2\over p_{_3}^2}\Big)^{n_{_1}}
\Big({p_{_3}^2\over p_{_1}^2}\Big)^{n_{_2}}
\nonumber\\
&&\hspace{3.2cm}\sim
{(p_{_3}^2)^{D/2-1}\over(p_{_3}^2-p_{_1}^2)(p_{_3}^2-p_{_2}^2)}
\;,\nonumber\\
&&\psi_{_{\{1,2,3,6\}}}^{(6)}(p_{_1}^2,p_{_2}^2,p_{_3}^2)
\sim(p_{_1}^2)^{D/2-3}
\sum\limits_{n_{_1},n_{_2}}^\infty\Big({p_{_2}^2\over p_{_1}^2}\Big)^{n_{_1}}
\Big({p_{_3}^2\over p_{_1}^2}\Big)^{n_{_2}}
\nonumber\\
&&\hspace{3.2cm}\sim
{(p_{_1}^2)^{D/2-1}\over(p_{_1}^2-p_{_2}^2)(p_{_1}^2-p_{_3}^2)}\;,
\label{1TR2-106a}
\end{eqnarray}
where the intersection of their convergent regions is a nonempty proper subset of the whole parameter space.
The fact indicates that those hypergeometric functions constitute a fundamental solution
system of the GKZ-system in the proper subset of the whole parameter space.

Corresponding to the integer lattice $n_{_1}E_{_{4}}^{(1)}$ $-n_{_2}E_{_{4}}^{(2)}$,
the exponents are
\begin{eqnarray}
&&\alpha_{_{1,4}}=1-{D\over2}\;,\;\alpha_{_{1,5}}={D\over2}-2\;,\;\alpha_{_{2,4}}=-1\;,\;
\alpha_{_{2,5}}=0\;,\;\alpha_{_{3,4}}=0,\;\alpha_{_{3,5}}=-1\;,
\label{1TR2-96}
\end{eqnarray}
whose geometric representation is drawn in Fig.\ref{fig10}(2)
with $\{a,b,c\}$ $=\{1,3,5\}$, $d=2$. The exponents of the
integer lattice $n_{_1}E_{_{4}}^{(1)}$ $-n_{_2}E_{_{4}}^{(4)}$ are
\begin{eqnarray}
&&\alpha_{_{1,4}}=0\;,\;\alpha_{_{1,5}}=-1\;,\;\alpha_{_{2,4}}=-{D\over2}\;,\;
\alpha_{_{2,5}}={D\over2}-1\;,\;\alpha_{_{3,4}}=0,\;\alpha_{_{3,5}}=-1\;,
\label{1TR2-100}
\end{eqnarray}
whose geometric representation is presented in Fig.\ref{fig10}(3)
with $a=2$, $\{b,c,d\}$ $=\{1,3,5\}$. As the integer lattice is chosen as
$n_{_1}E_{_{4}}^{(2)}$ $-n_{_2}E_{_{4}}^{(4)}$, the exponents are
\begin{eqnarray}
&&\alpha_{_{14}}=0\;,\;\alpha_{_{15}}=-1\;,\;\alpha_{_{24}}=-1\;,\;
\alpha_{_{25}}=0\;,\;\alpha_{_{34}}=1-{D\over2},\;\alpha_{_{35}}={D\over2}-2\;,
\label{1TR2-104}
\end{eqnarray}
whose geometric representation is presented in Fig.\ref{fig10}(2)
with $\{a,b,c\}$ $=\{1,3,5\}$, $d=2$. Those hypergeometric functions are
\begin{eqnarray}
&&\psi_{_{\{1,2,3,6\}}}^{(7)}(p_{_1}^2,p_{_2}^2,p_{_3}^2)
\sim{(p_{_2}^2)^{D/2-2}\over p_{_1}^2}
\sum\limits_{n_{_1},n_{_2}}^\infty\Big({p_{_3}^2\over p_{_2}^2}\Big)^{n_{_1}}
\Big({p_{_2}^2\over p_{_1}^2}\Big)^{n_{_2}}
\nonumber\\
&&\hspace{3.2cm}\sim
{(p_{_2}^2)^{D/2-1}\over(p_{_2}^2-p_{_1}^2)(p_{_2}^2-p_{_3}^2)}
\;,\nonumber\\
&&\psi_{_{\{1,2,3,6\}}}^{(8)}(p_{_1}^2,p_{_2}^2,p_{_3}^2)
\sim{(p_{_3}^2)^{D/2-1}\over p_{_1}^2p_{_2}^2}
\sum\limits_{n_{_1},n_{_2}}^\infty\Big({p_{_3}^2\over p_{_2}^2}\Big)^{n_{_1}}
\Big({p_{_3}^2\over p_{_1}^2}\Big)^{n_{_2}}
\nonumber\\
&&\hspace{3.2cm}\sim
{(p_{_3}^2)^{D/2-1}\over(p_{_3}^2-p_{_1}^2)(p_{_3}^2-p_{_2}^2)}
\;,\nonumber\\
&&\psi_{_{\{1,2,3,6\}}}^{(9)}(p_{_1}^2,p_{_2}^2,p_{_3}^2)
\sim{(p_{_1}^2)^{D/2-2}\over p_{_2}^2}
\sum\limits_{n_{_1},n_{_2}}^\infty\Big({p_{_1}^2\over p_{_2}^2}\Big)^{n_{_1}}
\Big({p_{_3}^2\over p_{_1}^2}\Big)^{n_{_2}}
\nonumber\\
&&\hspace{3.2cm}\sim
{(p_{_1}^2)^{D/2-1}\over(p_{_1}^2-p_{_2}^2)(p_{_1}^2-p_{_3}^2)}\;.
\label{1TR2-104a}
\end{eqnarray}
Although the intersection of the convergent regions of these series is
the empty subset of the whole parameter space, the sets of the hypergeometric functions
$\{\psi_{_{\{1,2,3,6\}}}^{(1)},\psi_{_{\{1,2,3,6\}}}^{(8)},\psi_{_{\{1,2,3,6\}}}^{(9)}\}$
and $\{\psi_{_{\{1,2,3,6\}}}^{(6)},\psi_{_{\{1,2,3,6\}}}^{(7)},\psi_{_{\{1,2,3,6\}}}^{(8)}\}$
constitute the fundamental solution systems of the nonempty proper subsets
of the whole parameter space, respectively.

The exponents of the integer lattice $-n_{_1}E_{_{4}}^{(1)}$ $+n_{_2}E_{_{4}}^{(2)}$ are
\begin{eqnarray}
&&\alpha_{_{1,4}}=1-{D\over2}\;,\;\alpha_{_{1,5}}={D\over2}-2\;,\;\alpha_{_{2,4}}=0\;,\;
\alpha_{_{2,5}}=-1\;,\;\alpha_{_{3,4}}=-1,\;\alpha_{_{3,5}}=0\;,
\label{1TR2-97}
\end{eqnarray}
whose geometric representation is presented in Fig.\ref{fig10}(2)
with $\{a,b,c\}$ $=\{1,2,5\}$, $d=3$. Similarly the exponents of
the integer lattice $-n_{_1}E_{_{4}}^{(1)}$ $+n_{_2}E_{_{4}}^{(4)}$ are
\begin{eqnarray}
&&\alpha_{_{1,4}}=-1\;,\;\alpha_{_{1,5}}=0\;,\;\alpha_{_{2,4}}=2-{D\over2}\;,\;
\alpha_{_{2,5}}={D\over2}-3\;,\;\alpha_{_{3,4}}=-1,\;\alpha_{_{3,5}}=0\;,
\label{1TR2-101}
\end{eqnarray}
whose geometric representation is plotted in Fig.\ref{fig10}(1)
with $(a,b)$ $=(2,5)$, $\{c,d\}$ $=\{1,3\}$. Corresponding to integer lattice
$-n_{_1}E_{_{4}}^{(2)}$ $+n_{_2}E_{_{4}}^{(4)}$, the exponents are
\begin{eqnarray}
&&\alpha_{_{1,4}}=-1\;,\;\alpha_{_{1,5}}=0\;,\;\alpha_{_{2,4}}=0\;,\;
\alpha_{_{2,5}}=-1\;,\;\alpha_{_{3,4}}=1-{D\over2},\;\alpha_{_{3,5}}={D\over2}-2\;,
\label{1TR2-105}
\end{eqnarray}
whose geometric representation is drawn in Fig.\ref{fig10}(2)
with $\{a,b,c\}$ $=\{2,3,5\}$, $d=1$.
The corresponding hypergeometric functions are
\begin{eqnarray}
&&\psi_{_{\{1,2,3,6\}}}^{(10)}(p_{_1}^2,p_{_2}^2,p_{_3}^2)\sim
{(p_{_2}^2)^{D/2-2}\over p_{_3}^2}
\sum\limits_{n_{_1},n_{_2}}^\infty\Big({p_{_2}^2\over p_{_3}^2}\Big)^{n_{_1}}
\Big({p_{_1}^2\over p_{_2}^2}\Big)^{n_{_2}}
\nonumber\\
&&\hspace{3.2cm}\sim
{(p_{_2}^2)^{D/2-1}\over(p_{_2}^2-p_{_1}^2)(p_{_2}^2-p_{_3}^2)}
\;,\nonumber\\
&&\psi_{_{\{1,2,3,6\}}}^{(11)}(p_{_1}^2,p_{_2}^2,p_{_3}^2)
\sim(p_{_3}^2)^{D/2-3}
\sum\limits_{n_{_1},n_{_2}}^\infty\Big({p_{_2}^2\over p_{_3}^2}\Big)^{n_{_1}}
\Big({p_{_1}^2\over p_{_3}^2}\Big)^{n_{_2}}
\nonumber\\
&&\hspace{3.2cm}\sim
{(p_{_3}^2)^{D/2-1}\over(p_{_3}^2-p_{_1}^2)(p_{_3}^2-p_{_2}^2)}
\;,\nonumber\\
&&\psi_{_{\{1,2,3,6\}}}^{(12)}(p_{_1}^2,p_{_2}^2,p_{_3}^2)
\sim{(p_{_1}^2)^{D/2-2}\over p_{_3}^2}
\sum\limits_{n_{_1},n_{_2}}^\infty\Big({p_{_2}^2\over p_{_1}^2}\Big)^{n_{_1}}
\Big({p_{_1}^2\over p_{_3}^2}\Big)^{n_{_2}}
\nonumber\\
&&\hspace{3.2cm}\sim
{(p_{_1}^2)^{D/2-1}\over(p_{_1}^2-p_{_2}^2)(p_{_1}^2-p_{_3}^2)}\;.
\label{1TR2-107}
\end{eqnarray}
Certainly the intersection of the convergent regions of these series is
the empty subset of the whole parameter space, the sets of the hypergeometric functions
$\{\psi_{_{\{1,2,3,6\}}}^{(3)},\psi_{_{\{1,2,3,6\}}}^{(10)},\psi_{_{\{1,2,3,6\}}}^{(11)}\}$
and $\{\psi_{_{\{1,2,3,6\}}}^{(4)},\psi_{_{\{1,2,3,6\}}}^{(11)},\psi_{_{\{1,2,3,6\}}}^{(12)}\}$
constitute the fundamental solution systems of the corresponding nonempty proper subsets
of the whole parameter space, respectively.

Because $\det(A_{_{\{1,2,5,6\}}}^{(1T)})=p_{_1}^2$ where the matrix $A_{_{\{1,2,5,6\}}}^{(1T)}$ denotes
the submatrix of $A^{(1T)}$ composed of the first, second, fifth and sixth columns, we have
\begin{eqnarray}
&&\Big(A_{_{\{1,2,5,6\}}}^{(1T)}\Big)^{-1}A^{(1T)}=\left(\begin{array}{ccccccc}
\;\;1\;\;&\;\;0\;\;&\;\;-{p_{_2}^2\over p_{_1}^2}\;\;&\;\;-1+{p_{_2}^2\over p_{_1}^2}\;\;
&\;\;0\;\;&\;\;0\;\;&\;\;0\;\;\\
\;\;0\;\;&\;\;1\;\;&\;\;-{p_{_3}^2\over p_{_1}^2}\;\;&\;\;-1+{p_{_3}^2\over p_{_1}^2}\;\;
&\;\;0\;\;&\;\;0\;\;&\;\;0\;\;\\
\;\;0\;\;&\;\;0\;\;&\;\;{1\over p_{_1}^2}\;\;&\;\;-{1\over p_{_1}^2}\;\;&\;\;1\;\;&\;\;0\;\;&\;\;0\;\;\\
\;\;0\;\;&\;\;0\;\;&\;\;0\;\;&\;\;1\;\;&\;\;0\;\;&\;\;1\;\;&\;\;1\;\;
\end{array}\right)\;.
\label{1TR2-109}
\end{eqnarray}
Obviously the matrix of exponents is written as
\begin{eqnarray}
&&\left(\begin{array}{ccccccc}
\;\;0\;\;&\;\;0\;\;&\;\;\alpha_{_{1,3}}\;\;&\;\;\alpha_{_{1,4}}\;\;&\;0\;\;&\;\;0\;\;&\;\;0\;\;\\
\;\;0\;\;&\;\;0\;\;&\;\;\alpha_{_{2,3}}\;\;&\;\;\alpha_{_{2,4}}\;\;&\;\;0\;\;&\;0\;\;&\;\;0\;\;\\
\;\;0\;\;&\;\;0\;\;&\;\;\alpha_{_{3,3}}\;\;&\;\;\alpha_{_{3,4}}\;\;&\;\;{D\over2}-3\;\;&\;0\;\;&\;\;0\;\;\\
\;\;0\;\;&\;\;0\;\;&\;\;0\;\;&\;\;3-{D\over2}\;\;&\;\;0\;\;&\;{D\over2}-3\;\;&\;\;-1\;\;
\end{array}\right)\;,
\label{1TR2-110}
\end{eqnarray}
where the matrix elements satisfy the relations
\begin{eqnarray}
&&\alpha_{_{1,3}}+\alpha_{_{1,4}}=-1\;,\;\alpha_{_{2,3}}+\alpha_{_{2,4}}=-1\;,\;
\alpha_{_{3,3}}+\alpha_{_{3,4}}=2-{D\over2}\;,
\nonumber\\
&&\alpha_{_{1,3}}+\alpha_{_{2,3}}+\alpha_{_{3,3}}=0\;,
\alpha_{_{1,4}}+\alpha_{_{2,4}}+\alpha_{_{3,4}}=-{D\over2}\;.
\label{1TR2-111}
\end{eqnarray}
Corresponding to the matrix of splitting local coordinates in Eq.(\ref{1TR2-109}),
we find 12 choices on the matrix of integer lattice whose submatrix composed of the third and
fourth columns is formulated as $\pm n_{_1}E_{_{4}}^{(i)}$ $\pm n_{_2}E_{_{4}}^{(j)}$,
with $n_{_{1,2}}\ge0$, $(i,\;j)\in\{(1,\;2),\;(1,\;4),\;(2,\;4)\}$,
and other elements are all zero. The exponents of the integer lattice
$n_{_1}E_{_{4}}^{(1)}$ $+n_{_2}E_{_{4}}^{(2)}$ are
\begin{eqnarray}
&&\alpha_{_{1,3}}={D\over2}-1\;,\;\alpha_{_{1,4}}=-{D\over2}\;,\;\alpha_{_{2,3}}=-1\;,\;
\alpha_{_{2,4}}=0\;,\;\alpha_{_{3,3}}=2-{D\over2},\;\alpha_{_{3,4}}=0\;,
\label{1TR2-112}
\end{eqnarray}
where the geometric representation is presented in Fig.\ref{fig10}(3)
with $a=1$, $\{b,c,d\}$ $=\{2,3,5\}$.
The exponents of the integer lattice $n_{_1}E_{_{4}}^{(1)}$ $+n_{_2}E_{_{4}}^{(4)}$ are
\begin{eqnarray}
&&\alpha_{_{1,3}}=0\;,\;\alpha_{_{1,4}}=-1\;,\;\alpha_{_{2,3}}={D\over2}-2\;,\;
\alpha_{_{2,4}}=1-{D\over2}\;,\;\alpha_{_{3,3}}=2-{D\over2},\;\alpha_{_{3,4}}=0\;,
\label{1TR2-116}
\end{eqnarray}
whose geometric representation is shown in Fig.\ref{fig10}(2)
with $\{a,b,c\}$ $=\{2,3,5\}$, $d=1$. In a similar way,
the exponents of the integer lattice $n_{_1}E_{_{4}}^{(2)}$ $+n_{_2}E_{_{4}}^{(4)}$ are
\begin{eqnarray}
&&\alpha_{_{1,3}}=0\;,\;\alpha_{_{1,4}}=-1\;,\;\alpha_{_{2,3}}=0\;,\;
\alpha_{_{2,4}}=-1\;,\;\alpha_{_{3,3}}=0,\;\alpha_{_{3,4}}=2-{D\over2}\;,
\label{1TR2-120}
\end{eqnarray}
whose geometric representation is plotted in Fig.\ref{fig10}(1)
with $(a,b)$ $=(3,5)$, $\{c,d\}$ $=\{1,2\}$.
The corresponding hypergeometric functions are evidently written as
\begin{eqnarray}
&&\psi_{_{\{1,2,5,6\}}}^{(1)}(p_{_1}^2,p_{_2}^2,p_{_3}^2)
\sim{(p_{_1}^2p_{_2}^2)^{D/2-1}\over p_{_3}^2(p_{_2}^2-p_{_1}^2)^{D/2}}
\sum\limits_{n_{_1},n_{_2}}^\infty{\Gamma({D\over2}-2+n_{_2})\over n_{_2}!}
\nonumber\\
&&\hspace{3.7cm}\times
\Big({p_{_2}^2(p_{_1}^2-p_{_3}^2)\over p_{_3}^2(p_{_1}^2-p_{_2}^2)}\Big)^{n_{_1}}
\Big({p_{_2}^2\over p_{_2}^2-p_{_1}^2}\Big)^{n_{_2}}
\nonumber\\
&&\hspace{3.2cm}\sim
{(p_{_2}^2)^{D/2-1}\over(p_{_2}^2-p_{_1}^2)(p_{_2}^2-p_{_3}^2)}
\;,\nonumber\\
&&\psi_{_{\{1,2,5,6\}}}^{(2)}(p_{_1}^2,p_{_2}^2,p_{_3}^2)
\sim{(p_{_1}^2)^{D/2-1}(p_{_3}^2)^{D/2-2}\over(p_{_1}^2-p_{_2}^2)(p_{_1}^2-p_{_3}^2)^{D/2-1}}
\sum\limits_{n_{_1},n_{_2}}^\infty{\Gamma({D\over2}-2+n_{_2})\over n_{_2}!}
\nonumber\\
&&\hspace{3.7cm}\times
\Big({p_{_2}^2(p_{_3}^2-p_{_1}^2)\over p_{_3}^2(p_{_2}^2-p_{_1}^2)}\Big)^{n_{_1}}
\Big({p_{_3}^2\over p_{_3}^2-p_{_1}^2}\Big)^{n_{_2}}
\nonumber\\
&&\hspace{3.2cm}\sim
{(p_{_3}^2)^{D/2-1}\over(p_{_3}^2-p_{_1}^2)(p_{_3}^2-p_{_2}^2)}
\;,\nonumber\\
&&\psi_{_{\{1,2,5,6\}}}^{(3)}(p_{_1}^2,p_{_2}^2,p_{_3}^2)
\sim{(p_{_1}^2)^{D/2-1}\over(p_{_1}^2-p_{_2}^2)(p_{_1}^2-p_{_3}^2)}
\sum\limits_{n_{_1},n_{_2}}^\infty{1\over\Gamma(3-{D\over2}+n_{_1}+n_{_2})}
\nonumber\\
&&\hspace{3.7cm}\times
{1\over\Gamma(1-n_{_1}-n_{_2})}
\Big({p_{_2}^2\over p_{_2}^2-p_{_1}^2}\Big)^{n_{_1}}
\Big({p_{_3}^2\over p_{_3}^2-p_{_1}^2}\Big)^{n_{_2}}
\nonumber\\
&&\hspace{3.2cm}\sim
{(p_{_1}^2)^{D/2-1}\over(p_{_1}^2-p_{_2}^2)(p_{_1}^2-p_{_3}^2)}\;,
\label{1TR2-120a}
\end{eqnarray}
where the intersection of their convergent regions is a nonempty proper subset of the whole parameter space.
In other words, those hypergeometric functions constitute a fundamental solution
system of the GKZ-system in the proper subset of the whole parameter space.

The matrix of exponents of the integer lattice
$-n_{_1}E_{_{4}}^{(1)}$ $-n_{_2}E_{_{4}}^{(2)}$ is
\begin{eqnarray}
&&\alpha_{_{1,3}}=0\;,\;\alpha_{_{1,4}}=-1\;,\;\alpha_{_{2,3}}=0\;,\;
\alpha_{_{2,4}}=-1\;,\;\alpha_{_{3,3}}=0,\;\alpha_{_{3,4}}=2-{D\over2}\;,
\label{1TR2-115}
\end{eqnarray}
whose geometric representation is presented in Fig.\ref{fig10}(1)
with $(a,b)$ $=(1,3)$, $\{c,d\}$ $=\{2,5\}$. Meanwhile the exponents of
the integer lattice $-n_{_1}E_{_{4}}^{(1)}$ $-n_{_2}E_{_{4}}^{(4)}$ are
\begin{eqnarray}
&&\alpha_{_{1,3}}=-1\;,\;\alpha_{_{1,4}}=0\;,\;\alpha_{_{2,3}}=1\;,\;
\alpha_{_{2,4}}=-2\;,\;\alpha_{_{3,3}}=0,\;\alpha_{_{3,4}}=2-{D\over2}\;,
\label{1TR2-119}
\end{eqnarray}
whose geometric representation is plotted in Fig.\ref{fig10}(2)
with $\{a,b,c\}$ $=\{1,2,3\}$, $d=5$. Similarly the exponents
of the integer lattice $-n_{_1}E_{_{4}}^{(2)}$ $-n_{_2}E_{_{4}}^{(4)}$ are
\begin{eqnarray}
&&\alpha_{_{1,3}}=-1\;,\;\alpha_{_{1,4}}=0\;,\;\alpha_{_{2,3}}=-1\;,\;
\alpha_{_{2,4}}=0\;,\;\alpha_{_{3,3}}=2,\;\alpha_{_{3,4}}=-{D\over2}\;,
\label{1TR2-123}
\end{eqnarray}
whose geometric representation is presented in Fig.\ref{fig10}(3)
with $a=5$, $\{b,c,d\}$ $=\{1,2,3\}$.
Three linear independent hypergeometric functions are
\begin{eqnarray}
&&\psi_{_{\{1,2,5,6\}}}^{(4)}(p_{_1}^2,p_{_2}^2,p_{_3}^2)
\sim{(p_{_1}^2)^{D/2-1}\over(p_{_1}^2-p_{_2}^2)(p_{_1}^2-p_{_3}^2)}
\sum\limits_{n_{_1},n_{_2}}^\infty{\Gamma({D\over2}-2+n_{_2})\over n_{_2}!}
\nonumber\\
&&\hspace{3.7cm}\times
\Big({p_{_3}^2(p_{_2}^2-p_{_1}^2)\over p_{_2}^2(p_{_3}^2-p_{_1}^2)}\Big)^{n_{_1}}
\Big({p_{_2}^2-p_{_1}^2\over p_{_2}^2}\Big)^{n_{_2}}
\nonumber\\
&&\hspace{3.2cm}\sim
{(p_{_2}^2)^{D/2-1}\over(p_{_2}^2-p_{_1}^2)(p_{_2}^2-p_{_3}^2)}
\;,\nonumber\\
&&\psi_{_{\{1,2,5,6\}}}^{(5)}(p_{_1}^2,p_{_2}^2,p_{_3}^2)
\sim{(p_{_1}^2)^{D/2-1}p_{_3}^2\over p_{_2}^2(p_{_1}^2-p_{_3}^2)^2}
\sum\limits_{n_{_1},n_{_2}}^\infty{\Gamma({D\over2}-2+n_{_2})\over n_{_2}!}
\nonumber\\
&&\hspace{3.7cm}\times
\Big({p_{_3}^2(p_{_2}^2-p_{_1}^2)\over p_{_2}^2(p_{_3}^2-p_{_1}^2)}\Big)^{n_{_1}}
\Big({p_{_3}^2-p_{_1}^2\over p_{_3}^2}\Big)^{n_{_2}}
\nonumber\\
&&\hspace{3.2cm}\sim
{(p_{_3}^2)^{D/2-1}\over(p_{_3}^2-p_{_1}^2)(p_{_3}^2-p_{_2}^2)}
\;,\nonumber\\
&&\psi_{_{\{1,2,5,6\}}}^{(6)}(p_{_1}^2,p_{_2}^2,p_{_3}^2)
\sim{(p_{_1}^2)^{D/2-1}\over p_{_2}^2p_{_3}^2}
\sum\limits_{n_{_1},n_{_2}}^\infty
{\Gamma({D\over2}+n_{_1}+n_{_2})\over\Gamma(2+n_{_1}+n_{_2})}
\nonumber\\
&&\hspace{3.7cm}\times
\Big({p_{_2}^2-p_{_1}^2\over p_{_2}^2}\Big)^{n_{_1}}
\Big({p_{_3}^2-p_{_1}^2\over p_{_3}^2}\Big)^{n_{_2}}\;,
\label{1TR2-123a}
\end{eqnarray}
where the intersection of their convergent regions is a nonempty proper subset of the whole parameter space.

Corresponding to the integer lattice $n_{_1}E_{_{4}}^{(1)}$ $-n_{_2}E_{_{4}}^{(2)}$,
the exponents are
\begin{eqnarray}
&&\alpha_{_{1,3}}=1\;,\;\alpha_{_{1,4}}=-2\;,\;\alpha_{_{2,3}}=-1\;,\;
\alpha_{_{2,4}}=0\;,\;\alpha_{_{3,3}}=0,\;\alpha_{_{3,4}}=2-{D\over2}\;,
\label{1TR2-113}
\end{eqnarray}
whose geometric representation is shown in Fig.\ref{fig10}(2)
with $\{a,b,c\}$ $=\{1,2,3\}$, $d=5$. Accordingly,
the exponents of the integer lattice of $n_{_1}E_{_{4}}^{(1)}$ $-n_{_2}E_{_{4}}^{(4)}$ are
\begin{eqnarray}
&&\alpha_{_{1,3}}=0\;,\;\alpha_{_{1,4}}=-1\;,\;\alpha_{_{2,3}}=0\;,\;
\alpha_{_{2,4}}=-1\;,\;\alpha_{_{3,3}}=0,\;\alpha_{_{3,4}}=2-{D\over2}\;,
\label{1TR2-117}
\end{eqnarray}
whose geometric representation is plotted in Fig.\ref{fig10}(1)
with $(a,b)$ $=(2,3)$, $\{c,d\}$ $=\{1,5\}$.
The matrix of exponents of the integer lattice
$n_{_1}E_{_{4}}^{(2)}$ $-n_{_2}E_{_{4}}^{(4)}$ is given by
\begin{eqnarray}
&&\alpha_{_{1,3}}=0\;,\;\alpha_{_{1,4}}=-1\;,\;\alpha_{_{2,3}}=-1\;,\;
\alpha_{_{2,4}}=0\;,\;\alpha_{_{3,3}}=1,\;\alpha_{_{3,4}}=1-{D\over2}\;,
\label{1TR2-121}
\end{eqnarray}
whose geometric representation is plotted in Fig.\ref{fig10}(2)
with $\{a,b,c\}$ $=\{2,3,5\}$, $d=1$.
The corresponding hypergeometric functions are presented as
\begin{eqnarray}
&&\psi_{_{\{1,2,5,6\}}}^{(7)}(p_{_1}^2,p_{_2}^2,p_{_3}^2)
\sim{(p_{_1}^2)^{D/2-1}p_{_2}^2\over p_{_3}^2(p_{_1}^2-p_{_2}^2)^2}
\sum\limits_{n_{_1},n_{_2}}^\infty{\Gamma({D\over2}-2+n_{_2})\over n_{_2}!}
\nonumber\\
&&\hspace{3.7cm}\times
\Big({p_{_2}^2(p_{_1}^2-p_{_3}^2)\over p_{_3}^2(p_{_1}^2-p_{_2}^2)}\Big)^{n_{_1}}
\Big({p_{_2}^2-p_{_1}^2\over p_{_2}^2}\Big)^{n_{_2}}
\nonumber\\
&&\hspace{3.2cm}\sim
{(p_{_2}^2)^{D/2-1}\over(p_{_2}^2-p_{_1}^2)(p_{_2}^2-p_{_3}^2)}
\;,\nonumber\\
&&\psi_{_{\{1,2,5,6\}}}^{(8)}(p_{_1}^2,p_{_2}^2,p_{_3}^2)
\sim{(p_{_1}^2)^{D/2-1}\over(p_{_1}^2-p_{_2}^2)(p_{_1}^2-p_{_3}^2)}
\sum\limits_{n_{_1},n_{_2}}^\infty{\Gamma({D\over2}-2+n_{_2})\over n_{_2}!}
\nonumber\\
&&\hspace{3.7cm}\times
\Big({p_{_2}^2(p_{_3}^2-p_{_1}^2)\over p_{_3}^2(p_{_2}^2-p_{_1}^2)}\Big)^{n_{_1}}
\Big({p_{_3}^2-p_{_1}^2\over p_{_3}^2}\Big)^{n_{_2}}
\nonumber\\
&&\hspace{3.2cm}\sim
{(p_{_3}^2)^{D/2-1}\over(p_{_3}^2-p_{_1}^2)(p_{_3}^2-p_{_2}^2)}
\;,\nonumber\\
&&\psi_{_{\{1,2,5,6\}}}^{(9)}(p_{_1}^2,p_{_2}^2,p_{_3}^2)
\sim{(p_{_1}^2)^{D/2-1}\over p_{_3}^2(p_{_1}^2-p_{_2}^2)}
\sum\limits_{n_{_1},n_{_2}}^\infty{\Gamma({D\over2}-1-n_{_1}+n_{_2})
\over\Gamma(2-n_{_1}+n_{_2})}
\nonumber\\
&&\hspace{3.7cm}\times
\Big({p_{_2}^2\over p_{_2}^2-p_{_1}^2}\Big)^{n_{_1}}
\Big({p_{_3}^2-p_{_1}^2\over p_{_3}^2}\Big)^{n_{_2}}
\nonumber\\
&&\hspace{3.2cm}\sim
{(p_{_1}^2)^{D/2-1}\over p_{_3}^2(p_{_1}^2-p_{_2}^2)}
\sum\limits_{n_{_1},n_{_2}}^\infty{\Gamma({D\over2}-2+n_{_2})\over n_{_2}!}
\nonumber\\
&&\hspace{3.7cm}\times
\Big({p_{_2}^2(p_{_3}^2-p_{_1}^2)\over p_{_3}^2(p_{_2}^2-p_{_1}^2)}\Big)^{n_{_1}}
\Big({p_{_3}^2-p_{_1}^2\over p_{_3}^2}\Big)^{n_{_2}}
\nonumber\\
&&\hspace{3.2cm}\sim
{(p_{_1}^2)^{D/2-1}\over(p_{_1}^2-p_{_2}^2)(p_{_1}^2-p_{_3}^2)}\;,
\label{1TR2-121a}
\end{eqnarray}
where the intersection of their convergent regions is a nonempty proper subset of the whole parameter space.
Furthermore, the sets of the hypergeometric functions
$\{\psi_{_{\{1,2,5,6\}}}^{(1)}$, $\psi_{_{\{1,2,5,6\}}}^{(8)}$, $\psi_{_{\{1,2,5,6\}}}^{(9)}\}$
and $\{\psi_{_{\{1,2,5,6\}}}^{(6)}$, $\psi_{_{\{1,2,5,6\}}}^{(7)}$, $\psi_{_{\{1,2,5,6\}}}^{(8)}\}$
constitute also the fundamental solution systems of the nonempty proper subsets
of the whole parameter space, respectively.

The matrix of exponents of the integer lattice
$-n_{_1}E_{_{4}}^{(1)}$ $+n_{_2}E_{_{4}}^{(2)}$ is
\begin{eqnarray}
&&\alpha_{_{1,3}}={D\over2}-2\;,\;\alpha_{_{1,4}}=1-{D\over2}\;,\;\alpha_{_{2,3}}=0\;,\;
\alpha_{_{2,4}}=-1\;,\;\alpha_{_{3,3}}=2-{D\over2},\;\alpha_{_{3,4}}=0\;,
\label{1TR2-114}
\end{eqnarray}
whose geometric representation is presented in Fig.\ref{fig10}(2)
with $\{a,b,c\}$ $=\{1,3,5\}$, $d=2$. Correspondingly the exponents of
the integer lattice $-n_{_1}E_{_{4}}^{(1)}$ $+n_{_2}E_{_{4}}^{(4)}$ are
\begin{eqnarray}
&&\alpha_{_{1,3}}=-1\;,\;\alpha_{_{1,4}}=0\;,\;\alpha_{_{2,3}}={D\over2}-1\;,\;
\alpha_{_{2,4}}=-{D\over2}\;,\;\alpha_{_{3,3}}=2-{D\over2},\;\alpha_{_{3,4}}=0\;,
\label{1TR2-118}
\end{eqnarray}
whose geometric representation is plotted in Fig.\ref{fig10}(3)
with $a=2$, $\{b,c,d\}$ $=\{1,3,5\}$. Finally the exponents
of the integer lattice $-n_{_1}E_{_{4}}^{(2)}$ $+n_{_2}E_{_{4}}^{(4)}$ are
\begin{eqnarray}
&&\alpha_{_{1,3}}=-1\;,\;\alpha_{_{1,4}}=0\;,\;\alpha_{_{2,3}}=0\;,\;
\alpha_{_{2,4}}=-1\;,\;\alpha_{_{3,3}}=1,\;\alpha_{_{3,4}}=1-{D\over2}\;,
\label{1TR2-122}
\end{eqnarray}
whose geometric representation is presented in Fig.\ref{fig10}(2)
with $\{a,b,c\}$ $=\{1,3,5\}$, $d=2$.
Those hypergeometric functions are formulated as
\begin{eqnarray}
&&\psi_{_{\{1,2,5,6\}}}^{(10)}(p_{_1}^2,p_{_2}^2,p_{_3}^2)\sim
{(p_{_1}^2)^{D/2-1}(p_{_2}^2)^{D/2-2}\over(p_{_3}^2-p_{_1}^2)(p_{_2}^2-p_{_1}^2)^{D/2-1}}
\sum\limits_{n_{_1},n_{_2}}^\infty{\Gamma({D\over2}-2+n_{_2})\over n_{_2}!}
\nonumber\\
&&\hspace{3.7cm}\times
\Big({p_{_3}^2(p_{_2}^2-p_{_1}^2)\over p_{_2}^2(p_{_3}^2-p_{_1}^2)}\Big)^{n_{_1}}
\Big({p_{_2}^2\over p_{_2}^2-p_{_1}^2}\Big)^{n_{_2}}
\nonumber\\
&&\hspace{3.2cm}\sim
{(p_{_2}^2)^{D/2-1}\over(p_{_2}^2-p_{_1}^2)(p_{_2}^2-p_{_3}^2)}
\;,\nonumber\\
&&\psi_{_{\{1,2,5,6\}}}^{(11)}(p_{_1}^2,p_{_2}^2,p_{_3}^2)
\sim{(p_{_1}^2p_{_3}^2)^{D/2-1}\over p_{_2}^2(p_{_3}^2-p_{_1}^2)^{D/2}}
\sum\limits_{n_{_1},n_{_2}}^\infty{\Gamma({D\over2}-2+n_{_2})\over n_{_2}!}
\nonumber\\
&&\hspace{3.7cm}\times
\Big({p_{_3}^2(p_{_2}^2-p_{_1}^2)\over p_{_2}^2(p_{_3}^2-p_{_1}^2)}\Big)^{n_{_1}}
\Big({p_{_3}^2\over p_{_3}^2-p_{_1}^2}\Big)^{n_{_2}}
\nonumber\\
&&\hspace{3.2cm}\sim
{(p_{_3}^2)^{D/2-1}\over(p_{_3}^2-p_{_1}^2)(p_{_3}^2-p_{_2}^2)}
\;,\nonumber\\
&&\psi_{_{\{1,2,5,6\}}}^{(12)}(p_{_1}^2,p_{_2}^2,p_{_3}^2)
\sim{(p_{_1}^2)^{D/2-2}(p_{_3}^2)^2\over(p_{_1}^2-p_{_3}^2)^2(p_{_2}^2-p_{_3}^2)}
\sum\limits_{n=0}^\infty{\Gamma(1+n)
\over\Gamma(4-{D\over2}+n)}
\Big({p_{_3}^2\over p_{_3}^2-p_{_1}^2}\Big)^{n}\;,
\label{1TR2-124}
\end{eqnarray}
where the intersection of their convergent regions is
a nonempty subset of the whole parameter space. The hypergeometric function of
the exponents of Eq.(\ref{1TR2-122}) is
\begin{eqnarray}
&&\psi_{_{\{1,2,5,6\}}}^{(12)\prime}(p_{_1}^2,p_{_2}^2,p_{_3}^2)
\sim{(p_{_1}^2)^{D/2-1}\over p_{_2}^2(p_{_1}^2-p_{_3}^2)}
\sum\limits_{n_{_1},n_{_2}}^\infty{\Gamma({D\over2}-1+n_{_1}-n_{_2})
\over\Gamma(2+n_{_1}-n_{_2})}
\nonumber\\
&&\hspace{3.7cm}\times
\Big({p_{_2}^2-p_{_1}^2\over p_{_2}^2}\Big)^{n_{_1}}
\Big({p_{_3}^2\over p_{_3}^2-p_{_1}^2}\Big)^{n_{_2}}
\nonumber\\
&&\hspace{3.2cm}\sim
\psi_{_{\{1,2,5,6\}}}^{(10)}(p_{_1}^2,p_{_2}^2,p_{_3}^2)
\nonumber\\
&&\hspace{3.7cm}
+\lim\limits_{\epsilon\rightarrow0}{\sin\pi\epsilon\over\sin\pi({D\over2}-3)}
\psi_{_{\{1,2,5,6\}}}^{(12)}(p_{_1}^2,p_{_2}^2,p_{_3}^2)\;.
\label{1TR2-124a}
\end{eqnarray}
The sets of the hypergeometric functions
$\{\psi_{_{\{1,2,5,6\}}}^{(3)}$, $\psi_{_{\{1,2,5,6\}}}^{(10)}$, $\psi_{_{\{1,2,5,6\}}}^{(11)}\}$
and $\{\psi_{_{\{1,2,5,6\}}}^{(4)}$, $\psi_{_{\{1,2,5,6\}}}^{(11)}$, $\psi_{_{\{1,2,5,6\}}}^{(12)}\}$
constitute the fundamental solution systems of the nonempty proper subsets
of the whole parameter space, respectively.

Because $\det(A_{_{1,3,5,6}})^{(1T)}=-p_{_3}^2$,
where the matrix $A_{_{\{1,3,5,6\}}}^{(1T)}$ denotes the submatrix of $A^{(1T)}$ composed
of the first, third, fifth and sixth columns, one finds
\begin{eqnarray}
&&\Big(A_{_{\{1,3,5,6\}}}^{(1T)}\Big)^{-1}\cdot A^{(1T)}=\left(\begin{array}{ccccccc}
\;\;1\;\;&\;\;-{p_{_2}^2\over p_{_3}^2}\;\;&\;\;0\;\;&\;\;-1+{p_{_2}^2\over p_{_3}^2}\;\;
&\;\;0\;\;&\;\;0\;\;&\;\;0\;\;\\
\;\;0\;\;&\;\;-{p_{_1}^2\over p_{_3}^2}\;\;&\;\;1\;\;&\;\;-1+{p_{_1}^2\over p_{_3}^2}\;\;&
\;\;0\;\;&\;\;0\;\;&\;\;0\;\;\\
\;\;0\;\;&\;\;{1\over p_{_3}^2}\;\;&\;\;0\;\;&\;\;-{1\over p_{_3}^2}\;\;&
\;\;1\;\;&\;\;0\;\;&\;\;0\;\;\\
\;\;0\;\;&\;\;0\;\;&\;\;0\;\;&\;\;1\;\;&\;\;0\;\;
&\;\;1\;\;&\;\;1\;\;
\end{array}\right)\;.
\label{1TR2-126}
\end{eqnarray}
Obviously the matrix of exponents is written as
\begin{eqnarray}
&&\left(\begin{array}{ccccccc}
\;\;0\;\;&\;\;\alpha_{_{1,2}}\;\;&\;\;0\;\;&\;\;\alpha_{_{1,4}}\;\;&\;\;0\;\;&\;0\;\;&\;\;0\;\;\\
\;\;0\;\;&\;\;\alpha_{_{2,2}}\;\;&\;\;0\;\;&\;\alpha_{_{2,4}}\;\;&\;\;0\;\;&\;\;0\;\;&\;0\;\;\\
\;\;0\;\;&\;\;\alpha_{_{3,2}}\;\;&\;\;0\;\;&\;\alpha_{_{3,4}}\;\;&\;\;{D\over2}-3\;\;&\;\;0\;\;&\;0\;\;\\
\;\;0\;\;&\;\;0\;\;&\;\;0\;\;&\;3-{D\over2}\;\;&\;\;0\;\;&\;\;{D\over2}-3\;\;&\;-1\;\;\\
\end{array}\right)\;,
\label{1TR2-127}
\end{eqnarray}
where the matrix elements satisfy the relations
\begin{eqnarray}
&&\alpha_{_{1,2}}+\alpha_{_{1,4}}=-1\;,\;\alpha_{_{2,2}}+\alpha_{_{2,4}}=-1\;,\;
\alpha_{_{3,2}}+\alpha_{_{3,4}}=2-{D\over2}\;,
\nonumber\\
&&\alpha_{_{1,2}}+\alpha_{_{2,2}}+\alpha_{_{3,2}}=0\;,
\alpha_{_{1,4}}+\alpha_{_{2,4}}+\alpha_{_{3,4}}=-{D\over2}\;.
\label{1TR2-128}
\end{eqnarray}
Correspondingly there are 12 choices on the matrix of integer lattice whose
submatrix composed of the second and fourth columns is formulated as
$\pm n_{_1}E_{_{4}}^{(i)}$ $\pm n_{_2}E_{_{4}}^{(j)}$,
where $n_{_{1,2}}\ge0$, $(i,\;j)\in\{(1,\;2),\;(1,\;4),\;(2,\;4)\}$,
and other elements are all zero.
Basing on the matrices of integer lattice, we have 12 choices on the
matrix of exponents whose geometric descriptions are obtained by the permutation
$\widehat{(23)}$ of that of $\psi_{_{\{1,2,5,6\}}}^{(i)}$.
Correspondingly the hypergeometric functions are
\begin{eqnarray}
&&\psi_{_{\{1,3,5,6\}}}^{(i)}(p_{_1}^2,p_{_2}^2,p_{_3}^2)=
\psi_{_{\{1,2,5,6\}}}^{(i)}(p_{_3}^2,p_{_2}^2,p_{_1}^2),\;\;i=1,\cdots,12.
\label{1TR2-129}
\end{eqnarray}

Because $\det(A_{_{2,3,5,6}})^{(1T)}=p_{_2}^2$,
where the matrix $A_{_{\{2,3,5,6\}}}^{(1T)}$ denotes the submatrix of $A^{(1T)}$ composed
of the second, third, fifth and sixth columns, one finds
\begin{eqnarray}
&&\Big(A_{_{\{2,3,5,6\}}}^{(1T)}\Big)^{-1}\cdot A^{(1T)}=\left(\begin{array}{ccccccc}
\;\;-{p_{_3}^2\over p_{_2}^2}\;\;&\;\;1\;\;&\;\;0\;\;&\;\;-1+{p_{_3}^2\over p_{_2}^2}\;\;
&\;\;0\;\;&\;\;0\;\;&\;\;0\;\;\\
\;\;-{p_{_1}^2\over p_{_2}^2}\;\;&\;\;0\;\;&\;\;1\;\;&\;\;-1+{p_{_1}^2\over p_{_2}^2}\;\;&
\;\;0\;\;&\;\;0\;\;&\;\;0\;\;\\
\;\;{1\over p_{_2}^2}\;\;&\;\;0\;\;&\;\;0\;\;&\;\;-{1\over p_{_2}^2}\;\;&
\;\;1\;\;&\;\;0\;\;&\;\;0\;\;\\
\;\;0\;\;&\;\;0\;\;&\;\;0\;\;&\;\;1\;\;&\;\;0\;\;
&\;\;1\;\;&\;\;1\;\;
\end{array}\right)\;.
\label{1TR2-130}
\end{eqnarray}
Obviously the matrix of exponents is written as
\begin{eqnarray}
&&\left(\begin{array}{ccccccc}
\;\;\alpha_{_{1,1}}\;\;&\;\;0\;\;&\;\;0\;\;&\;\;\alpha_{_{1,4}}\;\;&\;\;0\;\;&\;0\;\;&\;\;0\;\;\\
\;\;\alpha_{_{2,1}}\;\;&\;\;0\;\;&\;\;0\;\;&\;\alpha_{_{2,4}}\;\;&\;\;0\;\;&\;\;0\;\;&\;0\;\;\\
\;\;\alpha_{_{3,1}}\;\;&\;\;0\;\;&\;\;0\;\;&\;\alpha_{_{3,4}}\;\;&\;\;{D\over2}-3\;\;&\;\;0\;\;&\;0\;\;\\
\;\;0\;\;&\;\;0\;\;&\;\;0\;\;&\;3-{D\over2}\;\;&\;\;0\;\;&\;\;{D\over2}-3\;\;&\;-1\;\;\\
\end{array}\right)\;,
\label{1TR2-131}
\end{eqnarray}
where the matrix elements satisfy the relations
\begin{eqnarray}
&&\alpha_{_{1,1}}+\alpha_{_{1,4}}=-1\;,\;\alpha_{_{2,1}}+\alpha_{_{2,4}}=-1\;,\;
\alpha_{_{3,1}}+\alpha_{_{3,4}}=2-{D\over2}\;,
\nonumber\\
&&\alpha_{_{1,1}}+\alpha_{_{2,1}}+\alpha_{_{3,1}}=0\;,
\alpha_{_{1,4}}+\alpha_{_{2,4}}+\alpha_{_{3,4}}=-{D\over2}\;.
\label{1TR2-132}
\end{eqnarray}
Correspondingly there are 12 choices on the matrix of integer lattice
whose submatrix composed of the second and fourth columns is formulated
as $\pm n_{_1}E_{_{4}}^{(i)}$ $\pm n_{_2}E_{_{4}}^{(j)}$,
where $n_{_{1,2}}\ge0$, $(i,j)$ $\in\{(1,2)$, $(1,4)$, $(2,4)\}$,
and other elements are all zero.
Basing on the matrices of integer lattice, we have 12 choices on the
matrix of exponents whose geometric descriptions are obtained by the permutation
$\widehat{(123)}$ of that of $\psi_{_{\{1,2,5,6\}}}^{(i)}$.
Correspondingly the hypergeometric functions are
\begin{eqnarray}
&&\psi_{_{\{2,3,5,6\}}}^{(i)}(p_{_1}^2,p_{_2}^2,p_{_3}^2)=
\psi_{_{\{1,2,5,6\}}}^{(i)}(p_{_2}^2,p_{_3}^2,p_{_1}^2),\;\;i=1,\cdots,12.
\label{1TR2-133}
\end{eqnarray}

Since the Equation Eq.(\ref{1TR2-2+b}) does not depend on the splitting local
coordinate $z_{_{4,6}}$, we can embed the integral into the subvariety of the
Grassmannian $G_{_{4,7}}$ where the local coordinates are
\begin{eqnarray}
&&A^{(1T)^\prime}=\left(\begin{array}{ccccccc}
\;\;1\;\;&\;\;0\;\;&\;\;0\;\;&\;\;0\;\;&\;\;p_{_2}^2\;\;&\;\;1\;\;&\;\;1\;\;\\
\;\;0\;\;&\;\;1\;\;&\;\;0\;\;&\;\;0\;\;&\;\;p_{_3}^2\;\;&\;\;1\;\;&\;\;1\;\;\\
\;\;0\;\;&\;\;0\;\;&\;\;1\;\;&\;\;0\;\;&\;\;p_{_1}^2\;\;&\;\;1\;\;&\;\;1\;\;\\
\;\;0\;\;&\;\;0\;\;&\;\;0\;\;&\;\;1\;\;&\;\;0\;\;&\;\;0\;\;&\;\;1\;\;\\
\end{array}\right)\;.
\label{1TR2-3}
\end{eqnarray}
\begin{figure}[ht]
\setlength{\unitlength}{1cm}
\centering
\vspace{0.0cm}\hspace{-1.5cm}
\includegraphics[height=8cm,width=8.0cm]{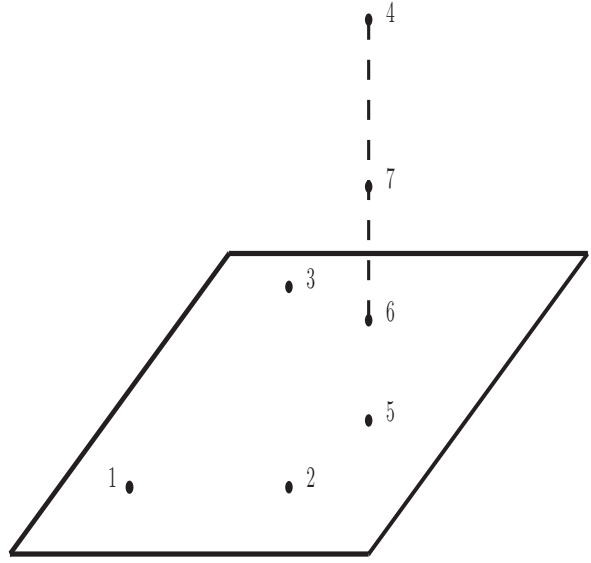}
\vspace{0cm}
\caption[]{The geometric description of the matroid $A^{(1T)\prime}$ in
Eq.(\ref{1TR2-3}) on the projective plane $CP^{4}$, where the points $1,\cdots,7$
denote the indices of columns of the $3\times7$ matrix. In the representation,
the points $5,6$ belong to the hyperplane $\{1,2,3\}$, the line \{4,6,7\}
intersects with the hyperplane $\{1,2,3\}$ at the point $6$, respectively.}
\label{fig5}
\end{figure}
Taking $\widehat{\sigma}_{_1}=\widehat{\sigma}_{_3}=\widehat{(56)}$,
$\widehat{\sigma}_{_2}=\widehat{\sigma}_{_4}=\widehat{\sigma}_{_5}
=\widehat{\sigma}_{_6}=\widehat{e}$ in Eq.(\ref{1TR2-2+a}), we have
\begin{eqnarray}
&&z_{_{1,6}}z_{_{2,5}}=p_{_3}^2\;,\;\;z_{_{1,5}}z_{_{3,6}}=p_{_2}^2\;,\;\;
z_{_{2,6}}z_{_{3,5}}=p_{_1}^2\;,
\nonumber\\
&&z_{_{1,5}}z_{_{4,6}}=0\;,
\;\;z_{_{2,5}}z_{_{4,6}}=0\;,\;\;z_{_{3,5}}z_{_{4,6}}=0\;.
\label{1TR2-3+b}
\end{eqnarray}
The solution of Eq.(\ref{1TR2-3+b})
\begin{eqnarray}
&&z_{_{1,5}}=p_{_2}^2\;,z_{_{2,5}}=p_{_3}^2\;,z_{_{3,5}}=p_{_1}^2\;,z_{_{4,5}}=z_{_{4,6}}=0\;,
\nonumber\\
&&z_{_{1,6}}=z_{_{2,6}}=z_{_{3,6}}=1\;,
\label{1TR2-3+c}
\end{eqnarray}
corresponds to the subvariety of the Grassmannian $G_{_{4,7}}$ in Eq.(\ref{1TR2-3}).
Accordingly we obtain the hypergeometric functions which are satisfy the PDEs
presented in Eq.(\ref{1TR2-94a4}) in combined local coordinates. To shorten the length
of text, we collect the hypergeometric functions in appendix~\ref{app1}.

The matrices of Eq.(\ref{1TR2-91}) and Eq.(\ref{1TR2-3}) belong to the embedding
of the same parametric representation Eq.(\ref{1TR2-2}) of the 1-loop triangle diagram with
zero virtual masses on different subvarieties of the Grassmannian $G_{_{4,7}}$.
Correspondingly, constructed fundamental solution systems either belong
to the equivalent fundamental solution systems of the neighborhoods of the same regular singularity,
or belong to the fundamental solution systems of neighborhoods of different regular singularities.

\section{The 2-loop vacuum\label{2Vac}}
\subsection{The fundamental system with one zero virtual mass}
\indent\indent
As one virtual mass $m_{_1}=0$, the scalar integral is simplified as
\begin{eqnarray}
&&A_{_{2Vac}}(0,\;m_{_2}^2,\;m_{_3}^2)=
-{\Gamma({D\over2}-1)\Gamma(3-D)\mu^{4\varepsilon}\over(4\pi)^{D}\Gamma({D\over2})}
\int \omega_{_3}(t)t_{_1}^{-1}(t_{_2}t_{_3})^{1-D/2}
\nonumber\\
&&\hspace{3.9cm}\times
{\delta(t_{_1}+t_{_2}+t_{_3})\over (t_{_2}m_{_2}^2+t_{_3}m_{_3}^2)^{3-D}}\;.
\label{2Vac-1-1}
\end{eqnarray}
The integral can be embedded in the subvariety of the Grassmannian $G_{_{3,5}}$
whose splitting local coordinates are written as
\begin{eqnarray}
&&A^{2V0}=\left(\begin{array}{ccccc}\;\;1\;\;&\;\;0\;\;&\;\;0\;\;&\;\;1\;\;&\;\;0\;\;\\
\;\;0\;\;&\;\;1\;\;&\;\;0\;\;&\;\;1\;\;&\;\;m_{_2}^2\;\;\\
\;\;0\;\;&\;\;0\;\;&\;\;1\;\;&\;\;1\;\;&\;\;m_{_3}^2\;\;\end{array}\right)\;,
\label{2Vac-1-2}
\end{eqnarray}
whose geometric description is presented in Fig.\ref{fig1}.
In the matrix of local coordinates, the first row corresponds to the integration variable $t_{_1}$,
the second row  corresponds to the integration variable $t_{_2}$,
the third row  corresponds to the integration variable $t_{_3}$, the first column represents
the power function $t_{_1}^{-1}$, the second column represents the power function $t_{_2}^{1-D/2}$,
the third column represents the power function $t_{_3}^{1-D/2}$,
the fourth column represents the function $\delta(t_{_1}+t_{_2}+t_{_3})$,
and the fifth column represents the power function $(t_{_2}m_{_2}^2+t_{_3}m_{_3}^2)^{D-3}$, respectively.

Accordingly, the Feynman integral satisfies the following GKZ-system
\begin{eqnarray}
&&\Big\{\vartheta_{_{1,1}}+\vartheta_{_{1,4}}\Big\}A_{_{2Vac}}(0,\;m_{_2}^2,\;m_{_3}^2)=
-A_{_{2Vac}}(0,\;m_{_2}^2,\;m_{_3}^2)
\;,\nonumber\\
&&\Big\{\vartheta_{_{2,2}}+\vartheta_{_{2,4}}+\vartheta_{_{2,5}}\Big\}A_{_{2Vac}}(0,\;m_{_2}^2,\;m_{_3}^2)=
-A_{_{2Vac}}(0,\;m_{_2}^2,\;m_{_3}^2)
\;,\nonumber\\
&&\Big\{\vartheta_{_{3,3}}+\vartheta_{_{3,4}}+\vartheta_{_{3,5}}\Big\}A_{_{2Vac}}(0,\;m_{_2}^2,\;m_{_3}^2)=
-A_{_{2Vac}}(0,\;m_{_2}^2,\;m_{_3}^2)
\;,\nonumber\\
&&\vartheta_{_{1,1}}A_{_{2Vac}}(0,\;m_{_2}^2,\;m_{_3}^2)=-A_{_{2Vac}}(0,\;m_{_2}^2,\;m_{_3}^2)
\;,\nonumber\\
&&\vartheta_{_{2,2}}A_{_{2Vac}}(0,\;m_{_2}^2,\;m_{_3}^2)=(1-{D\over2})A_{_{2Vac}}(0,\;m_{_2}^2,\;m_{_3}^2)
\;,\nonumber\\
&&\vartheta_{_{3,3}}A_{_{2Vac}}(0,\;m_{_2}^2,\;m_{_3}^2)=(1-{D\over2})A_{_{2Vac}}(0,\;m_{_2}^2,\;m_{_3}^2)
\;,\nonumber\\
&&\Big\{\vartheta_{_{1,4}}+\vartheta_{_{2,4}}+\vartheta_{_{3,4}}\Big\}A_{_{2Vac}}(0,\;m_{_2}^2,\;m_{_3}^2)=
-A_{_{2Vac}}(0,\;m_{_2}^2,\;m_{_3}^2)
\;,\nonumber\\
&&\Big\{\vartheta_{_{2,5}}+\vartheta_{_{3,5}}\Big\}A_{_{2Vac}}(0,\;m_{_2}^2,\;m_{_3}^2)=
(D-3)A_{_{2Vac}}(0,\;m_{_2}^2,\;m_{_3}^2)\;.
\label{2Vac-1-2a}
\end{eqnarray}
Certainly the exponent matrix is written as
\begin{eqnarray}
&&\left(\begin{array}{ccccc}\;-1\;&\;0\;&\;0\;&\;0\;&\;0\;\\
\;0&\;1-{D\over2}\;&\;0\;&\;\alpha_{_{2,4}}\;&\;\alpha_{_{2,5}}\;\\
\;0\;&\;0\;&\;1-{D\over2}\;&\;\alpha_{_{3,4}}\;&\;\alpha_{_{3,5}}\;\end{array}\right)\;,
\label{2Vac-1-3}
\end{eqnarray}
with
\begin{eqnarray}
&&\alpha_{_{2,4}}+\alpha_{_{2,5}}={D\over2}-2,\;\alpha_{_{3,4}}+\alpha_{_{3,5}}={D\over2}-2,
\nonumber\\
&&\alpha_{_{2,4}}+\alpha_{_{3,4}}=-1,\;\alpha_{_{2,5}}+\alpha_{_{3,5}}=D-3.
\label{2Vac-1-4}
\end{eqnarray}
Furthermore, the dual space of the GKZ-system of Eq.(\ref{2Vac-1-2a}) is spanned by
a $3\times5$ matrix $(0_{_{3\times3}}\Big|E_{_3}^{(1)})$.
The basis of the dual space implies the Feynman integral satisfying the PDE
\begin{eqnarray}
&&{\partial^2A_{_{2Vac}}\over\partial z_{_{2,4}}\partial z_{_{3,5}}}(0,\;m_{_2}^2,\;m_{_3}^2)
={\partial^2A_{_{2Vac}}\over\partial z_{_{2,5}}\partial z_{_{3,4}}}(0,\;m_{_2}^2,\;m_{_3}^2)\;.
\label{2Vac-1-4a}
\end{eqnarray}
When the exponents are given, the Feynman integral can be formally expressed as
\begin{eqnarray}
&&A_{_{2Vac}}(0,\;m_{_2}^2,\;m_{_3}^2)=\prod\limits_{i,j}z_{_{i,j}}^{\alpha_{_{i,j}}}
\varphi({m_{_3}^2\over m_{_2}^2})\;,
\label{2Vac-1-4b}
\end{eqnarray}
where $\varphi(x)$ satisfies the PDE in the combined coordinate
$x=z_{_{2,4}}z_{_{3,5}}/z_{_{2,5}}z_{_{3,4}}$ as
\begin{eqnarray}
&&x\Big[\alpha_{_{2,4}}\alpha_{_{3,5}}+(1-\alpha_{_{2,4}}-\alpha_{_{3,5}})x{d\over dx}
-x^2{d^2\over dx^2}\Big]\varphi(x)
\nonumber\\
&&\hspace{-0.5cm}=
\Big[\alpha_{_{2,5}}\alpha_{_{3,4}}+(1+\alpha_{_{2,5}}+\alpha_{_{3,4}})x{d\over dx}
-x^2{d^2\over dx^2}\Big]\varphi(x)\;.
\label{2Vac-1-4c}
\end{eqnarray}
At $\alpha_{_{2,5}}\alpha_{_{3,4}}=0$ this equation recovers the well-known Gauss equation.
The integer lattice $(0_{_{3\times3}}\Big|nE_{_3}^{(1)})$ ($n\ge0$) is compatible with two choices
of the exponents. The first choice is
\begin{eqnarray}
&&\alpha_{_{2,4}}=0,\;\alpha_{_{2,5}}={D\over2}-2,\;\alpha_{_{3,4}}=-1,\;\alpha_{_{3,5}}={D\over2}-1,
\label{2Vac-1-5}
\end{eqnarray}
whose geometric representation is plotted in Fig.\ref{fig2}(1) with
$a=3$, $\{(b,c)$, $(d,e)\}$ $=\{(1,4)$, $(2,5)\}$. The second choice is
\begin{eqnarray}
&&\alpha_{_{2,4}}=1-{D\over2},\;\alpha_{_{2,5}}=D-3,\;\alpha_{_{3,4}}={D\over2}-2,\;\alpha_{_{3,5}}=0,
\label{2Vac-1-6}
\end{eqnarray}
whose geometric representation is depicted in Fig.\ref{fig2}(2) with
$(a,b)$ $=(2,5)$, $\{c,d,e\}$ $=\{1,3,4\}$.
The corresponding hypergeometric functions are
\begin{eqnarray}
&&\psi_{_{\{1,2,3\}}}^{(1)}(m_{_2}^2,m_{_3}^2)\sim(m_{_2}^2)^{D/2-2}(m_{_3}^2)^{D/2-1}
\sum\limits_{n=0}^\infty{\Gamma(2-{D\over2}+n)\over\Gamma({D\over2}+n)}
\Big({m_{_3}^2\over m_{_2}^2}\Big)^n\;,
\nonumber\\
&&\psi_{_{\{1,2,3\}}}^{(2)}(m_{_2}^2,m_{_3}^2)\sim(m_{_2}^2)^{D-3}
\sum\limits_{n=0}^\infty{\Gamma(3-D+n)\over n!}
\Big({m_{_3}^2\over m_{_2}^2}\Big)^n
\nonumber\\
&&\hspace{2.8cm}\sim
(m_{_2}^2-m_{_3}^2)^{D-3}\;,
\label{2Vac-1-6a}
\end{eqnarray}
whose the intersection of their convergent regions is $|m_{_3}^2/m_{_2}^2|\le1$.
Two linearly independent Gauss functions constitute the fundamental
solution system of the GKZ-system of Eq.(\ref{1SE-9a}) in the region $|m_{_3}^2/m_{_2}^2|\le1$
which is consistent with the results in the literature. When $y_{_i}=0$ in Eq.(46)
of Ref.~\cite{Feng2020}, the fourth type Appell functions of the last two terms
are simplified as the Gauss functions above.

The third choice of the exponents is
\begin{eqnarray}
&&\alpha_{_{2,4}}={D\over2}-2,\;\alpha_{_{2,5}}=0,\;\alpha_{_{3,4}}=1-{D\over2},\;\alpha_{_{3,5}}=D-3,
\label{2Vac-1-7}
\end{eqnarray}
whose geometric representation is drawn in Fig.\ref{fig2}(2) with
$(a,b)$ $=(3,5)$, $\{c,d,e\}$ $=\{1,2,4\}$. Similarly the fourth choice of the exponents is
\begin{eqnarray}
&&\alpha_{_{2,4}}=-1,\;\alpha_{_{2,5}}={D\over2}-1,\;\alpha_{_{3,4}}=0,\;\alpha_{_{3,5}}={D\over2}-2,
\label{2Vac-1-8}
\end{eqnarray}
whose geometric representation is shown in Fig.\ref{fig2}(1) with
$a=2$, $\{(b,c)$, $(d,e)\}$ $=\{(1,4)$, $(3,5)\}$.
The hypergeometric functions are constructed accordingly
\begin{eqnarray}
&&\psi_{_{\{1,2,3\}}}^{(3)}(m_{_2}^2,m_{_3}^2)\sim(m_{_3}^2)^{D-3}
\sum\limits_{n=0}^\infty{\Gamma(3-D+n)\over n!}
\Big({m_{_2}^2\over m_{_3}^2}\Big)^n
\nonumber\\
&&\hspace{2.8cm}\sim
(m_{_2}^2-m_{_3}^2)^{D-3}\;,
\nonumber\\
&&\psi_{_{\{1,2,3\}}}^{(4)}(m_{_2}^2,m_{_3}^2)\sim(m_{_2}^2)^{D/2-1}(m_{_3}^2)^{D/2-2}
\sum\limits_{n=0}^\infty{\Gamma(2-{D\over2}+n)\over\Gamma({D\over2}+n)}
\Big({m_{_2}^2\over m_{_3}^2}\Big)^n\;,
\label{2Vac-1-9}
\end{eqnarray}
where the intersection of their convergent regions is $|m_{_2}^2/m_{_3}^2|\le1$.

Because $\det(A^{2V0}_{_{\{1,2,5\}}})=m_{_3}^2$, thus
\begin{eqnarray}
&&(A^{2V0}_{_{\{1,2,5\}}})^{-1}\cdot A^{2V0}=
\left(\begin{array}{ccccc}\;1\;&\;0\;&\;0\;&\;1\;&\;0\;\\
\;0\;&\;1\;&\;-{m_{_2}^2\over m_{_3}^2}\;&\;1-{m_{_2}^2\over m_{_3}^2}\;&\;0\;\\
\;0\;&\;0\;&\;{1\over m_{_3}^2}\;&\;{1\over m_{_3}^2}\;&\;1\;\end{array}\right)\;.
\label{2Vac-1-10}
\end{eqnarray}
The matrix of exponents is written as
\begin{eqnarray}
&&\left(\begin{array}{ccccc}\;-1\;&\;0\;&\;0\;&\;0\;&\;0\;\\
\;0&\;1-{D\over2}\;&\;\alpha_{_{2,3}}\;&\;\alpha_{_{2,4}}\;&\;0\;\\
\;0\;&\;0\;&\;\alpha_{_{3,3}}\;&\;\alpha_{_{3,4}}\;&\;D-3\;\end{array}\right)\;,
\label{2Vac-1-11}
\end{eqnarray}
with
\begin{eqnarray}
&&\alpha_{_{2,3}}+\alpha_{_{2,4}}={D\over2}-2,\;\alpha_{_{3,3}}+\alpha_{_{3,4}}=2-D,
\nonumber\\
&&\alpha_{_{2,3}}+\alpha_{_{3,3}}=1-{D\over2},\;\alpha_{_{2,4}}+\alpha_{_{3,4}}=-1.
\label{2Vac-1-12}
\end{eqnarray}
Corresponding to the matrix of local coordinates in Eq.(\ref{2Vac-1-10}), one has
two choices on the matrix of integer lattice whose submatrix composed of the third and
fourth columns is formulated as $\pm nE_{_{3}}^{(1)}$,
with $n\ge0$, and other elements are all zero. We have two
choices on the exponents of the integer lattice $nE_{_{3}}^{(1)}$. The first choice is
\begin{eqnarray}
&&\alpha_{_{2,3}}=0,\;\alpha_{_{2,4}}={D\over2}-2,\;\alpha_{_{3,3}}=1-{D\over2},\;\alpha_{_{3,4}}=1-{D\over2},
\label{2Vac-1-13}
\end{eqnarray}
where the geometric representation is presented in Fig.\ref{fig2}(2) with
$(a,b)$ $=(3,5)$, $\{c,d,e\}$ $=\{1,2,4\}$. The second choice of the exponents is
\begin{eqnarray}
&&\alpha_{_{2,3}}={D\over2}-1,\;\alpha_{_{2,4}}=-1,\;\alpha_{_{3,3}}=2-D,\;\alpha_{_{3,4}}=0,
\label{2Vac-1-14}
\end{eqnarray}
whose geometric representation is plotted in Fig.\ref{fig2}(1) with
$a=2$, $\{(b,c)$, $(d,e)\}$ $=\{(1,4)$, $(3,5)\}$.
The corresponding hypergeometric functions are
\begin{eqnarray}
&&\psi_{_{\{1,2,5\}}}^{(1)}(m_{_2}^2,m_{_3}^2)\sim(m_{_3}^2)^{D/2-1}(m_{_3}^2-m_{_2}^2)^{D/2-2}
\sum\limits_{n=0}^\infty{\Gamma({D\over2}-1+n)\over n!}
\Big({m_{_2}^2\over m_{_2}^2-m_{_3}^2}\Big)^n
\nonumber\\
&&\hspace{2.8cm}\sim
(m_{_3}^2-m_{_2}^2)^{D-3}\;,
\nonumber\\
&&\psi_{_{\{1,2,5\}}}^{(2)}(m_{_2}^2,m_{_3}^2)\sim
{(m_{_2}^2m_{_3}^2)^{D/2-1}\over m_{_3}^2-m_{_2}^2}
\sum\limits_{n=0}^\infty{\Gamma(D-2+n)\over\Gamma({D\over2}+n)}
\Big({m_{_2}^2\over m_{_2}^2-m_{_3}^2}\Big)^n\;,
\label{2Vac-1-14a}
\end{eqnarray}
where the intersection of their convergent regions is
$|m_{_2}^2/m_{_2}^2-m_{_3}^2|\le1$. In fact the basis $\{1,2,5\}$ of the projective
plane composed of the first, second and fifth columns of the matrix Eq.(\ref{2Vac-1-1})
is obtained from the basis $\{1,2,3\}$ through the permutation $\widehat{(35)}$,
accordingly the geometric description of the exponents Eq.(\ref{2Vac-1-13})
is derived from that of the exponents Eq.(\ref{2Vac-1-7}) through the permutation $\widehat{(35)}$.
Because the geometric description of Eq.(\ref{2Vac-1-13}) is same as that of Eq.(\ref{2Vac-1-7}),
$\psi_{_{\{1,2,5\}}}^{(1)}$ is proportional to
$\psi_{_{\{1,2,3\}}}^{(3)}$ in nonempty intersection of their convergent regions.
Similarly the geometric description of the exponents Eq.(\ref{2Vac-1-14})
is obtained through the permutation $\widehat{(35)}$ from that of Eq.(\ref{2Vac-1-8}).
Because the geometric description of Eq.(\ref{2Vac-1-14}) is same as that of Eq.(\ref{2Vac-1-8}),
$\psi_{_{\{1,2,5\}}}^{(2)}$ is proportional to
$\psi_{_{\{1,2,3\}}}^{(4)}$ in nonempty intersection of their convergent regions.
Actually $\psi_{_{\{1,2,5\}}}^{(2)}\sim \psi_{_{\{1,2,3\}}}^{(4)}$ is consistent
with the well-known relation~Eq.(\ref{1SE-24a}).

There are two choices on the exponents of the integer lattice
$-nE_{_{3}}^{(1)}$. The first choice is
\begin{eqnarray}
&&\alpha_{_{2,3}}={D\over2}-2,\;\alpha_{_{2,4}}=0,\;\alpha_{_{3,3}}=3-D,\;\alpha_{_{3,4}}=-1,
\label{2Vac-1-15}
\end{eqnarray}
where the geometric representation is drawn in Fig.\ref{fig2}(1) with
$a=5$, $\{(b,c)$, $(d,e)\}$ $=\{(1,4)$, $(2,3)\}$. The second choice is
\begin{eqnarray}
&&\alpha_{_{2,3}}=1-{D\over2},\;\alpha_{_{2,4}}=D-3,\;\alpha_{_{3,3}}=0,\;\alpha_{_{3,4}}=2-D,
\label{2Vac-1-16}
\end{eqnarray}
whose geometric representation is plotted in Fig.\ref{fig2}(2) with
$(a,b)$ $=(2,3)$, $\{c,d,e\}$ $=\{1,4,5\}$. Accordingly the hypergeometric
functions are
\begin{eqnarray}
&&\psi_{_{\{1,2,5\}}}^{(3)}(m_{_2}^2,m_{_3}^2)\sim(m_{_2}^2)^{D/2-2}(m_{_3}^2)^{D/2-1}
\sum\limits_{n=0}^\infty{\Gamma(2-{D\over2}+n)\over\Gamma(4-D+n)}
\Big({m_{_2}^2-m_{_3}^2\over m_{_2}^2}\Big)^n\;,
\nonumber\\
&&\psi_{_{\{1,2,5\}}}^{(4)}(m_{_2}^2,m_{_3}^2)\sim
{(m_{_3}^2)^{D/2-1}(m_{_3}^2-m_{_2}^2)^{D-3}\over(m_{_2}^2)^{D/2-1}}
\sum\limits_{n=0}^\infty{\Gamma({D\over2}-1+n)\over n!}
\Big({m_{_2}^2-m_{_3}^2\over m_{_2}^2}\Big)^n
\nonumber\\
&&\hspace{2.8cm}\sim
(m_{_3}^2-m_{_2}^2)^{D-3}\;.
\label{2Vac-1-17}
\end{eqnarray}

Similarly $\det(A^{2V0}_{_{\{1,3,5\}}})=-m_{_2}^2$, thus
\begin{eqnarray}
&&(A^{2V0}_{_{\{1,3,5\}}})^{-1}\cdot A^{2V0}=
\left(\begin{array}{ccccc}\;1\;&\;0\;&\;0\;&\;1\;&\;0\;\\
\;0\;&\;-{m_{_3}^2\over m_{_2}^2}\;&\;1\;&\;1-{m_{_3}^2\over m_{_2}^2}\;&\;0\;\\
\;0\;&\;{1\over m_{_2}^2}\;&\;0\;&\;{1\over m_{_2}^2}\;&\;1\;\end{array}\right)\;.
\label{2Vac-1-18}
\end{eqnarray}
The matrix of exponents is written as
\begin{eqnarray}
&&\left(\begin{array}{ccccc}\;-1\;&\;0\;&\;0\;&\;0\;&\;0\;\\
\;0&\;\alpha_{_{2,2}}\;&\;1-{D\over2}\;&\;\alpha_{_{2,4}}\;&\;0\;\\
\;0\;&\;\alpha_{_{3,2}}\;&\;0\;&\;\alpha_{_{3,4}}\;&\;D-3\;\end{array}\right)\;,
\label{2Vac-1-19}
\end{eqnarray}
where the entries satisfy the constraints of Eq.(\ref{2Vac-1-12}) with the replacement
$\alpha_{_{i,3}}\rightarrow\alpha_{_{i,2}}$, $i=2,3$. The geometric descriptions
are obtained from that of $\psi_{_{\{1,2,5\}}}^{(i)}$ by the permutation
$\widehat{(23)}$. The corresponding hypergeometric
functions are
\begin{eqnarray}
&&\psi_{_{\{1,3,5\}}}^{(i)}(m_{_2}^2,m_{_3}^2)\sim
\psi_{_{\{1,2,5\}}}^{(i)}(m_{_3}^2,m_{_2}^2),\;\;i=1,2,3,4.
\label{2Vac-1-20}
\end{eqnarray}

Because $\det(A^{2V0}_{_{\{2,3,4\}}})=1$, thus
\begin{eqnarray}
&&(A^{2V0}_{_{\{2,3,4\}}})^{-1}\cdot A^{2V0}=
\left(\begin{array}{ccccc}\;-1\;&\;1\;&\;0\;&\;0\;&\;m_{_2}^2\;\\
\;-1\;&\;0\;&\;1\;&\;0\;&\;m_{_3}^2\;\\
\;1\;&\;0\;&\;0\;&\;1\;&\;0\;\end{array}\right)\;.
\label{2Vac-1-21}
\end{eqnarray}
The matrix of exponents is written as
\begin{eqnarray}
&&\left(\begin{array}{ccccc}\;\alpha_{_{1,1}}\;&\;1-{D\over2}\;&\;0\;&\;0\;&\;\alpha_{_{1,5}}\;\\
\;\alpha_{_{2,3}}&\;0\;&\;1-{D\over2}\;&\;0\;&\;\alpha_{_{2,5}}\;\\
\;0\;&\;0\;&\;0\;&\;-1\;&\;0\;\end{array}\right)\;,
\label{2Vac-1-22}
\end{eqnarray}
with
\begin{eqnarray}
&&\alpha_{_{1,1}}+\alpha_{_{1,5}}={D\over2}-2,\;\alpha_{_{2,1}}+\alpha_{_{2,5}}={D\over2}-2,
\nonumber\\
&&\alpha_{_{1,1}}+\alpha_{_{2,1}}=-1,\;\alpha_{_{1,5}}+\alpha_{_{2,5}}=D-3.
\label{2Vac-1-23}
\end{eqnarray}
Corresponding to the local coordinates in Eq.(\ref{2Vac-1-21}), one finds
two choices on the matrix of integer lattice whose submatrix composed of the first and
fifth columns is formulated as $\pm nE_{_{3}}^{(1)}$,
where $n\ge0$, and other elements are all zero.
Basing on the matrices of integer lattice, we have 4 choices on the
exponents which are obtained that of $\psi_{_{\{1,2,3\}}}^{(i)}$
through the replacements $\alpha_{_{2,4}}\rightarrow\alpha_{_{1,1}}$,
$\alpha_{_{2,5}}\rightarrow\alpha_{_{1,5}}$, $\alpha_{_{3,4}}\rightarrow\alpha_{_{2,1}}$,
and $\alpha_{_{3,5}}\rightarrow\alpha_{_{2,5}}$,
respectively. Certainly the hypergeometric functions are
\begin{eqnarray}
&&\psi_{_{\{2,3,4\}}}^{(i)}(m_{_2}^2,m_{_3}^2)\sim
\psi_{_{\{1,2,3\}}}^{(i)}(m_{_2}^2,m_{_3}^2),\;\;i=1,2,3,4.
\label{2Vac-1-24}
\end{eqnarray}

\subsection{The fundamental solution systems in the general case}
\begin{figure}[ht]
\setlength{\unitlength}{1cm}
\centering
\vspace{0.0cm}\hspace{-1.5cm}
\includegraphics[height=8cm,width=8.0cm]{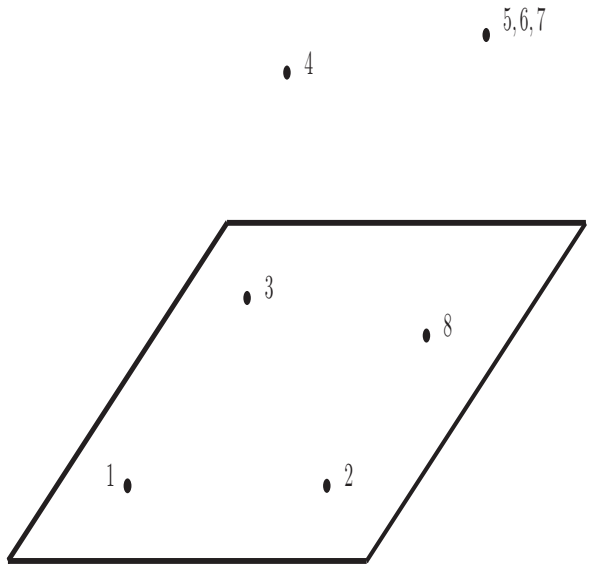}
\vspace{0cm}
\caption[]{The geometric description of the matroid $A^{(2V)}$ in
Eq.(\ref{2Vac-2-2}) on the projective plane $CP^{4}$, where the points $1,\cdots,8$
denote the indices of columns of the $4\times8$ matrix. In the representation,
the point $8$ belong to the hyperplane $\{1,2,3\}$, points \{5,6,7\} are
combined into a triple point, respectively.}
\label{fig7}
\end{figure}

\indent\indent
In the general case, the Feynman parametric representation is written as
\begin{eqnarray}
&&A_{_{2Vac}}(m_{_1}^2,\;m_{_2}^2,\;m_{_3}^2)=
{i^{4-D}\exp\{i\pi(2-D)/2\}\Gamma(3-D)\mu^{4\varepsilon}\over(4\pi)^D}
\int_0^\infty\prod\limits_{i=1}^3dt_{_i}
\nonumber\\
&&\hspace{3.9cm}\times
(t_{_1}t_{_2}t_{_3})^{1-D/2}
{\delta(t_{_1}t_{_2}t_{_3}-t_{_1}t_{_2}-t_{_1}t_{_3}-t_{_2}t_{_3})
\over(t_{_1}m_{_1}^2+t_{_2}m_{_2}^2+t_{_3}m_{_3}^2)^{3-D}}
\nonumber\\
&&\hspace{3.5cm}=
-{i^{4-D}\exp\{i\pi(2-D)/2\}\Gamma(3-D)\mu^{4\varepsilon}\over(4\pi)^D}
\nonumber\\
&&\hspace{3.9cm}\times
\int\omega(t)
{(t_{_1}t_{_2}t_{_3})^{1-D/2}t_{_4}^{D/2-1}
\over(t_{_1}m_{_1}^2+t_{_2}m_{_2}^2+t_{_3}m_{_3}^2)^{3-D}}
\nonumber\\
&&\hspace{3.9cm}\times
\delta(t_{_1}t_{_2}t_{_3}+t_{_1}t_{_2}t_{_4}+t_{_1}t_{_3}t_{_4}+t_{_2}t_{_3}t_{_4})
\label{2Vac-2-1}
\end{eqnarray}
The integral can be embedded in the subvariety of the Grassmannian $G_{_{4,8}}$
where the first row corresponds to the integration variable $t_{_1}$,
the second row  corresponds to the integration variable $t_{_2}$,
the third row  corresponds to the integration variable $t_{_3}$,
and the fourth row  corresponds to the integration variable $t_{_4}$, respectively.
Meanwhile the first column represents the power function $t_{_1}^{1-D/2}$,
the second column represents the power function $t_{_2}^{1-D/2}$, the third column
represents the power function $t_{_3}^{1-D/2}$, the fourth column
represents the power function $t_{_4}^{D/2-1}$, and the eighth column represents
the function $(t_{_1}m_{_1}^2+t_{_2}m_{_2}^2+t_{_3}m_{_3}^2)^{D-3}$, respectively.
In order to embed the homogeneous function
$\delta(t_{_1}t_{_2}t_{_3}+t_{_1}t_{_2}t_{_4}+t_{_1}t_{_3}t_{_4}+t_{_2}t_{_3}t_{_4})$
as the fifth, sixth, and seventh columns of the Grassmannian  $G_{_{4,8}}$, we rewrite
\begin{eqnarray}
&&t_{_1}t_{_2}t_{_3}+t_{_1}t_{_2}t_{_4}+t_{_1}t_{_3}t_{_4}+t_{_2}t_{_3}t_{_4}
\nonumber\\
&&\hspace{-0.5cm}=
z_{_{1,\widehat{\tau}_{_1}(5)}}z_{_{2,\widehat{\tau}_{_1}(6)}}z_{_{3,\widehat{\tau}_{_1}(7)}}t_{_1}t_{_2}t_{_3}
+z_{_{1,\widehat{\tau}_{_2}(5)}}z_{_{2,\widehat{\tau}_{_2}(6)}}z_{_{4,\widehat{\tau}_{_2}(7)}}t_{_1}t_{_2}t_{_4}
\nonumber\\
&&\hspace{0.0cm}
+z_{_{1,\widehat{\tau}_{_3}(5)}}z_{_{3,\widehat{\tau}_{_3}(6)}}z_{_{4,\widehat{\tau}_{_3}(7)}}t_{_1}t_{_3}t_{_4}
+z_{_{2,\widehat{\tau}_{_4}(5)}}z_{_{3,\widehat{\tau}_{_4}(6)}}z_{_{4,\widehat{\tau}_{_4}(7)}}t_{_2}t_{_3}t_{_4}\;,
\label{2Vac-2-1+a}
\end{eqnarray}
where $\widehat{\tau}_{_i},\;(i=1,\;2,\;3,\;4)$ are elements of the permutation
group $S_{_3}$ on the column indices $5,\;6,\;7$.
Taking $\widehat{\tau}_{_1}=\widehat{\tau}_{_4}=\hat{e}$, $\widehat{\tau}_{_2}=\widehat{(567)}$,
and $\widehat{\tau}_{_3}=\widehat{(576)}$, we have
\begin{eqnarray}
&&z_{_{1,5}}z_{_{2,6}}z_{_{3,7}}=z_{_{2,5}}z_{_{3,6}}z_{_{4,7}}=
z_{_{1,6}}z_{_{2,7}}z_{_{4,5}}=z_{_{1,7}}z_{_{3,5}}z_{_{4,6}}=1\;.
\label{2Vac-2-1+b}
\end{eqnarray}
The solution of Eq.(\ref{2Vac-2-1+b})
\begin{eqnarray}
&&z_{_{i,j}}=1\;,\;\;i=1,\cdots,4\;,j=5,\;6,\;7\;,
\label{2Vac-2-1+c}
\end{eqnarray}
indicates that the Feynman parametric representation can be embedded on
the subvariety of the Grassmannian $G_{_{4,8}}$
\begin{eqnarray}
&&A^{(2V)}=\left(\begin{array}{cccccccc}\;1\;&\;0\;&\;0\;&\;0\;&\;1\;&\;1\;&\;1\;&\;m_{_1}^2\;\\
\;0\;&\;1\;&\;0\;&\;0\;&\;1\;&\;1\;&\;1\;&\;m_{_2}^2\;\\
\;0\;&\;0\;&\;1\;&\;0\;&\;1\;&\;1\;&\;1\;&\;m_{_3}^2\;\\
\;0\;&\;0\;&\;0\;&\;1\;&\;1\;&\;1\;&\;1\;&\;0\;\end{array}\right)\;,
\label{2Vac-2-2}
\end{eqnarray}
whose geometric description is presented in Fig.\ref{fig7}.
Taking parametric representation of the 2-loop vacuum
as a function of the splitting local coordinates,
\begin{eqnarray}
&&A_{_{2Vac}}(m_{_1}^2,\;m_{_2}^2,\;m_{_3}^2)=\Psi^{2V}(z_{_1},\;\cdots,\;z_{_8})\;,
\label{2Vac-2-2+a}
\end{eqnarray}
we obviously have
\begin{eqnarray}
&&\Psi^{2V}(\chi_{_1}z_{_1},\;\cdots,\;\chi_{_8}z_{_8})=
(\chi_{_1}\chi_{_2}\chi_{_3})^{1-D/2}\chi_{_4}^{D/2-1}(\chi_{_5}\chi_{_6}\chi_{_7})^{-1}
\chi_{_8}^{D-3}\Psi^{2V}(z_{_1},\;\cdots,\;z_{_8})\;.
\label{2Vac-2-2+b}
\end{eqnarray}
Where $z_{_i}\;(i=1,\cdots,8)$ denotes the i-th column vector in the matrix Eq.(\ref{2Vac-2-2}),
and $\chi_{_i}$ is a nonzero constant, respectively.
If we choose the first, second, third, and fourth column vectors
of $A^{(2V)}$ as the basis of the projective plane,
the hypergeometric functions contains redundant interdependent arguments.
However $\det(A_{_{\{1,2,3,5\}}}^{(2V)})=1$ where the matrix $A_{_{\{1,2,3,5\}}}^{(2V)}$ denotes
the submatrix of $A^{(2V)}$ composed of the first, second, third and fifth columns, one derives
\begin{eqnarray}
&&\Big(A_{_{\{1,2,3,5\}}}^{(2V)}\Big)^{-1}\cdot A^{(2V)}=\left(\begin{array}{cccccccc}
\;\;1\;\;&\;\;0\;\;&\;\;0\;\;&\;\;-1\;\;&\;\;0\;\;&\;\;0\;\;&\;\;0\;\;&\;\;m_{_1}^2\;\;\\
\;\;0\;\;&\;\;1\;\;&\;\;0\;\;&\;\;-1\;\;&\;\;0\;\;&\;\;0\;\;&\;\;0\;\;&\;\;m_{_2}^2\;\;\\
\;\;0\;\;&\;\;0\;\;&\;\;1\;\;&\;\;-1\;\;&\;\;0\;\;&\;\;0\;\;&\;\;0\;\;&\;\;m_{_3}^2\;\;\\
\;\;0\;\;&\;\;0\;\;&\;\;0\;\;&\;\;1\;\;&\;\;1\;\;&\;\;1\;\;&\;\;1\;\;&\;\;0\;\;
\end{array}\right)\;.
\label{2Vac-2-3}
\end{eqnarray}
Accordingly, the Feynman integral satisfies the following GKZ-system in the splitting coordinates above
\begin{eqnarray}
&&\Big\{\vartheta_{_{1,1}}+\vartheta_{_{1,4}}+\vartheta_{_{1,8}}\Big\}A_{_{2Vac}}(m_{_1}^2,\;m_{_2}^2,\;m_{_3}^2)=
-A_{_{2Vac}}(m_{_1}^2,\;m_{_2}^2,\;m_{_3}^2)
\;,\nonumber\\
&&\Big\{\vartheta_{_{2,2}}+\vartheta_{_{2,4}}+\vartheta_{_{2,8}}\Big\}A_{_{2Vac}}(m_{_1}^2,\;m_{_2}^2,\;m_{_3}^2)=
-A_{_{2Vac}}(m_{_1}^2,\;m_{_2}^2,\;m_{_3}^2)
\;,\nonumber\\
&&\Big\{\vartheta_{_{3,3}}+\vartheta_{_{3,4}}+\vartheta_{_{3,8}}\Big\}A_{_{2Vac}}(m_{_1}^2,\;m_{_2}^2,\;m_{_3}^2)=
-A_{_{2Vac}}(m_{_1}^2,\;m_{_2}^2,\;m_{_3}^2)
\;,\nonumber\\
&&\Big\{\vartheta_{_{4,4}}+\vartheta_{_{4,5}}+\vartheta_{_{4,6}}
+\vartheta_{_{4,7}}\Big\}A_{_{2Vac}}(m_{_1}^2,\;m_{_2}^2,\;m_{_3}^2)=
-A_{_{2Vac}}(m_{_1}^2,\;m_{_2}^2,\;m_{_3}^2)
\;,\nonumber\\
&&\vartheta_{_{1,1}}A_{_{2Vac}}(m_{_1}^2,\;m_{_2}^2,\;m_{_3}^2)=
(1-{D\over2})A_{_{2Vac}}(m_{_1}^2,\;m_{_2}^2,\;m_{_3}^2)
\;,\nonumber\\
&&\vartheta_{_{2,2}}A_{_{2Vac}}(m_{_1}^2,\;m_{_2}^2,\;m_{_3}^2)=
(1-{D\over2})A_{_{2Vac}}(m_{_1}^2,\;m_{_2}^2,\;m_{_3}^2)
\;,\nonumber\\
&&\vartheta_{_{3,3}}A_{_{2Vac}}(m_{_1}^2,\;m_{_2}^2,\;m_{_3}^2)=
(1-{D\over2})A_{_{2Vac}}(m_{_1}^2,\;m_{_2}^2,\;m_{_3}^2)
\;,\nonumber\\
&&\Big\{\vartheta_{_{1,4}}+\vartheta_{_{2,4}}+\vartheta_{_{3,4}}+\vartheta_{_{4,4}}\Big\}
A_{_{2Vac}}(m_{_1}^2,\;m_{_2}^2,\;m_{_3}^2)=({D\over2}-1)A_{_{2Vac}}(m_{_1}^2,\;m_{_2}^2,\;m_{_3}^2)
\;,\nonumber\\
&&\vartheta_{_{4,5}}A_{_{2Vac}}(m_{_1}^2,\;m_{_2}^2,\;m_{_3}^2)=-A_{_{2Vac}}(m_{_1}^2,\;m_{_2}^2,\;m_{_3}^2)
\;,\nonumber\\
&&\vartheta_{_{4,6}}A_{_{2Vac}}(m_{_1}^2,\;m_{_2}^2,\;m_{_3}^2)=-A_{_{2Vac}}(m_{_1}^2,\;m_{_2}^2,\;m_{_3}^2)
\;,\nonumber\\
&&\vartheta_{_{4,7}}A_{_{2Vac}}(m_{_1}^2,\;m_{_2}^2,\;m_{_3}^2)=-A_{_{2Vac}}(m_{_1}^2,\;m_{_2}^2,\;m_{_3}^2)
\;,\nonumber\\
&&\Big\{\vartheta_{_{1,8}}+\vartheta_{_{2,8}}+\vartheta_{_{3,8}}\Big\}A_{_{2Vac}}(m_{_1}^2,\;m_{_2}^2,\;m_{_3}^2)=
(D-3)A_{_{2Vac}}(m_{_1}^2,\;m_{_2}^2,\;m_{_3}^2)\;.
\label{2Vac-2-3a}
\end{eqnarray}
Obviously the matrix of exponents is written as
\begin{eqnarray}
&&\left(\begin{array}{cccccccc}
\;\;1-{D\over2}\;\;&\;\;0\;\;&\;\;0\;\;&\;\;\alpha_{_{1,4}}\;\;&\;0\;\;&\;\;0\;\;&\;\;0\;\;&\;\;\alpha_{_{1,8}}\;\;\\
\;\;0\;\;&\;\;1-{D\over2}\;\;&\;0\;\;&\;\;\alpha_{_{2,4}}\;\;&\;\;0\;\;&\;0\;\;&\;\;0\;\;&\;\;\alpha_{_{2,8}}\;\;\\
\;\;0\;\;&\;\;0\;\;&\;1-{D\over2}\;\;&\;\;\alpha_{_{3,4}}\;\;&\;\;0\;\;&\;0\;\;&\;\;0\;\;&\;\;\alpha_{_{3,8}}\;\;\\
\;\;0\;\;&\;\;0\;\;&\;0\;\;&\;\;2\;\;&\;\;-1\;\;&\;-1\;\;&\;\;-1\;\;&\;\;0\;\;
\end{array}\right)\;,
\label{2Vac-2-4}
\end{eqnarray}
where the matrix elements satisfy the relations
\begin{eqnarray}
&&\alpha_{_{1,4}}+\alpha_{_{1,8}}={D\over2}-2\;,\;\alpha_{_{2,4}}+\alpha_{_{2,8}}={D\over2}-2\;,\;
\alpha_{_{3,4}}+\alpha_{_{3,8}}={D\over2}-2\;,
\nonumber\\
&&\alpha_{_{1,4}}+\alpha_{_{2,4}}+\alpha_{_{3,4}}={D\over2}-3\;,
\alpha_{_{1,8}}+\alpha_{_{2,8}}+\alpha_{_{3,8}}=D-3\;.
\label{2Vac-2-5}
\end{eqnarray}
\begin{figure}[ht]
\setlength{\unitlength}{1cm}
\centering
\vspace{0.0cm}\hspace{-1.5cm}
\includegraphics[height=8cm,width=8.0cm]{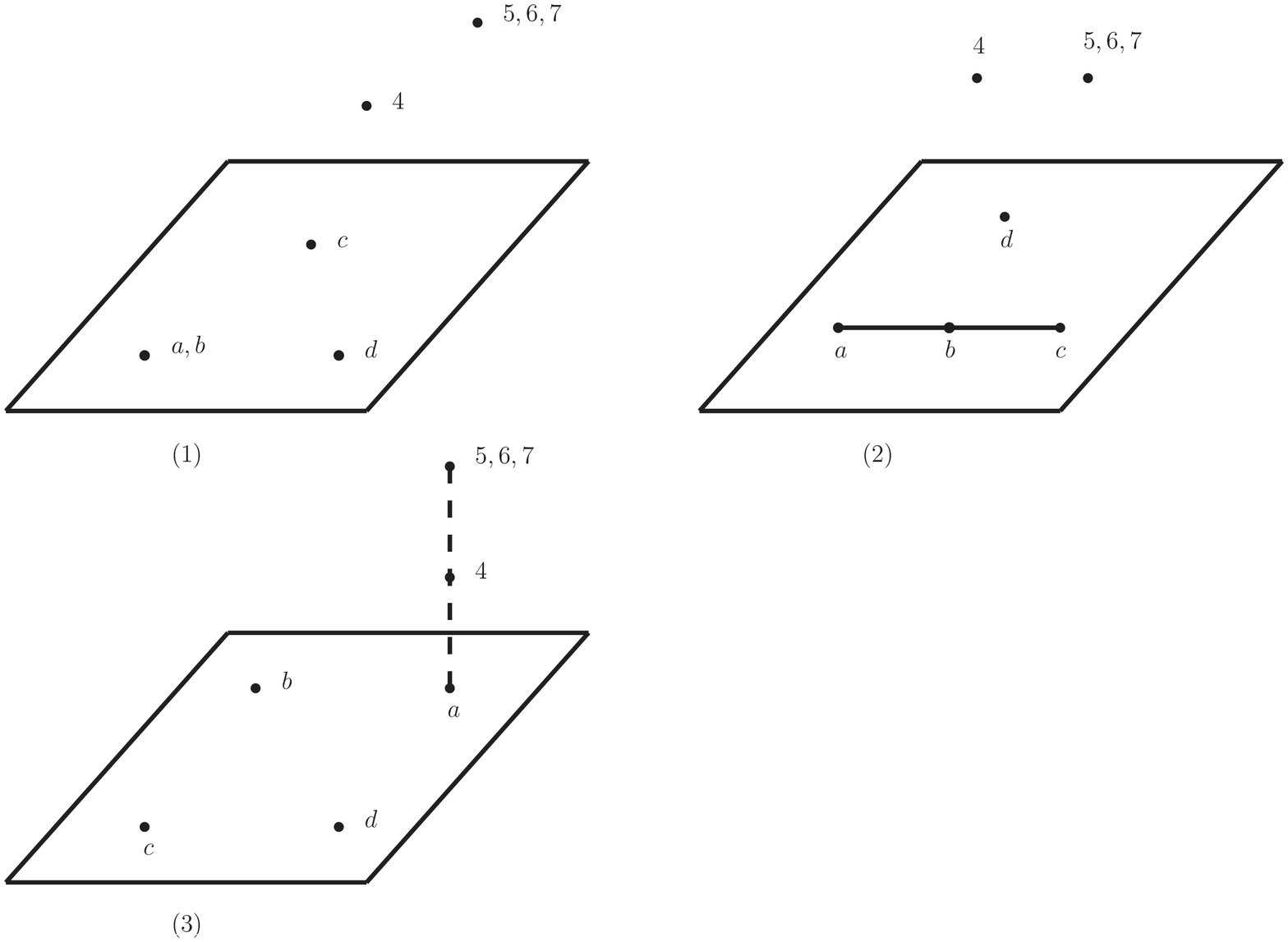}
\vspace{0cm}
\caption[]{The geometric descriptions of single orbits in
Eq.(\ref{2Vac-2-5}) on the projective plane $CP^{4}$, where single point 4
and triple point \{5,6,7\} lie outside the hyperplane $\{1,2,3,8\}$.}
\label{fig8}
\end{figure}
Corresponding to the matrix of local coordinates in Eq.(\ref{2Vac-2-3}),
we have 12 choices on the matrix of integer lattice whose submatrix composed of the fourth and
eighth columns is formulated as $\pm n_{_1}E_{_{4}}^{(i)}$ $\pm n_{_2}E_{_{4}}^{(j)}$,
where $n_{_{1,2}}\ge0$, $(i,j)$ $\in\{(1,2)$, $(1,4)$, $(2,4)\}$,
and other elements are all zero.
Note that $E_{_{4}}^{(1)}$ $=E_{_{4}}^{(2)}$ $-E_{_{4}}^{(4)}$, two linear independent
integer lattice matrices $E_{_{4}}^{(2)}$, $E_{_{4}}^{(4)}$ induce two independent PDEs as:
\begin{eqnarray}
&&{\partial^2A_{_{2Vac}}\over\partial z_{_{1,4}}\partial z_{_{3,8}}}(m_{_1}^2,\;m_{_2}^2,\;m_{_3}^2)
={\partial^2A_{_{2Vac}}\over\partial z_{_{1,8}}\partial z_{_{3,4}}}(m_{_1}^2,\;m_{_2}^2,\;m_{_3}^2)
\;,\nonumber\\
&&{\partial^2A_{_{2Vac}}\over\partial z_{_{2,4}}\partial z_{_{3,8}}}(m_{_1}^2,\;m_{_2}^2,\;m_{_3}^2)
={\partial^2A_{_{2Vac}}\over\partial z_{_{2,8}}\partial z_{_{3,4}}}(m_{_1}^2,\;m_{_2}^2,\;m_{_3}^2)\;.
\label{2Vac-2-5a}
\end{eqnarray}
When the exponents are given, the Feynman integral can be formally expressed as
\begin{eqnarray}
&&A_{_{2Vac}}(m_{_1}^2,\;m_{_2}^2,\;m_{_3}^2)=\prod\limits_{i,j}z_{_{i,j}}^{\alpha_{_{i,j}}}
\varphi({m_{_1}^2\over m_{_3}^2},{m_{_2}^2\over m_{_3}^2})\;,
\label{2Vac-2-5b}
\end{eqnarray}
where $(i,j)$ $\in\{(1,4)$, $(1,8)$, $(2,4)$, $(2,8)$, $(3,4)$, $(3,8)\}$,
and $\varphi(x_{_1},x_{_2})$ satisfies the PDEs in the combined coordinate
$x_{_1}$ $=z_{_{1,8}}z_{_{3,4}}/(z_{_{1,4}}z_{_{3,8}})$,
$x_{_2}$ $=z_{_{2,8}}z_{_{3,4}}/(z_{_{2,4}}z_{_{3,8}})$ as
\begin{eqnarray}
&&x_{_1}\Big[\alpha_{_{1,4}}\alpha_{_{3,8}}+(1-\alpha_{_{1,4}}
-\alpha_{_{3,8}})x_{_1}{\partial\over\partial x_{_1}}
-\alpha_{_{1,4}}x_{_2}{\partial\over\partial x_{_2}}
+x_{_1}x_{_2}{\partial^2\over\partial x_{_1}\partial x_{_2}}
+x_{_1}^2{\partial^2\over\partial x_{_1}^2}\Big]
\varphi(x_{_1},x_{_2})
\nonumber\\
&&\hspace{-0.5cm}=
\Big[\alpha_{_{1,8}}\alpha_{_{3,4}}+(1+\alpha_{_{1,8}}
+\alpha_{_{3,4}})x_{_1}{\partial\over\partial x_{_1}}
+\alpha_{_{1,8}}x_{_2}{\partial\over\partial x_{_2}}
+x_{_1}x_{_2}{\partial^2\over\partial x_{_1}\partial x_{_2}}
+x_{_1}^2{\partial^2\over\partial x_{_1}^2}\Big]
\varphi(x_{_1},x_{_2})
\;,\nonumber\\
&&x_{_2}\Big[\alpha_{_{2,4}}\alpha_{_{3,8}}
-\alpha_{_{2,4}}x_{_1}{\partial\over\partial x_{_1}}
+(1-\alpha_{_{2,4}}-\alpha_{_{3,8}})x_{_2}{\partial\over\partial x_{_2}}
+x_{_1}x_{_2}{\partial^2\over\partial x_{_1}\partial x_{_2}}
+x_{_2}^2{\partial^2\over\partial x_{_2}^2}\Big]
\varphi(x_{_1},x_{_2})
\nonumber\\
&&\hspace{-0.5cm}=
\Big[\alpha_{_{2,8}}\alpha_{_{3,4}}
+\alpha_{_{2,8}}x_{_1}{\partial\over\partial x_{_1}}
+(1+\alpha_{_{2,8}}+\alpha_{_{3,4}})x_{_2}{\partial\over\partial x_{_2}}
+x_{_1}x_{_2}{\partial^2\over\partial x_{_1}\partial x_{_2}}
+x_{_2}^2{\partial^2\over\partial x_{_2}^2}\Big]
\varphi(x_{_1},x_{_2})\;.
\label{2Vac-2-5c}
\end{eqnarray}
As $\alpha_{_{1,8}}=\alpha_{_{2,8}}=0$, the PDEs above are changed into
the first type Appell PDEs. The holonomic rank of the ideal composed by the
partial differential operators above is 3, which indicates three linear independent
hypergeometric functions constituting the fundamental solution systems
in neighborhoods of regular singularities.
It is known that the fourth type Appell function cannot be embedded into
Grassmannians~\cite{Gelfand1989}. Nevertheless, including the appropriate power factor,
the Feynman integral presented through
the fourth Appell functions~\cite{Feng2018,Feng2020} satisfies the above PDEs.
Those matrices of integer lattice permit 12 choices of the exponents.
Corresponding to the integer lattice $n_{_1}E_{_4}^{(1)}$ $+n_{_2}E_{_4}^{(2)}$,
the choice of exponents is written as
\begin{eqnarray}
&&\alpha_{_{1,4}}=1-{D\over2}\;,\;\alpha_{_{1,8}}=D-3\;,\;\alpha_{_{2,4}}={D\over2}-2\;,\;
\alpha_{_{2,8}}=0\;,\;\alpha_{_{3,4}}={D\over2}-2,\;\alpha_{_{3,8}}=0\;,
\label{2Vac-2-6}
\end{eqnarray}
whose geometric representation is presented in Fig.\ref{fig8}(1)
with $(a,b)$ $=(1,8)$, $\{c,d\}$ $=\{2,3\}$. In a similar way the exponents
of the integer lattice $n_{_1}E_{_4}^{(1)}$ $+n_{_2}E_{_4}^{(4)}$ are
\begin{eqnarray}
&&\alpha_{_{1,4}}=0\;,\;\alpha_{_{1,8}}={D\over2}-2\;,\;\alpha_{_{2,4}}=-1\;,\;
\alpha_{_{2,8}}={D\over2}-1\;,\;\alpha_{_{3,4}}={D\over2}-2,\;\alpha_{_{3,8}}=0\;,
\label{2Vac-2-10}
\end{eqnarray}
whose geometric representation is plotted in Fig.\ref{fig8}(2)
with $\{a,b,c\}$ $=\{1,2,8\}$, $d=3$. Corresponding to the integer lattice
$n_{_1}E_{_4}^{(2)}$ $+n_{_2}E_{_4}^{(4)}$, the exponents are
\begin{eqnarray}
&&\alpha_{_{1,4}}=0\;,\;\alpha_{_{1,8}}={D\over2}-2\;,\;\alpha_{_{2,4}}=0\;,\;
\alpha_{_{2,8}}={D\over2}-2\;,\;\alpha_{_{3,4}}={D\over2}-3,\;\alpha_{_{3,8}}=1\;,
\label{2Vac-2-14}
\end{eqnarray}
whose geometric representation is presented in Fig.\ref{fig8}(3)
with $a=3$, $(b,c,d)$ $=(1,2,8)$. The hypergeometric functions are evidently written as
\begin{eqnarray}
&&\psi_{_{\{1,2,3,5\}}}^{(1)}(m_{_1}^2,m_{_2}^2,m_{_3}^2)
\sim(m_{_1}^2)^{D-3}
\sum\limits_{n_{_1},n_{_2}}^\infty{\Gamma(2-{D\over2}+n_{_1})
\Gamma(2-{D\over2}+n_{_2})\over n_{_1}!n_{_2}!}
\nonumber\\
&&\hspace{4.2cm}\times
{\Gamma(3-D+n_{_1}+n_{_2})\over\Gamma(2-{D\over2}+n_{_1}+n_{_2})}
\Big({m_{_2}^2\over m_{_1}^2}\Big)^{n_{_1}}\Big({m_{_3}^2\over m_{_1}^2}\Big)^{n_{_2}}\;,
\nonumber\\
&&\psi_{_{\{1,2,3,5\}}}^{(2)}(m_{_1}^2,m_{_2}^2,m_{_3}^2)\sim
(m_{_1}^2)^{D/2-2}(m_{_2}^2)^{D/2-1}
\nonumber\\
&&\hspace{4.2cm}\times
\sum\limits_{n_{_1}=0}^\infty\sum\limits_{n_{_2}=0}^\infty
{\Gamma(2-{D\over2}+n_{_1}+n_{_2})\Gamma(1+n_{_1})\over
n_{_2}!\Gamma(1+n_{_1}+n_{_2})\Gamma({D\over2}+n_{_1})}
\nonumber\\
&&\hspace{4.2cm}\times
\Gamma(2-{D\over2}+n_{_2})
\Big({m_{_2}^2\over m_{_1}^2}\Big)^{n_{_1}}
\Big({m_{_3}^2\over m_{_1}^2}\Big)^{n_{_2}}\;,
\nonumber\\
&&\psi_{_{\{1,2,3,5\}}}^{(3)}(m_{_1}^2,m_{_2}^2,m_{_3}^2)
\sim(m_{_1}^2m_{_2}^2)^{D/2-2}m_{_3}^2
\sum\limits_{n_{_1},n_{_2}}^\infty{\Gamma(2-{D\over2}+n_{_1})
\Gamma(2-{D\over2}+n_{_2})\over n_{_1}!n_{_2}!}
\nonumber\\
&&\hspace{4.2cm}\times
{\Gamma(3-{D\over2}+n_{_1}+n_{_2})\over\Gamma(2+n_{_1}+n_{_2})}
\Big({m_{_3}^2\over m_{_1}^2}\Big)^{n_{_1}}
\Big({m_{_3}^2\over m_{_2}^2}\Big)^{n_{_2}}\;,
\label{2Vac-2-14a}
\end{eqnarray}
where the intersection of their convergent regions is a nonempty proper subset
of the whole parameter space.
In other words, three linear independent hypergeometric functions constitute a
fundamental solution system of the GKZ-system in the nonempty proper subset
of the whole parameter space.
Because non-positive integers are the first order zeros of
$\Gamma^{-1}(z)$, the exponents of Eq.(\ref{2Vac-2-10}) induce the following
hypergeometric function
\begin{eqnarray}
&&\psi_{_{\{1,2,3,5\}}}^{(2)\prime}(m_{_1}^2,m_{_2}^2,m_{_3}^2)
\sim{\sin^3{\pi D\over2}\over\pi^3}
(m_{_1}^2)^{D/2-2}(m_{_2}^2)^{D/2-1}
\nonumber\\
&&\hspace{4.2cm}\times
\sum\limits_{n_{_1}=0}^\infty\sum\limits_{n_{_2}=1+n_{_1}}^\infty
{\Gamma(1-{D\over2}-n_{_1}+n_{_2})\over
n_{_1}!n_{_2}!\Gamma(-n_{_1}+n_{_2})}
\nonumber\\
&&\hspace{4.2cm}\times
\Gamma(2-{D\over2}+n_{_1})\Gamma(2-{D\over2}+n_{_2})
\Big({m_{_2}^2\over m_{_1}^2}\Big)^{n_{_1}}
\Big({m_{_3}^2\over m_{_2}^2}\Big)^{n_{_2}}
\nonumber\\
&&\hspace{4.2cm}
+\lim\limits_{\epsilon\rightarrow0}
{\sin\pi\epsilon\sin^2{\pi D\over2}\over\pi^3}
(m_{_1}^2)^{D/2-2}(m_{_2}^2)^{D/2-1}
\nonumber\\
&&\hspace{4.2cm}\times
\sum\limits_{n_{_2}=0}^\infty\sum\limits_{n_{_1}=n_{_2}}^\infty
{\Gamma(1+n_{_1}-n_{_2})\over
n_{_1}!n_{_2}!\Gamma({D\over2}+n_{_1}-n_{_2})}
\nonumber\\
&&\hspace{4.2cm}\times
\Gamma(2-{D\over2}+n_{_1})\Gamma(2-{D\over2}+n_{_2})
\Big({m_{_2}^2\over m_{_1}^2}\Big)^{n_{_1}}
\Big({m_{_3}^2\over m_{_2}^2}\Big)^{n_{_2}}
\nonumber\\
&&\hspace{3.7cm}\sim
{\sin^3{\pi D\over2}\over\pi^3}
\psi_{_{\{1,2,3,5\}}}^{(3)}(m_{_1}^2,m_{_2}^2,m_{_3}^2)
\nonumber\\
&&\hspace{4.2cm}
+\lim\limits_{\epsilon\rightarrow0}
{\sin\pi\epsilon\sin^2{\pi D\over2}\over\pi^3}
\psi_{_{\{1,2,3,5\}}}^{(2)}(m_{_1}^2,m_{_2}^2,m_{_3}^2)\;.
\label{2Vac-2-14b}
\end{eqnarray}

The exponents of the integer lattice $-n_{_1}E_{_4}^{(1)}$
$-n_{_2}E_{_4}^{(2)}$ are
\begin{eqnarray}
&&\alpha_{_{1,4}}={D\over2}-3\;,\;\alpha_{_{1,8}}=1\;,\;\alpha_{_{2,4}}=0\;,\;
\alpha_{_{2,8}}={D\over2}-2\;,\;\alpha_{_{3,4}}=0,\;\alpha_{_{3,8}}={D\over2}-2\;,
\label{2Vac-2-9}
\end{eqnarray}
whose geometric representation is plotted in Fig.\ref{fig8}(3)
with $a=1$, $\{b,c,d\}$ $=\{2,3,8\}$. Similarly the exponents of
the integer lattice $-n_{_1}E_{_4}^{(1)}$ $-n_{_2}E_{_4}^{(4)}$ are given by
\begin{eqnarray}
&&\alpha_{_{1,4}}={D\over2}-2\;,\;\alpha_{_{1,8}}=0\;,\;\alpha_{_{2,4}}=-1\;,\;
\alpha_{_{2,8}}={D\over2}-1\;,\;\alpha_{_{3,4}}=0,\;\alpha_{_{3,8}}={D\over2}-2\;,
\label{2Vac-2-13}
\end{eqnarray}
whose geometric representation is presented in Fig.\ref{fig8}(2)
with $\{a,b,c\}$ $=\{2,3,8\}$, $d=1$. Corresponding to the integer lattice
$-n_{_1}E_{_4}^{(2)}$ $-n_{_2}E_{_4}^{(4)}$, the exponents are
\begin{eqnarray}
&&\alpha_{_{1,4}}={D\over2}-2\;,\;\alpha_{_{1,8}}=0\;,\;\alpha_{_{2,4}}={D\over2}-2\;,\;
\alpha_{_{2,8}}=0\;,\;\alpha_{_{3,4}}=1-{D\over2},\;\alpha_{_{3,8}}=D-3\;,
\label{2Vac-2-17}
\end{eqnarray}
whose geometric representation is depicted in Fig.\ref{fig8}(1)
with $(a,b)$ $=(3,8)$, $\{c,d\}$ $=\{1,2\}$. Accordingly, the hypergeometric
functions are
\begin{eqnarray}
&&\psi_{_{\{1,2,3,5\}}}^{(4)}(m_{_1}^2,m_{_2}^2,m_{_3}^2)
\sim m_{_1}^2(m_{_2}^2m_{_3}^2)^{D/2-2}
\sum\limits_{n_{_1},n_{_2}}^\infty{\Gamma(2-{D\over2}+n_{_1})
\Gamma(2-{D\over2}+n_{_2})\over n_{_1}!n_{_2}!}
\nonumber\\
&&\hspace{4.2cm}\times
{\Gamma(3-{D\over2}+n_{_1}+n_{_2})\over\Gamma(2+n_{_1}+n_{_2})}
\Big({m_{_1}^2\over m_{_2}^2}\Big)^{n_{_1}}
\Big({m_{_1}^2\over m_{_3}^2}\Big)^{n_{_2}}\;,
\nonumber\\
&&\psi_{_{\{1,2,3,5\}}}^{(5)}(m_{_1}^2,m_{_2}^2,m_{_3}^2)
\sim(m_{_2}^2)^{D/2-1}(m_{_3}^2)^{D/2-2}
\sum\limits_{n_{_1},n_{_2}}^\infty{\Gamma(2-{D\over2}+n_{_1}+n_{_2})
\over n_{_1}!\Gamma(1+n_{_1}+n_{_2})}
\nonumber\\
&&\hspace{4.2cm}\times
{\Gamma(2-{D\over2}+n_{_1})\Gamma(1+n_{_2})\over\Gamma({D\over2}+n_{_2})}
\Big({m_{_1}^2\over m_{_3}^2}\Big)^{n_{_1}}
\Big({m_{_2}^2\over m_{_3}^2}\Big)^{n_{_2}}
\;,\nonumber\\
&&\psi_{_{\{1,2,3,5\}}}^{(6)}(m_{_1}^2,m_{_2}^2,m_{_3}^2)
\sim(m_{_3}^2)^{D-3}
\sum\limits_{n_{_1},n_{_2}}^\infty
{\Gamma(2-{D\over2}+n_{_1})\Gamma(2-{D\over2}+n_{_2})\over n_{_1}!n_{_2}!}
\nonumber\\
&&\hspace{4.2cm}\times
{\Gamma(3-D+n_{_1}+n_{_2})\over\Gamma(2-{D\over2}+n_{_1}+n_{_2})}
\Big({m_{_1}^2\over m_{_3}^2}\Big)^{n_{_1}}
\Big({m_{_2}^2\over m_{_3}^2}\Big)^{n_{_2}}\;,
\label{2Vac-2-17a}
\end{eqnarray}
where the intersection of their convergent regions is a nonempty proper subset
of the whole parameter space.

Similarly the exponents of the integer lattice $n_{_1}E_{_4}^{(1)}$
$-n_{_2}E_{_4}^{(2)}$ are
\begin{eqnarray}
&&\alpha_{_{1,4}}=-1\;,\;\alpha_{_{1,8}}={D\over2}-1\;,\;\alpha_{_{2,4}}={D\over2}-2\;,\;
\alpha_{_{2,8}}=0\;,\;\alpha_{_{3,4}}=0,\;\alpha_{_{3,8}}={D\over2}-2\;,
\label{2Vac-2-7}
\end{eqnarray}
whose geometric representation is presented in Fig.\ref{fig8}(2)
with $\{a,b,c\}$ $=\{1,3,8\}$, $d=2$.
The exponents of the integer lattice
$n_{_1}E_{_4}^{(1)}$ $-n_{_2}E_{_4}^{(4)}$ are given as
\begin{eqnarray}
&&\alpha_{_{1,4}}=0\;,\;\alpha_{_{1,8}}={D\over2}-2\;,\;\alpha_{_{2,4}}={D\over2}-3\;,\;
\alpha_{_{2,8}}=1\;,\;\alpha_{_{3,4}}=0,\;\alpha_{_{3,8}}={D\over2}-2\;,
\label{2Vac-2-11}
\end{eqnarray}
whose geometric representation is drawn in Fig.\ref{fig8}(3)
with $a=2$, $\{b,c,d\}$ $=\{1,3,8\}$. Meanwhile the exponents of integer
lattice $n_{_1}E_{_4}^{(2)}-n_{_2}E_{_4}^{(4)}$ are
\begin{eqnarray}
&&\alpha_{_{1,4}}=0\;,\;\alpha_{_{1,8}}={D\over2}-2\;,\;\alpha_{_{2,4}}={D\over2}-2\;,\;
\alpha_{_{2,8}}=0\;,\;\alpha_{_{3,4}}=-1,\;\alpha_{_{3,8}}={D\over2}-1\;,
\label{2Vac-2-15}
\end{eqnarray}
whose geometric representation is presented in Fig.\ref{fig8}(2)
with $\{a,b,c\}$ $=\{1,3,8\}$, $d=2$. The corresponding hypergeometric functions are
\begin{eqnarray}
&&\psi_{_{\{1,2,3,5\}}}^{(7)}(m_{_1}^2,m_{_2}^2,m_{_3}^2)
\sim(m_{_1}^2)^{D/2-1}(m_{_3}^2)^{D/2-2}
\sum\limits_{n_{_1},n_{_2}}^\infty{\Gamma(2-{D\over2}+n_{_1}+n_{_2})
\over n_{_1}!\Gamma(1+n_{_1}+n_{_2})}
\nonumber\\
&&\hspace{4.2cm}\times
{\Gamma(2-{D\over2}+n_{_1})\Gamma(1+n_{_2})\over\Gamma({D\over2}+n_{_2})}
\Big({m_{_2}^2\over m_{_3}^2}\Big)^{n_{_1}}
\Big({m_{_1}^2\over m_{_3}^2}\Big)^{n_{_2}}\;,
\nonumber\\
&&\psi_{_{\{1,2,3,5\}}}^{(8)}(m_{_1}^2,m_{_2}^2,m_{_3}^2)
\sim(m_{_1}^2m_{_3}^2)^{D/2-2}m_{_2}^2
\sum\limits_{n_{_1},n_{_2}}^\infty{\Gamma(2-{D\over2}+n_{_1})
\Gamma(2-{D\over2}+n_{_2})\over n_{_1}!n_{_2}!}
\nonumber\\
&&\hspace{4.2cm}\times
{\Gamma(3-{D\over2}+n_{_1}+n_{_2})\over\Gamma(2+n_{_1}+n_{_2})}
\Big({m_{_2}^2\over m_{_1}^2}\Big)^{n_{_1}}
\Big({m_{_2}^2\over m_{_3}^2}\Big)^{n_{_2}}\;,
\nonumber\\
&&\psi_{_{\{1,2,3,5\}}}^{(9)}(m_{_1}^2,m_{_2}^2,m_{_3}^2)
\sim(m_{_1}^2)^{D/2-2}(m_{_3}^2)^{D/2-1}
\sum\limits_{n_{_1},n_{_2}}^\infty{\Gamma(2-{D\over2}+n_{_1}+n_{_2})\over
n_{_2}!\Gamma(1+n_{_1}+n_{_2})}
\nonumber\\
&&\hspace{4.2cm}\times
{\Gamma(1+n_{_1})\Gamma(2-{D\over2}+n_{_2})\over\Gamma({D\over2}+n_{_1})}
\Big({m_{_3}^2\over m_{_1}^2}\Big)^{n_{_1}}
\Big({m_{_2}^2\over m_{_1}^2}\Big)^{n_{_2}}\;,
\label{2Vac-2-15a}
\end{eqnarray}
where the intersection of their convergent regions is empty. However, the sets of the hypergeometric functions
$\{\psi_{_{\{1,2,3,5\}}}^{(1)}$, $\psi_{_{\{1,2,3,5\}}}^{(8)}$, $\psi_{_{\{1,2,3,5\}}}^{(9)}\}$
and $\{\psi_{_{\{1,2,3,5\}}}^{(6)}$, $\psi_{_{\{1,2,3,5\}}}^{(7)}$, $\psi_{_{\{1,2,3,5\}}}^{(8)}\}$
constitute the fundamental solution systems of the nonempty proper subsets
of the whole parameter space, respectively.

The exponents of the integer lattice $-n_{_1}E_{_4}^{(1)}$ $+n_{_2}E_{_4}^{(2)}$ are
\begin{eqnarray}
&&\alpha_{_{1,4}}=-1\;,\;\alpha_{_{1,8}}={D\over2}-1\;,\;\alpha_{_{2,4}}=0\;,\;
\alpha_{_{2,8}}={D\over2}-2\;,\;\alpha_{_{3,4}}={D\over2}-2,\;\alpha_{_{3,8}}=0\;,
\label{2Vac-2-8}
\end{eqnarray}
whose geometric representation is plotted in Fig.\ref{fig8}(2)
with $\{a,b,c\}$ $=\{1,2,8\}$, $d=3$.
The exponents of the integer lattice $-n_{_1}E_{_4}^{(1)}+n_{_2}E_{_4}^{(4)}$ are
\begin{eqnarray}
&&\alpha_{_{1,4}}={D\over2}-2\;,\;\alpha_{_{1,8}}=0\;,\;\alpha_{_{2,4}}=1-{D\over2}\;,\;
\alpha_{_{2,8}}=D-3\;,\;\alpha_{_{3,4}}={D\over2}-2,\;\alpha_{_{3,8}}=0\;,
\label{2Vac-2-12}
\end{eqnarray}
where the geometric representation is presented in Fig.\ref{fig8}(1)
with $(a,b)$ $=(2,8)$, $\{c,d\}$ $=\{1,3\}$. Corresponding to the integer
lattice $-n_{_1}E_{_4}^{(2)}$ $+n_{_2}E_{_4}^{(4)}$, the exponents are
\begin{eqnarray}
&&\alpha_{_{1,4}}={D\over2}-2\;,\;\alpha_{_{1,8}}=0\;,\;\alpha_{_{2,4}}=0\;,\;
\alpha_{_{2,8}}={D\over2}-2\;,\;\alpha_{_{3,4}}=-1,\;\alpha_{_{3,8}}={D\over2}-1\;,
\label{2Vac-2-16}
\end{eqnarray}
whose geometric representation is presented in Fig.\ref{fig8}(2)
with $\{a,b,c\}$ $=\{2,3,8\}$, $d=1$.
Accordingly the hypergeometric functions are evidently written as
\begin{eqnarray}
&&\psi_{_{\{1,2,3,5\}}}^{(10)}(m_{_1}^2,m_{_2}^2,m_{_3}^2)\sim
(m_{_1}^2)^{D/2-1}(m_{_2}^2)^{D/2-2}
\sum\limits_{n_{_1},n_{_2}}^\infty{\Gamma(2-{D\over2}+n_{_1}+n_{_2})
\over n_{_2}!\Gamma(1+n_{_1}+n_{_2})}
\nonumber\\
&&\hspace{4.2cm}\times
{\Gamma(1+n_{_1})\Gamma(2-{D\over2}+n_{_2})\over\Gamma({D\over2}+n_{_1})}
\Big({m_{_1}^2\over m_{_2}^2}\Big)^{n_{_1}}
\Big({m_{_3}^2\over m_{_2}^2}\Big)^{n_{_2}}\;,
\nonumber\\
&&\psi_{_{\{1,2,3,5\}}}^{(11)}(m_{_1}^2,m_{_2}^2,m_{_3}^2)
\sim(m_{_2}^2)^{D-3}
\sum\limits_{n_{_1},n_{_2}}^\infty
{\Gamma(2-{D\over2}+n_{_1})\Gamma(2-{D\over2}+n_{_2})\over n_{_1}!n_{_2}!}
\nonumber\\
&&\hspace{4.2cm}\times
{\Gamma(3-D+n_{_1}+n_{_2})\over\Gamma(2-{D\over2}+n_{_1}+n_{_2})}
\Big({m_{_1}^2\over m_{_2}^2}\Big)^{n_{_1}}
\Big({m_{_3}^2\over m_{_2}^2}\Big)^{n_{_2}}\;,
\nonumber\\
&&\psi_{_{\{1,2,3,5\}}}^{(12)}(m_{_1}^2,m_{_2}^2,m_{_3}^2)
\sim(m_{_2}^2)^{D/2-2}(m_{_3}^2)^{D/2-1}
\sum\limits_{n_{_1},n_{_2}}^\infty{\Gamma(2-{D\over2}+n_{_1}+n_{_2})\over
n_{_1}!\Gamma(1+n_{_1}+n_{_2})}
\nonumber\\
&&\hspace{4.2cm}\times
{\Gamma(2-{D\over2}+n_{_1})\Gamma(1+n_{_2})\over\Gamma({D\over2}+n_{_2})}
\Big({m_{_1}^2\over m_{_2}^2}\Big)^{n_{_1}}
\Big({m_{_3}^2\over m_{_2}^2}\Big)^{n_{_2}}\;,
\label{2Vac-2-18}
\end{eqnarray}
where the intersection of their convergent regions is empty. Nevertheless, the sets of the hypergeometric functions
$\{\psi_{_{\{1,2,3,5\}}}^{(3)}$, $\psi_{_{\{1,2,3,5\}}}^{(10)}$, $\psi_{_{\{1,2,3,5\}}}^{(11)}\}$
and $\{\psi_{_{\{1,2,3,5\}}}^{(4)}$, $\psi_{_{\{1,2,3,5\}}}^{(11)}$, $\psi_{_{\{1,2,3,5\}}}^{(12)}\}$
constitute the fundamental solution systems of the nonempty proper subsets
of the whole parameter space, respectively.

Because $\det(A_{_{\{1,2,5,8\}}})^{(2V)}=-m_{_3}^2$,
where the matrix $A_{_{\{1,2,5,8\}}}^{(2V)}$ denotes the submatrix of $A^{(2V)}$ composed
of the first, second, fifth and eighth columns, one finds
\begin{eqnarray}
&&\Big(A_{_{\{1,2,5,8\}}}^{(2V)}\Big)^{-1}\cdot A^{(2V)}=\left(\begin{array}{cccccccc}
\;\;1\;\;&\;\;0\;\;&\;\;-{m_{_1}^2\over m_{_3}^2}\;\;&\;\;-1+{m_{_1}^2\over m_{_3}^2}\;\;
&\;\;0\;\;&\;\;0\;\;&\;\;0\;\;&\;\;0\;\;\\
\;\;0\;\;&\;\;1\;\;&\;\;-{m_{_2}^2\over m_{_3}^2}\;\;&\;\;-1+{m_{_2}^2\over m_{_3}^2}\;\;&
\;\;0\;\;&\;\;0\;\;&\;\;0\;\;&\;\;0\;\;\\
\;\;0\;\;&\;\;0\;\;&\;\;0\;\;&\;\;1\;\;&\;\;1\;\;&\;\;1\;\;&\;\;1\;\;&\;\;0\;\;\\
\;\;0\;\;&\;\;0\;\;&\;\;{1\over m_{_3}^2}\;\;&\;\;-{1\over m_{_3}^2}\;\;&\;\;0\;\;
&\;\;0\;\;&\;\;0\;\;&\;\;1\;\;
\end{array}\right)\;.
\label{2Vac-2-20}
\end{eqnarray}
Obviously the matrix of exponents is written as
\begin{eqnarray}
&&\left(\begin{array}{cccccccc}
\;\;1-{D\over2}\;\;&\;\;0\;\;&\;\;\alpha_{_{1,3}}\;\;&\;\;\alpha_{_{1,4}}\;\;&\;\;0\;\;&\;0\;\;&\;\;0\;\;&\;\;0\;\;\\
\;\;0\;\;&\;\;1-{D\over2}\;\;&\;\;\alpha_{_{2,3}}\;\;&\;\alpha_{_{2,4}}\;\;&\;\;0\;\;&\;\;0\;\;&\;0\;\;&\;\;0\;\;\\
\;\;0\;\;&\;\;0\;\;&\;\;0\;\;&\;2\;\;&\;\;-1\;\;&\;\;-1\;\;&\;-1\;\;&\;\;0\;\;\\
\;\;0\;\;&\;\;0\;\;&\;\;\alpha_{_{4,3}}\;\;&\;\alpha_{_{4,4}}\;\;&\;\;0\;\;&\;\;0\;\;&\;0\;\;&\;\;D-3\;\;
\end{array}\right)\;,
\label{2Vac-2-21}
\end{eqnarray}
where the matrix elements satisfy the relations
\begin{eqnarray}
&&\alpha_{_{1,3}}+\alpha_{_{1,4}}={D\over2}-2\;,\;\alpha_{_{2,3}}+\alpha_{_{2,4}}={D\over2}-2\;,\;
\alpha_{_{4,3}}+\alpha_{_{4,4}}=2-D\;,
\nonumber\\
&&\alpha_{_{1,3}}+\alpha_{_{2,3}}+\alpha_{_{4,3}}=1-{D\over2}\;,
\alpha_{_{1,4}}+\alpha_{_{2,4}}+\alpha_{_{4,4}}={D\over2}-3\;.
\label{2Vac-2-22}
\end{eqnarray}
Corresponding to the matrix of splitting local coordinates in Eq.(\ref{2Vac-2-20}),
one finds 12 choices on the matrix of integer lattice whose submatrix composed of the third and
fourth columns is formulated as $\pm n_{_1}E_{_{4}}^{(i)}$ $\pm n_{_2}E_{_{4}}^{(j)}$,
where $n_{_{1,2}}\ge0$, $(i,j)$ $\in\{(1,3)$, $(1,5)$, $(3,5)\}$,
and other elements are all zero.

Corresponding to the integer lattice $n_{_1}E_{_{4}}^{(1)}$ $+n_{_2}E_{_{4}}^{(3)}$,
the exponents are written as
\begin{eqnarray}
&&\alpha_{_{1,3}}=1\;,\;\alpha_{_{1,4}}={D\over2}-3\;,\;\alpha_{_{2,3}}={D\over2}-2\;,\;
\alpha_{_{2,4}}=0\;,\;\alpha_{_{4,3}}=2-D,\;\alpha_{_{4,4}}=0\;,
\label{2Vac-2-23}
\end{eqnarray}
whose geometric representation is drawn in Fig.\ref{fig8}(3)
with $a=1$, $\{b,c,d\}$ $=\{2,3,8\}$.
In a similar way the exponents of the integer lattice $n_{_1}E_{_{4}}^{(1)}$
$+n_{_2}E_{_{4}}^{(5)}$ are
\begin{eqnarray}
&&\alpha_{_{1,3}}=0\;,\;\alpha_{_{1,4}}={D\over2}-2\;,\;\alpha_{_{2,3}}={D\over2}-1\;,\;
\alpha_{_{2,4}}=-1\;,\;\alpha_{_{4,3}}=2-D,\;\alpha_{_{4,4}}=0\;,
\label{2Vac-2-27}
\end{eqnarray}
whose geometric representation is presented in Fig.\ref{fig8}(2)
with $\{a,b,c\}$ $=\{2,3,8\}$, $d=1$. The exponents of the integer lattice
$n_{_1}E_{_{4}}^{(3)}$ $+n_{_2}E_{_{4}}^{(5)}$ are
\begin{eqnarray}
&&\alpha_{_{1,3}}=0\;,\;\alpha_{_{1,4}}={D\over2}-2\;,\;\alpha_{_{2,3}}=0\;,\;
\alpha_{_{2,4}}={D\over2}-2\;,\;\alpha_{_{4,3}}=1-{D\over2},\;\alpha_{_{4,4}}=1-{D\over2}\;,
\label{2Vac-2-31}
\end{eqnarray}
whose geometric representation is plotted in Fig.\ref{fig8}(1)
with $(a,b)$ $=(3,8)$, $\{c,d\}$ $=\{1,2\}$.
The corresponding hypergeometric functions are evidently written as
\begin{eqnarray}
&&\psi_{_{\{1,2,5,8\}}}^{(1)}(m_{_1}^2,m_{_2}^2,m_{_3}^2)
\sim{m_{_1}^2m_{_3}^2(m_{_2}^2)^{D/2-2}\over(m_{_1}^2-m_{_3}^2)^{3-D/2}}
\sum\limits_{n_{_1},n_{_2}}^\infty
{\Gamma(3-{D\over2}+n_{_1}+n_{_2})\over n_{_1}!n_{_2}!\Gamma(2+n_{_1}+n_{_2})}
\nonumber\\
&&\hspace{4.2cm}\times
\Gamma(2-{D\over2}+n_{_1})\Gamma(D-2+n_{_2})
\nonumber\\
&&\hspace{4.2cm}\times
\Big({m_{_1}^2(m_{_2}^2-m_{_3}^2)\over m_{_2}^2(m_{_1}^2-m_{_3}^2)}\Big)^{n_{_1}}
\Big({m_{_1}^2\over m_{_1}^2-m_{_3}^2}\Big)^{n_{_2}}\;,
\nonumber\\
&&\psi_{_{\{1,2,5,8\}}}^{(2)}(m_{_1}^2,m_{_2}^2,m_{_3}^2)
\sim{(m_{_2}^2)^{D/2-1}m_{_3}^2(m_{_1}^2-m_{_3}^2)^{D/2-2}\over m_{_2}^2-m_{_3}^2}
\nonumber\\
&&\hspace{4.2cm}\times
\sum\limits_{n_{_1},n_{_2}}^\infty{\Gamma(D-2+n_{_1}+n_{_2})\Gamma(2-{D\over2}+n_{_1})\over
n_{_1}!\Gamma(1+n_{_1}+n_{_2})}
\nonumber\\
&&\hspace{4.2cm}\times
{\Gamma(1+n_{_2})\over\Gamma({D\over2}+n_{_2})}
\Big({m_{_1}^2\over m_{_1}^2-m_{_3}^2}\Big)^{n_{_1}}
\Big({m_{_2}^2\over m_{_2}^2-m_{_3}^2}\Big)^{n_{_2}}\;,
\nonumber\\
&&\psi_{_{\{1,2,5,8\}}}^{(3)}(m_{_1}^2,m_{_2}^2,m_{_3}^2)
\sim\Big((m_{_1}^2-m_{_3}^2)(m_{_2}^2-m_{_3}^2)\Big)^{D/2-2}m_{_3}^2
\nonumber\\
&&\hspace{4.2cm}\times
\sum\limits_{n_{_1},n_{_2}}^\infty
{\Gamma(3-{D\over2}+n_{_1}+n_{_2})\over n_{_1}!n_{_2}!\Gamma(2+n_{_1}+n_{_2})}
\nonumber\\
&&\hspace{4.2cm}\times
\Gamma(2-{D\over2}+n_{_1})\Gamma(2-{D\over2}+n_{_2})
\nonumber\\
&&\hspace{4.2cm}\times
\Big({m_{_1}^2\over m_{_1}^2-m_{_3}^2}\Big)^{n_{_1}}
\Big({m_{_2}^2\over m_{_2}^2-m_{_3}^2}\Big)^{n_{_2}}\;,
\label{2Vac-2-31a}
\end{eqnarray}
where the intersection of their convergent regions is a nonempty proper subset
of the whole parameter space, and three linear independent hypergeometric
functions constitute a fundamental solution system of the GKZ-system in
the nonempty intersection.

The exponents of the integer lattice $-n_{_1}E_{_{4}}^{(1)}$
$-n_{_2}E_{_{4}}^{(3)}$ are
\begin{eqnarray}
&&\alpha_{_{1,3}}=1-{D\over2}\;,\;\alpha_{_{1,4}}=D-3\;,\;\alpha_{_{2,3}}=0\;,\;
\alpha_{_{2,4}}={D\over2}-2\;,\;\alpha_{_{4,3}}=0,\;\alpha_{_{4,4}}=2-D\;,
\label{2Vac-2-26}
\end{eqnarray}
whose geometric representation is presented in Fig.\ref{fig8}(1)
with $(a,b)$ $=(1,3)$, $\{c,d\}$ $=\{2,8\}$.
Similarly the exponents corresponding to the integer lattice
$-n_{_1}E_{_{4}}^{(1)}$ $-n_{_2}E_{_{4}}^{(5)}$ are given by
\begin{eqnarray}
&&\alpha_{_{1,3}}={D\over2}-2\;,\;\alpha_{_{1,4}}=0\;,\;\alpha_{_{2,3}}=3-D\;,\;
\alpha_{_{2,4}}={3D\over2}-5\;,\;\alpha_{_{4,3}}=0,\;\alpha_{_{4,4}}=2-D\;,
\label{2Vac-2-30}
\end{eqnarray}
whose geometric representation is shown in Fig.\ref{fig8}(2)
with $\{a,b,c\}$ $=\{1,2,3\}$, $d=8$. Finally the exponents of the integer
lattice $-n_{_1}E_{_{4}}^{(3)}$ $-n_{_2}E_{_{4}}^{(5)}$ are
\begin{eqnarray}
&&\alpha_{_{1,3}}={D\over2}-2\;,\;\alpha_{_{1,4}}=0\;,\;\alpha_{_{2,3}}={D\over2}-2\;,\;
\alpha_{_{2,4}}=0\;,\;\alpha_{_{4,3}}=5-{3D\over2},\;\alpha_{_{4,4}}={D\over2}-3\;,
\label{2Vac-2-34}
\end{eqnarray}
whose geometric representation is presented in Fig.\ref{fig8}(3)
with $a=8$, $\{b,c,d\}$ $=\{1,2,3\}$.
Accordingly the hypergeometric functions are summarized as
\begin{eqnarray}
&&\psi_{_{\{1,2,5,8\}}}^{(4)}(m_{_1}^2,m_{_2}^2,m_{_3}^2)
\sim{m_{_3}^2(m_{_2}^2m_{_3}^2)^{D/2-2}\over(m_{_1}^2)^{D/2-1}(m_{_1}^2-m_{_3}^2)^{3-D}}
\nonumber\\
&&\hspace{4.2cm}\times
\sum\limits_{n_{_1},n_{_2}}^\infty
{\Gamma({D\over2}-1+n_{_1}+n_{_2})
\Gamma(2-{D\over2}+n_{_1})\Gamma(D-2+n_{_2})
\over n_{_1}!n_{_2}!\Gamma(D-2+n_{_1}+n_{_2})}
\nonumber\\
&&\hspace{4.2cm}\times
\Big({m_{_2}^2(m_{_1}^2-m_{_3}^2)\over m_{_1}^2(m_{_2}^2-m_{_3}^2)}\Big)^{n_{_1}}
\Big({m_{_1}^2-m_{_3}^2\over m_{_1}^2}\Big)^{n_{_2}}\;,
\nonumber\\
&&\psi_{_{\{1,2,5,8\}}}^{(5)}(m_{_1}^2,m_{_2}^2,m_{_3}^2)
\sim{(m_{_1}^2)^{D/2-2}m_{_3}^2\over(m_{_2}^2)^{D-3}(m_{_2}^2-m_{_3}^2)^{5-3D/2}}
\nonumber\\
&&\hspace{4.2cm}\times
\sum\limits_{n_{_1},n_{_2}}^\infty{\Gamma(5-{3D\over2}-1+n_{_1}-n_{_2})
\Gamma(2-{D\over2}+n_{_1})\Gamma(D-2+n_{_2})
\over n_{_1}!n_{_2}!\Gamma(4-D+n_{_1}-n_{_2})}
\nonumber\\
&&\hspace{4.2cm}\times
\Big({m_{_2}^2(m_{_1}^2-m_{_3}^2)\over m_{_1}^2(m_{_2}^2-m_{_3}^2)}\Big)^{n_{_1}}
\Big({m_{_2}^2-m_{_3}^2\over m_{_2}^2}\Big)^{n_{_2}}\;,
\nonumber\\
&&\psi_{_{\{1,2,5,8\}}}^{(6)}(m_{_1}^2,m_{_2}^2,m_{_3}^2)
\sim m_{_3}^2(m_{_1}^2m_{_2}^2)^{D/2-2}
\sum\limits_{n_{_1},n_{_2}}^\infty
{\Gamma(2-{D\over2}+n_{_1})\Gamma(2-{D\over2}+n_{_2})\over n_{_1}!n_{_2}!}
\nonumber\\
&&\hspace{4.2cm}\times
{\Gamma(3-{D\over2}+n_{_1}+n_{_2})\over\Gamma(6-{3D\over2}+n_{_1}+n_{_2})}
\Big({m_{_1}^2-m_{_3}^2\over m_{_1}^2}\Big)^{n_{_1}}
\Big({m_{_2}^2-m_{_3}^2\over m_{_2}^2}\Big)^{n_{_2}}\;,
\label{2Vac-2-34a}
\end{eqnarray}
where the intersection of their convergent regions is a nonempty proper subset
of the whole parameter space. The hypergeometric functions constitute a
fundamental solution system of the GKZ-system in a nonempty proper subset
of the whole parameter space.

Similarly the exponents of the integer lattice $n_{_1}E_{_{4}}^{(1)}$
$-n_{_2}E_{_{4}}^{(3)}$ are written as
\begin{eqnarray}
&&\alpha_{_{1,3}}=3-D\;,\;\alpha_{_{1,4}}={3D\over2}-5\;,\;\alpha_{_{2,3}}={D\over2}-2\;,\;
\alpha_{_{2,4}}=0\;,\;\alpha_{_{4,3}}=0,\;\alpha_{_{4,4}}=2-D\;,
\label{2Vac-2-24}
\end{eqnarray}
whose geometric representation is presented in Fig.\ref{fig8}(2)
with $\{a,b,c\}$ $=\{1,2,3\}$, $d=8$.
The exponents of the integer lattice $n_{_1}E_{_{4}}^{(1)}$
$-n_{_2}E_{_{4}}^{(5)}$ are
\begin{eqnarray}
&&\alpha_{_{1,3}}=0\;,\;\alpha_{_{1,4}}={D\over2}-2\;,\;\alpha_{_{2,3}}=1-{D\over2}\;,\;
\alpha_{_{2,4}}=D-3\;,\;\alpha_{_{4,3}}=0,\;\alpha_{_{4,4}}=2-D\;,
\label{2Vac-2-28}
\end{eqnarray}
whose geometric representation is drawn in Fig.\ref{fig8}(1)
with $(a,b)$ $=(2,3)$, $\{c,d\}$ $=\{1,8\}$. Accordingly the exponents
of the integer lattice $n_{_1}E_{_{4}}^{(3)}$ $-n_{_2}E_{_{4}}^{(5)}$ are given by
\begin{eqnarray}
&&\alpha_{_{1,3}}=0\;,\;\alpha_{_{1,4}}={D\over2}-2\;,\;\alpha_{_{2,3}}={D\over2}-2\;,\;
\alpha_{_{2,4}}=0\;,\;\alpha_{_{4,3}}=3-D,\;\alpha_{_{4,4}}=-1\;,
\label{2Vac-2-32}
\end{eqnarray}
where the geometric representation is presented in Fig.\ref{fig8}(2)
with $\{a,b,c\}$ $=\{2,3,8\}$, $d=1$.
The corresponding hypergeometric functions are presented as
\begin{eqnarray}
&&\psi_{_{\{1,2,5,8\}}}^{(7)}(m_{_1}^2,m_{_2}^2,m_{_3}^2)
\sim{(m_{_1}^2)^{3-D}(m_{_2}^2)^{D/2-2}m_{_3}^2\over(m_{_1}^2-m_{_3}^2)^{5-3D/2}}
\sum\limits_{n_{_1},n_{_2}}^\infty
{\Gamma(2-{D\over2}+n_{_1})\Gamma(D-2+n_{_2})\over n_{_1}!n_{_2}!}
\nonumber\\
&&\hspace{4.2cm}\times
{\Gamma(5-{3D\over2}+n_{_1}-n_{_2})\over\Gamma(4-D+n_{_1}-n_{_2})}
\Big({m_{_1}^2(m_{_2}^2-m_{_3}^2)\over m_{_2}^2(m_{_1}^2-m_{_3}^2)}\Big)^{n_{_1}}
\Big({m_{_1}^2-m_{_3}^2\over m_{_1}^2}\Big)^{n_{_2}}\;,
\nonumber\\
&&\psi_{_{\{1,2,5,8\}}}^{(8)}(m_{_1}^2,m_{_2}^2,m_{_3}^2)
\sim{m_{_3}^2(m_{_1}^2-m_{_3}^2)^{D/2-2}\over(m_{_2}^2)^{D/2-1}(m_{_2}^2-m_{_3}^2)^{3-D}}
\sum\limits_{n_{_1},n_{_2}}^\infty
{\Gamma(2-{D\over2}+n_{_1})\Gamma(D-2+n_{_2})\over n_{_1}!n_{_2}!}
\nonumber\\
&&\hspace{4.2cm}\times
{\Gamma({D\over2}-1+n_{_1}+n_{_2})\over\Gamma(D-2+n_{_1}+n_{_2})}
\Big({m_{_1}^2(m_{_2}^2-m_{_3}^2)\over m_{_2}^2(m_{_1}^2-m_{_3}^2)}\Big)^{n_{_1}}
\Big({m_{_2}^2-m_{_3}^2\over m_{_2}^2}\Big)^{n_{_2}}\;,
\nonumber\\
&&\psi_{_{\{1,2,5,8\}}}^{(9)}(m_{_1}^2,m_{_2}^2,m_{_3}^2)
\sim{m_{_1}^2m_{_3}^2(m_{_2}^2)^{D/2-2}\over(m_{_1}^2-m_{_3}^2)^{3-D/2}}
\sum\limits_{n_{_1},n_{_2}}^\infty{\Gamma(3-{D\over2}+n_{_1}+n_{_2})\over
n_{_1}!n_{_2}!\Gamma(1+n_{_1}+n_{_2})}
\nonumber\\
&&\hspace{4.2cm}\times
\Gamma(D-2+n_{_1})\Gamma(2-{D\over2}+n_{_2})
\nonumber\\
&&\hspace{4.2cm}\times
\Big({m_{_1}^2\over m_{_1}^2-m_{_3}^2}\Big)^{n_{_1}}
\Big({m_{_1}^2(m_{_2}^2-m_{_3}^2)\over m_{_2}^2(m_{_1}^2-m_{_2}^2)}\Big)^{n_{_2}}\;,
\label{2Vac-2-32a}
\end{eqnarray}
where the intersection of their convergent regions is empty. However, the sets of the hypergeometric functions
$\{\psi_{_{\{1,2,5,8\}}}^{(1)}$, $\psi_{_{\{1,2,5,8\}}}^{(8)}$, $\psi_{_{\{1,2,5,8\}}}^{(9)}\}$
and $\{\psi_{_{\{1,2,5,8\}}}^{(6)}$, $\psi_{_{\{1,2,5,8\}}}^{(7)}$, $\psi_{_{\{1,2,5,8\}}}^{(8)}\}$
constitute the fundamental solution systems of the nonempty proper subsets
of the whole parameter space, respectively.

The exponents of the integer lattice $-n_{_1}E_{_{4}}^{(1)}$ $+n_{_2}E_{_{4}}^{(3)}$ are
\begin{eqnarray}
&&\alpha_{_{1,3}}={D\over2}-1\;,\;\alpha_{_{1,4}}=-1\;,\;\alpha_{_{2,3}}=0\;,\;
\alpha_{_{2,4}}={D\over2}-2\;,\;\alpha_{_{4,3}}=2-D,\;\alpha_{_{4,4}}=0\;,
\label{2Vac-2-25}
\end{eqnarray}
whose geometric representation is depicted in Fig.\ref{fig8}(2)
with $\{a,b,c\}$ $=\{1,2,8\}$, $d=3$.
The corresponding exponents of the integer lattice
$-n_{_1}E_{_{4}}^{(1)}$ $+n_{_2}E_{_{4}}^{(5)}$ are given as
\begin{eqnarray}
&&\alpha_{_{1,3}}={D\over2}-2\;,\;\alpha_{_{1,4}}=0\;,\;\alpha_{_{2,3}}=1\;,\;
\alpha_{_{2,4}}={D\over2}-3\;,\;\alpha_{_{4,3}}=2-D,\;\alpha_{_{4,4}}=0\;,
\label{2Vac-2-29}
\end{eqnarray}
whose geometric representation is shown in Fig.\ref{fig8}(3)
with $a=2$, $\{b,c,d\}$ $=\{1,3,8\}$. Finally, the exponents of the integer lattice
$-n_{_1}E_{_{4}}^{(3)}$ $+n_{_2}E_{_{4}}^{(5)}$ are
\begin{eqnarray}
&&\alpha_{_{1,3}}={D\over2}-2\;,\;\alpha_{_{1,4}}=0\;,\;\alpha_{_{2,3}}=0\;,\;
\alpha_{_{2,4}}={D\over2}-2\;,\;\alpha_{_{4,3}}=3-D,\;\alpha_{_{4,4}}=-1\;,
\label{2Vac-2-33}
\end{eqnarray}
whose geometric representation is presented in Fig.\ref{fig8}(2)
with $\{a,b,c\}$ $=\{1,3,8\}$, $d=2$.
Three hypergeometric functions are evidently written as
\begin{eqnarray}
&&\psi_{_{\{1,2,5,8\}}}^{(10)}(m_{_1}^2,m_{_2}^2,m_{_3}^2)\sim
{(m_{_1}^2)^{D/2-1}(m_{_2}^2)^{D/2-2}m_{_3}^2\over m_{_1}^2-m_{_3}^2}
\sum\limits_{n_{_1},n_{_2}}^\infty{\Gamma(D-2+n_{_1}+n_{_2})\over
\Gamma(1+n_{_1}+n_{_2})}
\nonumber\\
&&\hspace{4.2cm}\times
{\Gamma(2-{D\over2}+n_{_1})\Gamma(1+n_{_2})\over n_{_1}!\Gamma({D\over2}+n_{_2})}
\Big({m_{_2}^2\over m_{_2}^2-m_{_3}^2}\Big)^{n_{_1}}
\Big({m_{_1}^2\over m_{_1}^2-m_{_3}^2}\Big)^{n_{_2}}
\;,\nonumber\\
&&\psi_{_{\{1,2,5,8\}}}^{(11)}(m_{_1}^2,m_{_2}^2,m_{_3}^2)
\sim{m_{_2}^2m_{_3}^2(m_{_1}^2)^{D/2-2}\over(m_{_2}^2-m_{_3}^2)^{3-D/2}}
\sum\limits_{n_{_1},n_{_2}}^\infty{\Gamma(3-{D\over2}+n_{_1}+n_{_2})\over
\Gamma(2+n_{_1}+n_{_2})}
\nonumber\\
&&\hspace{4.2cm}\times
{\Gamma(2-{D\over2}+n_{_1})\Gamma(D-2+n_{_2})\over n_{_1}!n_{_2}!}
\nonumber\\
&&\hspace{4.2cm}\times
\Big({m_{_2}^2(m_{_1}^2-m_{_3}^2)\over m_{_1}^2(m_{_2}^2-m_{_3}^2)}\Big)^{n_{_1}}
\Big({m_{_2}^2\over m_{_2}^2-m_{_3}^2}\Big)^{n_{_2}}\;,
\nonumber\\
&&\psi_{_{\{1,2,5,8\}}}^{(12)}(m_{_1}^2,m_{_2}^2,m_{_3}^2)
\sim m_{_3}^2\Big(m_{_1}^2(m_{_2}^2-m_{_3}^2)\Big)^{D/2-2}
\sum\limits_{n_{_1},n_{_2}}^\infty{\Gamma(2-{D\over2}+n_{_1}+n_{_2})\over
\Gamma(1+n_{_1}+n_{_2})}
\nonumber\\
&&\hspace{4.2cm}\times
{\Gamma(1+n_{_1})\Gamma(2-{D\over2}+n_{_2})\over n_{_2}!\Gamma(4-D+n_{_1})}
\nonumber\\
&&\hspace{4.2cm}\times
\Big({m_{_1}^2-m_{_3}^2\over m_{_1}^2}\Big)^{n_{_1}}
\Big({m_{_2}^2(m_{_1}^2-m_{_3}^2)\over m_{_1}^2(m_{_2}^2-m_{_3}^2)}\Big)^{n_{_2}}\;,
\label{2Vac-2-35}
\end{eqnarray}
where the intersection of their convergent regions is empty. Nevertheless, the sets of the hypergeometric functions
$\{\psi_{_{\{1,2,5,8\}}}^{(3)}$, $\psi_{_{\{1,2,5,8\}}}^{(10)}$, $\psi_{_{\{1,2,5,8\}}}^{(11)}\}$
and $\{\psi_{_{\{1,2,5,8\}}}^{(4)}$, $\psi_{_{\{1,2,5,8\}}}^{(11)}$, $\psi_{_{\{1,2,5,8\}}}^{(12)}\}$
constitute the fundamental solution systems of the nonempty proper subsets
of the whole parameter space, respectively.

Because $\det(A_{_{\{1,3,5,8\}}})^{(2V)}=m_{_2}^2$,
where the matrix $A_{_{\{1,3,5,8\}}}^{(2V)}$ denotes the submatrix of $A^{(2V)}$ composed
of the first, third, fifth and eighth columns, one finds
\begin{eqnarray}
&&\Big(A_{_{\{1,3,5,8\}}}^{(2V)}\Big)^{-1}\cdot A^{(2V)}=\left(\begin{array}{cccccccc}
\;\;1\;\;&\;\;-{m_{_1}^2\over m_{_2}^2}\;\;&\;\;0\;\;&\;\;-1+{m_{_1}^2\over m_{_2}^2}\;\;
&\;\;0\;\;&\;\;0\;\;&\;\;0\;\;&\;\;0\;\;\\
\;\;0\;\;&\;\;-{m_{_3}^2\over m_{_2}^2}\;\;&\;\;1\;\;&\;\;-1+{m_{_3}^2\over m_{_2}^2}\;\;&
\;\;0\;\;&\;\;0\;\;&\;\;0\;\;&\;\;0\;\;\\
\;\;0\;\;&\;\;0\;\;&\;\;0\;\;&\;\;1\;\;&\;\;1\;\;&\;\;1\;\;&\;\;1\;\;&\;\;0\;\;\\
\;\;0\;\;&\;\;{1\over m_{_2}^2}\;\;&\;\;0\;\;&\;\;-{1\over m_{_2}^2}\;\;&\;\;0\;\;
&\;\;0\;\;&\;\;0\;\;&\;\;1\;\;
\end{array}\right)\;.
\label{2Vac-2-37}
\end{eqnarray}
Obviously the matrix of exponents is written as
\begin{eqnarray}
&&\left(\begin{array}{cccccccc}
\;\;1-{D\over2}\;\;&\;\;\alpha_{_{1,2}}\;\;&\;\;0\;\;&\;\;\alpha_{_{1,4}}\;\;&\;\;0\;\;&\;0\;\;&\;\;0\;\;&\;\;0\;\;\\
\;\;0\;\;&\;\;\alpha_{_{2,2}}\;\;&\;\;1-{D\over2}\;\;&\;\alpha_{_{2,4}}\;\;&\;\;0\;\;&\;\;0\;\;&\;0\;\;&\;\;0\;\;\\
\;\;0\;\;&\;\;0\;\;&\;\;0\;\;&\;2\;\;&\;\;-1\;\;&\;\;-1\;\;&\;-1\;\;&\;\;0\;\;\\
\;\;0\;\;&\;\;\alpha_{_{4,2}}\;\;&\;\;0\;\;&\;\alpha_{_{4,4}}\;\;&\;\;0\;\;&\;\;0\;\;&\;0\;\;&\;\;D-3\;\;
\end{array}\right)\;,
\label{2Vac-2-38}
\end{eqnarray}
where the matrix elements satisfy the relations
\begin{eqnarray}
&&\alpha_{_{1,2}}+\alpha_{_{1,4}}={D\over2}-2\;,\;\alpha_{_{2,2}}+\alpha_{_{2,4}}={D\over2}-2\;,\;
\alpha_{_{4,2}}+\alpha_{_{4,4}}=2-D\;,
\nonumber\\
&&\alpha_{_{1,2}}+\alpha_{_{2,2}}+\alpha_{_{4,2}}=1-{D\over2}\;,
\alpha_{_{1,4}}+\alpha_{_{2,4}}+\alpha_{_{4,4}}={D\over2}-3\;.
\label{2Vac-2-39}
\end{eqnarray}
Corresponding to the matrix of splitting local coordinates in Eq.(\ref{2Vac-2-37}),
we have 12 choices on the matrix of integer lattice whose submatrix composed of the second and
fourth columns is formulated as $\pm n_{_1}E_{_{4}}^{(i)}$ $\pm n_{_2}E_{_{4}}^{(j)}$,
where $n_{_{1,2}}\ge0$, $(i,j)$ $\in\{(1,3)$, $(1,5)$, $(3,5)\}$,
and other elements are all zero.
Basing on the matrices of integer lattice, we have 12 choices on the
matrix of exponents whose geometric descriptions are obtained from that
of $\psi_{_{\{1,2,5,8\}}}^{(i)}$ by the permutation $\widehat{(23)}$.
Correspondingly the hypergeometric functions are
\begin{eqnarray}
&&\psi_{_{\{1,3,5,8\}}}^{(i)}(m_{_1}^2,m_{_2}^2,m_{_3}^2)=
\psi_{_{\{1,2,5,8\}}}^{(i)}(m_{_1}^2,m_{_3}^2,m_{_2}^2)
\label{2Vac-2-40}
\end{eqnarray}

Similarly $\det(A_{_{\{2,3,5,8\}}})^{(2V)}=-m_{_1}^2$,
where the matrix $A_{_{\{2,3,5,8\}}}^{(2V)}$ denotes the submatrix of $A^{(2V)}$ composed
of the second, third, fifth and eighth columns, one finds
\begin{eqnarray}
&&\Big(A_{_{\{2,3,5,8\}}}^{(2V)}\Big)^{-1}\cdot A^{(2V)}=\left(\begin{array}{cccccccc}
\;\;-{m_{_2}^2\over m_{_1}^2}\;\;&\;\;1\;\;&\;\;0\;\;&\;\;-1+{m_{_2}^2\over m_{_1}^2}\;\;
&\;\;0\;\;&\;\;0\;\;&\;\;0\;\;&\;\;0\;\;\\
\;\;-{m_{_3}^2\over m_{_1}^2}\;\;&\;\;0\;\;&\;\;1\;\;&\;\;-1+{m_{_3}^2\over m_{_1}^2}\;\;&
\;\;0\;\;&\;\;0\;\;&\;\;0\;\;&\;\;0\;\;\\
\;\;0\;\;&\;\;0\;\;&\;\;0\;\;&\;\;1\;\;&\;\;1\;\;&\;\;1\;\;&\;\;1\;\;&\;\;0\;\;\\
\;\;{1\over m_{_1}^2}\;\;&\;\;0\;\;&\;\;0\;\;&\;\;-{1\over m_{_1}^2}\;\;&\;\;0\;\;
&\;\;0\;\;&\;\;0\;\;&\;\;1\;\;
\end{array}\right)\;.
\label{2Vac-2-41}
\end{eqnarray}
Obviously the matrix of exponents is written as
\begin{eqnarray}
&&\left(\begin{array}{cccccccc}
\;\;\alpha_{_{1,1}}\;\;&\;\;1-{D\over2}\;\;&\;\;0\;\;&\;\;\alpha_{_{1,4}}\;\;&\;\;0\;\;&\;0\;\;&\;\;0\;\;&\;\;0\;\;\\
\;\;\alpha_{_{2,1}}\;\;&\;\;0\;\;&\;\;1-{D\over2}\;\;&\;\alpha_{_{2,4}}\;\;&\;\;0\;\;&\;\;0\;\;&\;0\;\;&\;\;0\;\;\\
\;\;0\;\;&\;\;0\;\;&\;\;0\;\;&\;2\;\;&\;\;-1\;\;&\;\;-1\;\;&\;-1\;\;&\;\;0\;\;\\
\;\;\alpha_{_{4,1}}\;\;&\;\;0\;\;&\;\;0\;\;&\;\alpha_{_{4,4}}\;\;&\;\;0\;\;&\;\;0\;\;&\;0\;\;&\;\;D-3\;\;
\end{array}\right)\;,
\label{2Vac-2-42}
\end{eqnarray}
where the matrix elements satisfy the relations
\begin{eqnarray}
&&\alpha_{_{1,1}}+\alpha_{_{1,4}}={D\over2}-2\;,\;\alpha_{_{2,1}}+\alpha_{_{2,4}}={D\over2}-2\;,\;
\alpha_{_{4,1}}+\alpha_{_{4,4}}=2-D\;,
\nonumber\\
&&\alpha_{_{1,1}}+\alpha_{_{2,1}}+\alpha_{_{4,1}}=1-{D\over2}\;,
\alpha_{_{1,4}}+\alpha_{_{2,4}}+\alpha_{_{4,4}}={D\over2}-3\;.
\label{2Vac-2-43}
\end{eqnarray}
Correspondingly there are 12 choices on the matrix of integer lattice whose submatrix composed of the first and
fourth columns is formulated as $\pm n_{_1}E_{_{4}}^{(i)}$ $\pm n_{_2}E_{_{4}}^{(j)}$,
where $n_{_{1,2}}\ge0$, $(i,j)$ $\in\{(1,3)$, $(1,5)$, $(3,5)\}$,
and other elements are all zero.
Basing on the matrices of integer lattice, we have 12 choices on the
matrix of exponents whose geometric descriptions are obtained from that of
$\psi_{_{\{1,2,5,8\}}}^{(i)}$ by the permutation $\widehat{(13)}$.
Correspondingly the hypergeometric functions are
\begin{eqnarray}
&&\psi_{_{\{2,3,5,8\}}}^{(i)}(m_{_1}^2,m_{_2}^2,m_{_3}^2)=
\psi_{_{\{1,2,5,8\}}}^{(i)}(m_{_3}^2,m_{_2}^2,m_{_1}^2)\;.
\label{2Vac-2-44}
\end{eqnarray}

In view of combinatorial geometry~\cite{Oxley2011}, the points with the same
splitting local coordinates in the projective space can be taken as
one point whose exponent is the sum of the exponents of the coincident points.
Nevertheless, in the matrix of splitting local coordinates specifying a column
for each homogeneous factor in the integrand allows us to apply more combinatorial
information to efficiently obtain fundamental solution systems of neighborhoods
of different regular singularities.

\section{Parametric representations of the 2-loop self-erergies\label{2LSE}}

\begin{figure}[ht]
\setlength{\unitlength}{1cm}
\centering
\vspace{0.0cm}\hspace{-1.5cm}
\includegraphics[height=6cm,width=10.0cm]{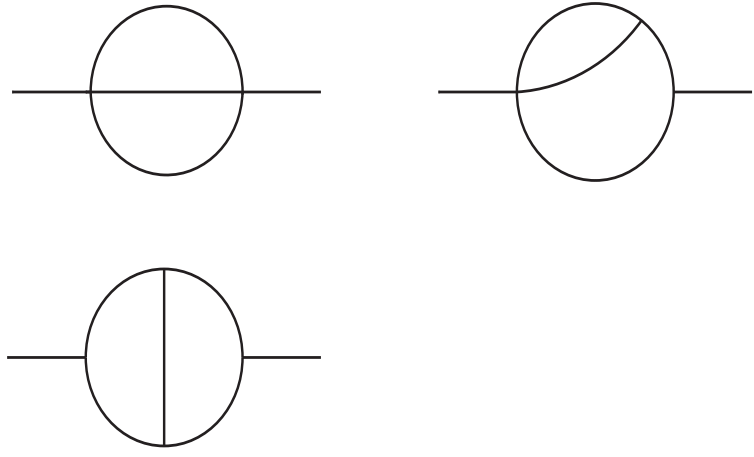}
\vspace{0cm}
\caption[]{The skeleton diagrams of two-loop self-energies.}
\label{fig11}
\end{figure}

To embed the Feynman integrals into the subvarieties of Grassmannians through
the Feynman parametric representation
is not practical because sizes of corresponding Grassmannians are too large to obtain
the hypergeometric functions efficiently. In order to construct hypergeometric functions
efficiently, we adopt the $\alpha$-parametric representation~\cite{V.A.Smirnov2012}
of the Feynman integrals for multiloop diagrams.
The diagrams of the 2-loop self-energies are drawn in Fig.\ref{fig11},
and the Feynman integral of the 2-loop self energy diagram
with 3 propagators can be expressed as
\begin{eqnarray}
&&A_{_{2L3}}(p^2,\;m_{_1}^2,\;m_{_2}^2,\;m_{_3}^2)=
-{i^{6-2D}\Gamma(3-D)\Big(\Lambda_{_{\rm RE}}^2\Big)^{3-D}\over(4\pi)^D}
\int\omega_{_4}(t)\Big(t_{_1}t_{_2}t_{_3}\Big)^{1-D/2}
\nonumber\\
&&\hspace{4.7cm}\times
t_{_4}^{D/2-1}
\delta(t_{_1}t_{_2}t_{_3}+t_{_1}t_{_2}t_{_4}+t_{_1}t_{_3}t_{_4}+t_{_2}t_{_3}t_{_4})
\nonumber\\
&&\hspace{4.7cm}\times
\Big[t_{_1}m_{_1}^2+t_{_2}m_{_2}^2+t_{_3}m_{_3}^2+t_{_4}p^2\Big]^{D-3}\;.
\label{2SunsetA}
\end{eqnarray}
The integral can be embedded in the subvariety of the Grassmannian $G_{_{4,8}}$
where the first row corresponds to the integration variable $t_{_1}$,
the second row corresponds to the integration variable $t_{_2}$,
the third row  corresponds to the integration variable $t_{_3}$,
and the fourth row  corresponds to the integration variable $t_{_4}$, respectively.
Meanwhile the first column represents the power function $t_{_1}^{1-D/2}$,
the second column represents the power function $t_{_2}^{1-D/2}$, the third column
represents the power function $t_{_3}^{1-D/2}$, the fourth column
represents the power function $t_{_4}^{D/2-1}$, the eighth column represents the function
$(t_{_1}m_{_1}^2+t_{_2}m_{_2}^2+t_{_3}m_{_3}^2+t_{_4}p^2)^{D-3}$. In order to
embed the homogeneous function
$\delta(t_{_1}t_{_2}t_{_3}+t_{_1}t_{_2}t_{_4}+t_{_1}t_{_3}t_{_4}+t_{_2}t_{_3}t_{_4})$
as the fifth, sixth, and seventh columns of the Grassmannian  $G_{_{4,8}}$, we rewrite
\begin{eqnarray}
&&t_{_1}t_{_2}t_{_3}+t_{_1}t_{_2}t_{_4}+t_{_1}t_{_3}t_{_4}+t_{_2}t_{_3}t_{_4}
\nonumber\\
&&\hspace{-0.5cm}=
z_{_{1,\widehat{\tau}_{_1}(5)}}z_{_{2,\widehat{\tau}_{_1}(6)}}z_{_{3,\widehat{\tau}_{_1}(7)}}t_{_1}t_{_2}t_{_3}
+z_{_{1,\widehat{\tau}_{_2}(5)}}z_{_{2,\widehat{\tau}_{_2}(6)}}z_{_{4,\widehat{\tau}_{_2}(7)}}t_{_1}t_{_2}t_{_4}
\nonumber\\
&&\hspace{0.0cm}
+z_{_{1,\widehat{\tau}_{_3}(5)}}z_{_{3,\widehat{\tau}_{_3}(6)}}z_{_{4,\widehat{\tau}_{_3}(7)}}t_{_1}t_{_3}t_{_4}
+z_{_{2,\widehat{\tau}_{_4}(5)}}z_{_{3,\widehat{\tau}_{_4}(6)}}z_{_{4,\widehat{\tau}_{_4}(7)}}t_{_2}t_{_3}t_{_4}\;,
\label{2SunsetA+d}
\end{eqnarray}
where $\widehat{\tau}_{_i},\;(i=1,\;2,\;3,\;4)$ are elements of the permutation
group $S_{_3}$ on the column indices $5,\;6,\;7$.
Taking $\widehat{\tau}_{_1}=\widehat{\tau}_{_4}=\hat{e}$, $\widehat{\tau}_{_2}=\widehat{(567)}$,
and $\widehat{\tau}_{_3}=\widehat{(576)}$, we have
\begin{eqnarray}
&&z_{_{1,5}}z_{_{2,6}}z_{_{3,7}}=z_{_{2,5}}z_{_{3,6}}z_{_{4,7}}=
z_{_{1,6}}z_{_{2,7}}z_{_{4,5}}=z_{_{1,7}}z_{_{3,5}}z_{_{4,6}}=1\;.
\label{2SunsetA+e}
\end{eqnarray}
The solution of Eq.(\ref{2Vac-2-1+b})
\begin{eqnarray}
&&z_{_{i,j}}=1\;,\;\;i=1,\cdots,4\;,j=5,\;6,\;7\;,
\label{2SunsetA+f}
\end{eqnarray}
The integral is embedded in the subvariety of the Grassmannian $G_{_{4,8}}$
whose splitting local coordinates are written as
\begin{eqnarray}
&&\left(\begin{array}{cccccccc}
\;\;1\;\;&\;\;0\;\;&\;\;0\;\;&\;\;0&\;\;1\;\;&\;\;1\;\;&\;\;1\;\;&\;\;m_{_1}^2\;\;\\
\;\;0\;\;&\;\;1\;\;&\;\;0\;\;&\;\;0\;\;&\;\;1\;\;&\;\;1\;\;&\;\;1\;\;&\;\;m_{_2}^2\;\;\\
\;\;0\;\;&\;\;0\;\;&\;\;1\;\;&\;\;0\;\;&\;\;1\;\;&\;\;1\;\;&\;\;1\;\;&\;\;m_{_3}^2\;\;\\
\;\;0\;\;&\;\;0\;\;&\;\;0\;\;&\;\;1\;\;&\;\;1\;\;&\;\;1\;\;&\;\;1\;\;&\;\;p^2\;\;
\end{array}\right)\;.
\label{2SunsetA-1}
\end{eqnarray}

Similarly the Feynman integral of the 2-loop diagram
with 4 propagators can be written as
\begin{eqnarray}
&&A_{_{2L4}}(p^2)=
-{i^{4-D}\Gamma(4-D)\exp\Big\{{i\pi(2-D)\over2}\Big\}
\Big(\Lambda_{_{\rm RE}}^2\Big)^{4-D}\over(4\pi)^D}
\int\prod\limits_{i=1}^4\omega_{_5}(t)\delta(t_{_1}t_{_2}t_{_3}
\nonumber\\
&&\hspace{2.2cm}
+t_{_1}t_{_2}t_{_4}+t_{_1}t_{_3}t_{_4}
+t_{_1}t_{_2}t_{_5}+t_{_1}t_{_3}t_{_5}
+t_{_2}t_{_3}t_{_5}+t_{_2}t_{_4}t_{_5}
+t_{_3}t_{_4}t_{_5})
\nonumber\\
&&\hspace{2.2cm}\times
\Big[t_{_1}t_{_2}+t_{_1}t_{_3}+t_{_2}t_{_3}
+t_{_2}t_{_4}+t_{_3}t_{_4}\Big]^{1-D/2}
\Big[\sum\limits_{i=1}^4t_{_i}m_{_i}^2+t_{_5}p^2\Big]^{D-4}\;,
\label{2LSE1234A}
\end{eqnarray}
and the Feynman integral of the 2-loop self energy with 5 propagators can
be accordingly expressed as
\begin{eqnarray}
&&A_{_{2L5}}(p^2)=
{\exp\Big\{{i\pi(8-2D)\over2}\Big\}\Gamma(5-D)
\Big(\Lambda_{_{\rm RE}}^2\Big)^{5-D}\over(4\pi)^D}
\int\omega_{_6}(t)
\nonumber\\
&&\hspace{2.3cm}\times
\Big[t_{_1}(t_{_2}+t_{_4}+t_{_5})
+(t_{_2}+t_{_4})(t_{_3}+t_{_5})
+t_{_3}t_{_5}\Big]^{1-D/2}
\nonumber\\
&&\hspace{2.3cm}\times
\delta\Big(t_{_1}t_{_2}(t_{_3}+t_{_4})
+(t_{_1}+t_{_2})(t_{_3}t_{_4}
+t_{_3}t_{_5}+t_{_4}t_{_5})
\nonumber\\
&&\hspace{2.3cm}
+t_{_1}t_{_6}(t_{_2}+t_{_4}+t_{_5})
+t_{_6}(t_{_2}+t_{_4})(t_{_3}+t_{_5})
+t_{_3}t_{_5}t_{_6}\Big)
\nonumber\\
&&\hspace{2.3cm}\times
\Big[\sum\limits_{i=1}^5t_{_i}m_{_i}^2+t_{_6}p^2\Big]^{D-5}\;,
\label{2LSE12345A}
\end{eqnarray}
respectively. Using the integrals in projective space above, we obtain
the fundamental solution systems in neighborhoods of different regular singularities.
Concrete results in detail are presented elsewhere.

\section{Summary\label{summary}}
\indent\indent
The Feynman integrals can be taken as functions on the subvarieties of Grassmannians
through homogenizing the parametric representation. Thus the GKZ-systems satisfied by
the Feynman integrals are obtained in splitting local coordinates accordingly.
The dual PDEs of GKZ-system determine the matrices of integer lattice
and corresponding exponent matrices in splitting local coordinates. Actually the dual
PDEs can be written as the PDEs satisfied by these scalar integrals in combined local coordinates.
Using those matrices of integer lattice and the corresponding exponent matrices,
we obtain the fundamental solution systems in neighborhoods of the regular singularities.

Taking the diagrams of the 1-loop self-energy, the 1-loop triangle with massless virtual
particles, and the 2-loop vacuum as examples, we elucidate how to construct the fundamental
solution systems of the corresponding Feynman integral in neighborhoods of the regular singularities.
For most multi-loop diagrams, the sizes of Grassmannians embedded by Feynman integrals
are too large to construct the fundamental systems in neighborhoods of the regular singularities
efficiently through the Feynman parametric representations.
To efficiently derive the fundamental solution system, we can embed the corresponding scalar
integrals into the subvarieties of Grassmannians using the $\alpha$-parametric representation.
In the last section, we present the parametric representations of diagrams of the 2-loop self energy
with 3 propagators, 4 propagators, and 5 propagators, respectively.
Generally the Feynman integrals can be written as linear combinations of
those hypergeometric functions of the fundamental solution systems in the
neighborhoods of regular singularities.
The integration constants, i.e. the combination coefficients, are determined
from the Feynman integral on an ordinary point or some regular singularities.
The concrete expressions of the Feynman integrals together with their partial
derivatives of some regular singularities can be evaluated by the
Gegenbauer polynomials~\cite{Chetyrkin1980} or Bessel functions~\cite{Feng2018,Gu2020}.

\begin{acknowledgments}
\indent\indent
The work has been supported partly by the National Natural
Science Foundation of China (NNSFC) with Grant No. 12075074, No. 11535002,
No. 11745006, No. 11675239, No. 11821505, No. 11705045, No. 11947302 and No. 12075301,
the Natural Science Foundation of Guangxi Autonomous Region with Grant No. 2022GXNSFDA035068,
and the Natural Science Foundation of Hebei Province with Grant No. A2022201017,
the youth top-notch talent support program of the Hebei Province.
Furthermore, the author (C.-H. Chang) is also supported by Key Research
program of Frontier Sciences, CAS, Grant No. QYZDY-SSW-SYS006.
\end{acknowledgments}

\appendix
\section{The fundamental solution systems of $A^{(1T)\prime}$\label{app1}}

When we choose the first, second, third, fourth column vectors of $A^{(1T)\prime}$
as the basis of the projective plane, the corresponding hypergeometric functions contain
redundant dependent indeterminates.
Because $\det(A_{_{\{1,2,4,6\}}}^{(1T)\prime})=-1$ where the matrix $A_{_{\{1,2,4,6\}}}^{(1T)\prime}$ denotes
the submatrix of $A^{(1T)\prime}$ composed of the first, second, fourth and sixth columns, one derives
\begin{eqnarray}
&&\Big(A_{_{\{1,2,4,6\}}}^{(1T)\prime}\Big)^{-1}\cdot A^{(1T)\prime}=\left(\begin{array}{ccccccc}
\;\;1\;\;&\;\;0\;\;&\;\;-1\;\;&\;\;0\;\;&\;\;p_{_2}^2-p_{_1}^2\;\;&\;\;0\;\;&\;\;0\;\;\\
\;\;0\;\;&\;\;1\;\;&\;\;-1\;\;&\;\;0\;\;&\;\;p_{_3}^2-p_{_1}^2\;\;&\;\;0\;\;&\;\;0\;\;\\
\;\;0\;\;&\;\;0\;\;&\;\;0\;\;&\;\;1\;\;&\;\;0\;\;&\;\;0\;\;&\;\;0\;\;\\
\;\;0\;\;&\;\;0\;\;&\;\;1\;\;&\;\;0\;\;&\;\;p_{_1}^2\;\;&\;\;1\;\;&\;\;1\;\;
\end{array}\right)\;.
\label{1TR2-4}
\end{eqnarray}
Obviously the matrix of exponents is written as
\begin{eqnarray}
&&\left(\begin{array}{ccccccc}
\;\;0\;\;&\;\;0\;\;&\;\;\alpha_{_{1,3}}\;\;&\;\;0\;\;&\;\alpha_{_{1,5}}\;\;&\;\;0\;\;&\;\;0\;\;\\
\;\;0\;\;&\;\;0\;\;&\;\alpha_{_{2,3}}\;\;&\;\;0\;\;&\;\;\alpha_{_{2,5}}\;\;&\;0\;\;&\;\;0\;\;\\
\;\;0\;\;&\;\;0\;\;&\;0\;\;&\;\;3-D\;\;&\;\;0\;\;&\;0\;\;&\;\;D-4\;\;\\
\;\;0\;\;&\;\;0\;\;&\;\alpha_{_{4,3}}\;\;&\;\;0\;\;&\;\;\alpha_{_{4,5}}\;\;&\;{D\over2}-3\;\;&\;\;3-D\;\;
\end{array}\right)\;,
\label{1TR2-5}
\end{eqnarray}
where the matrix elements satisfy the relations
\begin{eqnarray}
&&\alpha_{_{1,3}}+\alpha_{_{1,5}}=-1\;,\;\alpha_{_{2,3}}+\alpha_{_{2,5}}=-1\;,\;
\alpha_{_{4,3}}+\alpha_{_{4,5}}={D\over2}-1\;,
\nonumber\\
&&\alpha_{_{1,3}}+\alpha_{_{2,3}}+\alpha_{_{4,3}}=0\;,
\alpha_{_{1,5}}+\alpha_{_{2,5}}+\alpha_{_{4,5}}={D\over2}-3\;.
\label{1TR2-6}
\end{eqnarray}

Corresponding to the matrix of local coordinates in Eq.(\ref{1TR2-4}),
we find 12 choices on the matrix of integer lattice whose submatrix composed of the third and
fifth columns is formulated as $\pm n_{_1}E_{_{4}}^{(i)}$ $\pm n_{_2}E_{_{4}}^{(j)}$,
where $n_{_{1,2}}\ge0$, $(i,j)$ $\in\{(1,3)$, $(1,5)$, $(3,5)\}$,
and other elements are all zero.
\begin{figure}[ht]
\setlength{\unitlength}{1cm}
\centering
\vspace{0.0cm}\hspace{-1.5cm}
\includegraphics[height=8cm,width=8.0cm]{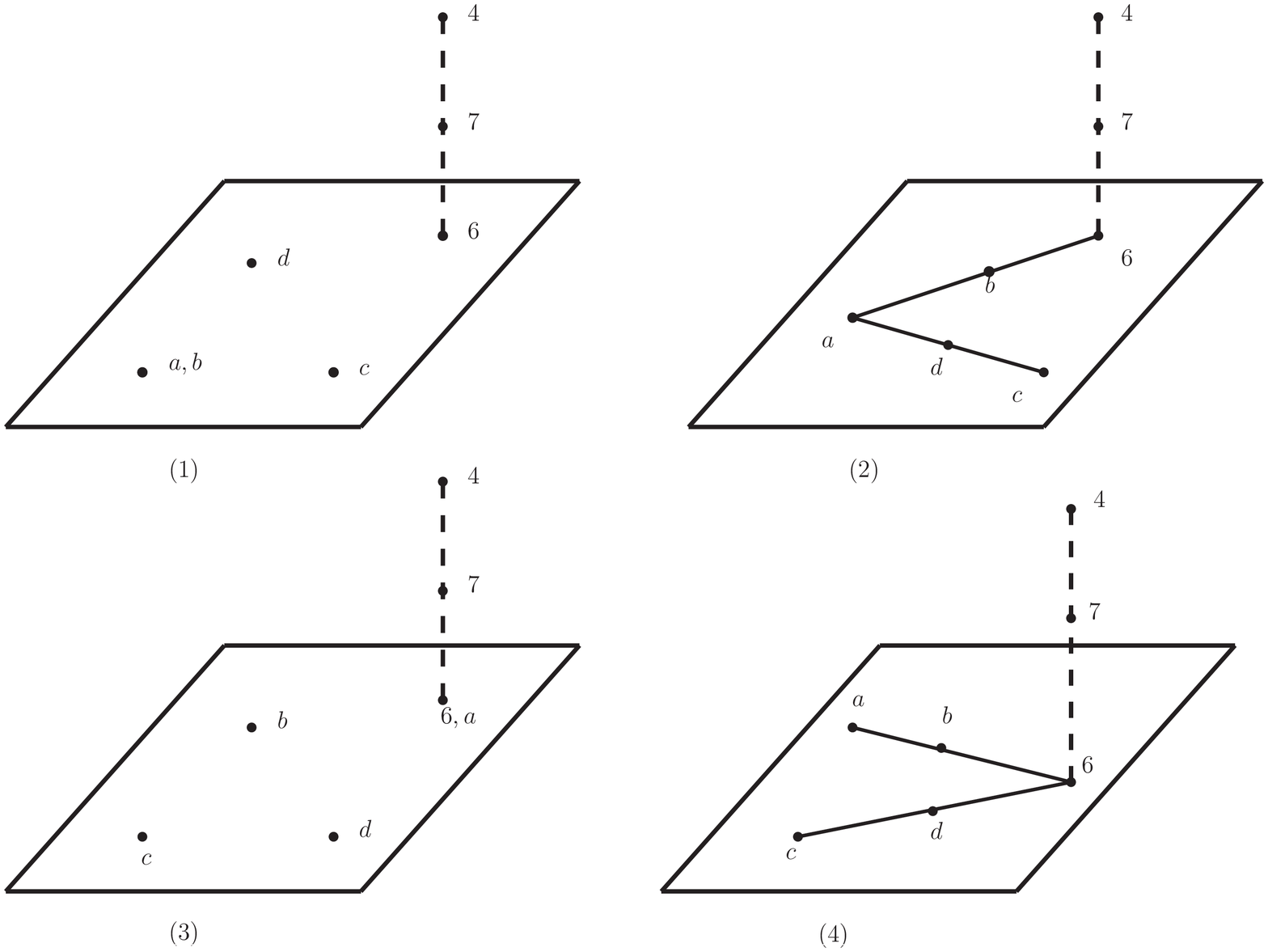}
\vspace{0cm}
\caption[]{The geometric descriptions of single orbits in
Eq.(\ref{1TR2-3}) on the projective plane $CP^{4}$, where the line \{4,6,7\}
intersects with the hyperplane $\{1,2,3\}$ at the point $6$.}
\label{fig6}
\end{figure}
Basing on the integer lattice $n_{_1}E_{_{4}}^{(1)}$ $+n_{_2}E_{_{4}}^{(3)}$,
we obtain the exponents as
\begin{eqnarray}
&&\alpha_{_{1,3}}=2-{D\over2}\;,\;\alpha_{_{1,5}}={D\over2}-3\;,\;\alpha_{_{2,3}}=-1\;,\;
\alpha_{_{2,5}}=0\;,\;\alpha_{_{4,3}}={D\over2}-1,\;\alpha_{_{4,5}}=0\;,
\label{1TR2-8}
\end{eqnarray}
where the geometric representation is presented in Fig.\ref{fig6}(1)
with $(a,b)$ $=(1,5)$, $\{c,d\}$ $=\{2,3\}$.
In a similar way the exponents of the integer lattice
$n_{_1}E_{_{4}}^{(1)}$ $+n_{_2}E_{_{4}}^{(5)}$ are
\begin{eqnarray}
&&\alpha_{_{1,3}}=0\;,\;\alpha_{_{1,5}}=-1\;,\;\alpha_{_{2,3}}=1-{D\over2}\;,\;
\alpha_{_{2,5}}={D\over2}-2\;,\;\alpha_{_{4,3}}={D\over2}-1,\;\alpha_{_{4,5}}=0\;,
\label{1TR2-12}
\end{eqnarray}
whose geometric representation is plotted in Fig.\ref{fig6}(2)
with $a=2$, $b=3$, $(c,d)$ $=(1,5)$.
The exponents corresponding to the integer lattice $n_{_1}E_{_{4}}^{(3)}$
$+n_{_2}E_{_{4}}^{(5)}$ are
\begin{eqnarray}
&&\alpha_{_{1,3}}=0\;,\;\alpha_{_{1,5}}=-1\;,\;\alpha_{_{2,3}}=0\;,\;
\alpha_{_{2,5}}=-1\;,\;\alpha_{_{4,3}}=0,\;\alpha_{_{4,5}}={D\over2}-1\;,
\label{1TR2-16}
\end{eqnarray}
whose geometric representation is drawn in Fig.\ref{fig6}(3)
with $a=3$, $(b,c,d)$ $=(1,2,5)$.
Three linear independent hypergeometric functions are evidently written as
\begin{eqnarray}
&&\psi_{_{\{1,2,4,6\}}}^{(1)}(p_{_1}^2,p_{_2}^2,p_{_3}^2)\sim
(p_{_2}^2-p_{_1}^2)^{D/2-3}\sum\limits_{n_{_1},n_{_2}}^\infty
{\Gamma(1-{D\over2}+n_{_2})\over n_{_2}!}
\nonumber\\
&&\hspace{3.7cm}\times
\Big({p_{_3}^2-p_{_1}^2\over p_{_2}^2-p_{_1}^2}\Big)^{n_{_1}}
\Big({p_{_1}^2\over p_{_1}^2-p_{_2}^2}\Big)^{n_{_2}}
\nonumber\\
&&\hspace{3.2cm}\sim
{(p_{_2}^2)^{D/2-1}\over(p_{_2}^2-p_{_1}^2)(p_{_2}^2-p_{_3}^2)}
\;,\nonumber\\
&&\psi_{_{\{1,2,4,6\}}}^{(2)}(p_{_1}^2,p_{_2}^2,p_{_3}^2)\sim
{(p_{_3}^2-p_{_1}^2)^{D/2-2}\over p_{_2}^2-p_{_1}^2}\sum\limits_{n_{_1},n_{_2}}^\infty
{\Gamma(1-{D\over2}+n_{_2})\over n_{_2}!}
\nonumber\\
&&\hspace{3.7cm}\times
\Big({p_{_3}^2-p_{_1}^2\over p_{_2}^2-p_{_1}^2}\Big)^{n_{_1}}
\Big({p_{_1}^2\over p_{_1}^2-p_{_3}^2}\Big)^{n_{_2}}
\nonumber\\
&&\hspace{3.2cm}\sim
{(p_{_3}^2)^{D/2-1}\over(p_{_3}^2-p_{_1}^2)(p_{_3}^2-p_{_2}^2)}
\;,\nonumber\\
&&\psi_{_{\{1,2,4,6\}}}^{(3)}(p_{_1}^2,p_{_2}^2,p_{_3}^2)\sim
{(p_{_1}^2)^{D/2-1}\over(p_{_2}^2-p_{_1}^2)(p_{_3}^2-p_{_1}^2)}
\sum\limits_{n_{_1},n_{_2}}^\infty
\nonumber\\
&&\hspace{3.7cm}\times
\Big({p_{_1}^2\over p_{_1}^2-p_{_2}^2}\Big)^{n_{_1}}
\Big({p_{_1}^2\over p_{_1}^2-p_{_3}^2}\Big)^{n_{_2}}
\nonumber\\
&&\hspace{3.7cm}\times
{1\over\Gamma(1-n_{_1}-n_{_2})\Gamma({D\over2}+n_{_1}+n_{_2})}
\nonumber\\
&&\hspace{3.2cm}\sim
{(p_{_1}^2)^{D/2-1}\over(p_{_2}^2-p_{_1}^2)(p_{_3}^2-p_{_1}^2)}\;,
\label{1TR2-16a}
\end{eqnarray}
where the intersection of their convergent regions is a nonempty proper subset of the whole parameter space.
In other words, those hypergeometric functions constitute a fundamental solution
system of the GKZ-system in a proper subset of the whole parameter space.
Although the embedded subvariety of Grassmannian Eq.(\ref{1TR2-3}) is different from
the subvariety of Eq.(\ref{1TR2-91}), the constructed hypergeometric function satisfies
the same PDEs of combined coordinates in Eq.(\ref{1TR2-94a4}).

The exponents of the integer lattice $-n_{_1}E_{_{4}}^{(1)}$ $-n_{_2}E_{_{4}}^{(3)}$
are presented as
\begin{eqnarray}
&&\alpha_{_{1,3}}=0\;,\;\alpha_{_{1,5}}=-1\;,\;\alpha_{_{2,3}}=0\;,\;
\alpha_{_{2,5}}=-1\;,\;\alpha_{_{4,3}}=0,\;\alpha_{_{4,5}}={D\over2}-1\;,
\label{1TR2-11}
\end{eqnarray}
whose geometric representation is presented in Fig.\ref{fig6}(1)
with $(a,b)$ $=(1,3)$, $\{c,d\}$ $=\{2,5\}$. Similarly the exponents of
the integer lattice $-n_{_1}E_{_{4}}^{(1)}$ $-n_{_2}E_{_{4}}^{(5)}$ are
\begin{eqnarray}
&&\alpha_{_{1,3}}=-1\;,\;\alpha_{_{1,5}}=0\;,\;\alpha_{_{2,3}}=1\;,\;
\alpha_{_{2,5}}=-2\;,\;\alpha_{_{4,3}}=0,\;\alpha_{_{4,5}}={D\over2}-1\;,
\label{1TR2-15}
\end{eqnarray}
whose geometric representation is drawn in Fig.\ref{fig6}(2)
with $a=2$, $b=5$, $(c,d)$ $=(1,3)$. Corresponding to the integer lattice
$-n_{_1}E_{_{4}}^{(3)}$ $-n_{_2}E_{_{4}}^{(5)}$, the exponents are
\begin{eqnarray}
&&\alpha_{_{1,3}}=-1\;,\;\alpha_{_{1,5}}=0\;,\;\alpha_{_{2,3}}=-1\;,\;
\alpha_{_{2,5}}=0\;,\;\alpha_{_{4,3}}=2,\;\alpha_{_{4,5}}={D\over2}-3\;,
\label{1TR2-19}
\end{eqnarray}
whose geometric representation is plotted in Fig.\ref{fig6}(3)
with $a=5$, $(b,c,d)$ $=(1,2,3)$.
The hypergeometric functions are evidently presented as
\begin{eqnarray}
&&\psi_{_{\{1,2,4,6\}}}^{(4)}(p_{_1}^2,p_{_2}^2,p_{_3}^2)\sim
{(p_{_1}^2)^{D/2-1}\over(p_{_2}^2-p_{_1}^2)(p_{_3}^2-p_{_1}^2)}
\sum\limits_{n_{_1},n_{_2}}^\infty
{\Gamma(1-{D\over2}+n_{_2})\over n_{_2}!}
\nonumber\\
&&\hspace{3.7cm}\times
\Big({p_{_2}^2-p_{_1}^2\over p_{_3}^2-p_{_1}^2}\Big)^{n_{_1}}
\Big({p_{_1}^2-p_{_2}^2\over p_{_1}^2}\Big)^{n_{_2}}
\nonumber\\
&&\hspace{3.2cm}\sim
{(p_{_2}^2)^{D/2-1}\over(p_{_2}^2-p_{_1}^2)(p_{_2}^2-p_{_3}^2)}
\;,\nonumber\\
&&\psi_{_{\{1,2,4,6\}}}^{(5)}(p_{_1}^2,p_{_2}^2,p_{_3}^2)\sim
{(p_{_1}^2)^{D/2-1}\over(p_{_3}^2-p_{_1}^2)^2}\sum\limits_{n_{_1},n_{_2}}^\infty
{\Gamma(1-{D\over2}+n_{_2})\over n_{_2}!}
\nonumber\\
&&\hspace{3.7cm}\times
\Big({p_{_2}^2-p_{_1}^2\over p_{_3}^2-p_{_1}^2}\Big)^{n_{_1}}
\Big({p_{_1}^2-p_{_3}^2\over p_{_1}^2}\Big)^{n_{_2}}
\nonumber\\
&&\hspace{3.2cm}\sim
{(p_{_3}^2)^{D/2-1}\over(p_{_3}^2-p_{_1}^2)(p_{_3}^2-p_{_2}^2)}
\;,\nonumber\\
&&\psi_{_{\{1,2,4,6\}}}^{(6)}(p_{_1}^2,p_{_2}^2,p_{_3}^2)\sim
(p_{_1}^2)^{D/2-3}\sum\limits_{n_{_1},n_{_2}}^\infty
{\Gamma(3-{D\over2}+n_{_1}+n_{_2})\over\Gamma(3+n_{_1}+n_{_2})}
\nonumber\\
&&\hspace{3.7cm}\times
\Big({p_{_1}^2-p_{_2}^2\over p_{_1}^2}\Big)^{n_{_1}}
\Big({p_{_1}^2-p_{_3}^2\over p_{_1}^2}\Big)^{n_{_2}}\;,
\label{1TR2-19a}
\end{eqnarray}
where the intersection of their convergent regions is a nonempty proper subset of the whole parameter space.
The hypergeometric functions constitute a fundamental solution
system of the GKZ-system of PDEs in a proper subset of the whole parameter space.

As the integer lattice is chosen as $n_{_1}E_{_{4}}^{(1)}$ $-n_{_2}E_{_{4}}^{(3)}$,
the exponents are
\begin{eqnarray}
&&\alpha_{_{1,3}}=1\;,\;\alpha_{_{1,5}}=-2\;,\;\alpha_{_{2,3}}=-1\;,\;
\alpha_{_{2,5}}=0\;,\;\alpha_{_{4,3}}=0,\;\alpha_{_{4,5}}={D\over2}-1\;,
\label{1TR2-9}
\end{eqnarray}
whose geometric representation is presented in Fig.\ref{fig6}(2)
with $a=1$, $b=5$, $\{c,d\}$ $=\{2,3\}$.
The exponents of the integer lattice $n_{_1}E_{_{4}}^{(1)}$ $-n_{_2}E_{_{4}}^{(5)}$
are given by
\begin{eqnarray}
&&\alpha_{_{1,3}}=0\;,\;\alpha_{_{1,5}}=-1\;,\;\alpha_{_{2,3}}=0\;,\;
\alpha_{_{2,5}}=-1\;,\;\alpha_{_{4,3}}=0,\;\alpha_{_{4,5}}={D\over2}-1\;,
\label{1TR2-13}
\end{eqnarray}
whose geometric representation is plotted in Fig.\ref{fig6}(1)
with $(a,b)$ $=(2,3)$, $\{c,d\}$ $=\{1,5\}$. The exponents of the integer lattice
$n_{_1}E_{_{4}}^{(3)}$ $-n_{_2}E_{_{4}}^{(5)}$ are
\begin{eqnarray}
&&\alpha_{_{1,3}}=0\;,\;\alpha_{_{1,5}}=-1\;,\;\alpha_{_{2,3}}=-1\;,\;
\alpha_{_{2,5}}=0\;,\;\alpha_{_{4,3}}=1,\;\alpha_{_{4,5}}={D\over2}-2\;,
\label{1TR2-17}
\end{eqnarray}
whose geometric representation is presented in Fig.\ref{fig6}(4)
with $\{(a,b)$, $(c,d)\}$ $=\{(1,5)$, $(2,3)\}$.
Three hypergeometric functions are obtained as
\begin{eqnarray}
&&\psi_{_{\{1,2,4,6\}}}^{(7)}(p_{_1}^2,p_{_2}^2,p_{_3}^2)\sim
{(p_{_1}^2)^{D/2-1}\over(p_{_2}^2-p_{_1}^2)^2}\sum\limits_{n_{_1},n_{_2}}^\infty
{\Gamma(1-{D\over2}+n_{_2})\over n_{_2}!}
\nonumber\\
&&\hspace{3.7cm}\times
\Big({p_{_3}^2-p_{_1}^2\over p_{_2}^2-p_{_1}^2}\Big)^{n_{_1}}
\Big({p_{_1}^2-p_{_2}^2\over p_{_1}^2}\Big)^{n_{_2}}
\nonumber\\
&&\hspace{3.2cm}\sim
{(p_{_2}^2)^{D/2-1}\over(p_{_2}^2-p_{_1}^2)(p_{_2}^2-p_{_3}^2)}
\;,\nonumber\\
&&\psi_{_{\{1,2,4,6\}}}^{(8)}(p_{_1}^2,p_{_2}^2,p_{_3}^2)\sim
{(p_{_1}^2)^{D/2-1}\over(p_{_1}^2-p_{_2}^2)(p_{_1}^2-p_{_3}^2)}\sum\limits_{n_{_1},n_{_2}}^\infty
{\Gamma(1-{D\over2}+n_{_2})\over n_{_2}!}
\nonumber\\
&&\hspace{3.7cm}\times
\Big({p_{_3}^2-p_{_1}^2\over p_{_2}^2-p_{_1}^2}\Big)^{n_{_1}}
\Big({p_{_1}^2-p_{_3}^2\over p_{_1}^2}\Big)^{n_{_2}}
\nonumber\\
&&\hspace{3.2cm}\sim
{(p_{_3}^2)^{D/2-1}\over(p_{_3}^2-p_{_1}^2)(p_{_3}^2-p_{_2}^2)}
\;,\nonumber\\
&&\psi_{_{\{1,2,4,6\}}}^{(9)}(p_{_1}^2,p_{_2}^2,p_{_3}^2)\sim
{(p_{_1}^2)^{D/2}\over(p_{_1}^2-p_{_2}^2)^2(p_{_3}^2-p_{_2}^2)}
\sum\limits_{n=0}^\infty
{\Gamma(1+n)\over\Gamma({D\over2}+1+n)}
\Big({p_{_1}^2\over p_{_1}^2-p_{_2}^2}\Big)^{n}\;,
\label{1TR2-17a}
\end{eqnarray}
where the intersection of their convergent regions is a nonempty proper subset of the whole parameter space.
The hypergeometric function of the exponents of Eq.(\ref{1TR2-17}) is
\begin{eqnarray}
&&\psi_{_{\{1,2,4,6\}}}^{(9)\prime}(p_{_1}^2,p_{_2}^2,p_{_3}^2)\sim
{(p_{_1}^2)^{D/2-2}\over p_{_1}^2-p_{_2}^2}
\sum\limits_{n_{_1},n_{_2}}^\infty
\Big({p_{_1}^2\over p_{_2}^2-p_{_1}^2}\Big)^{n_{_1}}
\Big({p_{_3}^2-p_{_1}^2\over p_{_1}^2}\Big)^{n_{_2}}
\nonumber\\
&&\hspace{3.7cm}\times
{1\over\Gamma(2-n_{_1}+n_{_2})\Gamma({D\over2}-1+n_{_1}-n_{_2})}
\nonumber\\
&&\hspace{3.2cm}\sim
-\psi_{_{\{1,2,4,6\}}}^{(8)}(p_{_1}^2,p_{_2}^2,p_{_3}^2)
\nonumber\\
&&\hspace{3.7cm}
+\lim\limits_{\epsilon\rightarrow0}{\sin\pi\epsilon\over\sin\pi({D\over2}+1)}
\psi_{_{\{1,2,4,6\}}}^{(9)}(p_{_1}^2,p_{_2}^2,p_{_3}^2)\;.
\label{1TR2-17b}
\end{eqnarray}
Furthermore, the sets of the hypergeometric functions
$\{\psi_{_{\{1,2,4,6\}}}^{(1)}$, $\psi_{_{\{1,2,4,6\}}}^{(8)}$, $\psi_{_{\{1,2,4,6\}}}^{(9)}\}$
and $\{\psi_{_{\{1,2,4,6\}}}^{(6)}$, $\psi_{_{\{1,2,4,6\}}}^{(7)}$, $\psi_{_{\{1,2,4,6\}}}^{(8)}\}$
constitute the fundamental solution systems of the nonempty proper subsets
of the whole parameter space, respectively.

The exponents of the integer lattice $-n_{_1}E_{_{4}}^{(1)}$ $+n_{_2}E_{_{4}}^{(3)}$ are
\begin{eqnarray}
&&\alpha_{_{1,3}}=1-{D\over2}\;,\;\alpha_{_{1,5}}={D\over2}-2\;,\;\alpha_{_{2,3}}=0\;,\;
\alpha_{_{2,5}}=-1\;,\;\alpha_{_{4,3}}={D\over2}-1,\;\alpha_{_{4,5}}=0\;,
\label{1TR2-10}
\end{eqnarray}
whose geometric representation is presented in Fig.\ref{fig6}(2)
with $a=1$, $b=3$, $\{c,d\}$ $=\{2,5\}$. Corresponding to the integer lattice
$-n_{_1}E_{_{4}}^{(1)}$ $+n_{_2}E_{_{4}}^{(5)}$, the exponents are
\begin{eqnarray}
&&\alpha_{_{1,3}}=-1\;,\;\alpha_{_{1,5}}=0\;,\;\alpha_{_{2,3}}=2-{D\over2}\;,\;
\alpha_{_{2,5}}={D\over2}-3\;,\;\alpha_{_{4,3}}={D\over2}-1,\;\alpha_{_{4,5}}=0\;,
\label{1TR2-14}
\end{eqnarray}
whose geometric representation is depicted in Fig.\ref{fig6}(1)
with $(a,b)$ $=(2,5)$, $\{c,d\}$ $=\{1,3\}$. Finally, the exponents
of the integer lattice $-n_{_1}E_{_{4}}^{(3)}$ $+n_{_2}E_{_{4}}^{(5)}$ are
\begin{eqnarray}
&&\alpha_{_{1,3}}=-1\;,\;\alpha_{_{1,5}}=0\;,\;\alpha_{_{2,3}}=0\;,\;
\alpha_{_{2,5}}=-1\;,\;\alpha_{_{4,3}}=1,\;\alpha_{_{4,5}}={D\over2}-2\;,
\label{1TR2-18}
\end{eqnarray}
whose geometric representation is depicted in Fig.\ref{fig6}(4)
with $\{(a,b)$, $(c,d)\}$ $=\{(1,3)$, $(2,5)\}$.
Three hypergeometric functions are evidently written as
\begin{eqnarray}
&&\psi_{_{\{1,2,4,6\}}}^{(10)}(p_{_1}^2,p_{_2}^2,p_{_3}^2)\sim
{(p_{_2}^2-p_{_1}^2)^{D/2-2}\over p_{_3}^2-p_{_1}^2}\sum\limits_{n_{_1},n_{_2}}^\infty
{\Gamma(1-{D\over2}+n_{_2})\over n_{_2}!}
\nonumber\\
&&\hspace{3.7cm}\times
\Big({p_{_2}^2-p_{_1}^2\over p_{_3}^2-p_{_1}^2}\Big)^{n_{_1}}
\Big({p_{_1}^2-p_{_2}^2\over p_{_1}^2}\Big)^{n_{_2}}
\nonumber\\
&&\hspace{3.2cm}\sim
{(p_{_2}^2)^{D/2-1}(p_{_2}^2-p_{_1}^2)^{D/2-2}\over(p_{_1}^2)^{D/2-1}(p_{_2}^2-p_{_3}^2)}
\;,\nonumber\\
&&\psi_{_{\{1,2,4,6\}}}^{(11)}(p_{_1}^2,p_{_2}^2,p_{_3}^2)\sim
(p_{_3}^2-p_{_1}^2)^{D/2-3}\sum\limits_{n_{_1},n_{_2}}^\infty
{\Gamma(1-{D\over2}+n_{_2})\over n_{_2}!}
\nonumber\\
&&\hspace{3.7cm}\times
\Big({p_{_2}^2-p_{_1}^2\over p_{_3}^2-p_{_1}^2}\Big)^{n_{_1}}
\Big({p_{_1}^2\over p_{_1}^2-p_{_3}^2}\Big)^{n_{_2}}
\nonumber\\
&&\hspace{3.2cm}\sim
{(p_{_3}^2)^{D/2-1}\over(p_{_3}^2-p_{_1}^2)(p_{_3}^2-p_{_2}^2)}
\;,\nonumber\\
&&\psi_{_{\{1,2,4,6\}}}^{(12)}(p_{_1}^2,p_{_2}^2,p_{_3}^2)\sim
{(p_{_1}^2)^{D/2-2}\over p_{_1}^2-p_{_2}^2}
\sum\limits_{n_{_1},n_{_2}}^\infty
\Big({p_{_2}^2-p_{_1}^2\over p_{_1}^2}\Big)^{n_{_1}}
\Big({p_{_1}^2\over p_{_3}^2-p_{_1}^2}\Big)^{n_{_2}}
\nonumber\\
&&\hspace{3.7cm}\times
{1\over\Gamma(2+n_{_1}-n_{_2})\Gamma({D\over2}-1-n_{_1}+n_{_2})}
\nonumber\\
&&\hspace{3.2cm}\sim
{(p_{_1}^2)^{D/2-1}\over(p_{_1}^2-p_{_2}^2)^2}
\sum\limits_{n_{_1}=0}^\infty\sum\limits_{n_{_2}=0}^\infty
{\Gamma(1-{D\over2}+n_{_1})\over n_{_1}!}
\nonumber\\
&&\hspace{3.7cm}\times
\Big({p_{_1}^2-p_{_2}^2\over p_{_1}^2}\Big)^{n_{_1}}
\Big({p_{_1}^2-p_{_2}^2\over p_{_1}^2-p_{_3}^2}\Big)^{n_{_2}}
\nonumber\\
&&\hspace{3.2cm}\sim
{(p_{_1}^2-p_{_3}^2)(p_{_2}^2)^{D/2-1}\over
(p_{_2}^2-p_{_1}^2)^2(p_{_2}^2-p_{_3}^2)}\;,
\label{1TR2-20}
\end{eqnarray}
where the intersection of their convergent regions is a nonempty proper subset of the whole parameter space.
In addition, the sets of the hypergeometric functions
$\{\psi_{_{\{1,2,4,6\}}}^{(3)}$, $\psi_{_{\{1,2,4,6\}}}^{(10)}$, $\psi_{_{\{1,2,4,6\}}}^{(11)}\}$
and $\{\psi_{_{\{1,2,4,6\}}}^{(4)}$, $\psi_{_{\{1,2,4,6\}}}^{(11)}$, $\psi_{_{\{1,2,4,6\}}}^{(12)}\}$
constitute the fundamental solution systems of the nonempty proper subsets
of the whole parameter space, respectively.

Because $\det(A_{_{\{1,3,4,6\}}}^{(1T)\prime})=1$ where the matrix $A_{_{\{1,3,4,6\}}}^{(1T)\prime}$ denotes
the submatrix of $A^{(1T)\prime}$ composed of the first, third, fourth and sixth column vectors, one derives
\begin{eqnarray}
&&\Big(A_{_{\{1,3,4,6\}}}^{(1T)\prime}\Big)^{-1}\cdot A^{(1T)\prime}=\left(\begin{array}{ccccccc}
\;\;1\;\;&\;\;-1\;\;&\;\;0\;\;&\;\;0\;\;&\;\;p_{_2}^2-p_{_3}^2\;\;&\;\;0\;\;&\;\;0\;\;\\
\;\;0\;\;&\;\;-1\;\;&\;\;0\;\;&\;\;0\;\;&\;\;p_{_1}^2-p_{_3}^2\;\;&\;\;0\;\;&\;\;0\;\;\\
\;\;0\;\;&\;\;0\;\;&\;\;0\;\;&\;\;1\;\;&\;\;0\;\;&\;\;0\;\;&\;\;0\;\;\\
\;\;0\;\;&\;\;1\;\;&\;\;0\;\;&\;\;0\;\;&\;\;p_{_3}^2\;\;&\;\;1\;\;&\;\;1\;\;
\end{array}\right)\;.
\label{1TR2-22}
\end{eqnarray}
Obviously the matrix of exponents is written as
\begin{eqnarray}
&&\left(\begin{array}{ccccccc}
\;\;0\;\;&\;\;\alpha_{_{1,2}}\;\;&\;\;0\;\;&\;\;0\;\;&\;\alpha_{_{1,5}}\;\;&\;\;0\;\;&\;\;0\;\;\\
\;\;0\;\;&\;\;\alpha_{_{2,2}}\;\;&\;0\;\;&\;\;0\;\;&\;\;\alpha_{_{2,5}}\;\;&\;0\;\;&\;\;0\;\;\\
\;\;0\;\;&\;\;0\;\;&\;0\;\;&\;\;3-D\;\;&\;\;0\;\;&\;0\;\;&\;\;D-4\;\;\\
\;\;0\;\;&\;\;\alpha_{_{4,2}}\;\;&\;0\;\;&\;\;0\;\;&\;\;\alpha_{_{4,5}}\;\;&\;{D\over2}-3\;\;&\;\;3-D\;\;
\end{array}\right)\;,
\label{1TR2-23}
\end{eqnarray}
where the matrix elements satisfy similar relations presented in Eq.(\ref{1TR2-7})
through the replacement $\alpha_{_{i3}}\rightarrow\alpha_{_{i2}}$, $i=1,2,4$.
Correspondingly there are 12 choices on the matrix of integer lattice
whose submatrix composed of the second and
fifth columns is formulated as $\pm n_{_1}E_{_{4}}^{(i)}$ $\pm n_{_2}E_{_{4}}^{(j)}$,
with $n_{_{1,2}}\ge0$, $(i,j)$ $\in\{(1,3)$, $(1,5)$, $(3,5)\}$,
and other elements are all zero.
Basing on the matrices of integer lattice, we have 12 choices on the
matrix of exponents which are obtained through replacements
$\alpha_{_{i3}}\rightarrow\alpha_{_{i2}}$ of the functions $\psi_{_{\{1,2,4,6\}}}^{(i)}$.
The geometric representations are also acquired through the permutation
$\widehat{(23)}$.
The corresponding hypergeometric functions are finally written as
\begin{eqnarray}
&&\psi_{_{\{1,3,4,6\}}}^{(i)}(p_{_1}^2,p_{_2}^2,p_{_3}^2)=
\psi_{_{\{1,2,4,6\}}}^{(i)}(p_{_3}^2,p_{_2}^2,p_{_1}^2),\;\;(i=1,\cdots,12).
\label{1TR2-24}
\end{eqnarray}

Because $\det(A_{_{\{2,3,4,6\}}}^{(1T)\prime})=-1$, we have
\begin{eqnarray}
&&\Big(A_{_{\{2,3,4,6\}}}^{(1T)\prime}\Big)^{-1}\cdot A^{(1T)\prime}=\left(\begin{array}{ccccccc}
\;\;-1\;\;&\;\;1\;\;&\;\;0\;\;&\;\;0\;\;&\;\;p_{_3}^2-p_{_2}^2\;\;&\;\;0\;\;&\;\;0\;\;\\
\;\;-1\;\;&\;\;0\;\;&\;\;0\;\;&\;\;0\;\;&\;\;p_{_1}^2-p_{_2}^2\;\;&\;\;0\;\;&\;\;0\;\;\\
\;\;0\;\;&\;\;0\;\;&\;\;0\;\;&\;\;1\;\;&\;\;0\;\;&\;\;0\;\;&\;\;0\;\;\\
\;\;1\;\;&\;\;0\;\;&\;\;0\;\;&\;\;0\;\;&\;\;p_{_2}^2\;\;&\;\;1\;\;&\;\;1\;\;
\end{array}\right)\;.
\label{1TR2-25}
\end{eqnarray}
Obviously the matrix of exponents is written as
\begin{eqnarray}
&&\left(\begin{array}{ccccccc}
\;\;\alpha_{_{1,1}}\;\;&\;\;0\;\;&\;\;0\;\;&\;\;0\;\;&\;\alpha_{_{1,5}}\;\;&\;\;0\;\;&\;\;0\;\;\\
\;\;\alpha_{_{2,1}}\;\;&\;\;0\;\;&\;0\;\;&\;\;0\;\;&\;\;\alpha_{_{2,5}}\;\;&\;0\;\;&\;\;0\;\;\\
\;\;0\;\;&\;\;0\;\;&\;0\;\;&\;\;3-D\;\;&\;\;0\;\;&\;0\;\;&\;\;D-4\;\;\\
\;\;\alpha_{_{4,1}}\;\;&\;\;0\;\;&\;0\;\;&\;\;0\;\;&\;\;\alpha_{_{4,5}}\;\;&\;{D\over2}-3\;\;&\;\;3-D\;\;
\end{array}\right)\;,
\label{1TR2-26}
\end{eqnarray}
where the matrix elements satisfy similar relations presented in Eq.(\ref{1TR2-7})
through the replacement $\alpha_{_{i3}}\rightarrow\alpha_{_{i1}}$, $i=1,2,4$.
Corresponding to the matrix of splitting local coordinates in Eq.(\ref{1TR2-4}),
we have 12 choices on the matrix of integer lattice whose submatrix composed of the first and
fifth columns is formulated as $\pm n_{_1}E_{_{4}}^{(i)}$ $\pm n_{_2}E_{_{4}}^{(j)}$,
where $n_{_{1,2}}\ge0$, $(i,j)$ $\in\{(1,3)$, $(1,5)$, $(3,5)\}$,
and other elements are all zero.
Basing on the matrices of integer lattice, we have 12 choices on the
matrix of exponents which are obtained through replacements
$\alpha_{_{i3}}\rightarrow\alpha_{_{i1}}$ of that of
the function $\psi_{_{\{1,2,4,6\}}}^{(i)}$.
The geometric representations are also acquired through the
permutation $\widehat{(13)}$ accordingly.
The hypergeometric functions are
\begin{eqnarray}
&&\psi_{_{\{2,3,4,6\}}}^{(i)}(p_{_1}^2,p_{_2}^2,p_{_3}^2)=
\psi_{_{\{1,2,4,6\}}}^{(i)}(p_{_2}^2,p_{_3}^2,p_{_1}^2),\;\;(i=1,\cdots,12).
\label{1TR2-27}
\end{eqnarray}

Obviously $\det(A_{_{\{1,2,6,7\}}}^{(1T)\prime})=1$ where the matrix $A_{_{\{1,2,6,7\}}}^{(1T)\prime}$ denotes
the submatrix of $A^{(1T)\prime}$ composed of the first, second, sixth and seventh columns, one has
\begin{eqnarray}
&&\Big(A_{_{\{1,2,6,7\}}}^{(1T)\prime}\Big)^{-1}\cdot A^{(1T)\prime}=\left(\begin{array}{ccccccc}
\;\;1\;\;&\;\;0\;\;&\;\;-1\;\;&\;\;0\;\;&\;\;p_{_2}^2-p_{_1}^2\;\;&\;\;0\;\;&\;\;0\;\;\\
\;\;0\;\;&\;\;1\;\;&\;\;-1\;\;&\;\;0\;\;&\;\;p_{_3}^2-p_{_1}^2\;\;&\;\;0\;\;&\;\;0\;\;\\
\;\;0\;\;&\;\;0\;\;&\;\;1\;\;&\;\;-1\;\;&\;\;p_{_1}^2\;\;&\;\;1\;\;&\;\;0\;\;\\
\;\;0\;\;&\;\;0\;\;&\;\;0\;\;&\;\;1\;\;&\;\;0\;\;&\;\;0\;\;&\;\;1\;\;
\end{array}\right)\;.
\label{1TR2-28}
\end{eqnarray}
Obviously the matrix of exponents is written as
\begin{eqnarray}
&&\left(\begin{array}{ccccccc}
\;\;0\;\;&\;\;0\;\;&\;\;\alpha_{_{1,3}}\;\;&\;\;0\;\;&\;\alpha_{_{1,5}}\;\;&\;\;0\;\;&\;\;0\;\;\\
\;\;0\;\;&\;\;0\;\;&\;\alpha_{_{2,3}}\;\;&\;\;0\;\;&\;\;\alpha_{_{2,5}}\;\;&\;0\;\;&\;\;0\;\;\\
\;\;0\;\;&\;\;0\;\;&\;\alpha_{_{3,3}}\;\;&\;\;3-D\;\;&\;\;\alpha_{_{3,5}}\;\;&\;{D\over2}-3\;\;&\;\;0\;\;\\
\;\;0\;\;&\;\;0\;\;&\;0\;\;&\;\;0\;\;&\;\;0\;\;&\;0\;\;&\;\;-1\;\;
\end{array}\right)\;,
\label{1TR2-29}
\end{eqnarray}
where the matrix elements satisfy the relations
\begin{eqnarray}
&&\alpha_{_{1,3}}+\alpha_{_{1,5}}=-1\;,\;\alpha_{_{2,3}}+\alpha_{_{2,5}}=-1\;,\;
\alpha_{_{3,3}}+\alpha_{_{3,5}}={D\over2}-1\;,
\nonumber\\
&&\alpha_{_{1,3}}+\alpha_{_{2,3}}+\alpha_{_{3,3}}=0\;,
\alpha_{_{1,5}}+\alpha_{_{2,5}}+\alpha_{_{3,5}}={D\over2}-3\;.
\label{1TR2-30}
\end{eqnarray}
Certainly there are 12 choices on the matrix of integer lattice whose submatrix composed of the third and
fifth columns is formulated as $\pm n_{_1}E_{_{4}}^{(i)}$ $\pm n_{_2}E_{_{4}}^{(j)}$,
where $n_{_{1,2}}\ge0$, $(i,j)$ $\in\{(1,2)$, $(1,4)$, $(2,4)\}$,
and other elements are all zero.
Basing on the matrices of integer lattice, we have 12 choices on the
matrix of exponents. The exponents of the integer lattice
$n_{_1}E_{_{4}}^{(1)}$ $+n_{_2}E_{_{4}}^{(2)}$ are written as
\begin{eqnarray}
&&\alpha_{_{1,3}}=2-{D\over2}\;,\;\alpha_{_{1,5}}={D\over2}-3\;,\;\alpha_{_{2,3}}=-1\;,\;
\alpha_{_{2,5}}=0\;,\;\alpha_{_{3,3}}={D\over2}-1,\;\alpha_{_{3,5}}=0\;,
\label{1TR2-32}
\end{eqnarray}
whose geometric representation is plotted in Fig.\ref{fig6}(1)
with $(a,b)$ $=(1,5)$, $\{c,d\}$ $=\{2,3\}$.
The exponents of the integer lattice $n_{_1}E_{_{4}}^{(1)}$ $+n_{_2}E_{_{4}}^{(4)}$ are
\begin{eqnarray}
&&\alpha_{_{1,3}}=0\;,\;\alpha_{_{1,5}}=-1\;,\;\alpha_{_{2,3}}=1-{D\over2}\;,\;
\alpha_{_{2,5}}={D\over2}-2\;,\;\alpha_{_{3,3}}={D\over2}-1,\;\alpha_{_{3,5}}=0\;,
\label{1TR2-36}
\end{eqnarray}
whose geometric representation is presented in Fig.\ref{fig6}(2)
with $a=2$, $b=3$, $(c,d)$ $=(1,5)$.
In a similar way, the exponents of the integer lattice
$n_{_1}E_{_{4}}^{(2)}$ $+n_{_2}E_{_{4}}^{(4)}$ are
\begin{eqnarray}
&&\alpha_{_{1,3}}=0\;,\;\alpha_{_{1,5}}=-1\;,\;\alpha_{_{2,3}}=0\;,\;
\alpha_{_{2,5}}=-1\;,\;\alpha_{_{3,3}}=0,\;\alpha_{_{3,5}}={D\over2}-1\;,
\label{1TR2-40}
\end{eqnarray}
whose geometric representation is depicted in Fig.\ref{fig6}(3)
with $a=3$, $(b,c,d)$ $=(1,2,5)$.

The exponents of the integer lattice $-n_{_1}E_{_{4}}^{(1)}$ $-n_{_2}E_{_{4}}^{(2)}$
are presented as
\begin{eqnarray}
&&\alpha_{_{1,3}}=0\;,\;\alpha_{_{1,5}}=-1\;,\;\alpha_{_{2,3}}=0\;,\;
\alpha_{_{2,5}}=-1\;,\;\alpha_{_{3,3}}=0,\;\alpha_{_{3,5}}={D\over2}-1\;,
\label{1TR2-35}
\end{eqnarray}
whose geometric representation is depicted in Fig.\ref{fig6}(1)
with $(a,b)$ $=(1,3)$, $\{c,d\}$ $=\{2,5\}$. Corresponding to the integer lattice
$-n_{_1}E_{_{4}}^{(1)}$ $-n_{_2}E_{_{4}}^{(4)}$, the exponents are given by
\begin{eqnarray}
&&\alpha_{_{1,3}}=-1\;,\;\alpha_{_{1,5}}=0\;,\;\alpha_{_{2,3}}=1\;,\;
\alpha_{_{2,5}}=-2\;,\;\alpha_{_{3,3}}=0,\;\alpha_{_{3,5}}={D\over2}-1\;,
\label{1TR2-39}
\end{eqnarray}
whose geometric representation is presented in Fig.\ref{fig6}(2)
with $a=2$, $b=5$, $(c,d)$ $=(1,3)$. Similarly the exponents of the integer
lattice $-n_{_1}E_{_{4}}^{(2)}$ $-n_{_2}E_{_{4}}^{(4)}$ are
\begin{eqnarray}
&&\alpha_{_{1,3}}=-1\;,\;\alpha_{_{1,5}}=0\;,\;\alpha_{_{2,3}}=-1\;,\;
\alpha_{_{2,5}}=0\;,\;\alpha_{_{3,3}}=2,\;\alpha_{_{3,5}}={D\over2}-3\;,
\label{1TR2-43}
\end{eqnarray}
whose geometric representation is plotted in Fig.\ref{fig6}(3)
with $a=5$, $(b,c,d)$ $=(1,2,3)$.

Corresponding to the integer lattice
$n_{_1}E_{_{4}}^{(1)}$ $-n_{_2}E_{_{4}}^{(2)}$, the exponents are
\begin{eqnarray}
&&\alpha_{_{1,3}}=1\;,\;\alpha_{_{1,5}}=-2\;,\;\alpha_{_{2,3}}=-1\;,\;
\alpha_{_{2,5}}=0\;,\;\alpha_{_{3,3}}=0,\;\alpha_{_{3,5}}={D\over2}-1\;,
\label{1TR2-33}
\end{eqnarray}
whose geometric representation is depicted in Fig.\ref{fig6}(2)
with $a=1$, $b=5$, $\{c,d\}$ $=\{2,3\}$.
The exponents of the integer lattice $n_{_1}E_{_{4}}^{(1)}$ $-n_{_2}E_{_{4}}^{(4)}$ are
\begin{eqnarray}
&&\alpha_{_{1,3}}=0\;,\;\alpha_{_{1,5}}=-1\;,\;\alpha_{_{2,3}}=0\;,\;
\alpha_{_{2,5}}=-1\;,\;\alpha_{_{3,3}}=0,\;\alpha_{_{3,5}}={D\over2}-1\;,
\label{1TR2-37}
\end{eqnarray}
whose geometric representation is presented in Fig.\ref{fig6}(1)
with $(a,b)$ $=(2,3)$, $\{c,d\}$ $=\{1,5\}$. Similarly the exponents of
the integer lattice $n_{_1}E_{_{4}}^{(2)}$ $-n_{_2}E_{_{4}}^{(4)}$
are given by
\begin{eqnarray}
&&\alpha_{_{1,3}}=0\;,\;\alpha_{_{1,5}}=-1\;,\;\alpha_{_{2,3}}=-1\;,\;
\alpha_{_{2,5}}=0\;,\;\alpha_{_{3,3}}=1,\;\alpha_{_{3,5}}={D\over2}-2\;,
\label{1TR2-41}
\end{eqnarray}
whose geometric representation is presented in Fig.\ref{fig6}(4)
with $\{(a,b)$, $(c,d)\}$ $=\{(1,5)$, $(2,3)\}$.
Meanwhile the exponents of the integer lattice
$-n_{_1}E_{_{4}}^{(1)}$ $+n_{_2}E_{_{4}}^{(2)}$ are
\begin{eqnarray}
&&\alpha_{_{1,3}}=1-{D\over2}\;,\;\alpha_{_{1,5}}={D\over2}-2\;,\;\alpha_{_{2,3}}=0\;,\;
\alpha_{_{2,5}}=-1\;,\;\alpha_{_{3,3}}={D\over2}-1,\;\alpha_{_{3,5}}=0\;,
\label{1TR2-34}
\end{eqnarray}
whose geometric representation is presented in Fig.\ref{fig6}(2)
with $a=1$, $b=3$, $\{c,d\}$ $=\{2,5\}$.
The matrix of exponents of the integer lattice
$-n_{_1}E_{_{4}}^{(1)}$ $+n_{_2}E_{_{4}}^{(4)}$ is formulated as
\begin{eqnarray}
&&\alpha_{_{1,3}}=-1\;,\;\alpha_{_{1,5}}=0\;,\;\alpha_{_{2,3}}=2-{D\over2}\;,\;
\alpha_{_{2,5}}={D\over2}-3\;,\;\alpha_{_{3,3}}={D\over2}-1,\;\alpha_{_{3,5}}=0\;,
\label{1TR2-38}
\end{eqnarray}
whose geometric representation is drawn in Fig.\ref{fig6}(1)
with $(a,b)$ $=(2,5)$, $\{c,d\}$ $=\{1,3\}$. Finally, the exponents of
the integer lattice $-n_{_1}E_{_{4}}^{(2)}$ $+n_{_2}E_{_{4}}^{(4)}$ are
\begin{eqnarray}
&&\alpha_{_{1,3}}=-1\;,\;\alpha_{_{1,5}}=0\;,\;\alpha_{_{2,3}}=0\;,\;
\alpha_{_{2,5}}=-1\;,\;\alpha_{_{3,3}}=1,\;\alpha_{_{3,5}}={D\over2}-2\;,
\label{1TR2-42}
\end{eqnarray}
whose geometric representation is depicted in Fig.\ref{fig6}(4)
with $\{(a,b)$, $(c,d)\}$ $=\{(1,3)$, $(2,5)\}$.
The corresponding hypergeometric functions are evidently written as
\begin{eqnarray}
&&\psi_{_{\{1,2,6,7\}}}^{(i)}(p_{_1}^2,p_{_2}^2,p_{_3}^2)
=\psi_{_{\{1,2,4,6\}}}^{(i)}(p_{_1}^2,p_{_2}^2,p_{_3}^2),\;\;(i=1,\cdots,12).
\label{1TR2-44}
\end{eqnarray}

Furthermore $\det(A_{_{\{1,3,6,7\}}}^{(1T)\prime})=-1$, $\det(A_{_{\{2,3,6,7\}}}^{(1T)\prime})=1$,
the calculation in detail implies
\begin{eqnarray}
&&\psi_{_{\{1,3,6,7\}}}^{(i)}(p_{_1}^2,p_{_2}^2,p_{_3}^2)
=\psi_{_{\{1,2,4,6\}}}^{(i)}(p_{_3}^2,p_{_2}^2,p_{_1}^2),
\nonumber\\
&&\psi_{_{\{2,3,6,7\}}}^{(i)}(p_{_1}^2,p_{_2}^2,p_{_3}^2)
=\psi_{_{\{1,2,4,6\}}}^{(i)}(p_{_2}^2,p_{_3}^2,p_{_1}^2).
\label{1TR2-45}
\end{eqnarray}

Because $\det(A_{_{1,4,5,6}}^{(1T)\prime})=p_{_3}^2-p_{_1}^2$,
where the matrix $A_{_{\{1,4,5,6\}}}^{(1T)\prime}$ denotes the submatrix of $A^{(1T)\prime}$ composed
of the first, fourth, fifth and sixth columns, we get
\begin{eqnarray}
&&\Big(A_{_{\{1,4,5,6\}}}^{(1T)\prime}\Big)^{-1}\cdot A^{(1T)\prime}=\left(\begin{array}{ccccccc}
\;\;1\;\;&\;\;-{p_{_2}^2-p_{_1}^2\over p_{_3}^2-p_{_1}^2}\;\;
&\;\;-{p_{_3}^2-p_{_2}^2\over p_{_3}^2-p_{_1}^2}\;\;&\;\;0\;\;&\;\;0\;\;&\;\;0\;\;&\;\;0\;\;\\
\;\;0\;\;&\;\;0\;\;&\;\;0\;\;&\;\;1\;\;&\;\;0\;\;&\;\;0\;\;&\;\;1\;\;\\
\;\;0\;\;&\;\;{1\over p_{_3}^2-p_{_1}^2}\;\;&\;\;-{1\over p_{_3}^2-p_{_1}^2}\;\;&
\;\;0\;\;&\;\;1\;\;&\;\;0\;\;&\;\;0\;\;\\
\;\;0\;\;&\;\;-{p_{_1}^2\over p_{_3}^2-p_{_1}^2}\;\;&
\;\;{p_{_3}^2\over p_{_3}^2-p_{_1}^2}\;\;&\;\;0\;\;&\;\;0\;\;&\;\;1\;\;&\;\;1\;\;
\end{array}\right)\;.
\label{1TR2-46}
\end{eqnarray}
Obviously the matrix of exponents is written as
\begin{eqnarray}
&&\left(\begin{array}{ccccccc}
\;\;0\;\;&\;\;\alpha_{_{1,2}}\;\;&\;\;\alpha_{_{1,3}}\;\;&\;\;0\;\;&\;0\;\;&\;\;0\;\;&\;\;0\;\;\\
\;\;0\;\;&\;\;0\;\;&\;0\;\;&\;\;3-D\;\;&\;\;0\;\;&\;0\;\;&\;\;D-4\;\;\\
\;\;0\;\;&\;\;\alpha_{_{3,2}}\;\;&\;\alpha_{_{3,3}}\;\;&\;\;0\;\;&\;\;{D\over2}-3\;\;&\;0\;\;&\;\;0\;\;\\
\;\;0\;\;&\;\;\alpha_{_{4,2}}\;\;&\;\alpha_{_{4,3}}\;\;&\;\;0\;\;&\;\;0\;\;&\;{D\over2}-3\;\;&\;\;3-D\;\;
\end{array}\right)\;,
\label{1TR2-47}
\end{eqnarray}
where the matrix elements satisfy the relations
\begin{eqnarray}
&&\alpha_{_{1,2}}+\alpha_{_{1,3}}=-1\;,\;\alpha_{_{3,2}}+\alpha_{_{3,3}}=2-{D\over2}\;,\;
\alpha_{_{4,2}}+\alpha_{_{4,3}}={D\over2}-1\;,
\nonumber\\
&&\alpha_{_{1,2}}+\alpha_{_{3,2}}+\alpha_{_{4,2}}=0\;,
\alpha_{_{1,3}}+\alpha_{_{3,3}}+\alpha_{_{4,3}}=0\;.
\label{1TR2-48}
\end{eqnarray}
Corresponding to the matrix of splitting local coordinates in Eq.(\ref{1TR2-46}),
one obtains 12 choices on the matrix of integer lattice whose submatrix composed of the third and
fifth columns is formulated as $\pm n_{_1}E_{_{4}}^{(i)}$ $\pm n_{_2}E_{_{4}}^{(j)}$,
where $n_{_{1,2}}\ge0$, $(i,j)$ $\in\{(2,3)$, $(2,6)$, $(3,6)\}$,
and other elements are all zero.

The exponents of the integer lattice $n_{_1}E_{_{4}}^{(2)}+n_{_2}E_{_{4}}^{(3)}$ are
\begin{eqnarray}
&&\alpha_{_{1,2}}=-1\;,\;\alpha_{_{1,3}}=0\;,\;\alpha_{_{3,2}}=2-{D\over2}\;,\;
\alpha_{_{3,3}}=0\;,\;\alpha_{_{4,2}}={D\over2}-1,\;\alpha_{_{4,3}}=0\;,
\label{1TR2-49}
\end{eqnarray}
whose geometric representation is presented in Fig.\ref{fig6}(1)
with $(a,b)$ $=(1,3)$, $\{c,d\}$ $=\{2,5\}$.
The exponents of the integer lattice $n_{_1}E_{_{4}}^{(2)}$ $+n_{_2}E_{_{4}}^{(6)}$ are
\begin{eqnarray}
&&\alpha_{_{1,2}}=0\;,\;\alpha_{_{1,3}}=-1\;,\;\alpha_{_{3,2}}=1-{D\over2}\;,\;
\alpha_{_{3,3}}=1\;,\;\alpha_{_{4,2}}={D\over2}-1,\;\alpha_{_{4,3}}=0\;,
\label{1TR2-53}
\end{eqnarray}
whose geometric representation is depicted in Fig.\ref{fig6}(2)
with $a=5$, $b=2$, $(c,d)$ $=(1,3)$.
The exponents of the integer lattice $n_{_1}E_{_{4}}^{(3)}$ $+n_{_2}E_{_{4}}^{(6)}$
are presented by
\begin{eqnarray}
&&\alpha_{_{1,2}}=0\;,\;\alpha_{_{1,3}}=-1\;,\;\alpha_{_{3,2}}=0\;,\;
\alpha_{_{3,3}}=2-{D\over2}\;,\;\alpha_{_{4,2}}=0,\;\alpha_{_{4,3}}={D\over2}-1\;,
\label{1TR2-57}
\end{eqnarray}
whose geometric representation is presented in Fig.\ref{fig6}(3)
with $a=2$, $(b,c,d)$ $=(1,3,5)$.
The corresponding hypergeometric functions are obviously written as
\begin{eqnarray}
&&\psi_{_{\{1,4,5,6\}}}^{(1)}(p_{_1}^2,p_{_2}^2,p_{_3}^2)
\sim{(p_{_1}^2)^{D/2-1}\over(p_{_2}^2-p_{_1}^2)(p_{_3}^2-p_{_1}^2)}
\sum\limits_{n_{_1},n_{_2}}^\infty{\Gamma({D\over2}-2+n_{_1})\Gamma(1-{D\over2}+n_{_2})
\over n_{_1}!n_{_2}!}
\nonumber\\
&&\hspace{3.7cm}\times
\Big({p_{_1}^2-p_{_2}^2\over p_{_3}^2-p_{_2}^2}\Big)^{n_{_1}}
\Big({p_{_3}^2(p_{_1}^2-p_{_2}^2)\over p_{_1}^2(p_{_3}^2-p_{_2}^2)}\Big)^{n_{_2}}
\nonumber\\
&&\hspace{3.3cm}\sim
{(p_{_2}^2)^{D/2-1}\over(p_{_1}^2-p_{_2}^2)(p_{_3}^2-p_{_2}^2)}\;,
\nonumber\\
&&\psi_{_{\{1,4,5,6\}}}^{(2)}(p_{_1}^2,p_{_2}^2,p_{_3}^2)
\sim{(p_{_1}^2)^{D/2-3}(p_{_3}^2)^2\over(p_{_3}^2-p_{_1}^2)(p_{_3}^2-p_{_2}^2)}
\sum\limits_{n_{_1},n_{_2}}^\infty{\Gamma(3-{D\over2}+n_{_1}+n_{_2})
\over\Gamma(3+n_{_1}+n_{_2})}
\nonumber\\
&&\hspace{3.7cm}\times
{\Gamma(1+n_{_2})\over\Gamma(4-{D\over2}+n_{_2})}
\Big({p_{_3}^2(p_{_1}^2-p_{_2}^2)\over p_{_1}^2(p_{_3}^2-p_{_2}^2)}\Big)^{n_{_1}}
\Big({p_{_3}^2\over p_{_1}^2}\Big)^{n_{_2}}\;,
\nonumber\\
&&\psi_{_{\{1,4,5,6\}}}^{(3)}(p_{_1}^2,p_{_2}^2,p_{_3}^2)
\sim{(p_{_3}^2)^{D/2-1}\over(p_{_3}^2-p_{_1}^2)(p_{_3}^2-p_{_2}^2)}
\sum\limits_{n_{_1},n_{_2}}^\infty{\Gamma({D\over2}-2+n_{_2})
\over n_{_2}!\Gamma(1-n_{_1}-n_{_2})\Gamma({D\over2}+n_{_1}+n_{_2})}
\nonumber\\
&&\hspace{3.7cm}\times
(-)^{n_{_1}+n_{_2}}\Big({p_{_3}^2(p_{_1}^2-p_{_2}^2)\over p_{_1}^2(p_{_3}^2-p_{_2}^2)}\Big)^{n_{_1}}
\Big({p_{_3}^2\over p_{_1}^2}\Big)^{n_{_2}}
\nonumber\\
&&\hspace{3.3cm}\sim
{(p_{_3}^2)^{D/2-1}\over(p_{_3}^2-p_{_1}^2)(p_{_3}^2-p_{_2}^2)}\;,
\label{1TR2-57a}
\end{eqnarray}
where the intersection of their convergent regions is a nonempty proper subset of the whole parameter space.
In addition the hypergeometric function of the exponents of Eq.(\ref{1TR2-53}) is expressed as
\begin{eqnarray}
&&\psi_{_{\{1,4,5,6\}}}^{(2)\prime}(p_{_1}^2,p_{_2}^2,p_{_3}^2)
\sim{(p_{_1}^2)^{D/2-1}\over(p_{_3}^2-p_{_1}^2)(p_{_3}^2-p_{_2}^2)}
\sum\limits_{n_{_1},n_{_2}}^\infty{\Gamma({D\over2}-1+n_{_1}-n_{_2})
\over\Gamma(2+n_{_1}-n_{_2})}
\nonumber\\
&&\hspace{3.7cm}\times
{\Gamma(1-{D\over2}+n_{_2})\over n_{_2}!}
\Big({p_{_1}^2-p_{_2}^2\over p_{_3}^2-p_{_2}^2}\Big)^{n_{_1}}
\Big({p_{_3}^2\over p_{_1}^2}\Big)^{n_{_2}}
\nonumber\\
&&\hspace{3.3cm}\sim
\psi_{_{\{1,4,5,6\}}}^{(1)}(p_{_1}^2,p_{_2}^2,p_{_3}^2)
\nonumber\\
&&\hspace{3.7cm}
+\lim\limits_{\epsilon\rightarrow0}{\sin\pi\epsilon\over\sin\pi(4-{D\over2})}
\psi_{_{\{1,4,5,6\}}}^{(2)}(p_{_1}^2,p_{_2}^2,p_{_3}^2)\;.
\label{1TR2-57b}
\end{eqnarray}

The exponents of the integer lattice $-n_{_1}E_{_{4}}^{(2)}-n_{_2}E_{_{4}}^{(3)}$
are presented as
\begin{eqnarray}
&&\alpha_{_{1,2}}=0\;,\;\alpha_{_{1,3}}=-1\;,\;\alpha_{_{3,2}}=0\;,\;
\alpha_{_{3,3}}=2-{D\over2}\;,\;\alpha_{_{4,2}}=0,\;\alpha_{_{4,3}}={D\over2}-1\;,
\label{1TR2-52}
\end{eqnarray}
whose geometric representation is plotted in Fig.\ref{fig6}(1)
with $(a,b)$ $=(1,2)$, $\{c,d\}$ $=\{3,5\}$. Similarly the exponents of the integer
lattice $-n_{_1}E_{_{4}}^{(2)}$ $-n_{_2}E_{_{4}}^{(6)}$ are
\begin{eqnarray}
&&\alpha_{_{1,2}}=-1\;,\;\alpha_{_{1,3}}=0\;,\;\alpha_{_{3,2}}=1\;,\;
\alpha_{_{3,3}}=1-{D\over2}\;,\;\alpha_{_{4,2}}=0,\;\alpha_{_{4,3}}={D\over2}-1\;,
\label{1TR2-56}
\end{eqnarray}
whose geometric representation is presented in Fig.\ref{fig6}(2)
with $a=5$, $b=3$, $(c,d)$ $=(1,2)$. Corresponding to the integer lattice
$-n_{_1}E_{_{4}}^{(3)}$ $-n_{_2}E_{_{4}}^{(6)}$, the exponents are
\begin{eqnarray}
&&\alpha_{_{1,2}}=-1\;,\;\alpha_{_{1,3}}=0\;,\;\alpha_{_{3,2}}=2-{D\over2}\;,\;
\alpha_{_{3,3}}=0\;,\;\alpha_{_{4,2}}={D\over2}-1,\;\alpha_{_{4,3}}=0\;,
\label{1TR2-60}
\end{eqnarray}
whose geometric representation is plotted in Fig.\ref{fig6}(3)
with $a=3$, $(b,c,d)$ $=(1,2,5)$.
The corresponding hypergeometric functions are obviously written as
\begin{eqnarray}
&&\psi_{_{\{1,4,5,6\}}}^{(4)}(p_{_1}^2,p_{_2}^2,p_{_3}^2)
\sim{(p_{_3}^2)^{D/2-1}\over(p_{_3}^2-p_{_1}^2)(p_{_3}^2-p_{_2}^2)}
\sum\limits_{n_{_1},n_{_2}}^\infty{\Gamma({D\over2}-2+n_{_1})\Gamma(1-{D\over2}+n_{_2})
\over n_{_1}!n_{_2}!}
\nonumber\\
&&\hspace{3.7cm}\times
\Big({p_{_3}^2-p_{_2}^2\over p_{_1}^2-p_{_2}^2}\Big)^{n_{_1}}
\Big({p_{_1}^2(p_{_3}^2-p_{_2}^2)\over p_{_3}^2(p_{_1}^2-p_{_2}^2)}\Big)^{n_{_2}}
\nonumber\\
&&\hspace{3.3cm}\sim
{(p_{_2}^2)^{D/2-1}\over
(p_{_1}^2-p_{_2}^2)(p_{_3}^2-p_{_2}^2)}\;,
\nonumber\\
&&\psi_{_{\{1,4,5,6\}}}^{(5)}(p_{_1}^2,p_{_2}^2,p_{_3}^2)
\sim{(p_{_1}^2)^2(p_{_3}^2)^{D/2-3}\over(p_{_2}^2-p_{_1}^2)(p_{_3}^2-p_{_1}^2)}
\sum\limits_{n_{_1},n_{_2}}^\infty{\Gamma(3-{D\over2}+n_{_1}+n_{_2})
\over\Gamma(3+n_{_1}+n_{_2})}
\nonumber\\
&&\hspace{3.7cm}\times
{\Gamma(1+n_{_2})\over\Gamma(4-{D\over2}+n_{_2})}
\Big({p_{_1}^2(p_{_3}^2-p_{_2}^2)\over p_{_3}^2(p_{_1}^2-p_{_2}^2)}\Big)^{n_{_1}}
\Big({p_{_1}^2\over p_{_3}^2}\Big)^{n_{_2}}\;,
\nonumber\\
&&\psi_{_{\{1,4,5,6\}}}^{(6)}(p_{_1}^2,p_{_2}^2,p_{_3}^2)
\sim{(p_{_1}^2)^{D/2-1}\over(p_{_2}^2-p_{_1}^2)(p_{_3}^2-p_{_1}^2)}
\sum\limits_{n_{_1},n_{_2}}^\infty{\Gamma({D\over2}-2+n_{_2})
\over n_{_2}!\Gamma(1-n_{_1}-n_{_2})\Gamma({D\over2}+n_{_1}+n_{_2})}
\nonumber\\
&&\hspace{3.7cm}\times
(-)^{n_{_1}+n_{_2}}\Big({p_{_1}^2(p_{_3}^2-p_{_2}^2)\over p_{_3}^2(p_{_1}^2-p_{_2}^2)}\Big)^{n_{_1}}
\Big({p_{_1}^2\over p_{_3}^2}\Big)^{n_{_2}}
\nonumber\\
&&\hspace{3.3cm}\sim
{(p_{_1}^2)^{D/2-1}\over(p_{_2}^2-p_{_1}^2)(p_{_3}^2-p_{_1}^2)}\;,
\label{1TR2-60a}
\end{eqnarray}
where the intersection of their convergent regions is a nonempty proper subset of the whole parameter space.
The fact indicates that those linear independent hypergeometric functions
constitute a fundamental solution system of the GKZ-system in the
proper subset of the whole parameter space. Actually the hypergeometric function
of the exponents of Eq.(\ref{1TR2-56}) is
\begin{eqnarray}
&&\psi_{_{\{1,4,5,6\}}}^{(5)\prime}(p_{_1}^2,p_{_2}^2,p_{_3}^2)
\sim{(p_{_3}^2)^{D/2-1}\over(p_{_2}^2-p_{_1}^2)(p_{_3}^2-p_{_1}^2)}
\sum\limits_{n_{_1},n_{_2}}^\infty{\Gamma({D\over2}-1+n_{_1}-n_{_2})\Gamma(1-{D\over2}+n_{_2})
\over n_{_2}!\Gamma(2+n_{_1}-n_{_2})}
\nonumber\\
&&\hspace{3.7cm}\times
\Big({p_{_3}^2-p_{_2}^2\over p_{_1}^2-p_{_2}^2}\Big)^{n_{_1}}
\Big({p_{_1}^2\over p_{_3}^2}\Big)^{n_{_2}}
\nonumber\\
&&\hspace{3.3cm}\sim
\psi_{_{\{1,4,5,6\}}}^{(4)}(p_{_1}^2,p_{_2}^2,p_{_3}^2)
\nonumber\\
&&\hspace{3.7cm}
+\lim\limits_{\epsilon\rightarrow0}{\sin\pi\epsilon\over\sin\pi({D\over2}-3)}
\psi_{_{\{1,4,5,6\}}}^{(5)}(p_{_1}^2,p_{_2}^2,p_{_3}^2)\;.
\label{1TR2-60b}
\end{eqnarray}

Similarly the exponents of the integer lattice $n_{_1}E_{_{4}}^{(2)}$
$-n_{_2}E_{_{4}}^{(3)}$ are
\begin{eqnarray}
&&\alpha_{_{1,2}}={D\over2}-2\;,\;\alpha_{_{1,3}}=1-{D\over2}\;,\;\alpha_{_{3,2}}=2-{D\over2}\;,\;
\alpha_{_{3,3}}=0\;,\;\alpha_{_{4,2}}=0,\;\alpha_{_{4,3}}={D\over2}-1\;,
\label{1TR2-50}
\end{eqnarray}
whose geometric representation is drawn in Fig.\ref{fig6}(2)
with $a=1$, $b=3$, $\{c,d\}$ $=\{2,5\}$.
The matrix of exponents of the integer lattice
$n_{_1}E_{_{4}}^{(2)}$ $-n_{_2}E_{_{4}}^{(6)}$ is formulated as
\begin{eqnarray}
&&\alpha_{_{1,2}}=0\;,\;\alpha_{_{1,3}}=-1\;,\;\alpha_{_{3,2}}=0\;,\;
\alpha_{_{3,3}}=2-{D\over2}\;,\;\alpha_{_{4,2}}=0,\;\alpha_{_{4,3}}={D\over2}-1\;,
\label{1TR2-54}
\end{eqnarray}
whose geometric representation is plotted in Fig.\ref{fig6}(1)
with $(a,b)$ $=(2,5)$, $\{c,d\}$ $=\{1,3\}$.
The exponents of the integer lattice $n_{_1}E_{_{4}}^{(3)}$ $-n_{_2}E_{_{4}}^{(6)}$
are given by
\begin{eqnarray}
&&\alpha_{_{1,2}}=0\;,\;\alpha_{_{1,3}}=-1\;,\;\alpha_{_{3,2}}=2-{D\over2}\;,\;
\alpha_{_{3,3}}=0\;,\;\alpha_{_{4,2}}={D\over2}-2,\;\alpha_{_{4,3}}=1\;,
\label{1TR2-58}
\end{eqnarray}
where the geometric representation is presented in Fig.\ref{fig6}(4)
with $\{(a,b)$, $(c,d)\}$ $=\{(1,3)$, $(2,5)\}$.
The corresponding hypergeometric functions are obviously written as
\begin{eqnarray}
&&\psi_{_{\{1,4,5,6\}}}^{(7)}(p_{_1}^2,p_{_2}^2,p_{_3}^2)
\sim{(p_{_3}^2)^{D/2-1}(p_{_2}^2-p_{_1}^2)^{D/2-2}\over
(p_{_3}^2-p_{_1}^2)(p_{_3}^2-p_{_2}^2)^{D/2-1}}
\nonumber\\
&&\hspace{3.7cm}\times
\sum\limits_{n_{_1},n_{_2}}^\infty{\Gamma({D\over2}-2+n_{_1})\Gamma(1-{D\over2}+n_{_2})
\over n_{_1}!n_{_2}!}
\nonumber\\
&&\hspace{3.7cm}\times
\Big({p_{_1}^2-p_{_2}^2\over p_{_3}^2-p_{_2}^2}\Big)^{n_{_1}}
\Big({p_{_1}^2(p_{_3}^2-p_{_2}^2)\over p_{_3}^2(p_{_1}^2-p_{_2}^2)}\Big)^{n_{_2}}
\nonumber\\
&&\hspace{3.3cm}\sim
{(p_{_2}^2)^{D/2-1}\over
(p_{_1}^2-p_{_2}^2)(p_{_3}^2-p_{_2}^2)}\;,
\nonumber\\
&&\psi_{_{\{1,4,5,6\}}}^{(8)}(p_{_1}^2,p_{_2}^2,p_{_3}^2)
\sim{(p_{_3}^2)^{D/2-1}\over(p_{_3}^2-p_{_1}^2)(p_{_3}^2-p_{_2}^2)}
\nonumber\\
&&\hspace{3.7cm}\times
\sum\limits_{n_{_1},n_{_2}}^\infty{\Gamma(1-{D\over2}+n_{_2})
\over n_{_2}!\Gamma(1-n_{_1}-n_{_2})\Gamma(3-{D\over2}+n_{_1}+n_{_2})}
\nonumber\\
&&\hspace{3.7cm}\times
(-)^{n_{_1}+n_{_2}}\Big({p_{_1}^2-p_{_2}^2\over p_{_3}^2-p_{_2}^2}\Big)^{n_{_1}}
\Big({p_{_1}^2\over p_{_3}^2}\Big)^{n_{_2}}
\nonumber\\
&&\hspace{3.3cm}\sim
{(p_{_3}^2)^{D/2-1}\over(p_{_3}^2-p_{_1}^2)(p_{_3}^2-p_{_2}^2)}\;,
\nonumber\\
&&\psi_{_{\{1,4,5,6\}}}^{(9)}(p_{_1}^2,p_{_2}^2,p_{_3}^2)
\sim{(p_{_1}^2)^{D/2}p_{_3}^2\over p_{_3}^2(p_{_3}^2-p_{_1}^2)(p_{_3}^2-p_{_2}^2)}
\nonumber\\
&&\hspace{3.7cm}\times
\sum\limits_{n_{_1},n_{_2}}^\infty{\Gamma({D\over2}+n_{_1}+n_{_2})\Gamma(1+n_{_2})
\over\Gamma(3+n_{_1}+n_{_2})\Gamma({D\over2}+1+n_{_2})}
\nonumber\\
&&\hspace{3.7cm}\times
\Big({p_{_1}^2-p_{_2}^2\over p_{_3}^2-p_{_2}^2}\Big)^{n_{_1}}
\Big({p_{_1}^2\over p_{_3}^2}\Big)^{n_{_2}}\;,
\label{1TR2-58a}
\end{eqnarray}
where the intersection of their convergent regions is a nonempty proper subset of the whole parameter space.
The hypergeometric function of the exponents in Eq.(\ref{1TR2-58}) is
\begin{eqnarray}
&&\psi_{_{\{1,4,5,6\}}}^{(9)\prime}(p_{_1}^2,p_{_2}^2,p_{_3}^2)
\sim{(p_{_1}^2)^{D/2-2}p_{_3}^2\over(p_{_3}^2-p_{_1}^2)(p_{_3}^2-p_{_2}^2)}
\nonumber\\
&&\hspace{3.7cm}\times
\sum\limits_{n_{_1},n_{_2}}^\infty{\Gamma(2-{D\over2}+n_{_1}-n_{_2})\Gamma({D\over2}-2+n_{_2})
\over n_{_2}!\Gamma(2+n_{_1}-n_{_2})}
\nonumber\\
&&\hspace{3.7cm}\times
\Big({p_{_3}^2(p_{_1}^2-p_{_2}^2)\over p_{_1}^2(p_{_3}^2-p_{_2}^2)}\Big)^{n_{_1}}
\Big({p_{_1}^2\over p_{_3}^2}\Big)^{n_{_2}}
\nonumber\\
&&\hspace{3.3cm}\sim
{(p_{_2}^2)^{D/2-1}\over
(p_{_1}^2-p_{_2}^2)(p_{_3}^2-p_{_2}^2)}
\nonumber\\
&&\hspace{3.7cm}
+\lim\limits_{\epsilon\rightarrow0}{\sin\pi\epsilon\over\sin\pi({D\over2}+1)}
\psi_{_{\{1,4,5,6\}}}^{(9)}(p_{_1}^2,p_{_2}^2,p_{_3}^2)\;,
\label{1TR2-58b}
\end{eqnarray}
Furthermore, the sets of the hypergeometric functions
$\{\psi_{_{\{1,4,5,6\}}}^{(1)}$, $\psi_{_{\{1,4,5,6\}}}^{(8)}$, $\psi_{_{\{1,4,5,6\}}}^{(9)}\}$
and $\{\psi_{_{\{1,4,5,6\}}}^{(6)}$, $\psi_{_{\{1,4,5,6\}}}^{(7)}$, $\psi_{_{\{1,4,5,6\}}}^{(8)}\}$
constitute the fundamental solution systems of the nonempty proper subsets
of the whole parameter space, respectively.

Meanwhile the exponents of the integer lattice $-n_{_1}E_{_{4}}^{(2)}$ $+n_{_2}E_{_{4}}^{(3)}$ are
\begin{eqnarray}
&&\alpha_{_{1,2}}=1-{D\over2}\;,\;\alpha_{_{1,3}}={D\over2}-2\;,\;\alpha_{_{3,2}}=0\;,\;
\alpha_{_{3,3}}=2-{D\over2}\;,\;\alpha_{_{4,2}}={D\over2}-1,\;\alpha_{_{4,3}}=0\;,
\label{1TR2-51}
\end{eqnarray}
whose geometric representation is depicted in Fig.\ref{fig6}(2)
with $a=1$, $b=2$, $\{c,d\}$ $=\{3,5\}$. The exponents of the integer lattice
$-n_{_1}E_{_{4}}^{(2)}$ $+n_{_2}E_{_{4}}^{(6)}$ are
\begin{eqnarray}
&&\alpha_{_{1,2}}=-1\;,\;\alpha_{_{1,3}}=0\;,\;\alpha_{_{3,2}}=2-{D\over2}\;,\;
\alpha_{_{3,3}}=0\;,\;\alpha_{_{4,2}}={D\over2}-1,\;\alpha_{_{4,3}}=0\;,
\label{1TR2-55}
\end{eqnarray}
whose geometric representation is plotted in Fig.\ref{fig6}(1)
with $(a,b)$ $=(3,5)$, $\{c,d\}$ $=\{1,2\}$. Finally, the exponents
of the integer lattice $-n_{_1}E_{_{4}}^{(3)}$ $+n_{_2}E_{_{4}}^{(6)}$ are
\begin{eqnarray}
&&\alpha_{_{1,2}}=-1\;,\;\alpha_{_{1,3}}=0\;,\;\alpha_{_{3,2}}=0\;,\;
\alpha_{_{3,3}}=2-{D\over2}\;,\;\alpha_{_{4,2}}=1,\;\alpha_{_{4,3}}={D\over2}-2\;,
\label{1TR2-59}
\end{eqnarray}
whose geometric representation is shown in Fig.\ref{fig6}(4)
with $\{(a,b)$, $(c,d)\}$ $=\{(1,2)$, $(3,5)\}$.
The corresponding hypergeometric functions are obviously written as
\begin{eqnarray}
&&\psi_{_{\{1,4,5,6\}}}^{(10)}(p_{_1}^2,p_{_2}^2,p_{_3}^2)
\sim{(p_{_1}^2)^{D/2-1}(p_{_3}^2-p_{_2}^2)^{D/2-2}\over
(p_{_2}^2-p_{_1}^2)^{D/2-1}(p_{_3}^2-p_{_1}^2)}
\nonumber\\
&&\hspace{3.7cm}\times
\sum\limits_{n_{_1},n_{_2}}^\infty{\Gamma({D\over2}-2+n_{_1})\Gamma(1-{D\over2}+n_{_2})
\over n_{_1}!n_{_2}!}
\nonumber\\
&&\hspace{3.7cm}\times
\Big({p_{_3}^2-p_{_2}^2\over p_{_1}^2-p_{_2}^2}\Big)^{n_{_1}}
\Big({p_{_3}^2(p_{_1}^2-p_{_2}^2)\over p_{_1}^2(p_{_3}^2-p_{_2}^2)}\Big)^{n_{_2}}
\nonumber\\
&&\hspace{3.3cm}\sim
{(p_{_2}^2)^{D/2-1}\over(p_{_1}^2-p_{_2}^2)(p_{_3}^2-p_{_2}^2)}\;,
\nonumber\\
&&\psi_{_{\{1,4,5,6\}}}^{(11)}(p_{_1}^2,p_{_2}^2,p_{_3}^2)
\sim{(p_{_1}^2)^{D/2-1}\over(p_{_2}^2-p_{_1}^2)(p_{_3}^2-p_{_1}^2)}
\nonumber\\
&&\hspace{3.7cm}\times
\sum\limits_{n_{_1},n_{_2}}^\infty{\Gamma(1-{D\over2}+n_{_2})
\over n_{_2}!\Gamma(1-n_{_1}-n_{_2})\Gamma(3-{D\over2}+n_{_1}+n_{_2})}
\nonumber\\
&&\hspace{3.7cm}\times
(-)^{n_{_1}+n_{_2}}\Big({p_{_3}^2-p_{_2}^2\over p_{_1}^2-p_{_2}^2}\Big)^{n_{_1}}
\Big({p_{_3}^2\over p_{_1}^2}\Big)^{n_{_2}}
\nonumber\\
&&\hspace{3.3cm}\sim
{(p_{_1}^2)^{D/2-1}\over(p_{_2}^2-p_{_1}^2)(p_{_3}^2-p_{_1}^2)}\;,
\nonumber\\
&&\psi_{_{\{1,4,5,6\}}}^{(12)}(p_{_1}^2,p_{_2}^2,p_{_3}^2)
\sim{(p_{_3}^2)^{D/2}\over p_{_1}^2(p_{_2}^2-p_{_1}^2)(p_{_3}^2-p_{_1}^2)}
\nonumber\\
&&\hspace{3.7cm}\times
\sum\limits_{n_{_1},n_{_2}}^\infty{\Gamma({D\over2}+n_{_1}+n_{_2})\Gamma(1+n_{_2})
\over\Gamma(3+n_{_1}+n_{_2})\Gamma({D\over2}+1+n_{_2})}
\nonumber\\
&&\hspace{3.7cm}\times
\Big({p_{_3}^2-p_{_2}^2\over p_{_1}^2-p_{_2}^2}\Big)^{n_{_1}}
\Big({p_{_3}^2\over p_{_1}^2}\Big)^{n_{_2}}\;,
\label{1TR2-61}
\end{eqnarray}
where the intersection of their convergent regions is a nonempty proper subset of the whole parameter space.
The hypergeometric function of the exponents of Eq.(\ref{1TR2-59}) is
\begin{eqnarray}
&&\psi_{_{\{1,4,5,6\}}}^{(12)\prime}(p_{_1}^2,p_{_2}^2,p_{_3}^2)
\sim{p_{_1}^2(p_{_3}^2)^{D/2-2}\over(p_{_2}^2-p_{_1}^2)(p_{_3}^2-p_{_1}^2)}
\nonumber\\
&&\hspace{3.7cm}\times
\sum\limits_{n_{_1},n_{_2}}^\infty{\Gamma(2-{D\over2}+n_{_1}-n_{_2})\Gamma({D\over2}-2+n_{_2})
\over n_{_2}!\Gamma(2+n_{_1}-n_{_2})}
\nonumber\\
&&\hspace{3.7cm}\times
\Big({p_{_1}^2(p_{_3}^2-p_{_2}^2)\over p_{_3}^2(p_{_1}^2-p_{_2}^2)}\Big)^{n_{_1}}
\Big({p_{_3}^2\over p_{_1}^2}\Big)^{n_{_2}}
\nonumber\\
&&\hspace{3.3cm}\sim
{(p_{_2}^2)^{D/2-1}\over(p_{_3}^2-p_{_2}^2)(p_{_1}^2-p_{_2}^2)}
\nonumber\\
&&\hspace{3.7cm}
+\lim\limits_{\epsilon\rightarrow0}{\sin\pi\epsilon\over\sin\pi
({D\over2}+1)}\psi_{_{\{1,4,5,6\}}}^{(12)}(p_{_1}^2,p_{_2}^2,p_{_3}^2)\;,
\label{1TR2-61a}
\end{eqnarray}
Additionally the sets of the hypergeometric functions
$\{\psi_{_{\{1,4,5,6\}}}^{(3)}$, $\psi_{_{\{1,4,5,6\}}}^{(10)}$, $\psi_{_{\{1,4,5,6\}}}^{(11)}\}$
and $\{\psi_{_{\{1,4,5,6\}}}^{(4)}$, $\psi_{_{\{1,4,5,6\}}}^{(11)}$, $\psi_{_{\{1,4,5,6\}}}^{(12)}\}$
constitute the fundamental solution systems of the nonempty proper subsets
of the whole parameter space, respectively.

Similarly $\det(A_{_{2,4,5,6}}^{(1T)\prime})=p_{_1}^2-p_{_2}^2$, and
\begin{eqnarray}
&&\Big(A_{_{\{2,4,5,6\}}}^{(1T)\prime}\Big)^{-1}\cdot A^{(1T)\prime}=\left(\begin{array}{ccccccc}
\;\;-{p_{_3}^2-p_{_1}^2\over p_{_2}^2-p_{_1}^2}\;\;&\;\;1\;\;
&\;\;-{p_{_2}^2-p_{_3}^2\over p_{_2}^2-p_{_1}^2}\;\;&\;\;0\;\;&\;\;0\;\;&\;\;0\;\;&\;\;0\;\;\\
\;\;0\;\;&\;\;0\;\;&\;\;0\;\;&\;\;1\;\;&\;\;0\;\;&\;\;0\;\;&\;\;1\;\;\\
\;\;{1\over p_{_2}^2-p_{_1}^2}\;\;&\;\;0\;\;&\;\;-{1\over p_{_2}^2-p_{_1}^2}\;\;&
\;\;0\;\;&\;\;1\;\;&\;\;0\;\;&\;\;0\;\;\\
\;\;-{p_{_1}^2\over p_{_2}^2-p_{_1}^2}\;\;&\;\;0\;\;&
\;\;{p_{_2}^2\over p_{_2}^2-p_{_1}^2}\;\;&\;\;0\;\;&\;\;0\;\;&\;\;1\;\;&\;\;1\;\;
\end{array}\right)\;.
\label{1TR2-63}
\end{eqnarray}
Obviously the matrix of exponents is written as
\begin{eqnarray}
&&\left(\begin{array}{ccccccc}
\;\;\alpha_{_{1,1}}\;\;&\;\;0\;\;&\;\;\alpha_{_{1,3}}\;\;&\;\;0\;\;&\;0\;\;&\;\;0\;\;&\;\;0\;\;\\
\;\;0\;\;&\;\;0\;\;&\;0\;\;&\;\;3-D\;\;&\;\;0\;\;&\;0\;\;&\;\;D-4\;\;\\
\;\;\alpha_{_{3,1}}\;\;&\;\;0\;\;&\;\alpha_{_{3,3}}\;\;&\;\;0\;\;&\;\;{D\over2}-3\;\;&\;0\;\;&\;\;0\;\;\\
\;\;\alpha_{_{4,1}}\;\;&\;\;0\;\;&\;\alpha_{_{4,3}}\;\;&\;\;0\;\;&\;\;0\;\;&\;{D\over2}-3\;\;&\;\;3-D\;\;
\end{array}\right)\;,
\label{1TR2-64}
\end{eqnarray}
where the matrix elements satisfy the relations which are obtained by the replacement
$\alpha_{_{i2}}\rightarrow\alpha_{_{i1}}$, with $(i=1,3,4)$ in Eq.(\ref{1TR2-48}).
Correspondingly geometric representations of those exponents are
obtained from that of $\psi_{_{\{1,4,5,6\}}}^{(i)}$
through the permutation $\widehat{(12)}$.
The constructed hypergeometric functions are
\begin{eqnarray}
&&\psi_{_{\{2,4,5,6\}}}^{(i)}(p_{_1}^2,p_{_2}^2,p_{_3}^2)=
\psi_{_{\{1,4,5,6\}}}^{(i)}(p_{_1}^2,p_{_3}^2,p_{_2}^2),\;\;(i=1,\cdots,12)\;.
\label{1TR2-65}
\end{eqnarray}

Because $\det(A_{_{3,4,5,6}}^{(1T)\prime})=p_{_2}^2-p_{_3}^2$, and
\begin{eqnarray}
&&\Big(A_{_{\{3,4,5,6\}}}^{(1T)\prime}\Big)^{-1}\cdot A^{(1T)\prime}=\left(\begin{array}{ccccccc}
\;\;-{p_{_1}^2-p_{_3}^2\over p_{_2}^2-p_{_3}^2}\;\;&\;\;-{p_{_2}^2-p_{_1}^2\over p_{_2}^2-p_{_3}^2}\;\;
&\;\;1\;\;&\;\;0\;\;&\;\;0\;\;&\;\;0\;\;&\;\;0\;\;\\
\;\;0\;\;&\;\;0\;\;&\;\;0\;\;&\;\;1\;\;&\;\;0\;\;&\;\;0\;\;&\;\;1\;\;\\
\;\;{1\over p_{_2}^2-p_{_3}^2}\;\;&\;\;-{1\over p_{_2}^2-p_{_3}^2}\;\;&\;\;0\;\;&
\;\;0\;\;&\;\;1\;\;&\;\;0\;\;&\;\;0\;\;\\
\;\;-{p_{_3}^2\over p_{_2}^2-p_{_3}^2}\;\;&\;\;{p_{_2}^2\over p_{_2}^2-p_{_3}^2}\;\;&
\;\;0\;\;&\;\;0\;\;&\;\;0\;\;&\;\;1\;\;&\;\;1\;\;
\end{array}\right)\;,
\label{1TR2-66}
\end{eqnarray}
the matrix of exponents is written as
\begin{eqnarray}
&&\left(\begin{array}{ccccccc}
\;\;\alpha_{_{1,1}}\;\;&\;\;\alpha_{_{1,2}}\;\;&\;\;0\;\;&\;\;0\;\;&\;0\;\;&\;\;0\;\;&\;\;0\;\;\\
\;\;0\;\;&\;\;0\;\;&\;0\;\;&\;\;3-D\;\;&\;\;0\;\;&\;0\;\;&\;\;D-4\;\;\\
\;\;\alpha_{_{3,1}}\;\;&\;\;\alpha_{_{3,2}}\;\;&\;0\;\;&\;\;0\;\;&\;\;{D\over2}-3\;\;&\;0\;\;&\;\;0\;\;\\
\;\;\alpha_{_{4,1}}\;\;&\;\;\alpha_{_{4,2}}\;\;&\;0\;\;&\;\;0\;\;&\;\;0\;\;&\;{D\over2}-3\;\;&\;\;3-D\;\;
\end{array}\right)\;,
\label{1TR2-67}
\end{eqnarray}
where the matrix elements satisfy the relations which are obtained by the replacement
$\alpha_{_{i,3}}\rightarrow\alpha_{_{i,1}}$, with $(i=1,3,4)$ in Eq.(\ref{1TR2-48}).
Correspondingly the geometric representations of those matrices of exponents are
obtained from that of $\psi_{_{\{1,4,5,6\}}}^{(i)}$ through the permutation $\widehat{(132)}$.
The constructed hypergeometric functions are
\begin{eqnarray}
&&\psi_{_{\{3,4,5,6\}}}^{(i)}(p_{_1}^2,p_{_2}^2,p_{_3}^2)=
\psi_{_{\{1,4,5,6\}}}^{(i)}(p_{_3}^2,p_{_1}^2,p_{_2}^2),\;\;(i=1,\cdots,12)\;.
\label{1TR2-68}
\end{eqnarray}

Because $\det(A_{_{1,5,6,7}}^{(1T)\prime})=p_{_3}^2-p_{_1}^2$,
where the matrix $A_{_{\{1,5,6,7\}}}^{(1T)\prime}$ denotes the submatrix of $A^{(1T)\prime}$ composed
of the first, fifth, sixth and seventh columns, one finds
\begin{eqnarray}
&&\Big(A_{_{\{1,5,6,7\}}}^{(1T)\prime}\Big)^{-1}\cdot A^{(1T)\prime}=\left(\begin{array}{ccccccc}
\;\;1\;\;&\;\;-{p_{_2}^2-p_{_1}^2\over p_{_3}^2-p_{_1}^2}\;\;
&\;\;-{p_{_3}^2-p_{_2}^2\over p_{_3}^2-p_{_1}^2}\;\;&\;\;0\;\;&\;\;0\;\;&\;\;0\;\;&\;\;0\;\;\\
\;\;0\;\;&\;\;{1\over p_{_3}^2-p_{_1}^2}\;\;&\;\;-{1\over p_{_3}^2-p_{_1}^2}\;\;&
\;\;0\;\;&\;\;1\;\;&\;\;0\;\;&\;\;0\;\;\\
\;\;0\;\;&\;\;-{p_{_1}^2\over p_{_3}^2-p_{_1}^2}\;\;&
\;\;{p_{_3}^2\over p_{_3}^2-p_{_1}^2}\;\;&\;\;-1\;\;&\;\;0\;\;&\;\;1\;\;&\;\;0\;\;\\
\;\;0\;\;&\;\;0\;\;&\;\;0\;\;&\;\;1\;\;&\;\;0\;\;&\;\;0\;\;&\;\;1\;\;
\end{array}\right)\;.
\label{1TR2-69}
\end{eqnarray}
Obviously the matrix of exponents is written as
\begin{eqnarray}
&&\left(\begin{array}{ccccccc}
\;\;0\;\;&\;\;\alpha_{_{1,2}}\;\;&\;\;\alpha_{_{1,3}}\;\;&\;\;0\;\;&\;0\;\;&\;\;0\;\;&\;\;0\;\;\\
\;\;0\;\;&\;\;\alpha_{_{2,2}}\;\;&\;\alpha_{_{2,3}}\;\;&\;\;0\;\;&\;\;{D\over2}-3\;\;&\;0\;\;&\;\;0\;\;\\
\;\;0\;\;&\;\;\alpha_{_{3,2}}\;\;&\;\alpha_{_{3,3}}\;\;&\;\;3-D\;\;&\;\;0\;\;&\;{D\over2}-3\;\;&\;\;0\;\;\\
\;\;0\;\;&\;\;0\;\;&\;0\;\;&\;\;0\;\;&\;\;0\;\;&\;0\;\;&\;\;-1\;\;
\end{array}\right)\;,
\label{1TR2-70}
\end{eqnarray}
where the matrix elements satisfy the relations
\begin{eqnarray}
&&\alpha_{_{1,2}}+\alpha_{_{1,3}}=-1\;,\;\alpha_{_{2,2}}+\alpha_{_{2,3}}=2-{D\over2}\;,\;
\alpha_{_{3,2}}+\alpha_{_{3,3}}={D\over2}-1\;,
\nonumber\\
&&\alpha_{_{1,2}}+\alpha_{_{2,2}}+\alpha_{_{3,2}}=0\;,
\alpha_{_{1,3}}+\alpha_{_{2,3}}+\alpha_{_{3,3}}=0\;.
\label{1TR2-71}
\end{eqnarray}
Corresponding to the matrix of local coordinates in Eq.(\ref{1TR2-46}),
one obtains 12 choices on the matrix of integer lattice whose submatrix composed of the third and
fifth columns is formulated as $\pm n_{_1}E_{_{4}}^{(i)}$ $\pm n_{_2}E_{_{4}}^{(j)}$,
where $n_{_{1,2}}\ge0$, $(i,j)$ $\in\{(1,2)$, $(1,4)$, $(2,4)\}$,
and other elements are all zero.
Basing on the matrices of integer lattice, we have 12 choices on the
matrix of exponents.
The exponents of the integer lattice $n_{_1}E_{_{4}}^{(1)}$ $+n_{_2}E_{_{4}}^{(2)}$
are written as
\begin{eqnarray}
&&\alpha_{_{1,2}}=-1\;,\;\alpha_{_{1,3}}=0\;,\;\alpha_{_{2,2}}=2-{D\over2}\;,\;
\alpha_{_{2,3}}=0\;,\;\alpha_{_{3,2}}={D\over2}-1,\;\alpha_{_{3,3}}=0\;,
\label{1TR2-72}
\end{eqnarray}
whose geometric representation is presented in Fig.\ref{fig6}(1)
with $(a,b)$ $=(1,3)$, $\{c,d\}$ $=\{2,5\}$.
In a similar way the exponents of the integer lattice
$n_{_1}E_{_{4}}^{(1)}$ $+n_{_2}E_{_{4}}^{(4)}$ are
\begin{eqnarray}
&&\alpha_{_{1,2}}=0\;,\;\alpha_{_{1,3}}=-1\;,\;\alpha_{_{2,2}}=1-{D\over2}\;,\;
\alpha_{_{2,3}}=1\;,\;\alpha_{_{3,2}}={D\over2}-1,\;\alpha_{_{3,3}}=0\;,
\label{1TR2-76}
\end{eqnarray}
whose geometric representation is plotted in Fig.\ref{fig6}(2)
with $a=5$, $b=2$, $(c,d)$ $=(1,3)$.
The corresponding exponents of the integer lattice $n_{_1}E_{_{4}}^{(2)}$
$+n_{_2}E_{_{4}}^{(4)}$ are presented by
\begin{eqnarray}
&&\alpha_{_{1,2}}=0\;,\;\alpha_{_{1,3}}=-1\;,\;\alpha_{_{2,2}}=0\;,\;
\alpha_{_{2,3}}=2-{D\over2}\;,\;\alpha_{_{3,2}}=0,\;\alpha_{_{3,3}}={D\over2}-1\;,
\label{1TR2-80}
\end{eqnarray}
whose geometric representation is shown in Fig.\ref{fig6}(3)
with $a=2$, $(b,c,d)$ $=(1,3,5)$.

The exponents of the integer lattice $-n_{_1}E_{_{4}}^{(1)}$ $-n_{_2}E_{_{4}}^{(2)}$
are presented as
\begin{eqnarray}
&&\alpha_{_{1,2}}=0\;,\;\alpha_{_{1,3}}=-1\;,\;\alpha_{_{2,2}}=0\;,\;
\alpha_{_{2,3}}=2-{D\over2}\;,\;\alpha_{_{3,2}}=0,\;\alpha_{_{3,3}}={D\over2}-1\;,
\label{1TR2-75}
\end{eqnarray}
whose geometric representation is depicted in Fig.\ref{fig6}(1)
with $(a,b)$ $=(1,2)$, $\{c,d\}$ $=\{3,5\}$. Corresponding to the integer lattice
$-n_{_1}E_{_{4}}^{(1)}$ $-n_{_2}E_{_{4}}^{(4)}$, the exponents are given by
\begin{eqnarray}
&&\alpha_{_{1,2}}=-1\;,\;\alpha_{_{1,3}}=0\;,\;\alpha_{_{2,2}}=1\;,\;
\alpha_{_{2,3}}=1-{D\over2}\;,\;\alpha_{_{3,2}}=0,\;\alpha_{_{3,3}}={D\over2}-1\;,
\label{1TR2-79}
\end{eqnarray}
whose geometric representation is drawn in Fig.\ref{fig6}(2)
with $a=5$, $b=3$, $(c,d)$ $=(1,2)$. Similarly the exponents of the
integer lattice $-n_{_1}E_{_{4}}^{(2)}$ $-n_{_2}E_{_{4}}^{(4)}$ are
\begin{eqnarray}
&&\alpha_{_{1,2}}=-1\;,\;\alpha_{_{1,3}}=0\;,\;\alpha_{_{2,2}}=2-{D\over2}\;,\;
\alpha_{_{2,3}}=0\;,\;\alpha_{_{3,2}}={D\over2}-1,\;\alpha_{_{3,3}}=0\;,
\label{1TR2-83}
\end{eqnarray}
whose geometric representation is presented in Fig.\ref{fig6}(3)
with $a=3$, $(b,c,d)$ $=(1,2,5)$.

Meanwhile the exponents of the integer lattice $n_{_1}E_{_{4}}^{(1)}$ $-n_{_2}E_{_{4}}^{(2)}$
are written as
\begin{eqnarray}
&&\alpha_{_{1,2}}={D\over2}-2\;,\;\alpha_{_{1,3}}=1-{D\over2}\;,\;\alpha_{_{2,2}}=2-{D\over2}\;,\;
\alpha_{_{2,3}}=0\;,\;\alpha_{_{3,2}}=0,\;\alpha_{_{3,3}}={D\over2}-1\;,
\label{1TR2-73}
\end{eqnarray}
whose geometric representation is presented in Fig.\ref{fig6}(2)
with $a=1$, $b=3$, $\{c,d\}$ $=\{2,5\}$.
The corresponding exponents of integer lattice $n_{_1}E_{_{4}}^{(1)}$ $-n_{_2}E_{_{4}}^{(4)}$ are
\begin{eqnarray}
&&\alpha_{_{1,2}}=0\;,\;\alpha_{_{1,3}}=-1\;,\;\alpha_{_{2,2}}=0\;,\;
\alpha_{_{2,3}}=2-{D\over2}\;,\;\alpha_{_{3,2}}=0,\;\alpha_{_{3,3}}={D\over2}-1\;,
\label{1TR2-77}
\end{eqnarray}
whose geometric representation is drawn in Fig.\ref{fig6}(1)
with $(a,b)$ $=(2,5)$, $\{c,d\}$ $=\{1,3\}$.
The exponents corresponding to the integer lattice $n_{_1}E_{_{4}}^{(2)}$ $-n_{_2}E_{_{4}}^{(4)}$
are given by
\begin{eqnarray}
&&\alpha_{_{1,2}}=0\;,\;\alpha_{_{1,3}}=-1\;,\;\alpha_{_{2,2}}=2-{D\over2}\;,\;
\alpha_{_{2,3}}=0\;,\;\alpha_{_{3,2}}={D\over2}-2,\;\alpha_{_{3,3}}=1\;,
\label{1TR2-81}
\end{eqnarray}
whose geometric representation is depicted in Fig.\ref{fig6}(4)
with $\{(a,b)$, $(c,d)\}$ $=\{(1,3)$, $(2,5)\}$.

The exponents of the integer lattice $-n_{_1}E_{_{4}}^{(1)}$ $+n_{_2}E_{_{4}}^{(2)}$ are
\begin{eqnarray}
&&\alpha_{_{1,2}}=1-{D\over2}\;,\;\alpha_{_{1,3}}={D\over2}-2\;,\;\alpha_{_{2,2}}=0\;,\;
\alpha_{_{2,3}}=2-{D\over2}\;,\;\alpha_{_{3,2}}={D\over2}-1,\;\alpha_{_{3,3}}=0\;,
\label{1TR2-74}
\end{eqnarray}
whose geometric representation is presented in Fig.\ref{fig6}(2)
with $a=1$, $b=2$, $\{c,d\}$ $=\{3,5\}$.
The corresponding exponents of the integer lattice $-n_{_1}E_{_{4}}^{(1)}$ $+n_{_2}E_{_{4}}^{(4)}$
are presented by
\begin{eqnarray}
&&\alpha_{_{1,2}}=-1\;,\;\alpha_{_{1,3}}=0\;,\;\alpha_{_{2,2}}=2-{D\over2}\;,\;
\alpha_{_{2,3}}=0\;,\;\alpha_{_{3,2}}={D\over2}-1,\;\alpha_{_{3,3}}=0\;,
\label{1TR2-78}
\end{eqnarray}
whose geometric representation is drawn in Fig.\ref{fig6}(1)
with $(a,b)$ $=(3,5)$, $\{c,d\}$ $=\{1,2\}$. Finally the exponents
of the integer lattice $-n_{_1}E_{_{4}}^{(2)}$ $+n_{_2}E_{_{4}}^{(4)}$ are
\begin{eqnarray}
&&\alpha_{_{1,2}}=-1\;,\;\alpha_{_{1,3}}=0\;,\;\alpha_{_{2,2}}=0\;,\;
\alpha_{_{2,3}}=2-{D\over2}\;,\;\alpha_{_{3,2}}=1,\;\alpha_{_{3,3}}={D\over2}-2\;,
\label{1TR2-82}
\end{eqnarray}
whose geometric representation is plotted in Fig.\ref{fig6}(4)
with $\{(a,b)$, $(c,d)\}$ $=\{(1,2)$, $(3,5)\}$.
The corresponding hypergeometric functions are evidently written as
\begin{eqnarray}
&&\psi_{_{\{1,5,6,7\}}}^{(i)}(p_{_1}^2,p_{_2}^2,p_{_3}^2)
=\psi_{_{\{1,4,5,6\}}}^{(i)}(p_{_1}^2,p_{_2}^2,p_{_3}^2),\;\;(i=1,\cdots,12).
\label{1TR2-84}
\end{eqnarray}

Similarly $\det(A_{_{2,5,6,7}}^{(1T)\prime})=p_{_1}^2-p_{_2}^2$, and
\begin{eqnarray}
&&\Big(A_{_{\{2,5,6,7\}}}^{(1T)\prime}\Big)^{-1}\cdot A^{(1T)\prime}=\left(\begin{array}{ccccccc}
\;\;-{p_{_3}^2-p_{_1}^2\over p_{_2}^2-p_{_1}^2}\;\;&\;\;1\;\;
&\;\;-{p_{_2}^2-p_{_3}^2\over p_{_2}^2-p_{_1}^2}\;\;&\;\;0\;\;&\;\;0\;\;&\;\;0\;\;&\;\;0\;\;\\
\;\;{1\over p_{_2}^2-p_{_1}^2}\;\;&\;\;0\;\;&\;\;-{1\over p_{_2}^2-p_{_1}^2}\;\;&
\;\;0\;\;&\;\;1\;\;&\;\;0\;\;&\;\;0\;\;\\
\;\;-{p_{_1}^2\over p_{_2}^2-p_{_1}^2}\;\;&\;\;0\;\;&
\;\;{p_{_2}^2\over p_{_2}^2-p_{_1}^2}\;\;&\;\;-1\;\;&\;\;0\;\;&\;\;1\;\;&\;\;0\;\;\\
\;\;0\;\;&\;\;0\;\;&\;\;0\;\;&\;\;1\;\;&\;\;0\;\;&\;\;0\;\;&\;\;1\;\;
\end{array}\right)\;.
\label{1TR2-85}
\end{eqnarray}
Obviously the matrix of exponents is written as
\begin{eqnarray}
&&\left(\begin{array}{ccccccc}
\;\;\alpha_{_{1,1}}\;\;&\;\;0\;\;&\;\;\alpha_{_{1,3}}\;\;&\;\;0\;\;&\;0\;\;&\;\;0\;\;&\;\;0\;\;\\
\;\;\alpha_{_{2,1}}\;\;&\;\;0\;\;&\;\alpha_{_{2,3}}\;\;&\;\;0\;\;&\;\;{D\over2}-3\;\;&\;0\;\;&\;\;0\;\;\\
\;\;\alpha_{_{3,1}}\;\;&\;\;0\;\;&\;\alpha_{_{3,3}}\;\;&\;\;3-D\;\;&\;\;0\;\;&\;{D\over2}-3\;\;&\;\;0\;\;\\
\;\;0\;\;&\;\;0\;\;&\;0\;\;&\;\;0\;\;&\;\;0\;\;&\;0\;\;&\;\;-1\;\;
\end{array}\right)\;,
\label{1TR2-86}
\end{eqnarray}
where the matrix elements satisfy the relations which are obtained by the replacement
$\alpha_{_{i,2}}\rightarrow\alpha_{_{i,1}}$, with $(i=1,3,4)$ in Eq.(\ref{1TR2-71}).
Correspondingly the geometric representations of those exponent matrices are
obtained from that of exponents of $\psi_{_{\{1,4,5,6\}}}^{(i)}$
through the permutation $\widehat{(12)}$.
The constructed hypergeometric functions are
\begin{eqnarray}
&&\psi_{_{\{2,5,6,7\}}}^{(i)}(p_{_1}^2,p_{_2}^2,p_{_3}^2)=
\psi_{_{\{1,4,5,6\}}}^{(i)}(p_{_1}^2,p_{_3}^2,p_{_2}^2),\;\;(i=1,\cdots,12)\;.
\label{1TR2-87}
\end{eqnarray}

Because $\det(A_{_{3,5,6,7}}^{(1T)\prime})=p_{_2}^2-p_{_3}^2$, and
\begin{eqnarray}
&&\Big(A_{_{\{3,5,6,7\}}}^{(1T)\prime}\Big)^{-1}\cdot A^{(1T)\prime}=\left(\begin{array}{ccccccc}
\;\;-{p_{_1}^2-p_{_3}^2\over p_{_2}^2-p_{_3}^2}\;\;&\;\;-{p_{_2}^2-p_{_1}^2\over p_{_2}^2-p_{_3}^2}\;\;
&\;\;1\;\;&\;\;0\;\;&\;\;0\;\;&\;\;0\;\;&\;\;0\;\;\\
\;\;{1\over p_{_2}^2-p_{_3}^2}\;\;&\;\;-{1\over p_{_2}^2-p_{_3}^2}\;\;&\;\;0\;\;&
\;\;0\;\;&\;\;1\;\;&\;\;0\;\;&\;\;0\;\;\\
\;\;-{p_{_3}^2\over p_{_2}^2-p_{_3}^2}\;\;&\;\;{p_{_2}^2\over p_{_2}^2-p_{_3}^2}\;\;&
\;\;0\;\;&\;\;-1\;\;&\;\;0\;\;&\;\;1\;\;&\;\;0\;\;\\
\;\;0\;\;&\;\;0\;\;&\;\;0\;\;&\;\;1\;\;&\;\;0\;\;&\;\;0\;\;&\;\;1\;\;
\end{array}\right)\;,
\label{1TR2-88}
\end{eqnarray}
the matrix of exponents is written as
\begin{eqnarray}
&&\left(\begin{array}{ccccccc}
\;\;\alpha_{_{1,1}}\;\;&\;\;\alpha_{_{1,2}}\;\;&\;\;0\;\;&\;\;0\;\;&\;0\;\;&\;\;0\;\;&\;\;0\;\;\\
\;\;\alpha_{_{3,1}}\;\;&\;\;\alpha_{_{3,2}}\;\;&\;0\;\;&\;\;0\;\;&\;\;{D\over2}-3\;\;&\;0\;\;&\;\;0\;\;\\
\;\;\alpha_{_{4,1}}\;\;&\;\;\alpha_{_{4,2}}\;\;&\;0\;\;&\;\;3-D\;\;&\;\;0\;\;&\;{D\over2}-3\;\;&\;\;0\;\;\\
\;\;0\;\;&\;\;0\;\;&\;0\;\;&\;\;0\;\;&\;\;0\;\;&\;0\;\;&\;\;-1\;\;
\end{array}\right)\;,
\label{1TR2-89}
\end{eqnarray}
where the matrix elements satisfy the relations which are obtained by the replacement
$\alpha_{_{i,3}}\rightarrow\alpha_{_{i,1}}$, with $(i=1,3,4)$ in Eq.(\ref{1TR2-71}).
Correspondingly the geometric representations of those exponent matrices are
obtained from that of exponents of $\psi_{_{\{1,4,5,6\}}}^{(i)}$
through the permutation $\widehat{(132)}$.
The constructed hypergeometric functions are
\begin{eqnarray}
&&\psi_{_{\{3,5,6,7\}}}^{(i)}(p_{_1}^2,p_{_2}^2,p_{_3}^2)=
\psi_{_{\{1,4,5,6\}}}^{(i)}(p_{_3}^2,p_{_1}^2,p_{_2}^2),\;\;(i=1,\cdots,12)\;.
\label{1TR2-90}
\end{eqnarray}

\end{document}